\documentclass[sigconf, nonacm]{acmart}
\usepackage{array,color,url}
\usepackage[ruled,linesnumbered]{algorithm2e}
\usepackage{indentfirst}
\usepackage{algpseudocode}
\usepackage{tabu}
\usepackage{amsthm}
\usepackage{bm}
\usepackage{bbm}
\usepackage{graphicx}
\usepackage{float}
\usepackage{epsfig}
\usepackage{epstopdf}
\usepackage{subfigure}
\usepackage{color,xcolor}
\usepackage{comment}
\usepackage{booktabs}
\usepackage{multirow}
\usepackage{url}
\usepackage{caption}
\usepackage{verbatimbox}
\usepackage{comment}
\usepackage{diagbox}
\usepackage{enumitem}
\usepackage{setspace}
\usepackage[explicit]{titlesec}
\setlist[itemize]{leftmargin=*}
\usepackage{appendix}
\usepackage{color}
\usepackage{makecell}
\usepackage{hyperref}

\begin{document}
\title{Fine-grained Manipulation Attack to Local Differential Privacy Protocols for Data Streams}

\author{Xinyu Li}
\affiliation{
  \institution{Xi'an Jiaotong University}
  \city{Xi'an}
  \state{China}
}
\email{2213311049@stu.xjtu.edu.cn}

\author{Xuebin Ren}
\affiliation{
  \institution{Xi'an Jiaotong University}
  \city{Xi'an}
  \state{China}
}
\email{xuebinren@mail.xjtu.edu.cn}

\author{Shusen Yang}
\affiliation{
  \institution{Xi'an Jiaotong University}
  \city{Xi'an}
  \state{China}
}
\email{shusenyang@mail.xjtu.edu.cn}

\author{Liang Shi}
\affiliation{
  \institution{Xi'an Jiaotong University}
  \city{Xi'an}
  \state{China}
}
\email{sl1624@stu.xjtu.edu.cn}

\author{Chia-Mu Yu}
\affiliation{
  \institution{National Yang Ming Chiao Tung University}
  \city{Hsinchu}
  \state{Taiwan}
}
\email{chiamuyu@gmail.com}


\titlespacing*{\section}{0pt}{0.7ex plus .0ex minus .0ex}{.0ex plus .0ex}
\titlespacing*{\subsection}{0pt}{0.6ex plus .0ex minus .0ex}{.0ex plus .0ex}
\titlespacing*{\subsubsection}{0pt}{0.6ex plus .0ex minus .0ex}{.3ex plus .0ex}

\textfloatsep 0.1mm plus 0.1mm \intextsep 0.1mm plus 0.1mm

\begin{abstract}
Local Differential Privacy (LDP) enables massive data collection and analysis while protecting end users' privacy against untrusted aggregators. It has been applied to various data types (e.g., categorical, numerical, and graph data) and application settings (e.g., static and streaming). Recent findings indicate that LDP protocols can be easily disrupted by poisoning or manipulation attacks, which leverage injected/corrupted fake users to send crafted data conforming to the LDP reports. However, current attacks primarily target static protocols, neglecting the security of LDP protocols in the streaming settings. Our research fills the gap by developing novel fine-grained manipulation attacks to LDP protocols for data streams. By reviewing the attack surfaces in existing algorithms, we introduce a unified attack framework with composable modules, which can manipulate the LDP estimated stream toward a target stream. Our attack framework can adapt to state-of-the-art streaming LDP algorithms with different analytic tasks (e.g., frequency and mean) and LDP models (event-level, user-level, $w$-event level). 
We validate our attacks theoretically and through extensive experiments on both synthetic and real-world datasets, and finally explore a possible defense mechanism for mitigating these attacks.
\end{abstract}

\maketitle

\section{Introduction}
Local Differential Privacy (LDP)~\cite{duchi2013local,duchi2014privacy,kasiviswanathan2011can} enables massive data collection and analytics while ensuring end-users' privacy without relying on a trusted aggregator. Due to the rigorous guarantee and easy implementation, it has been widely deployed at major companies like Google~\cite{Erlingsson2014RAPPORRA}, Microsoft~\cite{ding2017collecting}, and Apple~\cite{AppleDP}. 
Early studies on LDP focus on various static analytic tasks like frequency estimation~\cite{arcolezi2021random,Erlingsson2014RAPPORRA,kairouz2016extremal,Wang2017LocallyDPFE,arcolezi2022frequency,arcolezi2022improving,ohrimenko2022randomize}, mean/variance estimation~\cite{Wang2019CollectingAA}, key-value data collection~\cite{ye2019privkv}, frequent itemset mining~\cite{Qin2016HeavyHE,li2022frequent,wang2018privset,wang2018locally} and graph data mining~\cite{Imola2021LDPgraph,Ye2020LDPgraph}. Recent work proposes to adapt LDP protocols for more complicated streaming settings~\cite{wang2020continuous,baocgm,LDP_IDS,schaler2023benchmarking,li2023privsketch}, which can realize continual data collection and analysis over streams. These work adopts different stream LDP models, including event-level LDP~\cite{joseph2018local,wang2020continuous}, user-level LDP~\cite{baocgm,Erlingsson2019AmplificationBS}, and $w$-event LDP~\cite{LDP_IDS}. 
In particular, $w$-event LDP can be easily extended to both event-level and (approximate) user-level LDP, thus being a popular paradigm~\cite{schaler2023benchmarking}. 

Recently, data poisoning attacks, arising in general data community~\cite{10.1145/3639292,10.1145/3514221.3517867}, have also emerged as a threat to LDP protocols~\cite{Cao2021poisonFE,Wu2022poisonKey,Li2023Finegrained}. Research indicates that LDP aggregators are vulnerable to manipulated user data distributions. Cao \emph{et al.}~\cite{Cao2021poisonFE} initiated attacks on LDP frequency estimation and heavy-hitter detection by inflating target item frequencies. Wu \emph{et al.}~\cite{Wu2022poisonKey} targeted LDP for key-value data, aiming to boost both frequencies and mean values of selected keys using data from fictitious users. Li \emph{et al.}~\cite{Li2023Finegrained} developed a fine-grained attack on LDP for mean and variance estimation, enabling precise manipulation of statistical estimates and demonstrating that a larger LDP privacy budget enhances attack effectiveness. However, these studies are all limited to \textit{one-shot} attacks in static settings, and none of them consider the streaming data. Despite being an essential analytic setting and extensively used in various applications, the vulnerability of streaming data LDP protocols to data poisoning attacks remains unexplored.

In this work, we explore data poisoning attacks on streaming LDP protocols. Specifically, 
we focus on fine-grained manipulation attacks~\cite{Li2023Finegrained} of the stream of estimated statistics (e.g., frequency and mean), to persistently match a target stream of intended statistics at each timestamp, allowing for varied targets over time. We propose a novel framework to show how fake users can continuously manipulate the estimation over streams by submitting carefully crafted data to the aggregator at each timestamp, thus minimizing the overall gap between the released and intended statistics throughout the whole stream. These attacks can lead the online estimate sequence over massive users' streams close to a sequence with intended statistics. For example, a malicious company wants to manipulate seasonal consumer interest or preferences over time. It can leverage some fake users to send crafted data continuously, thus manipulating the preference trends to follow desired time-varying patterns. 
We identify three technical challenges of the attacks:

\begin{itemize}[itemsep=2pt,topsep=0pt,parsep=0pt]
\item \noindent \textbf{C1: Complicated optimization over streams.}
Unlike poisoning attacks for static (non-streaming) LDP that only perform one-shot manipulation, attacks for streaming LDP require attackers to reconcile the correlations across timestamps and formulate attack strategies to optimize the attack performance over the whole stream. Directly applying the one-shot attacks into streaming settings cannot achieve the overall optimality of the attack.

\item \noindent \textbf{C2: Fine-grained manipulation of statistics.}
We focus on fine-grained manipulation of common statistics like frequency and mean estimation.
On one hand, existing attacks for frequency estimation like MGA~\cite{Cao2021poisonFE} are not fine-grained with specific targets. 
On the other hand, existing fine-grained attacks~\cite{Li2023Finegrained} only work for mean and variance estimation in static settings, lacking consideration of frequency estimation and Challenge C1.

\item \noindent \textbf{C3: Sophisticated mechanisms within LDP protocols.} 
Streaming LDP protocols typically consist of multiple coupled phases, choosing different submechanisms (i.e., publication or approximation) for statistical data release at different timestamps. It is challenging to design corresponding attack modules for different phases that can coordinate the choice of different strategies toward a specific target in a unified manner.







	
\end{itemize}

To address the above challenges, we propose a novel fine-grained manipulation attack framework against streaming LDP protocols, by formulating it as an optimization problem. We first summarize existing streaming LDP protocols into three phases with inherent attack surfaces. Based on the surfaces, we then propose two theory-driven attack modules for manipulating the publication and approximation strategies respectively, with consideration of both input and output poisoning. To coordinate the two attacks toward the overall optimization, we also propose a manipulation strategy determination module by mimicking the adaptive LDP protocol themselves. To demonstrate the effectiveness of our attack framework, we further propose both baseline and adaptive attacking algorithms against the state-of-the-art LDP protocols. Besides theoretical analysis, we conduct extensive experiments for performance evaluation. Finally, we discuss a possible defense with validations. Our contributions can be summarized as follows:

\begin{itemize}[itemsep=2pt,topsep=0pt,parsep=0pt]
	\item To the best of our knowledge, we are the first to explore fine-grained data poisoning attacks against state-of-the-art LDP protocols for infinite data streams, which achieves not only arbitrary targets-driven LDP poisoning but also nearly-optimal attacking performance \textit{on-the-fly} over the whole streams.

	\item We propose a general attack framework against LDP protocols for data streams, which considers both different knowledge assumptions (full and partial knowledge) and attacking modes (input and output poisoning). The unified framework with composable modules can effectively adapt to various analytical tasks (frequency and mean estimation etc) and streaming LDP models ($w$-event, event-level, and user-level LDP).

	\item We study the proposed attacks both theoretically and empirically. We present theoretical analysis of the attack performance, and discuss the sufficient conditions. We also implement all the proposed attacks and conduct experiments on both synthetic and real-world datasets. The results show that compared to the baseline attacks, our proposed attacks can achieve significant improvement in attack effectiveness. In addition, we explore a possible defense method against our attacks.

\end{itemize}

\section{Background Knolwedge}\label{sec:background}
We outline the background knowledge, defining Local Differential Privacy (LDP), and introduce Local Differential Privacy over Data Streams, our attacking targets.

\subsection{Local Differential Privacy (LDP)}
In LDP, $\mathcal{M}$ is a randomized mechanism that perturbs each user's input $v$ before being aggregated for data analytics.

\begin{definition}[Local Differential Privacy]
A mechanism $\mathcal{M}$ with $\mathcal{O}$ denotes the set of all possible outputs of $\mathcal{M}$ satisfies $\epsilon$-Local Differential Privacy (i.e., $\epsilon$-LDP) if and only if, for $\forall~v, v^\prime \in Dom\left(\mathcal{M}\right)$ and $\forall S\subseteq \mathcal{O}$ there is $Pr\left[\mathcal{M}\left(v\right)\in S\right]\le e^\epsilon Pr\left[\mathcal{M}\left(v^\prime\right)\in S\right]$.
\end{definition}


LDP also applies to the streaming setting with the similar definitions of event-level, user-level and $w$-event LDP, which are defined over different adjacency of stream prefixes. Event-level adjacency has at most one timeslot of difference while user-level adjacency can differ at all the user's contributed slots. \textit{w-neighboring} means they share identical elements within a window of up to $w$ timestamps. More details can refer to \cite{Kellaris2014DifferentiallyPE} and \cite{LDP_IDS}.




\begin{definition}
Let $\mathcal{M}$ be a mechanism that takes as input a stream prefix $V_t=\left(v_1,v_2,\ldots,v_t\right)$ consisting of an arbitrary number of consecutive input values $v_t$ of a single user, and $\mathcal{O}$ be the set of all possible outputs of $\mathcal{M}$. $\mathcal{M}$ satisfies $w$-event (event-level, user-level resp.) $\epsilon$-LDP if, for any $w$-neighboring (event-level, user-level adjacent resp.) stream prefixes $V_t$,$V_t^\prime$ with arbitrary $t$, and $\forall S\subseteq \mathcal{O}$ it holds $\mathcal{O},Pr\left[\mathcal{M}\left(V_t\right) \in 
 S\right]\le e^\epsilon Pr\left[\mathcal{M}\left(V_t^\prime\right)\in S\right]$.
\end{definition}



\subsection{Frequency and Mean Estimation under LDP}\label{sec: Frequency Oracle Under LDP}
Frequency and mean estimation are two common analytic tasks in streaming LDP. Frequency estimation is based on Frequency Oracles (FOs)~\cite{Wang2017LocallyDPFE}, which estimate the frequency of any item $\omega_k$ in a domain $\Omega=\{\omega_1,...,\omega_d\}$ of size $d$. Three commonly used FO protocols are kRR~\cite{duchi2013local}, OUE~\cite{Wang2017LocallyDPFE} and Ada~\cite{ZhangWLHC18}. For every $j\in [n]$ user, a FO first perturbs its input item $v_{j}$ to $y_{j}$ and sends the output value $y_{j}$ to the aggregator. Then the aggregator calculates the frequency of each distinct item $k$, denoted as $\mathbf{\hat{f}}[k]$, as follows:
\begin{small}\vspace{-1mm}
    \begin{align}\label{f hat}
    \mathbf{\hat{f}}\left[k\right]=\frac{\frac{1}{n}\sum_{j=1}^{n}\mathbb{I}_{S\left(y_{j}\right)}^{(k)}-q}{p-q}, 
\end{align}\vspace{-0mm}
\end{small} where $n$ is the total number of users, $y_{j}$ is the output value from the $j$-th user, $p$ and $q$ are two perturbation parameters of FOs, and $S(y_{j})$ is the support set of $y_{j}$ (i.e., the input values that can produce $y_{j}$ in FO). $\mathbb{I}_{S\left(y_{j}\right)}^{\left(k\right)}$ is a characteristic function, which is defined as 1 if $k\ \in S\left(y_{j}\right)$ and 0 otherwise. FOs provides unbiased estimation of the actual item frequencies $\mathbf{f}[k]$, i.e., $\mathbb{E}(\mathbf{\hat{f}}[k])=\mathbf{f}[k]$. \cite{Wang2017LocallyDPFE} shows the variance of estimated frequency. For kRR, it is $\text{Var}\left(\mathbf{\hat{f}}\left[k\right],\epsilon,n\right)=\frac{d-2+e^\epsilon}{n\left(e^\epsilon-1\right)^2}+\frac{\mathbf{f}\left[k\right]\left(d-2\right)}{n\left(e^\epsilon-1\right)}$. Since $\sum_{k=1}^{d}\mathbf{f}\left[k\right]=1$, we denote $\frac{1}{d}\sum_{k=1}^{d}\text{Var}\left(\mathbf{\hat{f}}\left[k\right],\epsilon,n\right)$ as $\text{Var}\left(n,\epsilon\right)=\frac{d-2+e^\epsilon}{n\left(e^\epsilon-1\right)^2}+\frac{d-2}{nd\left(e^\epsilon-1\right)}$. For OUE, $\text{Var}(n,\epsilon)=\frac{1}{d}\sum_{k=1}^{d}\text{Var}\left(\mathbf{\hat{f}}\left[k\right],\epsilon,n\right)=\frac{4e^\epsilon}{n\left(e^\epsilon-1\right)^2}+\frac{1}{nd}$. The variance of Ada is the smaller one of kRR and OUE.

Mean estimation usually adopts the Hybrid Mechanism \cite{Wang2019CollectingAA}, which merges Stochastic Rounding (SR) \cite{Duchi2016MinimaxOP} and Piecewise Mechanism (PM) \cite{Wang2019CollectingAA} for minimal error. For $\epsilon>0.61$, it employs PM with probability $1-e^{-\epsilon/2}$ and SR with $e^{-\epsilon/2}$. Below $\epsilon\le 0.61$, it solely uses SR. 
Like FOs, the variance of HM can also be denoted as the function of population $n$ and privacy budget $\epsilon$, i.e., $\text{Var}(n,\epsilon)$.


\subsection{LDP over Data Streams}\label{subsec:LDP for Infinite Data Streams}
\subsubsection{System Models.}
We consider the stream data analytics, under the definition of local differential privacy. We assume there are $n$ users and a central aggregator. At each timestamp $t$, each user $j \in [n]$ holds a value $v_t^j$ from a domain $\Omega$, either categorical $\Omega=\{\omega_1,\ldots,\omega_d\}$ with the cardinality of $|\Omega|=d$ or numerical $\Omega=[0,B]$ with a maximal differential bound $B$. Meanwhile, the aggregator aims to analyze the aggregate statistics (e.g., frequency or mean estimation) over all $n$ users' data $v^1_t, v^2_t, \ldots, v^n_t$ at each time $t$. 
With time evolves, each user actually has an infinite data stream $V^j=(v_1^j, v_2^j, \ldots, v_t^j, \ldots)$. Different LDP model (i.e., event-level, $w$-event LDP, user-level) can be adopted to provide specific DP guarantees to these users. Then LDP streaming analytic aims to derive a stream of aggregate statistics $\mathbf{\hat{f}}=(\mathbf{\hat{f}}_1, \mathbf{\hat{f}}_2, \ldots, \mathbf{\hat{f}}_t,\ldots)$ as close to the actual stream $\mathbf{f}=(\mathbf{{f}}_1, \mathbf{{f}}_2, \ldots, \mathbf{f}_t...)$ as possible \textit{on-the-fly}, according to massive end users' LDP perturbed report stream. 
For any given stream with a length of $T$ timestamps, it seeks to minimize the average estimate error as:
\begin{small}
	\begin{align} \label{LDP-IDS problem}
		\min~~~\frac{1}{d}\frac{1}{T}\sum\nolimits_{t=1}^{T}\sum\nolimits_{k=1}^{d}(\mathbf{\hat{f}}_t[k]-\mathbf{f}_t[k])^2.
	\end{align}
\end{small} 
In particular, two typical LDP analytic tasks can be considered: 
\textit{Frequency estimation for categorical data} aims to estimate the frequency histogram $\mathbf{f}_t=<\mathbf{f}_t[1],\mathbf{f}_t[2],\ldots,\mathbf{f}_t[d]>$ over a domain $\Omega$ of size $d$. 
\textit{Mean-value estimation for numerical data} estimates the mean value $f_t=\frac{1}{n}\sum_{j=1}^{n}v_t^j$ over all $n$ users' data $v_t^j$ in a transformed domain $[0,B]$ where $B$ is the maximum value. 


As $w$-event LDP can be easily extended to event-level and user-level ones, we primarily introduce the streaming protocols with $w$-event LDP and briefly discuss those with the other two models.

\begin{figure}[t]
    \centering
    \includegraphics[width=220pt] {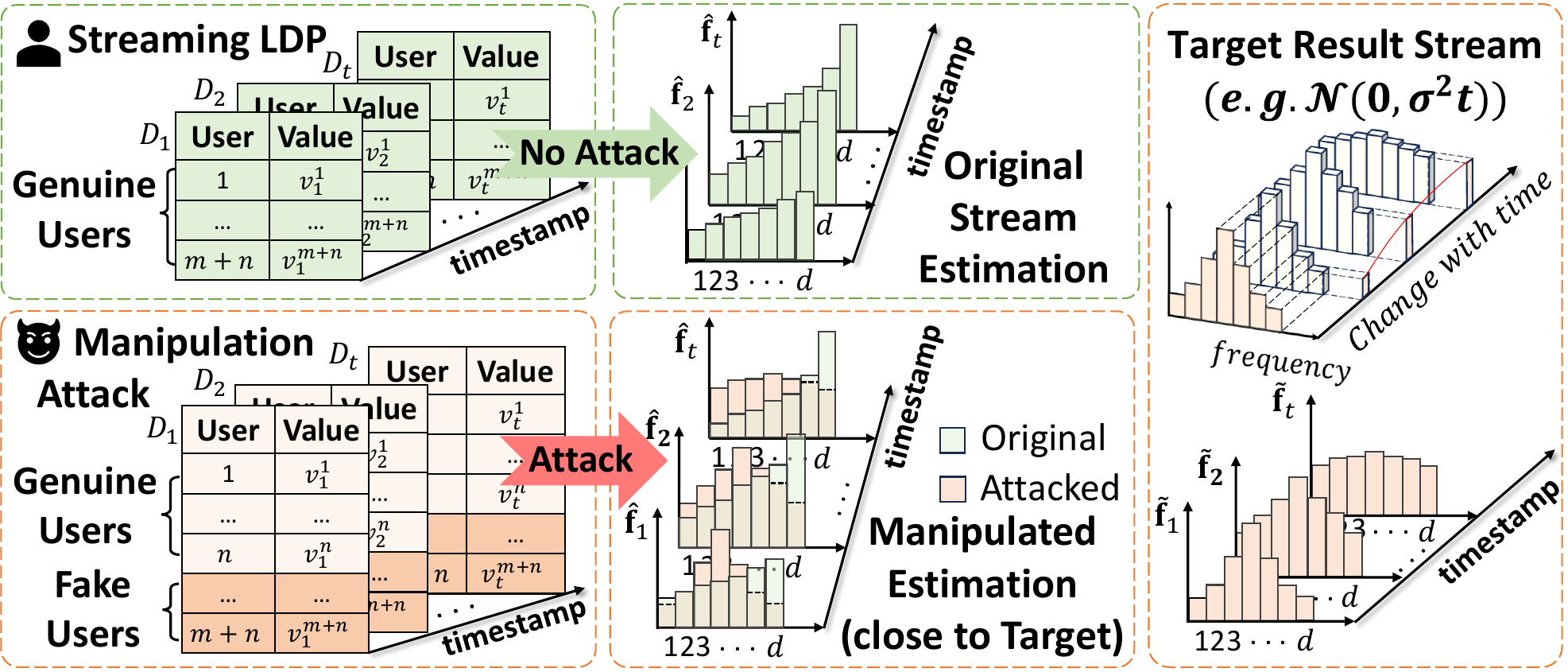}\vspace{-3.5mm}
    \caption{\small Manipulation to LDP Frequency Estimation over Streams.}
    \label{fig:Manipulation attack in a streaming setting}
\end{figure}

\subsubsection{$w$-event Adaptive Budget-Division and Population-Division.}
\begin{itemize}[itemsep=2pt,topsep=0pt,parsep=0pt]
	\item \textbf{Private Dissimilarity Calculation.} At time $t$, LDP-IDS~\cite{LDP_IDS} invokes the FO to derive an initial estimation $\mathbf{\bar{f}}_t$ using budget $\epsilon_{t,1}=\frac{\epsilon}{2w}$ with whole population or population $|U_{t,1}|=\frac{n}{2w}$ with whole budget. Then, a private dissimilarity $\overline{dis}$ can be calculated 
		\begin{small}\vspace{-2mm}
		\begin{align}\label{dis}
			\overline{dis}=\frac{1}{d}\sum\nolimits_{k=1}^{d}\left(\mathbf{\bar{f}}_t\left[k\right]-\mathbf{\hat{f}}_{t-1}\left[k\right]\right)^2-\frac{1}{d}\sum\nolimits_{k=1}^{d}\text{Var}\left(\mathbf{\hat{f}}_t\left[k\right]\right).
	\end{align} \vspace{-5mm}
 \end{small}
	\item \textbf{Private Strategy Determination.} Some publication budget $\epsilon_{t,2}$ or population $|U_{t,2}|$ is assigned to derive a potential publication error $err$ (equivalent to noise variance $\text{Var}(n,\epsilon_{t,2})$ for budget-division and $\text{Var}(|U_{t,2}|,\epsilon)$ for population-division) for possible publication strategy. Then $\overline{dis}$ and $err$ are compared to choose a strategy with less error. That is, if $\overline{dis} > err$, the publication is chosen; otherwise, approximation without spending budget.  
	\item \textbf{Publication Budget Allocation.} 
 If publication is chosen, LDP-IDS will consume $\epsilon_{t,2}$ budget with whole population or $|U_{t,2}|$ population with whole budget and release the estimates as $\mathbf{\hat{f}}_t$; otherwise, it will directly publish the last released statistics as an approximation and save budget or population for next publication.
 $\epsilon_{t,2}$ and $|U_{t,2}|$ is assigned based on different rules. In LDP budget distribution (LBD) and LDP population distribution (LPD), $\epsilon_{t,2}$ and $|U_{t,2}|$ is distributed in an exponentially decaying way to the timestamps when a publication is chosen. In LDP budget absorption (LBA) and LDP population absorption (LPA), $\epsilon_{t,2}$ and $|U_{t,2}|$ is uniformly assigned as $\epsilon_{t,2}=\frac{\epsilon}{2w}$ and $|U_{t,2}|=\frac{n}{2w}$ first and then unused budget is absorbed at timestamps where approximation is chosen. 
	
\end{itemize}

\subsubsection{Other $w$-event LDP Methods.}
LDP-IDS also provides three baselines, LBU (LDP Budget Uniform), LPU (LDP Population Uniform), and LSP (LDP Sampling). Furthermore, RescueDP~\cite{Wang2016RescueDPRS} and $\text{DSAT}_w$~\cite{10.1145/2806416.2806441} can be extended to $w$-event LDP, as discussed in \cite{LDP_IDS}.


\begin{itemize}[itemsep=2pt,topsep=0pt,parsep=0pt]
	\item \textbf{Private Dissimilarity Calculation.} $\text{DSAT}_w$ calculates $\overline{dis}$ using budget $\epsilon_{t,1}$ with whole population or population $|U_{t,1}|$ with whole budget. RescueDP calculates feedback and PID errors without allocating budget or population. Other methods do not apply.
	\item \textbf{Private Strategy Determination.} They adopt a similar budget-division or population-division framework, except that $\text{DSAT}_w$ modifies $err$ as a dynamic threshold. LBU/LPU publishes always. LSP selects one timestamp in a window for publication and approximates other $w-1$ timestamps. RescueDP calculates the next sampling intervals according to PID error, publishes at the next sampling timestamps and approximates at other timestamps.
	\item \textbf{Publication Budget/Population Allocation.} $\text{DSAT}_w$ dynamically allocates $\epsilon_{t,1}$ (or $|U_{t,1}|$) and $\epsilon_{t,2}$ (or $|U_{t,2}|$) by a PID controller. LBU/LPU allocates fixed budget $\epsilon/w$ or population $n/w$ across timestamps. LSP allocates the entire $\epsilon$ and population at the sampling timestamp. RescueDP dynamically allocates budget or population at sampling timestamps and approximates with a Kalman filter.
	
\end{itemize}

\subsubsection{Event-level and User-Level Methods.} 
We briefly discuss event-level protocols, including ToPL~\cite{wang2020continuous} and (LDP extended) PeGaSus~\cite{Chen2017PeGaSusDD}, and user-level protocols, including CGM~\cite{baocgm} and (LDP extended) FAST~\cite{Fan2014AnAA}. 
\begin{itemize}[itemsep=2pt,topsep=0pt,parsep=0pt]
	\item \textbf{Private Dissimilarity Calculation.} FAST calculates the feedback error and PID error. PeGaSus, ToPL and CGM do not apply.
	\item \textbf{Private Strategy Determination.} PeGaSus, ToPL and CGM publish at each timestamp. FAST calculates the next sampling interval according to PID error, publishes at next sampling timestamp and approximates the rest timestamps.
	\item \textbf{Publication Budget/Population Allocation.} 
    PeGaSus allocates all privacy and population at each timestamps to satisfy event-level LDP and post-processes the estimates with Grouper and Smoother. ToPL consumes all privacy and population for mean estimation at each timestamp. FAST allocates budget for publication equally at each sampling timestamp. CGM distributes budget to one user's data stream for user-level LDP.

\end{itemize}

\section{Threat Model}\label{sec: threat model}
We detail attackers' abilities and goals in our proposed attacks.

\subsection{Assumptions}
We assume that the attacker can inject or compromise multiple users to manipulate the aggregator's estimation by sending crafted data. Specifically, the attacker may inject $m$ fake users and blend them with $n$ genuine users (or the attacker may compromise $m$ out of $n+m$ as fake users), all participating in the streaming LDP protocols. 
Similar to~\cite{Li2023Finegrained}, we assume a sufficient number of fake users is available, i.e., $m$ is large enough. Existing study~\cite{thomas2013trafficking} indicates the low cost of acquiring fake accounts on platforms like Twitter, Google, and Hotmail (approximately $0.0004-0.03$ US dollar per account).
Besides, we follow \cite{Li2023Finegrained} to assume that the attacker can estimate the number of real users $n$ as $n^e$ from public sources. 
We also assume that attackers can estimate the input distribution $\mathbf{f}_t$ as $\mathbf{f}^e_t$ at every timestamp, where we further consider two major types of attackers who have different levels of knowledge about the information: \textit{full-knowledge} and \textit{partial-knowledge} as discussed in~\cite{tong2024data}. The former means attackers know the true statistics while the latter means attackers can compromise a subset of users and know their inputs for $\mathbf{f}^e_t$ estimation. A variant of the partial-knowledge is the \textit{man-in-the-middle} (MITM) attack, where the attacker can only estimate the frequencies of items based on intercepted perturbed reports. 
Additionally, the attacker accesses LDP protocol parameters like $\epsilon$ and $w$ from public documents~\cite{AppleDP} and detailed FO implementation information~\cite{Cao2021poisonFE,Wu2022poisonKey,Li2023Finegrained}.

\begin{table}
\centering
\caption{\label{tab:notations}Notations.}\vspace{-2mm}
\scalebox{0.9}{
\begin{tabular}{|c|l|}
\hline
\textbf{Notation} & \textbf{Description}\\
\hline
$\epsilon$, $w$& \makecell[l]{Privacy budget, size of sliding windows} \\
\hline
$n$, $n^e$& \makecell[l]{Number of genuine users, attacker-estimated $n$} \\
\hline
$\mathbf{f}$, $\mathbf{f}^e$& \makecell[l]{Genuine distribution, attacker-estimated $\mathbf{f}$} \\
\hline
$m$& \makecell[l]{Number of fake users} \\
\hline
$d$& \makecell[l]{Domain size of FOs} \\
\hline
$\mathbf{\hat{f}}_t$, $\mathbf{\hat{f}}_t[k]$& \makecell[l]{Released distribution $t$, $k$-th item frequency in $\mathbf{\hat{f}}_t$} \\
\hline
$\mathbf{\tilde{f}}_t$, $\mathbf{\tilde{f}}_t[k]$& \makecell[l]{Target distribution at $t$, $k$-th item frequency in $\mathbf{\tilde{f}}_t$} \\
\hline
$\overline{dis}$, $err$& \makecell[l]{Private dissimilarity, potential publication error} \\
\hline
$\mathbf{\bar{f}}_t$, $\mathbf{\bar{f}}_t[k]$& \makecell[l]{Frequency for $\overline{dis}$ calculation, $k$-th item in $\mathbf{\bar{f}}_t$} \\
\hline
$G_{I,t}$, $G_{O,t}$& \makecell[l]{Manipulation gap of IPMA/OPMA} \\
\hline

\end{tabular}
}
\end{table}

\subsection{Attacker's Goal}
We consider the attacker aim to modify the estimated statistical result $\mathbf{\hat{f}}_t$ (e.g., frequency distribution or mean value) of streaming LDP protocols at each timestamp $t$, such that the estimated result stream $\mathbf{\hat{f}}=\{\mathbf{\hat{f}}_1,\mathbf{\hat{f}}_2,\ldots,\mathbf{\hat{f}}_t\ldots\}$ to be as close to the target result stream $\mathbf{\tilde{f}}=\{\mathbf{\tilde{f}}_1,\mathbf{\tilde{f}}_2,\ldots,\mathbf{\tilde{f}}_t\ldots\}$ as possible. 
Fig.~\ref{fig:Manipulation attack in a streaming setting} illustrates the attack scenario for frequency estimation. As shown, the attacker can set a target frequency distribution (e.g., $\mathbf{\tilde{f}}_t=\mathcal{N}(0,\sigma^2t)$) at each timestamp, which forms a target stream. Without attack, the streaming LDP protocol continuously releases the estimated frequency histogram from the genuine users at each timestamp. Once attacked by some fake users, the released histogram is manipulated to be close to the target at each timestamp. 
To measure how successful the fine-grained attack is at each timestamp, we define the \textit{manipulation gap} $G_t=\sum_{k=1}^{d}(\mathbf{\hat{f}}_t[k]-\mathbf{\tilde{f}}_t[k])^2/d$ as the distance between the estimated result $\mathbf{\hat{f}}_t$ and the target result $\mathbf{\tilde{f}}_t$ at timestamp $t$. Then, \textit{average manipulation gap} $G_{avg}=\frac{1}{T}\sum\nolimits_{t=1}^T G_t$ can be used as the performance metric over the whole stream. A smaller gap $G_{avg}$ implies a more successful fine-grained attack.
For any given streams of $T$ timestamps and $d$-dimensional statistics, the attacker seeks to minimize the average gap between them
\begin{small}
     \begin{equation}\label{overall goal}
     \min ~~~\frac{1}{d}\frac{1}{T}\sum\nolimits_{t=1}^{T}\sum\nolimits_{k=1}^{d}(\mathbf{\hat{f}}_t[k]-\mathbf{\tilde{f}}_t[k])^2.
 \end{equation}
\end{small}
Note that, 
the target $\mathbf{\tilde{f}}_t$ changes by time $t$, manipulating the post-attack estimated result to follow specific time-varying patterns. Table~\ref{tab:notations} summarizes key notations.


\section{Attack Framework and Modules} \label{sec: attack details}

\subsection{Overview}
\begin{figure}[t]
\centering	
\includegraphics[width=0.48\textwidth]{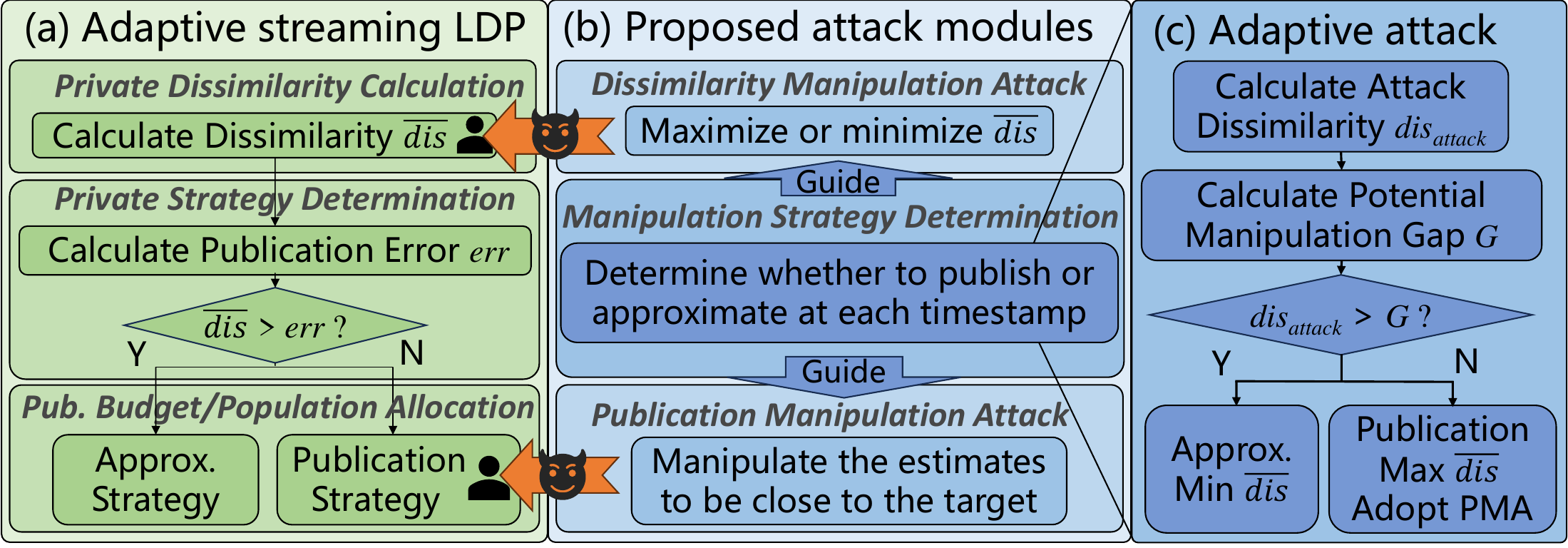}\vspace{-3mm}
	\caption{\small Overview of adaptive attack framework: (a) attack surfaces; (b) proposed attack modules; (c) proposed adaptive attack strategy.
 }\centering
	\label{fig:overview} 
\end{figure}

\textbf{Attack Surfaces.} As introduced in Sec.~\ref{subsec:LDP for Infinite Data Streams}, existing streaming LDP protocols can be summarized into three phases, shown in Fig.~\ref{fig:overview}(a), which also leaves surfaces for manipulation attacks. 


\textsf{(1) Private Dissimilarity Calculation:} This step estimates a dissimilarity value $\overline{dis}$ via FOs to measure the stream change at the current timestamp, which is compared with the potential publication error $err$ in the following step of private strategy determination. Therefore, the attacker can manipulate $\overline{dis}$ to control the private strategy determination, which can be leveraged to amplify the attack. Specifically, the dissimilarity $\overline{dis}$ is calculated based on $\mathbf{\bar{f}}_t$ which is derived by FOs, and can be manipulated to be increased or decreased. 



\textsf{(2) Private Strategy Determination:} This step tries to select between publication and approximation at each timestamp to adaptively minimize the total noise introduced over the non-deterministic data stream, as illustrated in Sec.~\ref{Sec: Optimization View of LDP-IDS}. The attacker also need a guideline to manipulate the strategy choice at each timestamp to minimize the overall gap throughout the stream. Interestingly, we found that the optimization problem of minimizing the average manipulation gap is very similar to that of minimizing the total noise in streaming LDP protocols. Therefore, the attacker can mimic the optimization strategy in streaming LDP protocols to adaptively manipulate the strategy choice in fine-grained attacks. Specifically, the manipulation gap of publication or approximation at each timestamp is compared for greedily choosing a better strategy. 


\textsf{(3) Publication Budget/Population Allocation:} This step distributes the LDP budget/population and invokes FOs for the publication strategy. The attacker cannot directly affect the distribution process of publication budget/population since it is independent of user data. However, the invoked FOs can be manipulated to make the estimate $\mathbf{\hat{f}}_t$ at publication timestamps close to the target $\mathbf{\tilde{f}}_t$. Note that, despite no direct manipulation on approximation timestamps, the manipulated result at a publication timestamp would be set as the approximate value on later approximation timestamps. 

\textbf{Attack Modules.} 
Thus, we introduce three attack modules: Publication Manipulation Attack (PMA), Dissimilarity Manipulation Attack (DMA), and Manipulation Strategy Determination (MSD). At each timestamp, the attacker performs these modules to mount the attack. Specifically, based on current knowledge, the attacker first adopts MSD to determine whether publication or approximation is more beneficial in reducing the manipulation gap between $\mathbf{\hat{f}}_t$ and $\mathbf{\tilde{f}}_t$. If MSD chooses the publication strategy, the attacker will invoke DMA to manipulate $\overline{dis}$ and steer the LDP aggregator to choose publication. Otherwise, DMA will minimize $\overline{dis}$ to make the aggregator choose the approximation strategy. If the publication strategy is chosen as expected, PMA will be further invoked to manipulate the released statistics $\mathbf{\hat{f}}_t$ to approach the target $\mathbf{\tilde{f}}_t$. Fig.~\ref{fig:overview}(b) summarizes the attack modules and interactions with the exposed attack surfaces of the adaptive streaming LDP protocols.
Note that, 
these three attack modules can be applied to any streaming LDP algorithms with the aforementioned attack surfaces. For DMA and PMA, we also consider both input and output manipulation methods.  


In what follows, we introduce PMA and DMA in Sec.~\ref{subsec: Publication Manipulation Attack} and Sec.~\ref{subsec: Dissimilarity Manipulation Attack}, respectively. Then we design MSD in Sec.~\ref{subsec: Arttack Strategy Determination}. 

\subsection{Publication Manipulation Attacks (PMAs)}\label{subsec: Publication Manipulation Attack}
\subsubsection{The Goal of PMAs.}
PMAs aim to manipulate the estimated statistic $\mathbf{\hat{f}}_t$ (e.g., frequency and mean estimate) via FOs to be as close to a target $\mathbf{\tilde{f}}_t$ as possible (at any timestamp $t$), i.e., minimizing the manipulation gap $G_t$. It can be formulated as follows.
\begin{small}
\begin{align}\label{eq: PMA goal}
    \min G_t=\frac{1}{d}\sum\nolimits_{k=1}^{d}\left(\mathbf{\hat{f}}_t\left[k\right]-\mathbf{\tilde{f}}_t\left[k\right]\right)^2
\end{align}
\end{small}

Eq.~(\ref{eq: PMA goal}) aligns with our main objective (Eq.~(\ref{overall goal})), allowing the attacker to precisely target a specific distribution $\mathbf{\tilde{f}}_t$ with PMAs to manipulate LDP estimates at publication timestamps. Note that existing poisoning or manipulation attacks cannot directly apply here. Maximum gain attack (MGA)~\cite{Cao2021poisonFE} works for frequency estimation but is not fine-grained to precisely manipulate each item, while existing fine-grained attacks~\cite{Li2023Finegrained} are only proposed for mean and variance estimation. 


\subsubsection{Input Publication Manipulation Attack (IPMA).} 
We first introduce the Input Publication Manipulation Attack (IPMA), aimed at modifying $\mathbf{\hat{f}}_t$ by adjusting the inputs from fake users to approximate the target distribution. 

IPMA can minimize the manipulation gap $G_t$ by crafting the input values for the $m$ controllable fake users. According to the FO protocols, once the input distribution $\mathbf{f}_t$ of $n+m$ total users' data can be manipulated to be close to the target distribution $\mathbf{\tilde{f}}_t$, the estimates $\mathbf{\hat{f}}_t$ derived from FOs will also approach $\mathbf{\tilde{f}}_t$. Therefore, the optimization in Eq. (\ref{eq: PMA goal}) can be transformed into the minimization of the gap between $\mathbf{f}_t$ and $\mathbf{\tilde{f}}_t$. Besides, for $\forall k \in [1,..,d]$, suppose $m[k]$ represents the number of fake users whose input is the $k$-th item $\omega_k \in \Omega$, then the true frequency $\mathbf{f}_t[k]$ $\forall k \in [1,...,d]$ can be estimated as $\mathbf{f}_t[k] \approx\left(m[k]+n^e\cdot\mathbf{f}_t^e[k]\right)/\left(m+n^e\right)$. Specifically, we formulate IPMA as finding $m[k]~\forall k \in [1,..,d]$ such that $\mathbf{f}_t[k]=\mathbf{\tilde{f}}_t[k]$ and the attacker can determine $m[k]$ for different $k$ by solving the following convex optimization problem.
\begin{small}
    \begin{align}
\label{IPMA_optimization}
\nonumber
&\min~~~ \frac{1}{d}\sum\limits_{k=1}^{d}\left(\mathbf{f}_t\left[k\right]-\mathbf{\tilde{f}}_t\left[k\right]\right)^2=\frac{1}{d}\sum\limits_{k=1}^{d}\left(\frac{m\left[k\right]+n^e\cdot\mathbf{f}^e_t\left[k\right]}{n^e+m}-\mathbf{\tilde{f}}_t\left[k\right]\right)^2  \\
&s.t.~~~\sum\nolimits_{k=1}^{d}m\left[k\right]=m,0\le m\left[k\right]\le m 
\end{align}
\end{small}
And the input values of fake users at timestamp $t$ can be set as an arbitrary combination of $[v_1,\ldots,v_m]$ ($\forall j: v_{j} \in \Omega=\{\omega_1, \ldots, \omega_d\}$) that satisfies $\sum\nolimits_{j=1}^m \mathbb{I}_{v_j}^{(k)} = m[k], \forall k \in [1,\ldots,d]$
where $\mathbb{I}_{v_j}^{(k)}$ equals to $1$ if $v_j = \omega_k$ and $0$ otherwise.

\noindent\textbf{Analysis of IPMA.} Here we analyze the properties and sufficient conditions of IPMA. 

\begin{theorem}\label{Theorem:G_{I,t}}
\vspace{-2mm}
    The expected manipulation gap for IPMA $G_{I,t}$ can be calculated as 
    \begin{small}
    \begin{align}
\label{G_{I,t}}
\nonumber
&G_{I,t}\left(m,n,\epsilon\right)=\mathbb{E}\left(\frac{1}{d}\sum\nolimits_{k=1}^{d}\left(\mathbf{\hat{f}}_t\left[k\right]-\mathbf{\tilde{f}}_t\left[k\right]\right)^2\right) \\
&=\frac{1}{d}\sum\nolimits_{k=1}^{d}\left(\frac{m^\ast\left[k\right]+n\cdot\mathbf{f}_t\left[k\right]}{n+m}-\mathbf{\tilde{f}}_t\left[k\right]\right)^2+\text{Var}\left(n+m,\epsilon\right),
\end{align}
\end{small}
where $(m^\ast[1],m^\ast[2],\ldots,m^\ast[d])$ is the solution of the above optimization problem, $\text{Var}(n+m,\epsilon)=\frac{1}{d}\sum_{k=1}^d \text{Var}(\mathbf{\hat{f}}_t[k])$ and $\epsilon$ is the privacy budget of the attacked FO.
\end{theorem}

\begin{proof}
\vspace{-2mm}
    See Appendix~\ref{Proof G_{I,t}}.
\end{proof}

\begin{corollary}
\vspace{-2mm}
    When $\epsilon$ is larger, $G_{I,t}\left(m,n,\epsilon\right)$ becomes smaller, implying a more successful attack. 
    \label{Theorem:IPMA_epsilon}
\end{corollary}

\begin{proof}
\vspace{-2mm}
    See Appendix~\ref{Proof IPMA_epsilon}.
    \vspace{-2mm}
\end{proof}

Also, the relation between $G_{I,t}\left(m,n,\epsilon\right)$ and the number of users $n$ and $m$ could be illustrated as follows. 

\begin{corollary}
\vspace{-2mm}
For $\forall \alpha>1$, $G_{I,t}\left(m,n,\epsilon\right)>G_{I,t}\left(\alpha m,\alpha n,\epsilon\right)$.
    \label{Theorem:IPMA_mn}
\end{corollary}
\begin{proof}
\vspace{-2mm}
See Appendix~\ref{sec: Proof of Theorem 5.2}. 
\vspace{-2mm}
\end{proof}

Corollaries~\ref{Theorem:IPMA_epsilon} 
and~\ref{Theorem:IPMA_mn} show the relationship between the manipulation gap $G_{I,t}$ and the budget $\epsilon$ or population $n$.

\begin{corollary}\label{Theorem:condition_IPMA}
\vspace{-2mm}
For $\forall k\in\left[d\right]$, if the attacker has 
    \begin{align}
\label{condition_IPMA}
m\geq \max\left(\frac{n \cdot \mathbf{f}_t\left[k\right]}{\mathbf{\tilde{f}}_t\left[k\right]}-n,\frac{n\cdot\mathbf{\tilde{f}}_t\left[k\right]-n\cdot\mathbf{f}_t\left[k\right]}{1-\mathbf{\tilde{f}}_t\left[k\right]}\right),
\end{align}
fake users and the accurate knowledge of $n$ and $\mathbf{f}_t$, i.e., $n^e=n$ and $\mathbf{f}_t^e=\mathbf{f}_t$, then the expected manipulation gap is
\begin{align}
\label{G^*_I}
G_{I,t}^\ast\left(m,n,\epsilon\right)=\text{Var}\left(n+m,\epsilon\right).
\end{align}
\end{corollary}

\begin{proof}
\vspace{-2mm}
See Appendix~\ref{Proof condition_IPMA}. 
\end{proof}


\subsubsection{Output Publication Manipulation Attack (OPMA).}
We then present the Output Publication Manipulation Attack (OPMA), which utilizes the LDP implementation details to manipulate the outputs sent to the aggregator directly.

We consider the attacker targets the FO at a publication timestamp $t$ with privacy budget $\epsilon$. The $n$ real users' output are denoted as $\left(y_1,y_2,\ldots,y_n\right)$, and the $m$ fake users' outputs $\left(z_1,z_2,\ldots,z_m\right)$.
We use $m[k]$ to denote the number of fake users who send the $k$-th item to the aggregator (when using kRR) or who set the $k$-th bit of the encoding binary strings as 1 (when using OUE). That is, $m[k]=\mathbb{E}\left(\sum_{j=1}^{m}\mathbb{I}_{S\left(z_j\right)}^{\left(k\right)}\right)$.
Using FOs' unbiased estimation, there is $\sum_{j=1}^{n}\mathbb{E}\left(\mathbb{I}_{S\left(y_j\right)}^{\left(k\right)}\right)=n\left(\mathbf{f}_t\left[k\right]\left(p-q\right)+q\right)\approx n^e\left(\mathbf{f}_t^e\left[k\right]\left(p-q\right)+q\right)$, where $p$ and $q$ are FO parameters. So, the expected estimate $\mathbb{E}(\mathbf{\hat{f}}_t[k])$ after the attack can be calculated as
\begin{small}
    \begin{align}
\nonumber
\mathbb{E}\left(\frac{\sum_{j=1}^{n}\mathbb{I}_{S\left(y_j\right)}^{\left(k\right)}+\sum_{j=1}^{m}\mathbb{I}_{S\left(z_j\right)}^{\left(k\right)}-q\left(m+n^e\right)}{\left(m+n^e\right)\left(p-q\right)}\right) =\frac{A_k+m\left[k\right]}{\left(p-q\right)\left(m+n^e\right)},
\end{align}
\end{small}where $A_k=n^e\left(p-q\right) \mathbf{f}_t^e\left[k\right]-mq$. 

Also, $\mathbb{E}(\frac{1}{d}{\sum_{k=1}^{d}(\mathbf{\hat{f}}_t\left[k\right]-\mathbf{\tilde{f}}_t\left[k\right])^2})$ can be computed by \\ 
${\frac{1}{d}\sum\nolimits_{k=1}^{d}\left(\mathbb{E}\left(\mathbf{\hat{f}}_t\left[k\right]\right)-\mathbf{\tilde{f}}_t\left[k\right]\right)^2}+\frac{1}{d}\sum\nolimits_{k=1}^{d}\text{Var}\left(\mathbf{\hat{f}}_t\left[k\right]\right)$. 

However, unlike IPMA perturbs all inputs (including those of fake users), only the input of genuine users is perturbed in FOs. Considering the estimate variance \\ $\text{Var}\left(\mathbf{\hat{f}}_t\left[k\right]\right)=\text{Var}\left(\frac{\sum_{j=1}^{n}\mathbb{I}_{S\left(y_j\right)}^{\left(k\right)}+\sum_{j=1}^{m}\mathbb{I}_{S\left(z_j\right)}^{\left(k\right)}- q\left(m+n^e\right)}{\left(m+n^e\right)\left(p-q\right)}\right)$ is usually small when there are a large number of users.
In OPMA, we focus on minimizing the first term, i.e. $\min~\frac{1}{d}{\sum_{k=1}^{d}\left(\mathbb{E}(\mathbf{\hat{f}}_t\left[k\right])-\mathbf{\tilde{f}}_t\left[k\right]\right)^2}$

There is $0\le m\left[k\right]\le m$, and $\sum_{k=1}^{d}m\left[k\right]=m$ when kRR is used as the FO and only $0\le m\left[k\right]\le m$ when using OUE. To simplify the computation, we try to minimize $\frac{1}{d}\sum_{k=1}^{d}\left|\mathbb{E}(\mathbf{\hat{f}}_t\left[k\right])-\mathbf{\tilde{f}}_t\left[k\right]\right|$, instead of ${\frac{1}{d}\sum_{k=1}^{d}\left(\mathbb{E}\left(\mathbf{\hat{f}}_t\left[k\right]\right)-\mathbf{\tilde{f}}_t\left[k\right]\right)}^2$. Specifically, there is
    \vspace{-1mm}
    \begin{small}
    \begin{align}
\nonumber
\sum_{k=1}^{d}\left|\mathbb{E}\left(\mathbf{\hat{f}}_t\left[k\right]\right)-\mathbf{\tilde{f}}_t\left[k\right]\right|
= \frac{\sum_{k=1}^{d} \left|m\left[k\right]+C_k \right|}{\left(p-q\right)\left(m+n^e\right)},
\end{align}
    \end{small}
where $C_k=A_k-\mathbf{\tilde{f}}_t\left[k\right]\left(p-q\right)\left(m+n^e\right)$.
Particularly, this simplifies the problem into a linear optimization problem, solvable in polynomial time \cite{huangfu2018parallelizing}.


Let $u_k=\frac{\left|m\left[k\right]+C_k\right|+\left(m\left[k\right]+C_k\right)}{2}$ and $v_k=\frac{\left|m\left[k\right]+C_k\right|-\left(m\left[k\right]+C_k\right)}{2}$, so the minimization of $\sum_{k=1}^{d}\left|m\left[k\right]+C_k\right|$ could be converted to 
\begin{small}
    \begin{align}\label{OPMA_optimization}
\min \sum\nolimits_{k=1}^{d}u_k+v_k \ \ \ s.t.~ C_k\le u_k-v_k\le m+C_k.
\end{align}
\end{small}
When attacking kRR, $u_k$, $v_k$ also satisfy $\sum_{k=1}^{d}\left(u_k-v_k-C_k\right)=m$. We use HiGHS~\cite{huangfu2018parallelizing} to solve the optimization problem. After obtaining $u_k$ and $v_k$, we calculate $m[k]$ as $u_k-v_k-C_k$.

\noindent\textbf{Analysis of OPMA.} Here we also analyze the properties and sufficient conditions of OPMA. 
\begin{theorem}
\vspace{-2mm}
\label{Theorem:G_{O,t}}
    The expected manipulation gap for OPMA $G_{O,t}$ can be calculated as 
    \begin{small}
        \begin{align}\label{G_{O,t}}
            \nonumber
 &G_{O,t}\left(m,n,\epsilon\right)=\mathbb{E}\left(\frac{1}{d}\sum_{k=1}^{d}\left(\mathbf{\hat{f}}_t\left[k\right]-\mathbf{\tilde{f}}_t\left[k\right]\right)^2\right)\\&=\frac{1}{d}\sum_{k=1}^{d}{\left( \frac{n\mathbf{f}_t\left[k\right]\left(p-q\right)+m^\ast\left[k\right]-mq}{\left(m+n\right)\left(p-q\right)}-\mathbf{\tilde{f}}_t\left[k\right]\right)^2}+\frac{n^2 \cdot \text{Var}\left(n,\epsilon\right)}{(m+n)^2},
        \end{align}
    \end{small}where $\left(m^\ast\left[1\right],m^\ast\left[2\right],\ldots,m^\ast\left[d\right]\right)$ is the solution of the above optimization problem, $p$,$q$ are parameters of FO and $\epsilon$ is the privacy budget of the attacked FO.
\end{theorem}

\begin{proof}
\vspace{-2mm}
    See Appendix~\ref{Proof G_{O,t}}.
\end{proof}

\begin{corollary}\label{Theorem:condition_OPMA}
\vspace{-2mm}
For $\forall k\in\left[d\right]$, if the attacker has 
\begin{small}
\begin{align}
\label{condition_OPMA}
m\geq \max\left(\frac{n\mathbf{\tilde{f}}_t\left[k\right]-n\mathbf{f}_t\left[k\right]}{\frac{1-q}{p-q}-\mathbf{\tilde{f}}_t\left[k\right]},\frac{n\mathbf{f}_t\left[k\right]-n \mathbf{\tilde{f}}_t\left[k\right]}{\frac{q}{p-q}+\mathbf{\tilde{f}}_t\left[k\right]}\right).
\end{align}
\end{small}
fake users and the accurate knowledge of $n$ and $\mathbf{f}_t$, i.e., $n^e=n$ and $\mathbf{f}_t^e=\mathbf{f}_t$, then the expected manipulation gap is
\begin{small}\vspace{-0.5mm}
    \begin{align}
\label{G^*_O}
G_{O,t}^\ast\left(m,n,\epsilon\right)=\left(\frac{n}{m+n}\right)^2\text{Var}\left(n,\epsilon\right)=\frac{n}{m+n}\text{Var}\left(n+m,\epsilon\right).
\end{align}
\end{small}
\end{corollary}

\begin{proof}
\vspace{-1mm}
   See Appendix~\ref{Proof condition_OPMA}.
   \vspace{-2mm}
\end{proof}


For OPMA, we also explore its relation between manipulation gap and privacy budget $\epsilon$.
\begin{corollary}\label{Theorem:OPMA_epsilon}
\vspace{-2mm}
    When~(\ref{condition_OPMA}), $n^e=n$ and $\mathbf{f}_t^e=\mathbf{f}_t$ holds, $G_{O,t}$ becomes smaller as $\epsilon$ becomes larger, implying more successful attack.
\end{corollary}

\begin{proof}
\vspace{-2mm}
    See Appendix~\ref{Proof OPMA_epsilon}.
    \vspace{-2mm}
\end{proof}

Unlike IPMA, OPMA's optimization and $m^\ast[k]$ depend on $\epsilon$, making it impossible to determine the relationship between $G_{O,t}(m,n,\epsilon)$ and $\epsilon$ without enough fake users. However, the impact of user numbers $n$ and $m$ on attack effectiveness is similar in OPMA and IPMA. 

\begin{corollary}
\vspace{-2mm}
    For $\forall \alpha>1$, $G_{O,t}\left(m,n,\epsilon\right)>G_{O,t}\left(\alpha m,\alpha n,\epsilon\right)$.
    \label{Theorem:OPMA_mn}
\end{corollary}
\begin{proof}
\vspace{-2mm}
    See Appendix~\ref{Proof OPMA_mn}.
    \vspace{-2mm}
\end{proof}

Corollaries~\ref{Theorem:OPMA_epsilon} and~\ref{Theorem:OPMA_mn} also show the relation between the manipulation gap and the $\epsilon$ (or population) used when attacking adaptive streaming LDP protocols.

The manipulation gaps of PMAs, regardless of OPMA or IPMA, decrease with the increase of $m+n$ and/or $\epsilon$. This implies the similarity between PMAs and FOs. For FOs, a larger $\epsilon$ or population $n$ means a better estimation with a smaller variance. Similarly, for PMAs, a larger $\epsilon$ or population $m+n$ also means a better attack with a smaller manipulation gap. This inspires us that the strategy for minimizing the estimation error of streaming LDP protocols (i.e., Eq~(\ref{LDP-IDS problem})), can be referenced for achieving our attack goal (i.e., Eq~(\ref{overall goal})). We will detail it in Sec.~\ref{subsec: Arttack Strategy Determination}.

Based on the above introduction, one naive idea is manipulating the FOs at each publication timestamp by PMAs. However, it fails to achieve the optimal. Consider a simple situation where target $\mathbf{\tilde{f}}_t$ keeps as a constant across timestamps and there are enough fake users with accurate knowledge $n^e=n$, $\mathbf{f}^e_t=\mathbf{f}_t$. Then, at publication timestamps, PMAs can achieve the manipulation gap as Eq.~(\ref{G^*_I}) using IPMA or Eq.~(\ref{G^*_O}) using OPMA. At approximation timestamps, the manipulation gap equals to that of the last publication manipulation gap. In such a case, the optimal attack strategy is to find the minimal manipulation gap and try to keep it approximate the timestamps after it. Recall that a larger budget or population helps PMAs to achieve a smaller manipulation gap. Thus, PMAs achieve the optimal when one publication timestamp has the largest budget or population that the procotol can allocate and the rest timestamps directly approximate it. 
Unfortunately, PMAs can not control the strategy determination within LDP protocol. Specifically, the strategy is chosen according to the calculated dissimilarity $\overline{dis}$, which depends on $\mathbf{\bar{f}}_t$ and is uncontrolled by PMAs. 

\subsection{Dissimilarity Manipulation Attacks (DMAs)} \label{subsec: Dissimilarity Manipulation Attack}
\subsubsection{The Goal of DMA.} The adaptive streaming LDP protocol selects a per-timestamp strategy by comparing dissimilarity and potential publication error $err$. The aggregator calculates $err$ using user population and privacy budget. For dissimilarity, it first aggregates users' LDP data to estimate $\mathbf{\bar{f}}_t[k]$, then computes dissimilarity as $\overline{dis}=\frac{1}{d}\sum_{k=1}^{d}(\mathbf{\bar{f}}_t[k]-\mathbf{\hat{f}}_{t-1}[k])^2-\frac{1}{d}\sum_{k=1}^{d}\text{Var}(\mathbf{\bar{f}}_t[k])$. This data aggregation allows attackers to manipulate $\mathbf{\bar{f}}_t[k]$.

The objective of DMAs is to manipulate $\overline{dis}$ via $\mathbf{\bar{f}}_t[k]$ to steer LDP protocol towards a desired strategy for enhanced attack outcomes. If attackers want the protocol to choose the publication strategy, they should increase the $\overline{dis}$ ($\max\:\overline{dis}$) so it surpasses the publication error. Conversely, to trigger the approximation strategy, they should reduce $\overline{dis}$ ($\min\:\overline{dis}$) to make it lower than the publication error. This manipulation allows attackers to control the protocol's Private Strategy Determination phase. And the goal of DMAs can be formulated as follows:
\begin{small}
\begin{align}
\max/\min ~\frac{1}{d}\sum\nolimits_{k=1}^{d}(\mathbf{\bar{f}}_t[k]-\mathbf{\hat{f}}_{t-1}[k])^2-\frac{1}{d}\sum\nolimits_{k=1}^{d}\text{Var}(\mathbf{\bar{f}}_t[k]).
\end{align}
\end{small}

\subsubsection{Input Dissimilarity Manipulation Attack (IDMA).} 
We also introduce the Input Dissimilarity Manipulation Attack (IDMA) first.
Similar to the idea of manipulating the distribution of fake users' inputs of FOs in IPMA, IDMA can adjust the input distribution of FOs to either maximize or minimize the estimated dissimilarity between timestamps.
According to \cite{LDP_IDS}, dissimilarity is an unbiased estimate of the genuine dissimilarity $dis=\frac{1}{d}\sum_{k=1}^{d}\left(\mathbf{f}_t[k]-\mathbf{\hat{f}}_{t-1}[k]\right)^2$. The attacker can manipulate $\mathbf{f}_t[k]$ to either align closely or deviate significantly from $\mathbf{\hat{f}}_{t-1}$, by controlling the number of fake users $m[k]$ per $k$-th element, to manipulate $dis$. This can be formulated as the following optimization problem.
\begin{small}
    \begin{align}\label{eq: IDMA goal}
\nonumber
&\max / \min~~~\frac{1}{d}\sum\nolimits_{k=1}^{d}\left(\mathbf{f}_t\left[k\right]-\mathbf{\hat{f}}_{t-1}\left[k\right]\right)^2 \\
&s.t.\sum\nolimits_{k=1}^{d}m\left[k\right]=m,0\le m\left[k\right]\le m 
\end{align}
\end{small}where $\mathbf{f}_t\left[k\right]$ approximates to $\frac{m\left[k\right]+n^e\mathbf{f}^e_t\left[k\right]}{m+n^e}$.

In Eq.~(\ref{eq: IDMA goal}), the minimization of the estimated dissimilarity can be solved similarly as Eq.~(\ref{IPMA_optimization}). In particular, the attacker only has to replace $\mathbf{\tilde{f}}_t[k]$ with $\mathbf{\hat{f}}_{t-1}[k]$. On the other hand, the maximization of the estimated dissimilarity, we use the following theorem.

\begin{theorem}\label{Theorem:IDMA}
\vspace{-2mm}
    Assume $k^\ast=\arg\max_{k\in[1,..,d]}(n^e\mathbf{f}^e_t[k]/(m+n^e)-\mathbf{\hat{f}}^{t-1}[k])$, the maximum of 
    \begin{small}
        \begin{align}\nonumber
            \frac{1}{d}\sum\nolimits_{k=1}^{d}\left(\mathbf{f}_t\left[k\right]-\mathbf{\tilde{f}}_t\left[k\right]\right)^2=\frac{1}{d}\sum\nolimits_{k=1}^{d}\left(\frac{m\left[k\right]+n^e\cdot\mathbf{f}^e_t\left[k\right]}{n^e+m}-\mathbf{\tilde{f}}_t\left[k\right]\right)^2
        \end{align}
    \end{small}
     is reached when $m\left[k^\ast\right]=m$ and $m\left[k\right]\ =0\ (k\neq\ k^\ast)$.
\end{theorem}

\begin{proof}
\vspace{-2mm}
    See Appendix~\ref{Proof IDMA}.
\vspace{-2mm}
\end{proof}

\subsubsection{Output Dissimilarity Manipulation Attack (ODMA).} IDMA can evolve into Output Dissimilarity Manipulation Attack (ODMA), where fake values are sent directly to the aggregator to control dissimilarity. According to \cite{LDP_IDS}, dissimilarity at time $t$ is $\overline{dis}=\frac{1}{d}\sum_{k=1}^{d}(\mathbf{\bar{f}}_t[k] - \mathbf{\hat{f}}_{t-1}[k])^2 - \frac{1}{d}\sum_{k=1}^{d}\text{Var}(\mathbf{\bar{f}}_t[k])$. The aggregator computes the variance term as a constant $\text{Var}(n,\epsilon)$. To manipulate dissimilarity, only the first term is adjusted, i.e.,
\begin{small}
    \begin{align}
        \max / \min~~\mathbb{E}\left(\frac{1}{d}\sum\nolimits_{k=1}^{d}(\mathbf{\bar{f}}_t[k] - \mathbf{\hat{f}}_{t-1}[k])^2\right).
    \end{align}
\end{small}

OPMA can be adapted as ODMA to reduce dissimilarity by substituting $\mathbf{\tilde{f}}_t[k]$ with $\mathbf{\hat{f}}_{t-1}[k]$. For maximizing dissimilarity, we use the following theorm.

\begin{theorem}\label{Theorem:ODMA}
\vspace{-2mm}
    Assume $k^\ast=\arg\min_{k\in[1,..,d]}(\mathbf{\hat{f}}_{t-1}[k])$, the lower bound of 
    \begin{small}
        \begin{align}\nonumber
            \nonumber
&\mathbb{E}\left(\frac{1}{d}\sum\nolimits_{k=1}^{d}\left(\mathbf{\bar{f}}_t\left[k\right]-\mathbf{\hat{f}}_{t-1}\left[k\right]\right)^2\right) \\ \nonumber
&=\frac{1}{d}\sum\nolimits_{k=1}^{d}\left(\frac{n\mathbf{f}_t\left[k\right]\left(p-q\right)+m\left[k\right]-mq}{\left(m+n\right)\left(p-q\right)}-\mathbf{\hat{f}}_{t-1}\left[k\right]\right)^2 \\ \nonumber
&+\frac{1}{d}\sum\nolimits_{k=1}^{d}\text{Var}\left(\frac{\sum_{j=1}^{n}\mathbb{I}_{S\left(y_j\right)}^{\left(k\right)}+\sum_{j=1}^{m}\mathbb{I}_{S\left(z_j\right)}^{\left(k\right)}-q\left(m+n\right)}{\left(m+n\right)\left(p-q\right)}\right)
        \end{align}
    \end{small}
     can be maximized when $m\left[k^\ast\right]=m$ and $m\left[k\right]\ =0\ (k\neq\ k^\ast)$, where $(y_1, y_2, \ldots, y_n)$ are real users' outputs, $(z_1, z_2, \ldots, z_m)$ are fake users' outputs and $m[k]=\mathbb{E}(\sum_{j=1}^{m}\mathbb{I}_{S(z_j)}^{(k)})$.
\end{theorem}


\begin{proof}
\vspace{-2mm}
    See Appendix~\ref{Proof ODMA}.
\vspace{-2mm}
\end{proof}

According to Theorem~\ref{Theorem:ODMA}, to maximize the dissimilarity, the attacker can first identify the least frequent item (denoted as $k^*$-th item) in the last release $\mathbf{\hat{f}}_{t-1}$, and then control all the fake users to send the data representing $k^*$-th item to the aggregator, i.e, send $k^*$-th item when using kRR or set the $k^*$-th bit of the encoding binary strings as 1 and other bits as 0 when using OUE.

\subsection{Manipulation Strategy Determination}\label{subsec: Arttack Strategy Determination}
PMAs and DMAs provide composable modules for manipulating the per-timestamp estimate processes in streaming LDP protocols. With them, we first propose two baselines applicable to two special target streams.
\subsubsection{Baseline Strategy for Special Targets\\}

\quad\textbf{Baseline 1: Sampling Attack.}\label{SP baseline} Let us reconsider a simple situation where the target $\mathbf{\tilde{f}}_t$ remains as a constant with $t$. With DMAs, an attacker could steer LDP protocols into a desired strategy at each timestamp. Recall that for a constant target, the optimal attack can be achieved when the protocol chooses publication at one timestamp with allocating the largest budget or population while approximating at other timestamps. In a window of size $w$, the attacker needs to steer the protocol into approximation for $w-1$ timestamps and then into publication to make it allocate the largest budget or population. Then the attacker only needs to approximate until the end of the stream. With loss of attack effectiveness, we modify the above idea by publishing at the sampling timestamp in the window, and approximating at other timestamps, termed Sampling Attack, since only one publication with continuous approximation may raise aggregator's concern. This attack can achieve high attacking effectiveness against LSP by solely manipulating sampling timestamps.

\textbf{Baseline 2: Uniform Attack.}\label{BD baseline} In contrast to the constant target, here we consider another special target stream $\mathbf{\tilde{f}}_t$ with huge fluctuations at each timestamp. 
Similar to the above analysis, at each publication timestamp, the attacker can achieve a manipulation gap as $G^*_{I,t}$ or $G^*_{O,t}$. 
Since the huge fluctuations of target, meaning large distance between two adjacent timestamps, the manipulation gap for approximation will be larger than that of the publication timestamp. Thus, always publication is more beneficial for the attack. We term this Uniform Attack, meaning attacking every publication timestamp uniformly. The uniform attack works well for LBU/LPU, which publishes at every timestamp. 

The above baselines assume special targets.
However, for arbitrary targets, adopting Uniform Attack to publish always will result in a publication budget or population not exceeding $\epsilon/2w$ or $\left(m+n\right)/2w$. When the budget or population is exactly or below $\epsilon/2w$ or $\left(m+n\right)/2w$, $G_t$ meets or exceeds $G_{I(O),t}(m,n,\epsilon/2w)$ or $G_{I(O),t}(m/2w,n/2w,\epsilon)$. As analysed in Sec.~\ref{subsec: Publication Manipulation Attack}, with a large $w$, $G_t$ will become very large,
which contradicts the objective of minimizing $G_{avg}$. When adopting the Sampling Attack for arbitrary targets, the protocol allocates at most a publication budget of $\epsilon/2$ or population $(m+n)/2$ at the sampling timestamp and zero at others. The average manipulation gap $G_{avg}$ becomes $(G_{I(O),t}\left(m,n,\frac{\epsilon}{2}\right)+\frac{1}{d}\sum_{t=2}^{w}\sum_{k=1}^d(\mathbf{\tilde{f}}_t[k]-\mathbf{\hat{f}}_1[k])^2)/w$ for budget-division methods and $(G_{I(O),t}\left(\frac{m}{2},\frac{n}{2},\epsilon\right)+\frac{1}{d}\sum_{t=2}^{w}\sum_{k=1}^d(\mathbf{\tilde{f}}_t[k]-\mathbf{\hat{f}}_1[k])^2)/w$ for population-division methods, assuming the first timestamp is the sampling time. The attack effectiveness diminishes unless $\mathbf{\tilde{f}}_t$ remains constant.

\subsubsection{Adaptive Strategy Chosen for Arbitrary Targets\\}
Minimizing the time-average manipulation gap for an arbitrary target stream requires adaptive coordination of the two modules across timestamps. We propose a MSD module to achieve this by mimicking the idea of adaptively optimizing the estimation error in streaming LDP protocols as follows.


\textbf{Optimization View of Streaming LDP.}\label{Sec: Optimization View of LDP-IDS} 
Adaptive streaming LDP protocols seek to minimize the average estimate error as an optimization problem in Eq.~(\ref{LDP-IDS problem}).
Specifically, the estimation error $error_t=\sum_{k=1}^d(\mathbf{\hat{f}}_t[k]-\mathbf{f}_t[k])^2/d$ per timestamp can be denoted as a binary-state variable $error_t(i)$ concerning the \textit{privacy strategy} of publication or approximation. It equals to either publication error $err$ in the publication (denoted as $i=1$) or approximation error (dissimilarity $dis$, where $\mathbb{E}(\overline{dis})=dis$) in the approximation ($i=0$). The former $err$ equals the noise variance of FO (i.e., $\text{Var}(n_t,\epsilon_t)$) which is decided by the allocated user population $n_t$ and budget $\epsilon_t$. The latter $dis$ can be denoted as $\sum_{k=1}^d(\mathbf{\hat{f}}_{t-1}[k]-\mathbf{f}_t[k])^2/d$, where $\mathbf{\hat{f}}_{t-1}$ is the last release at $t-1$. So,
\begin{small}
	\begin{align}\nonumber
error_t(i)=\left\{\begin{matrix}dis=\sum_{k=1}^d(\mathbf{\hat{f}}_{t-1}[k]-\mathbf{f}_t[k])^2/d,~i=0 ~(\text{approximation})\\err=\text{Var}\left(n_t,\epsilon_t\right),~i=1 ~(\text{publication})\\\end{matrix}\right.
	\end{align}
\end{small}
Then, the optimization problem in Eq.~(\ref{LDP-IDS problem}) is a strategy-chosen problem of finding a binary sequence $Q$ for minimizing the time-average estimation error, i.e. $\arg\min_{Q\in{\{0,1\}}^T} \frac{1}{T}\sum\nolimits_{t=1}^{T}error_t(Q[t])$.
Adaptive streaming LDP protocols in Sec.~\ref{subsec:LDP for Infinite Data Streams} provide different online adaptive solutions to this problem by comparing $err$ with $dis$ and greedily choosing a \textit{private strategy} with a smaller error.

\textbf{Optimization View of Attack.} The overall goal of our attack is formulated in Eq.~(\ref{overall goal}). The \textit{manipulation gap} $G_t=\sum_{k=1}^d(\mathbf{\hat{f}}_t[k]-\mathbf{\tilde{f}}_t[k])^2/d$ at timestamp $t$ varies with the aggregator's choice between publication and approximation. If the aggregator chooses the approximation strategy, it will publish the previously estimated frequency $\mathbf{\hat{f}}_{t-1}$, so $G_t=\sum_{k=1}^d(\mathbf{\hat{f}}_{t-1}[k]-\mathbf{\tilde{f}}_t[k])^2/d$. Otherwise, it chooses the publication strategy, the aggregator will invoke the FO to publish a fresh estimate. 
The attacker will also initiate PMAs to manipulate the estimate to approach the target. Overall, when publication, $G_t$ can be computed as $G_{I,t}(m_t^a,n_t^a,\epsilon_t)$ when using IPMA or $G_{O,t}(m_t^a,n_t^a,\epsilon_t)$ when using OPMA. $m_t^a$, $n_t^a$, and $\epsilon_t$ are the number of fake users and genuine users, and the privacy budget used for publication, respectively. 
Considering the aggregator's choice over two \textit{privacy strategies} can be manipulated by DMAs, there are also two \textit{attack strategies} from the attacker's view. 
They are approximation-based and publication-based strategies, i.e., manipulating the dissimilarity calculation to steer the aggregator to choose approximation and publication, respectively. 
And the \textit{manipulation gap} $G_t$ can be formulated as the following variable, where $i=0$ and $1$ represent the approximation and publication-based attack strategies, respectively.
\begin{small}
    \begin{align}\nonumber
    G_t\left(i\right)=\left\{\begin{matrix}\sum_{k=1}^d(\mathbf{\hat{f}}_{t-1}[k]-\mathbf{\tilde{f}}_t[k])^2/d,&i=0 ~(\text{approximation})\\G_{I(O),t}(m_t^a,n_t^a,\epsilon_t),&i=1 ~(\text{publication})\\\end{matrix}\right.
\end{align}
\end{small}

Therefore, the optimization problem in Eq.~(\ref{overall goal}) can also be seen as \textit{a strategy-chosen problem}, i.e., seeking a binary vector $Q=\{0,1\}^T$ of length $T$ which can minimize the average manipulation gap $G_{avg}=\frac{1}{T}\sum_{t=1}^T G_t$ over the whole stream:
\begin{small}
    \begin{align}
\nonumber
    \arg\min_{Q\in{\{0,1\}}^T} &\frac{1}{T}\sum\nolimits_{t=1}^{T}G_t(Q[t]).
\end{align}
\end{small}


\textbf{MSD: Adaptive Strategy Chosen.}\label{subsec:Solving the Strategy-Chosen Problem}
Comparing the strategy-chosen problems of the adaptive streaming LDP protocol to adaptive attack for arbitrary targets, it is easy to find that they share a similar problem structure. 
For publication, both $error_t(1)$ and $G_t(1)$ are determined similarly by privacy budget and population size; larger values result in smaller errors or manipulation gaps, enhancing estimation or attack effectiveness. 
For approximation, both $error_t(0)$ and $G_t(0)$ represents approximating the results of current timestamp (i.e.,$\mathbf{f}_t$ or $\mathbf{\tilde{f}}_t$) with the last one. The similar problem structure inspires us the design of MSD module, which solves the attack strategy-chosen problem in an online adaptive way as the private strategy-chosen problem of adaptive streaming LDP protocols.
In particular,
MSD guides the attacker to greedily choose between approximation and publication-based strategy with a better attack effect (i.e., smaller manipulation gap). Specifically, if $G_t(0)$ is less than $G_t(1)$, MSD chooses the approximation-based attack strategy; otherwise, MSD chooses the publication-based attack strategy. 


\begin{algorithm}[t]\footnotesize\setstretch{0.2}
	\SetAlgoLined
	\caption{Adaptive Attack to LDP-IDS}        \label{alg:Adaptive Attack}
	\KwIn{Target stream $\mathbf{\tilde{f}}=\{\mathbf{\tilde{f}}_1, \mathbf{\tilde{f}}_2,...,\mathbf{\tilde{f}}_t,...\}$}
        \SetKwInOut{KwOut}{Output}
        \KwOut{Manipulated released statistics $\mathbf{\hat{f}}=\{\mathbf{\hat{f}}_1,\mathbf{\hat{f}}_2,...,\mathbf{\hat{f}}_t,...\}$}
	Initialize Queue of $\epsilon_{t,2}$ or $|U_{t,2}|$, $Q_m=\left[0,..,0\right]^{w-1}$; \\
        Estimate the number of genuine users as $n^e$; \\
            
	\For{each timestamp $t$}
	{
             Get $\mathbf{f}^e_t$ from fake users and estimate $\epsilon_{t,2}$ or $|U_{t,2}|$ using $Q_m$;\\
            Compute $m^*$ for PMAs' optimization problem (e.g. Eq.~(\ref{OPMA_optimization}));\\
            Calculate attack dissimilarity $dis_{attack}=\sum\nolimits_{k=1}^d(\mathbf{\hat{f}}_{t-1}[k]-\mathbf{\tilde{f}}_t[k])^2/d$;\\
            Calculate potential manipulation gap $G(m,n^e,\epsilon_{t,2})$ for budget-division or $G\left(\frac{m\left|U_{t,2}\right|}{\left(m+n^e\right)},\frac{n^e\left|U_{t,2}\right|}{\left(m+n^e\right)}, \epsilon\right)$ for population-division; \\
            \tcp{Manipulation Strategy Determination (MSD)}
            \eIf{$dis_{attack}$ $>$ $G$}
            {
                Adopt DMAs to maximize the dissimilarity $\overline{dis}$;
            }
            {
                Adopt DMAs to minimize the dissimilarity $\overline{dis}$;
            }
            \eIf{LDP Aggregator chooses publication}
            {
                Launch PMAs (using $m^*$ in line 5) to derive $\mathbf{\hat{f}}_{attack}$;\\
            \Return{$\mathbf{\hat{f}}_t=\mathbf{\hat{f}}_{attack}$;}
            }
            {
                \Return{$\mathbf{\hat{f}}_t=\mathbf{\hat{f}}_{t-1}$};
            }
            Pop the first item from $Q_m$, enqueue new $\epsilon_{t,2}$ or $|U_{t,2}|$;
        }
        
\end{algorithm}

\section{Attacking Streaming LDP Protocols}
In the following, we present the details of our Adaptive Attack against the adaptive streaming LDP protocols by integrating the above proposed MSD, PMAs, and DMAs. 

\subsection{Attacking Adaptive Streaming LDP}\label{BD and PD AA}
For LDP estimation, a larger privacy budget or population enhances data utility. Similarly, for attacks, a larger budget (Corollaries~\ref{Theorem:IPMA_epsilon} and~\ref{Theorem:OPMA_epsilon}) or population (Corollaries~\ref{Theorem:IPMA_mn} and ~\ref{Theorem:OPMA_mn}) reduces the manipulation gap, thus increasing the attack effectiveness. This suggests adopting a streaming-LDP-like adaptive budget/population allocation could enhance attacks, as discussed in Sec.~\ref{subsec:Solving the Strategy-Chosen Problem}.

\subsubsection{Algorithm details.}Algorithm~\ref{alg:Adaptive Attack} shows the detailed framework of our proposed Adaptive Attack to LDP-IDS (adaptive framework of streaming LDP protocol~\cite{LDP_IDS}) with an arbitrary target stream $\mathbf{\tilde{f}}=\{\mathbf{\tilde{f}}_1, \mathbf{\tilde{f}}_2,...,\mathbf{\tilde{f}}_t,...\}$. 
At each timestamp $t$, the attacker first gets the inputs (or outputs in the MITM attack) of FOs from a portion of compromised users to estimate $\mathbf{f}^e_t$ (line 4). 
Then, the attacker computes $\left(m^\ast\left[1\right],\ldots,m^\ast\left[d\right]\right)$ via solving the optimization problem of PMAs (line 5). 
After that, the attacker can calculate $dis_{attack}=\sum_{k=1}^d(\mathbf{\hat{f}}_{t-1}[k]-\mathbf{\tilde{f}}_t[k])^2/d$ as the manipulation gap for the approximation-based attack strategy (line 6), and the \textit{potential manipulation gap} $G$ (i.e., $G_{I,t}$ in IPMA or $G_{O,t}$ in OPMA) for the publication-based attack strategy (line 7). For budget-division protocols LBD/LBA, by estimating $n$ and $\mathbf{f}_t$ as $n^e$ and $\mathbf{f}^e_t$, $G$ is calculated as $G(m,n^e,\epsilon_{t,2})$ since $\epsilon_{t,2}$ budget with all users ($n$ genuine and $m$ fake users) are used. For population-division protocols LPD/LPA, $G$ equals to $G\left(\frac{m\left|U_{t,2}\right|}{\left(m+n^e\right)},\frac{n^e\left|U_{t,2}\right|}{\left(m+n^e\right)}, \epsilon\right)$ as the whole budget with $|U_{t,2}|$ users ($\frac{n|U_{t,2}|}{m+n}$ genuine and $\frac{m|U_{t,2}|}{m+n}$ fake users) are used. 

Then, MSD chooses a beneficial strategy by comparing $dis_{attack}$ with $G$ (line 8). If $dis_{attack}>G$, MSD will choose the publication-based attack strategy. Thus, DMAs will be launched to maximize $\overline{dis}$ in the streaming LDP protocol and enforce the LDP aggregator to choose publication (line 9). If the aggregator does so, PMAs will be further mounted to manipulate the FO in publication (line 14). If $dis_{attack}\leq G$, DMAs instead of PMAs will then be called to minimize $\overline{dis}$ (line 11). Fig.~\ref{fig:overview}(c) illustrates the strategy selection-based adaptive attack, which interestingly mimics that of the strategy selection in the adaptive LDP estimation in Fig.~\ref{fig:overview}(a). 

Note that, the historical usage of budget $\epsilon_{t,2}$ or population $|U_{t,2}|$ for the previous $w-1$ timestamps is needed for calculating $G$ (line 4). So, the attacker can store them in a first-in-first-out queue $Q_m$ (line 19). For example, when attacking LBD, $\epsilon_{t,2}=(\epsilon/2-\sum_{i=1}^{w-1}Q_m[i])/2$.
Current $\epsilon_{t,2}$ or $|U_{t,2}|$ at $t$ can be obtained as follows. For the publication timestamp, the fake users will receive $\epsilon_{t,2}$ from the aggregator, $|U_{t,2}|$ can be estimated as $\frac{m_t(n^e+m)}{m}$ where $m_t$ is the number of fake users sampled. For the approximation timestamp, $\epsilon_{t,2}$ or $|U_{t,2}|$ is 0. 

\subsubsection{Theoretical Analysis.}\label{sec:analysis}
Tables~\ref{tab:attack_effect} and~\ref{tab:manipulation gap of baseline} summarize the effects of all proposed attacking methods. For LBU and LPU, only the Uniform Attack (including both Input and Output poisoning) is used. For LSP, only the Sampling Attack (also including Input and Output) is applied, and we abbreviated $G_{I(O),t}\left(m,n,\epsilon\right)+\sum_{t=\left(i-1\right)w+2}^{iw}\sum_{k=1}^{d}(\mathbf{\tilde{f}}_t[k]-\mathbf{\hat{f}}_{\left(i-1\right)w+1}[k])^2/d$ as $G_{I(O),t}^S\left(m, n,\epsilon,i\right)$. In each attack, the same type of LDP poisoning (i.e., Input or Output) is assumed for both DMAs and PMAs. These methods employ Uniform, Sampling, and Adaptive Attacks, each available in input and output forms. Particularly, it is assumed that there are enough $m$ fake users so that DMAs can always succeed in inducing the LDP aggregator into the desired strategy. Since the adaptive allocation of privacy budget and population results in varying sequences of $\epsilon_{t,2}$ and $U_{t,2}$, it is hard to calculate the exact average manipulation gap. Therefore, we estimate the upper and lower bounds of the manipulation gap based on the minimum and maximum privacy budget and population that can be allocated. For LBD, the privacy budget ranges from $\epsilon/2^{w+1}$ to $\epsilon/4$. For LBA, it ranges from $\epsilon/2w$ to $\epsilon/2$. For LPD, the user population is between $(m+n)/2^{w+1}$ and $(m+n)/4$. For LPA, the range is from $(m+n)/2w$ to $(m+n)/2$.

Tables~\ref{tab:attack_effect} and~\ref{tab:manipulation gap of baseline} manifest that the Adaptive Attacks' performance always equals to or exceeds Uniform Attacks'. It is hard to compare Adaptive Attacks and Sampling Attacks directly while experiments show that Adaptive Attacks also outperform Sampling Attacks. Interestingly, we find that the baseline LDP algorithms are more robust to our attacks than the adaptive ones when adopting the same type of attacks. For example, attacking LBA has less manipulation gap than attacking LBU with Uniform Attacks. This raises concerns that, despite achieving higher utility, adaptive LDP methods may be less robust to fine-grained manipulation.

\begin{table}[tb]
    \centering
    \caption{Average manipulation gap of attacks against LDP-IDS baselines. $t_a$ is the total number of attack timestamps.}\vspace{-3mm}
    \scalebox{0.65}{
    \begin{tabular}{|c|c|c|}
\hline
& LBU & LPU \\ \hline
\makecell[c]{Input Uniform \\Attack(Sec.~\ref{BD baseline})} & $G_{I,t}\left(m,n,\frac{\epsilon}{w}\right)$ & $G_{I,t}\left(\frac{m}{w},\frac{n}{w},\epsilon\right)$  \\ \hline
\makecell[c]{Output Uniform \\Attack(Sec.~\ref{BD baseline})}	& $G_{O,t}\left(m,n,\frac{\epsilon}{w}\right)$ & $G_{O,t}\left(\frac{m}{w},\frac{n}{w},\epsilon\right)$  \\ \hline
\end{tabular}
\begin{tabular}{|c|c|}
\hline
  & LSP   \\ \hline
\makecell[c]{Input Sampling \\Attack(Sec.~\ref{SP baseline})} & $\left(\sum_{i}{G_{I,t}^S\left(m,n,\epsilon,i\right)}\right)/t_a$ \\ \hline
\makecell[c]{Output Sampling \\Attack(Sec.~\ref{SP baseline})}& $\left(\sum_{i}{G_{O,t}^S\left(m,n,\epsilon,i\right)}\right)/t_a$ \\ \hline
\end{tabular}
}
    \label{tab:manipulation gap of baseline}
\end{table}

\begin{table*}[t]
\centering
\caption{Average manipulation gap of attacks against adaptive LDP-IDS. $t_a$ is the total number of attack timestamps.}
\label{tab:attack_effect}


\scalebox{0.750}{
\begin{tabular}{|c|c|c|c|c|}
\hline
& LBD & LBA & LPD & LPA \\ \hline
\makecell[c]{Input Uniform Attack\\(Sec.~\ref{BD baseline})} & $\left(G_{I,t}\left(m,n,\frac{\epsilon}{4}\right),G_{I,t}\left(m,n,\frac{\epsilon}{2^{w+1}}\right)\right)$ & $G_{I,t}\left(m,n,\frac{\epsilon}{2w}\right)$ & $\left(G_{I,t}\left(\frac{m}{4},\frac{n}{4},\epsilon\right),G_{I,t}\left(\frac{m}{2^{w+1}},\frac{n}{2^{w+1}},\epsilon\right)\right)$ & $G_{I,t}\left(\frac{m}{2w},\frac{n}{2w},\epsilon\right)$ \\ \hline
\makecell[c]{Output Uniform Attack\\(Sec.~\ref{BD baseline})}	&  $\left(G_{O,t}\left(m,n,\frac{\epsilon}{4}\right),G_{O,t}\left(m,n,\frac{\epsilon}{2^{w+1}}\right)\right)$ & $G_{O,t}\left(m,n,\frac{\epsilon}{2w}\right)$ & $\left(G_{O,t}\left(\frac{m}{4},\frac{n}{4},\epsilon\right),G_{O,t}\left(\frac{m}{2^{w+1}},\frac{n}{2^{w+1}},\epsilon\right)\right)$ & $G_{O,t}\left(\frac{m}{2w},\frac{n}{2w},\epsilon\right)$\\ \hline

\makecell[c]{Input Sampling Attack\\(Sec.~\ref{SP baseline})} & $\left(\sum_{i}{G_{I,t}^S\left(m,n,\frac{\epsilon}{4},i\right)}\right)/t_a$ & $\left(\sum_{i}{G_{I,t}^S\left(m,n,\frac{\epsilon}{2},i\right)}\right)/t_a$ & $\left(\sum_{i}{G_{I,t}^S\left(\frac{m}{4},\frac{n}{4},\epsilon,i\right)}\right)/t_a$ & $\left(\sum_{i}{G_{I,t}^S\left(\frac{m}{2},\frac{n}{2},\epsilon,i\right)}\right)/t_a$ \\ \hline

\makecell[c]{Output Sampling Attack\\(Sec.~\ref{SP baseline})} & $\left(\sum_{i}{G_{O,t}^S\left(m,n,\frac{\epsilon}{4},i\right)}\right)/t_a$ & $\left(\sum_{i}{G_{O,t}^S\left(m,n,\frac{\epsilon}{2},i\right)}\right)/t_a$ & $\left(\sum_{i}{G_{O,t}^S\left(\frac{m}{4},\frac{n}{4},\epsilon,i\right)}\right)/t_a$ & $\left(\sum_{i}{G_{O,t}^S\left(\frac{m}{2},\frac{n}{2},\epsilon,i\right)}\right)/t_a$ \\ \hline

\makecell[c]{Input Adaptive Attack\\(Sec.~\ref{BD and PD AA})} & $\left(G_{I,t}\left(m,n,\frac{\epsilon}{4}\right),G_{I,t}\left(m,n,\frac{\epsilon}{2^{w+1}}\right)\right)$ & $\left(G_{I,t}\left(m,n,\frac{\epsilon}{2}\right),G_{I,t}\left(m,n,\frac{\epsilon}{2w}\right)\right)$ & $\left(G_{I,t}\left(\frac{m}{4},\frac{n}{4},\epsilon\right),G_{I,t}\left(\frac{m}{2^{w+1}},\frac{n}{2^{w+1}},\epsilon\right)\right)$ & $\left(G_{I,t}\left(\frac{m}{2},\frac{n}{2},\epsilon\right),G_{I,t}\left(\frac{m}{2w},\frac{n}{2w},\epsilon\right)\right)$ \\ \hline
\makecell[c]{Output Adaptive Attack\\(Sec.~\ref{BD and PD AA})} & $\left(G_{O,t}\left(m,n,\frac{\epsilon}{4}\right),G_{O,t}\left(m,n,\frac{\epsilon}{2^{w+1}}\right)\right)$ & $\left(G_{O,t}\left(m,n,\frac{\epsilon}{2}\right),G_{O,t}\left(m,n,\frac{\epsilon}{2w}\right)\right)$ & $\left(G_{O,t}\left(\frac{m}{4},\frac{n}{4},\epsilon\right),G_{O,t}\left(\frac{m}{2^{w+1}},\frac{n}{2^{w+1}},\epsilon\right)\right)$ & $\left(G_{O,t}\left(\frac{m}{2},\frac{n}{2},\epsilon\right),G_{O,t}\left(\frac{m}{2w},\frac{n}{2w},\epsilon\right)\right)$ \\ \hline
\end{tabular}
}


\vspace{-4mm}
\end{table*}

\subsection{Attacking Other LDP Tasks and Protocols}\label{sec:Attacking Other LDP Tasks and Protocols}

\begin{itemize}
    \item \textbf{Attacking Mean Estimation.} Besides frequency estimation, our methods can also apply to continual mean estimation over data streams \cite{wang2020continuous, LDP_IDS} with only replacing PMA with OPA~\cite{Li2023Finegrained} since both PMA and OPA benefits from a larger budget and population. For further information, refer to Appendix~\ref{sec: Attack Mean Estimation over Numerical Domain}.
    
    \item \textbf{Attacking Other Streaming LDP Protocols.} Our attack methods are applicable to other streaming LDPs like PeGaSus~\cite{Chen2017PeGaSusDD}, FAST~\cite{Fan2014AnAA}, $\text{DSAT}_w$~\cite{10.1145/2806416.2806441}, RescueDP~\cite{Wang2016RescueDPRS}, CGM~\cite{baocgm} and ToPL~\cite{wang2020continuous}. To summarize, since $\text{DSAT}_w$ adopts budget-division or population-division like LDP-IDS, all our proposed attacks are applicable. PeGasus, CGM and ToPL choose to publish at every timestamp like LBU/LPU, thus Uniform Attack are adopted. FAST and RecueDP samples timestamps for publication like LSP, so the Sampling Attack is considered. The post-processing of event-level LDP does not affect our attacks. Both LDP estimation and attacks benefit from a larger budget population (Corollaries~\ref{Theorem:IPMA_epsilon}, ~\ref{Theorem:IPMA_mn}, ~\ref{Theorem:OPMA_epsilon}, and ~\ref{Theorem:OPMA_mn}), sharing similar properties. Thus, the post-processing for better LDP estimation can be directly exploited to enhance our attacks. More details are in Appendix~\ref{sec: Existing LDP Finite Data Stream Algorithms}. 
\end{itemize}

\section{Performance Evaluation}\label{sec:Experiments}

\subsection{Experimental Setup}
\subsubsection{Datasets.} Three real-world datasets were used as follows. Results on synthetic datasets are shown in Appendix~\ref{sec: Experiment Results on Synthetic Datasets}.
\begin{itemize}
\item\textbf{\textsf{Taxi}}\footnote{\url{https://www.microsoft.com/en-us/research/publication/t-drive-trajectory-data-sample/}} comprises taxi trajectories in Beijing from Feb. 2nd to Feb. 8th, 2008. We extracted data from $N=10,357$ taxis, each with $T=885$ timestamps at 10-minute intervals, across $6$ grid partitions ($d=6$). We also utilize longitude data of Taxi dataset for numerical domain attacks, denoted as \textbf{Taxi-Longitude}.
\item\textbf{\textsf{Foursquare}}\footnote{\url{https://sites.google.com/site/yangdingqi/home/foursquare-dataset}} includes Foursquare check-ins from Apr. 2012 to Sep. 2013, detailing time, place, and user ID. It's transformed into $N=266,909$ data streams, each with $T=456$ timestamps, and places categorized into 100 types ($d=100$).
\item\textbf{\textsf{Taobao}}\footnote{\url{https://tianchi.aliyun.com/dataset/dataDetail?dataId=56}} includes AD click logs from 1.14 million Taobao customers across 12,973 categories. For simplicity, AD commodities were grouped into $d=150$ categories. The extracted click streams of $N=728,745$ customers, each representing the category of the last click every ten minutes over three days, totaling $T=432$ timestamps. For numerical domains, we obtain the labeled price of each AD commodities in Taobao, referred to as \textbf{Taobao-Price}.
\end{itemize}

We also synthesized four types of datasets and run experiments on them. The results can be found in Appendix~\ref{sec: Experiment Results on Synthetic Datasets}.

\subsubsection{Targets.} We used four different target streams $\mathbf{\tilde{f}}=\left(\mathbf{\tilde{f}}_1,\mathbf{\tilde{f}}_2,\ldots,\mathbf{\tilde{f}}_{T}\right)$ for frequency and mean with the following time-varying patterns. 

\begin{itemize}
    \item \textbf{Uniform:} $\mathbf{\tilde{f}}_t$ at each time $t$ was set as a uniform distribution for frequency estimation and a constant for mean-value estimation to simulate a stream with no fluctuation.
    \item \textbf{Pulse:} To simulate huge fluctuation stream, for frequency estimation, at each time $t$, we selected one item $k\in\left[d\right]$ and set its frequency $\mathbf{\tilde{f}}_t\left[k\right]=1$ and others $\mathbf{\tilde{f}}_t\left[k'\right]=0~(k' \neq k)$. We set the target as an extreme value for mean-value estimation.
    \item \textbf{Gaussian:}  Gaussian distribution is a common distribution that always appears in natural and social sciences. We set $\mathbf{\tilde{f}}_t$ at each time $t$ as a discrete Gaussian distribution $\mathcal{N}\left(0,\sigma^2t\right)$, where $\sigma=0.5$. For mean-value estimation, we set the target as the value sampled from the Gaussian distribution.
    \item \textbf{Sigmoid:} To simulate frequency increasing pattern, for each $\mathbf{\tilde{f}}_t$, we chose one item $k\in\left[d\right]$ and set its frequency as $\mathbf{\tilde{f}}_t\left[k\right]=2 \cdot \text{Sigmoid}\left(0.01t\right)-1$. For mean estimation, we set the target increasing according to Sigmoid pattern.
\end{itemize}

\subsubsection{Metrics.} We measured the attack performance using the time and dimension averaged MSE between manipulated statistical stream $\hat{\textbf{f}}$ and the target one $\tilde{\textbf{f}}$, i.e., $MSE=\frac{1}{dT}\sum_{t=1}^{T}\sum_{k=1}^{d}(\mathbf{\hat{f}}_t[k]-\mathbf{\tilde{f}}_t[k])^2$, aligning with Eq.~(\ref{overall goal}). A smaller MSE means better attack effectiveness. We also used the $Success\ Rate=\frac{\text{\#success}}{\text{\#total}}$ to evaluate the effectiveness of DMAs, where $\text{\#success}$ is the number of DMAs that successfully induce the LDP protocol into beneficial strategies for the attack, and $\text{\#total}$ is the total number of launched DMAs.

\subsubsection{Parameter Settings.} We set default $\epsilon=1$ and $w=20$ and used Ada as the default FO. We evaluated attacks against other FOs with larger domain in Sec.~\ref{sec: Attack different FO}. We chose the default estimated user number $n^e$ based on a common observation that online reports tend to publish round numbers instead of precise values~\cite{Erlingsson2014RAPPORRA}. We selected 1000 fake users per dataset for $\mathbf{f}^e$ information. Here, we mainly consider a partial-knowledge attacker and compare different levels of knowledge in Sec.~\ref{sec: Compare different levels of knowledge}. Given the difficulty of achieving special targets like Pulse, we adjusted $\beta=m/(m+n)$ variably for different targets. For budget-division attacks with OPMA, sufficient fake users are necessary due to Corollary~\ref{Theorem:OPMA_epsilon}, setting $\beta$ at 0.2 for uniform and Gaussian, and 0.3 for pulse and sigmoid per Eq.~(\ref{condition_OPMA}). IPMA inherently negates the need for sufficient fake users as per Eq.~(\ref{condition_IPMA}) due to Corollaries~\ref{Theorem:IPMA_epsilon} and~\ref{Theorem:IPMA_mn}. Attacks on population-division are based on Corollaries~\ref{Theorem:IPMA_mn} and~\ref{Theorem:OPMA_mn}, without requiring for sufficient fake users. Reasons for different $\beta$ settings are detailed in Sec.~\ref{sec: Conditions for sufficient fake users}.

\begin{figure*}[t]
	\centering	
	\subfigure[Uniform $\mathbf{\tilde{f}}$]{
		\includegraphics[width=0.24\textwidth]{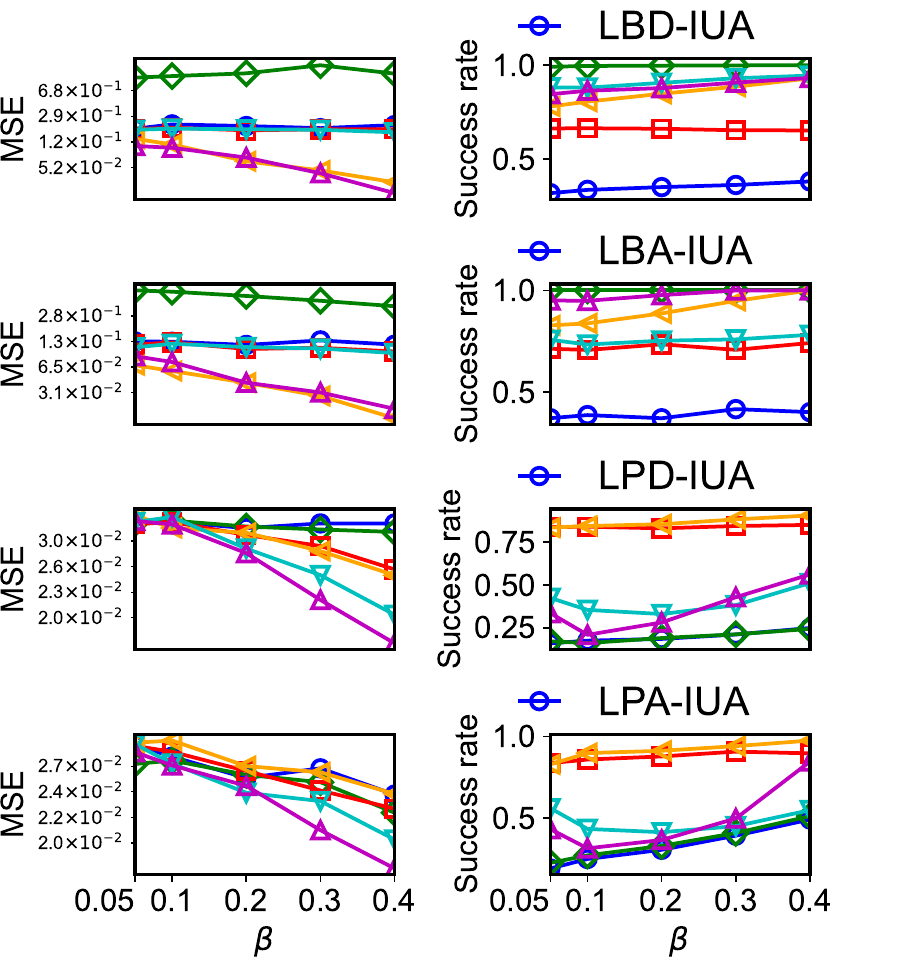}}\vspace{-0.15cm}\hspace{-4mm}
	\subfigure[Pulse $\mathbf{\tilde{f}}$]{
		\includegraphics[width=0.24\textwidth]{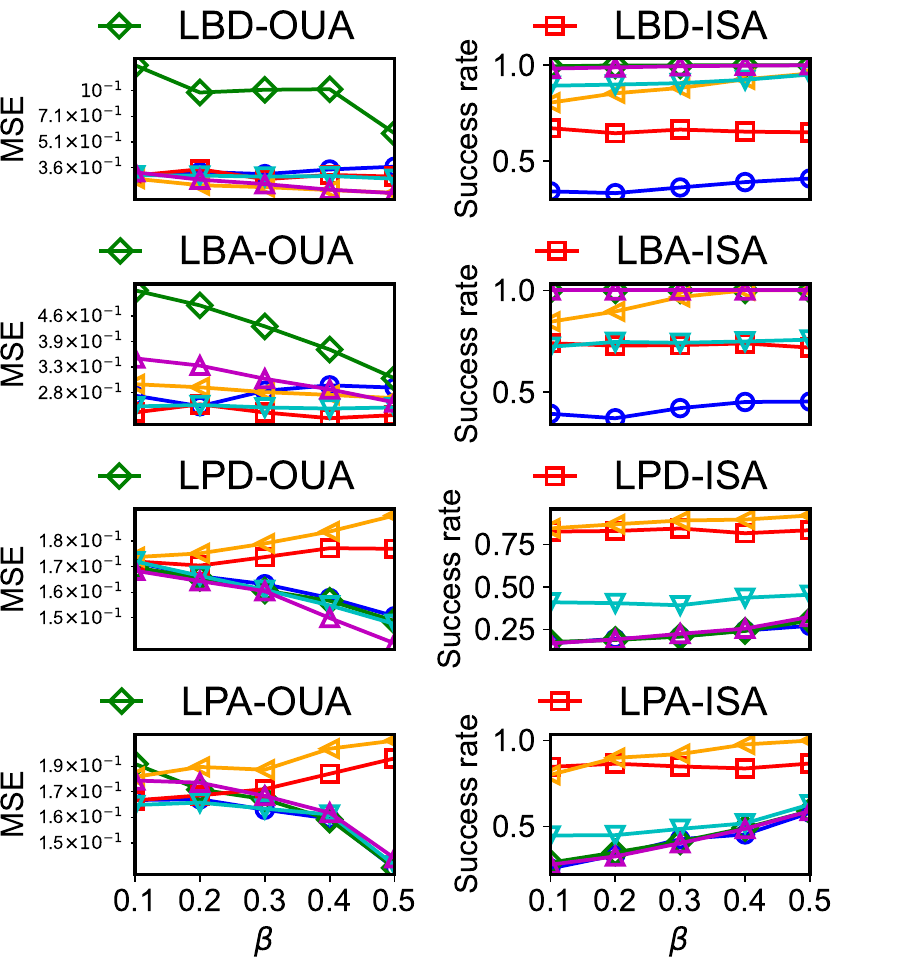}}\vspace{-0.15cm}\hspace{-4mm}
	\subfigure[Gaussian $\mathbf{\tilde{f}}$]{
		\includegraphics[width=0.24\textwidth]{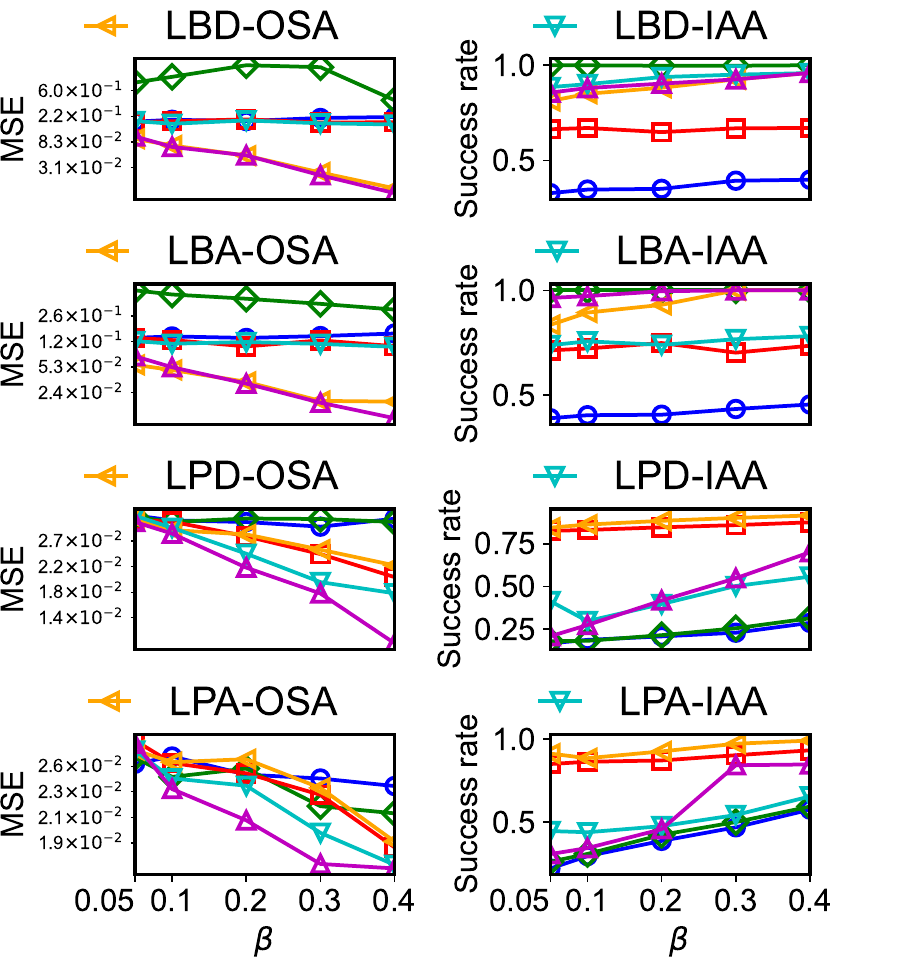}}\vspace{-0.15cm}\hspace{-4mm}
	\subfigure[Sigmoid $\mathbf{\tilde{f}}$]{
		\includegraphics[width=0.24\textwidth]{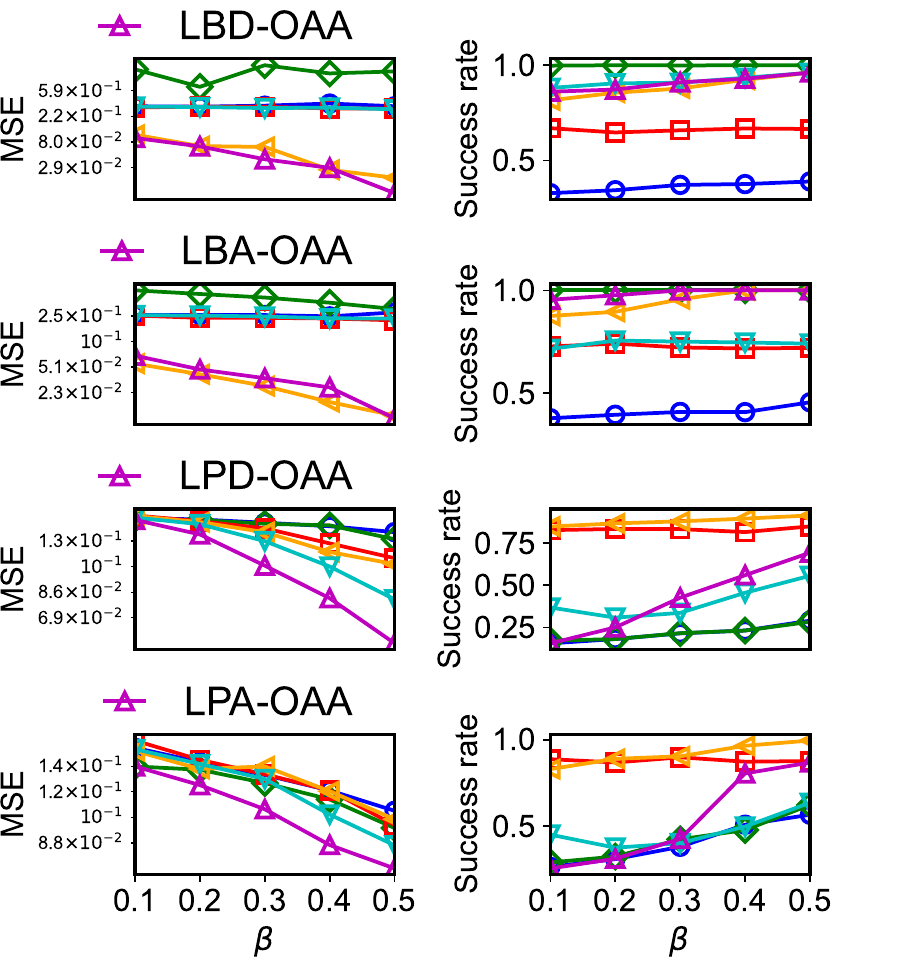}}\vspace{-0.15cm}
	\caption{\small Attacking effectiveness against LBD, LBA, LPD and LPA, varying fake user ratio $\beta$ (\textsf{Taxi} with different targets). 
	}\centering
	\label{fig:varying beta} 
	\vspace{-0.6cm}
\end{figure*}

\begin{figure*}[t]
	\centering	
	\subfigure[Varying $\epsilon$.]{
		\includegraphics[width=0.24\textwidth]{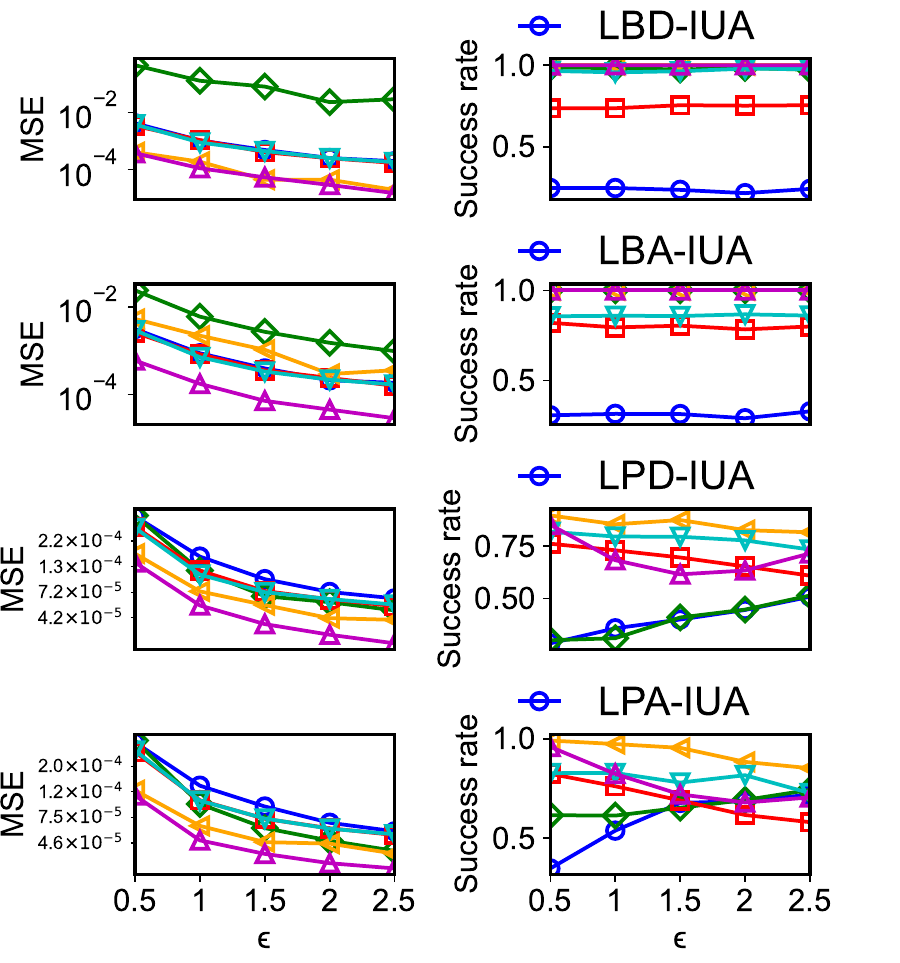}
          \label{fig:varyinig epsilon}  
        }\vspace{-0.15cm}\hspace{-4mm}
	\subfigure[Varying $w$.]{
		\includegraphics[width=0.24\textwidth]{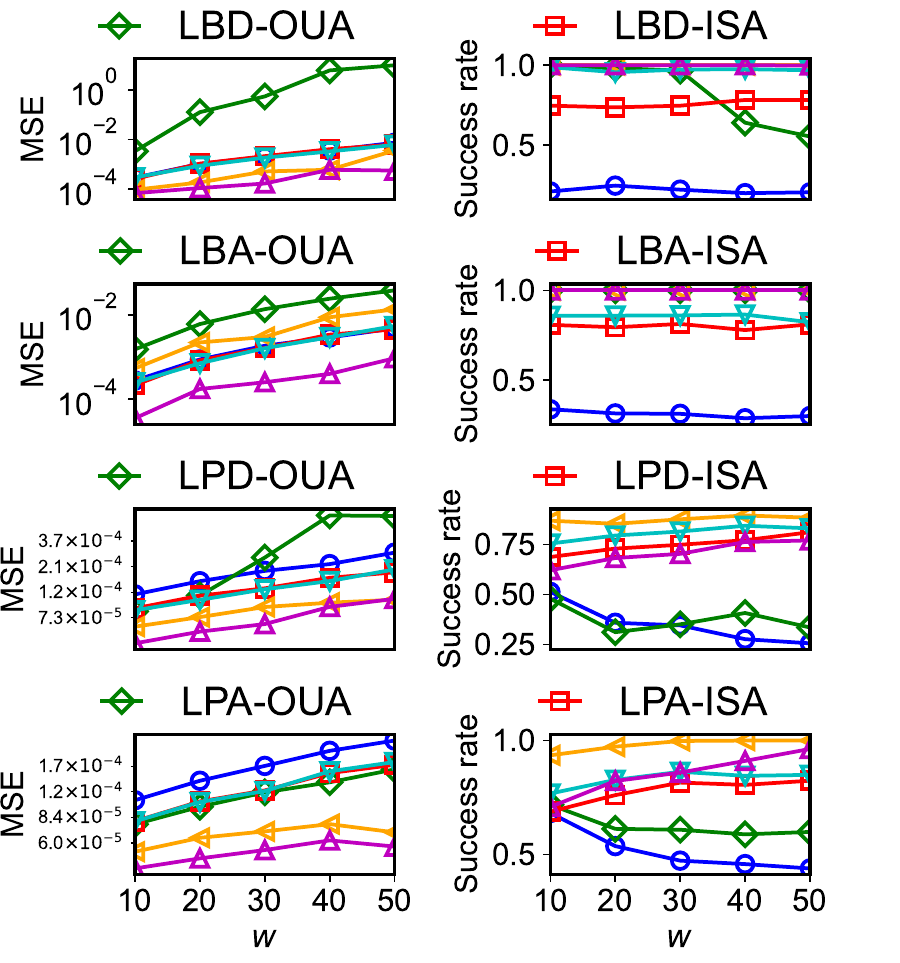}
         \label{fig:varyinig w}  
        }\vspace{-0.15cm}\hspace{-4mm}
	\subfigure[Varying $n^e$.]{
		\includegraphics[width=0.24\textwidth]{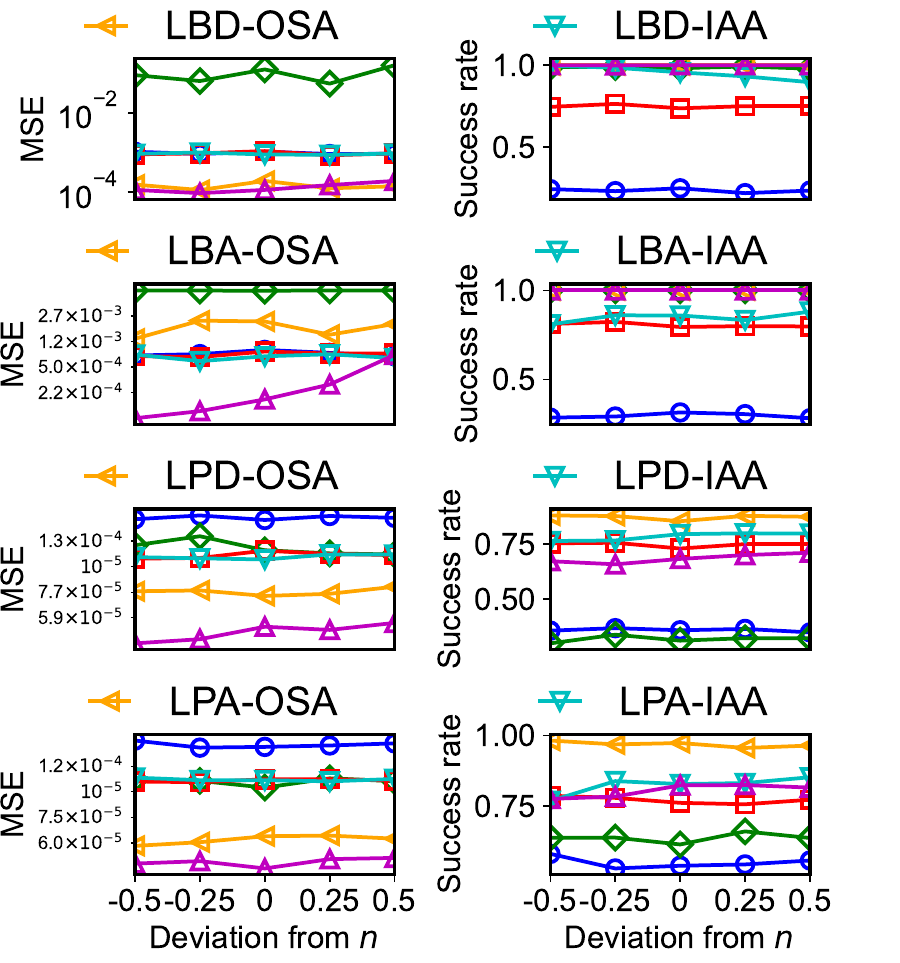}
  \label{fig:varyinig n^e}  
    }\vspace{-0.15cm}\hspace{-4mm}
  \subfigure[\centering Varying $\mathbf{f}^e$.]{
		\includegraphics[width=0.24\textwidth]{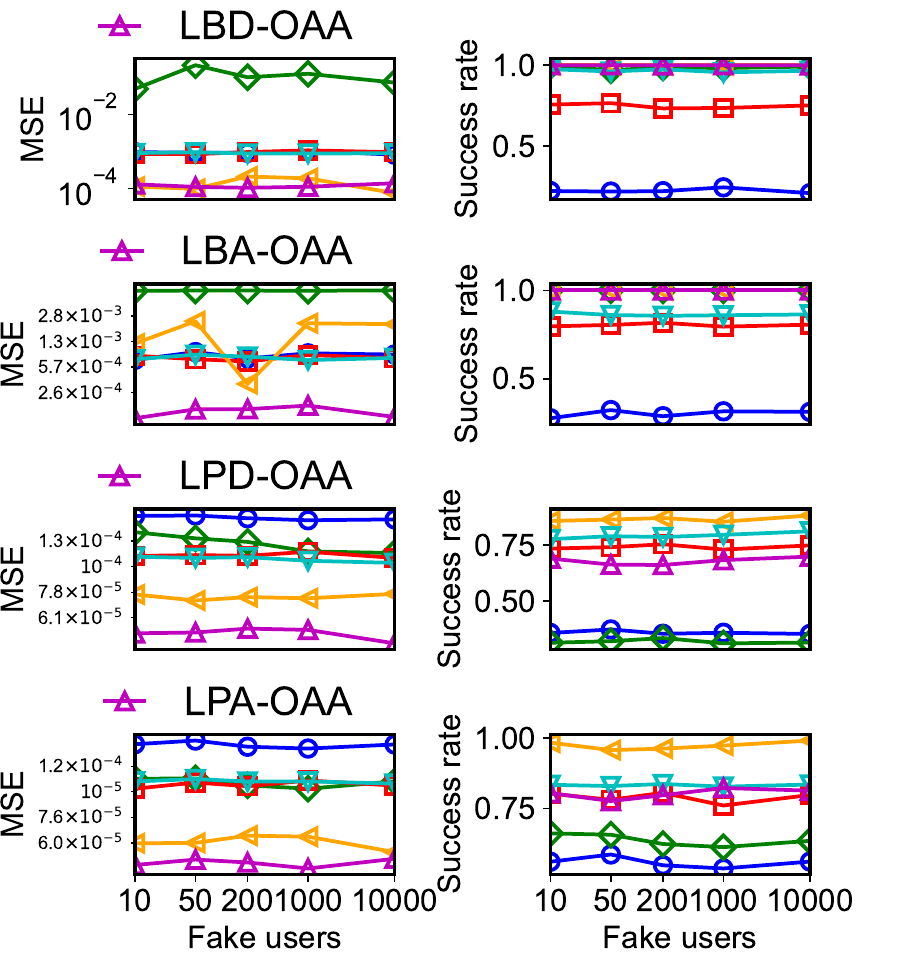}
  \label{fig:varyinig f^e} }\vspace{-0.1cm}
  \caption{\small Attack effectiveness w.r.t. different parameters (\textsf{Taobao} with Sigmoid target). 
  \label{fig:varying parameters}}\centering
  \vspace{-3mm}
\end{figure*}


\begin{figure}[htbp]
	\centering	
	\includegraphics[width=0.24\textwidth]{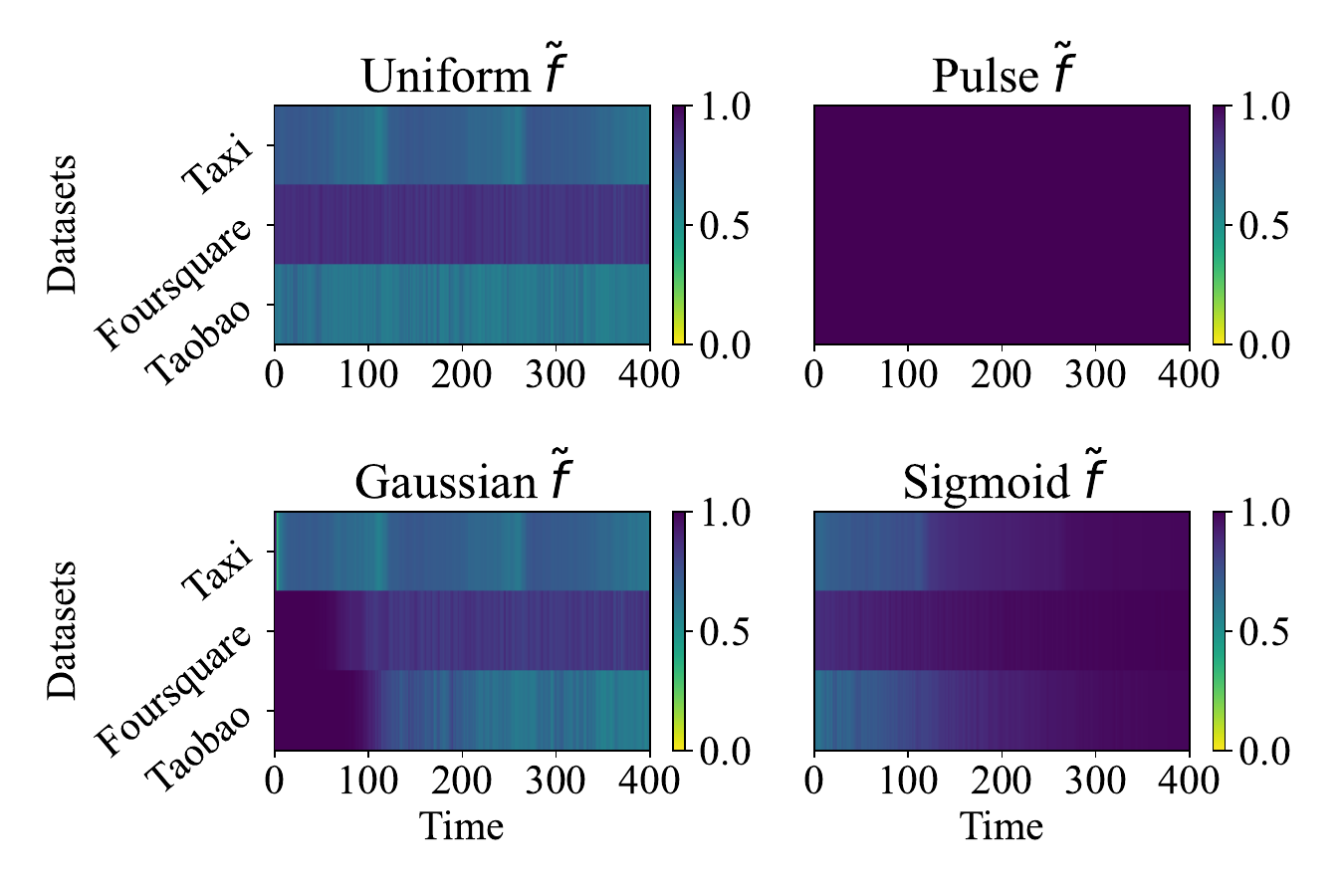}\hspace{-0.4cm}\hfill\vspace{-1mm}
 \includegraphics[width=0.24\textwidth]{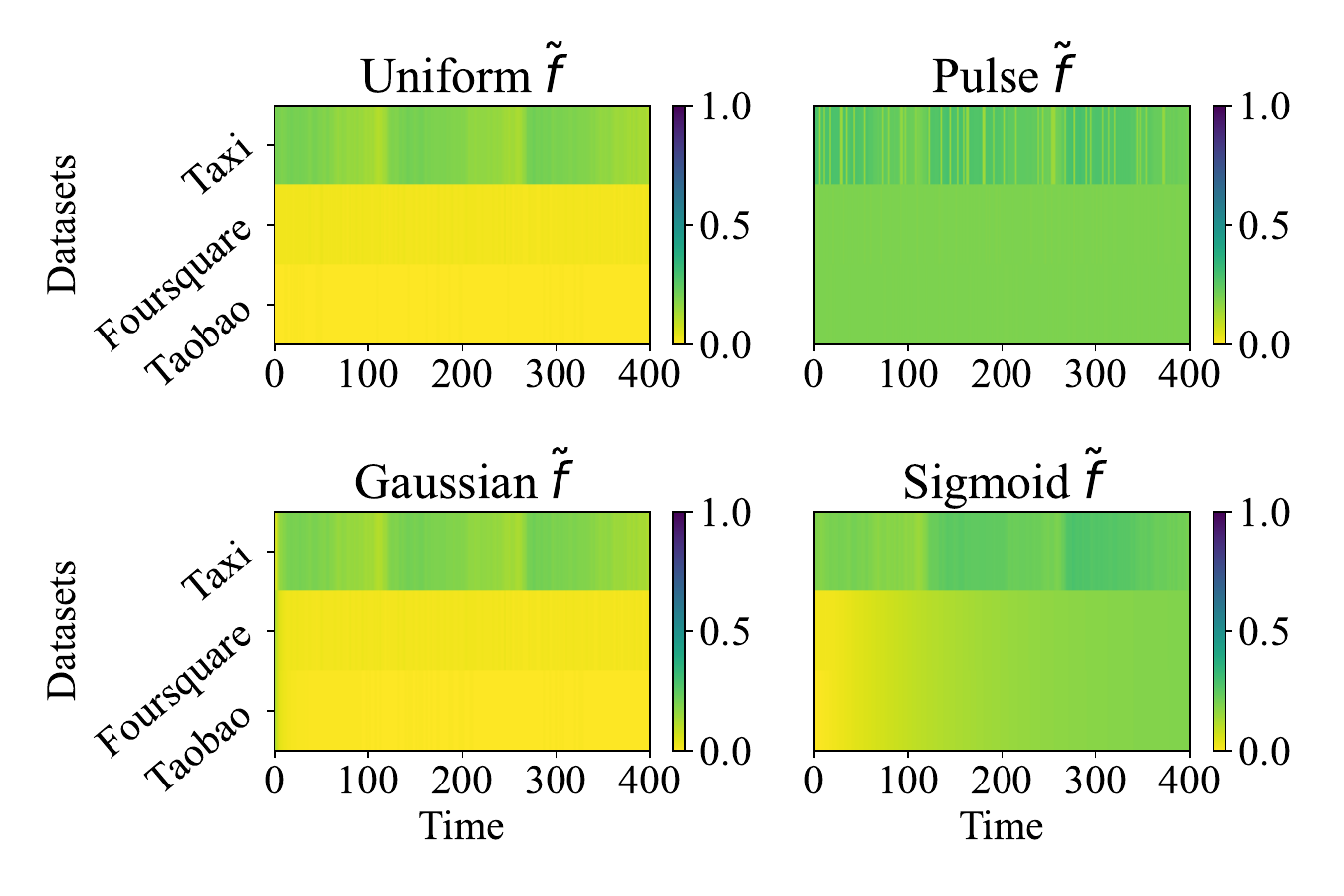}\vspace{-1mm}
 \text{\footnotesize (a) IPMA \quad\quad\quad\quad\quad\quad\quad\quad\quad\quad\quad\quad\quad (b) OPMA}\vspace{-3mm}
	\caption{The minimum $\beta$ needed for IPMA/OPMA with varying targets $\mathbf{\tilde{f}}$ on different datasets (Ada FO with $\epsilon=0.5$).}
	\label{fig:condition} 
\end{figure}

\subsubsection{Compared Algorithms.} Due to no existing study, we compared our proposed Adaptive Attack with the baselines by applying them to the following algorithms in the stream setting.

\noindent $\bullet$\textbf{$w$-event Level:} LDP-IDS~\cite{LDP_IDS}, $\text{DSAT}_w$~\cite{10.1145/2806416.2806441} and RescueDP~\cite{Wang2016RescueDPRS};

\noindent $\bullet$\textbf{Event Level:} PeGaSus~\cite{Chen2017PeGaSusDD} 
 and ToPL~\cite{wang2020continuous};
 
\noindent $\bullet$\textbf{User Level:} FAST~\cite{Fan2014AnAA} and CGM~\cite{baocgm}.

We assume the same type of attack for DMAs and PMAs (e.g., IDMA with IPMA). We abbreviated attacks (e.g., IUA for Input Uniform Attack) and denoted attacking format as ``X-Y'' (e.g., LBD-IUA means attacking LBD with IUA).



\subsection{Experimental Results}
We present our findings on the sufficient conditions for effective attacks and how different factors impact the attack's performance. 

\subsubsection{Overall Results.}\label{sec: Overall effect} 
Fig.~\ref{fig:varying beta} shows the attack effectiveness of all proposed attacks with different targets while varying the fake user ratio $\beta$. As shown, for target streams with minor fluctuations like Uniform (Fig.~\ref{fig:varying beta}(a)), Adaptive and Sampling Attacks outperform Uniform Attacks, which is consistent with the analysis in Sec.~\ref{sec:analysis}. For those with significant fluctuations like Pulse (Fig.~\ref{fig:varying beta}(b)), Adaptive and Uniform Attacks outperform Sampling Attacks. 
Generally, we can observe that, Output Attacks (OUA, OSA and OAA) outperform Input Attacks (IUA, ISA and IAA). Adaptive Attacks (IAA and OAA) surpass Uniform Attacks (IUA and OUA) and Sampling Attacks (ISA and OSA). Among all, OAA achieves the best performance. A larger $\beta$ (i.e., more fake users) generally enhances the attack effectiveness and DMA's success rate. However, for OSA with the Pulse target, the effectiveness reduces when $\beta$ is larger since the improvement of DMA's success rate causes LDP-IDS to choose approximation more, which is contrary to the way to achieve Pulse target, i.e., requiring more publication.
Appendix~\ref{sec: Experiment Results on Real-World Datasets} shows more results on different datasets.

\subsubsection{Sufficient Conditions for Attacks.}\label{sec: Conditions for sufficient fake users} We examined the minimum number of fake users for achieving targets in IPMA and OPMA. Given $\mathbf{\tilde{f}}$, $\epsilon$, and the dataset, the minimum number of fake users could be derived by Eqs.~(\ref{condition_IPMA}) and (\ref{condition_OPMA}). Then, the minimum $\beta$ for attacks can be obtained. Figs.~\ref{fig:condition}(a) and~\ref{fig:condition}(b) show the minimum $\beta$ required for successful IPMA and OPMA attacks respectively, during the first 400 timestamps for each dataset. We can observe that OPMA requires less faker users when achieving the same attack effect.



\subsubsection{Impact of Other Factors\\}


\textbf{Impact of $\epsilon$.} \label{subsubsection:Impact of epsilon}
Fig.~\ref{fig:varyinig epsilon} shows the attack effectiveness with different privacy budget $\epsilon$. Attacks improve when $\epsilon$ is larger. Success rates vary with $\epsilon$. Large $\epsilon$ typically boosts the success rate of Uniform Attacks since a larger $\epsilon$ reduces the noise injected into dissimilarity calculation, which allows DMAs to succeed with a higher probability. The success rate for Sampling Attacks either drops or remains stable. This is because Sampling Attacks always need to minimize $\overline{dis}$ via DMAs to make it smaller than $err$ but a larger $\epsilon$ also makes $err$ smaller, leading worse successful rate for DMAs.
Appendix~\ref{sec: Experiment Results on Real-World Datasets} gives more results on different datasets and different targets.

\noindent\textbf{Impact of $w$.} Fig.~\ref{fig:varyinig w} shows the attack effectiveness on varying sliding window size $w$. Overall, the attack effectiveness improves with smaller $w$. Also, Output Adaptive Attack always achieves better performance. Different attacks show different relations between their success rates and $w$. For Uniform Attacks, it decreases with $w$ as large $w$ increases the result variance in DMAs. For Sampling Attacks, the success rate is basically unchanged since the budget or population allocated at the sampling timestamp is independent of $w$.
Appendix~\ref{sec: Experiment Results on Real-World Datasets} shows more results.

\noindent\textbf{Impact of Attacker’s knowledge.} Figs.~\ref{fig:varyinig n^e} and~\ref{fig:varyinig f^e} demonstrate the effects of varying $n^e$ and $\mathbf{f}^e$ on the attacks. We varied $n^e$ by $\pm50\%$ and $\pm25\%$ from $n$ to assess the impact. For $\mathbf{f}^e$, we assume the attacker can access different numbers of fake users for $\mathbf{f}^e$ estimation. Generally, more accurate estimation of $n^e$ and $\mathbf{f}^e$ lead to a smaller manipulation gap and higher success rate. Appendix~\ref{sec: Experiment Results on Real-World Datasets} gives more results.

\begin{figure}[htbp]
	\centering 
	\includegraphics[width=240pt]    {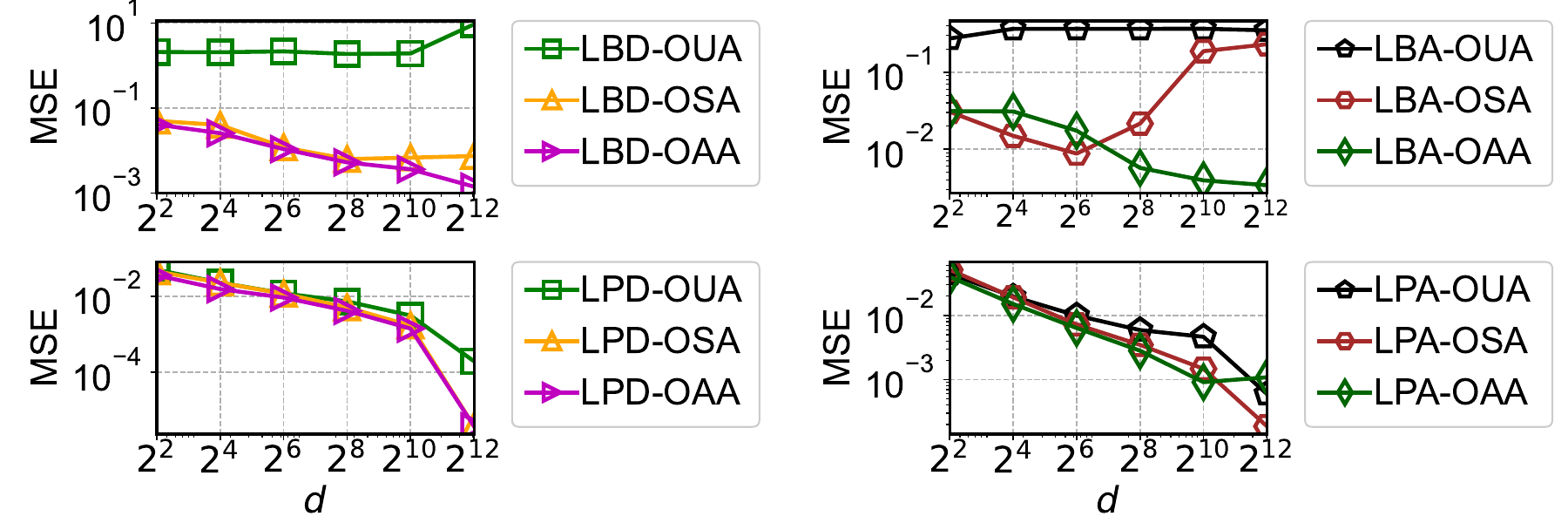}
	\vspace{-6mm}
	\caption{\small Effectiveness w.r.t $d$ (\textsf{Taobao} with Gaussian target).}\centering
	\label{fig:attack different FO}
\end{figure}

\noindent\textbf{Impact of dimension $d$.}\label{sec: Attack different FO} 
Fig.~\ref{fig:attack different FO} illustrates the attack effectiveness on Ada, with varying domain size $d$. We sorted Taobao categories and varied $d$ by grouping item IDs into different numbers of categories (from $2^2$ to $2^{12}$). Results indicate that the attack enhances as the domain size increases, highlighting the effectiveness of our attacks in large domains. Results on different FOs (kRR and OUE), are in Appendix~\ref{appendix:different FOs}.

\noindent\textbf{Impact of Attack Mismatch.} Table~\ref{tab:mismatch} illustrates the mismatch scenarios where the attacker launches an unmatched attacking algorithm against an LDP protocol, e.g., using an LBA-based adaptive attack on an aggregator operating with LBD. Mismatched attacks mainly result in the wrong calculation of $G_{I,t}$ or $G_{O,t}$, leading the aggregator to choose a strategy less beneficial for attacks. Each entry $(X,Y)$ in Table~\ref{tab:mismatch} means using OAA designed for LDP protocol in column $Y$ to attack row $X$. As shown, despite the matched attack gains the smallest gap, the mismatch often has a slight impact if they adopt the same budget or population division framework. Otherwise, mismatched attacks with different division frameworks often lead to much poorer attack performances since the calculation of $G_{I,t}$ or $G_{O,t}$ is quite different in the two division frameworks.  


\begin{table}[htbp]
\centering
\caption{\label{tab:mismatch} Impact of Mismatched Attacks (Gaussian target).}\vspace{-3.5mm}
\scalebox{0.55}{

\begin{tabular}{|cc|c|c|c|c|}
\hline
\multicolumn{2}{|c|}{\multirow{2}{*}{\makecell[c]{MSE}}}& \multicolumn{4}{c|}{\makecell[c]{Attack Methods}} \\ \cline{3-6}
&&\makecell[c]{OAA-BD} & \makecell[c]{OAA-BA} &\makecell[c]{OAA-PD}&\makecell[c]{OAA-PA}\\ \hline

\multirow{3}{*}{\makecell[c]{\raisebox{-18pt}[0pt][0pt]{\rotatebox{90}{LDP Alg.}}}}&\multicolumn{1}{|c|}{LBD}&\textbf{0.0178}&0.0213	&7.9034	&	7.4535		\\ \cline{2-6}											
&\multicolumn{1}{|c|}{LBA}&0.4736&	\textbf{0.0152}&3.537	&		3.1310		\\ \cline{2-6}												
&\multicolumn{1}{|c|}{LPD}&0.0021&	0.00166&	\textbf{0.0004}&0.0012		\\ \cline{2-6}		

&\multicolumn{1}{|c|}{LPA}&0.0083&0.0038&	0.0011&	\textbf{0.0003}		\\ \hline						
\end{tabular}

\begin{tabular}{|cc|c|c|c|c|}
\hline
\multicolumn{2}{|c|}{\multirow{2}{*}{\makecell[c]{MSE}}}& \multicolumn{4}{c|}{\makecell[c]{Attack Methods}} \\ \cline{3-6}
&&\makecell[c]{OAA-BD} & \makecell[c]{OAA-BA} &\makecell[c]{OAA-PD}&\makecell[c]{OAA-PA}\\ \hline	
\multirow{3}{*}{\makecell[c]{\raisebox{-18pt}[0pt][0pt]{\rotatebox{90}{LDP Alg.}}}}&\multicolumn{1}{|c|}{LBD}&\textbf{0.0011}&0.0013	&2.3891	&2.9994		\\ \cline{2-6}								
&\multicolumn{1}{|c|}{LBA}&0.0029&	\textbf{0.0009}&1.1676	&	1.2000	\\ \cline{2-6}							
&\multicolumn{1}{|c|}{LPD}&0.0004&	0.0004&	\textbf{0.0002}&	0.0004	\\ \cline{2-6}							
&\multicolumn{1}{|c|}{LPA}&0.0017	&0.0017	&0.0004&	\textbf{0.0002}	 \\ \hline							

\end{tabular}
}
\text{\footnotesize (a) \textsf{Foursquare.} \quad\quad\quad\quad\quad\quad\quad\quad\quad\quad\quad (b) \textsf{Taobao.}}
\end{table}

\begin{figure}[htbp]
	\centering	
		\includegraphics[width=240pt,height=80pt] {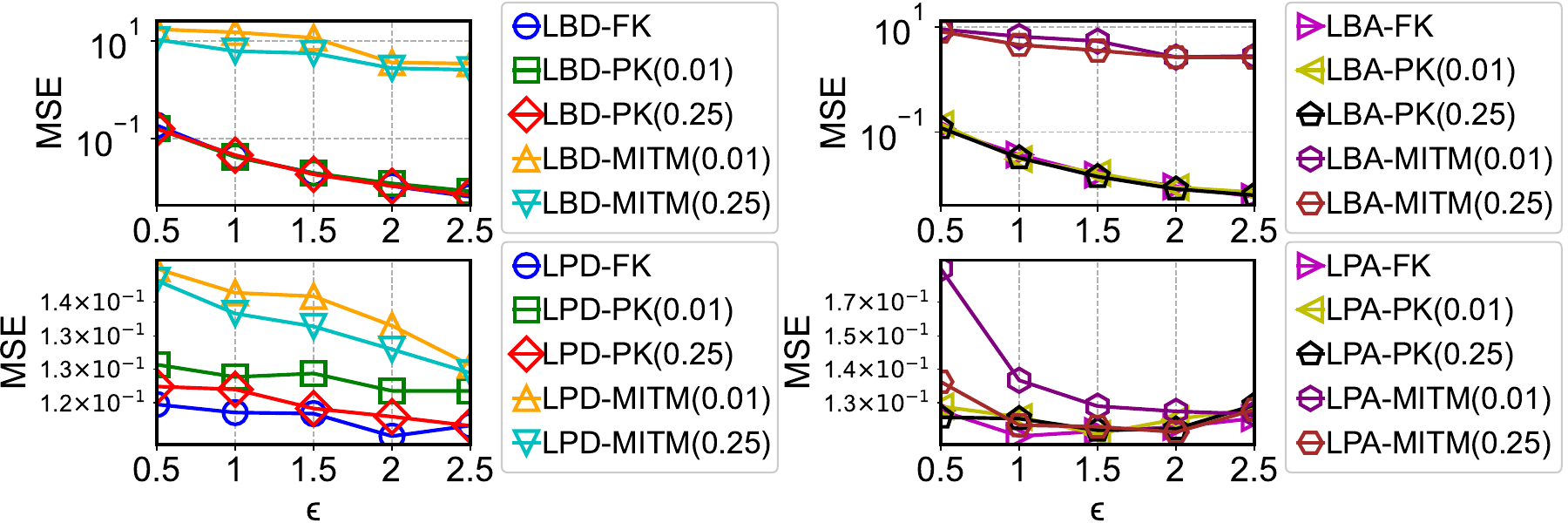}\vspace{-2.5mm}
	\caption{\small Attack effectiveness using OAA with different levels of knowledge, varying $\epsilon$ (\textsf{Taxi} with Sigmoid target).}\centering
 \label{fig:different level knowledge}
\end{figure}

\noindent\textbf{Different Levels of Knowledge.}\label{sec: Compare different levels of knowledge}
Fig.~\ref{fig:different level knowledge} compares the performance of the full-knowledge (FK), partial-knowledge (PK) and the man-in-the-middle attack (MITM). In these scenarios, the PK attackers obtain $1\%$ or $25\%$ of the knowledge about frequencies, denoted by PK(0.01) and PK(0.25). MITM attackers intercept $1\%$ or $25\%$ of the communication between users and aggregator and recover the frequencies with FO, denoted by MITM(0.01) and MITM(0.25). Results show that FK attackers perform better than PK attackers. The performance of PK attackers with $25\%$ information is almost as good as that of FK attackers. MITM attackers perform the worst because recovering frequencies with FO often introduce more noise. More results are in Appendix~\ref{appendix:different knowledge}.

\subsubsection{Attacking Baseline Methods.}\label{subsubsec:attack baseline methods} 
\begin{figure}[htbp]
	\centering	
        \includegraphics[width=0.235\textwidth] {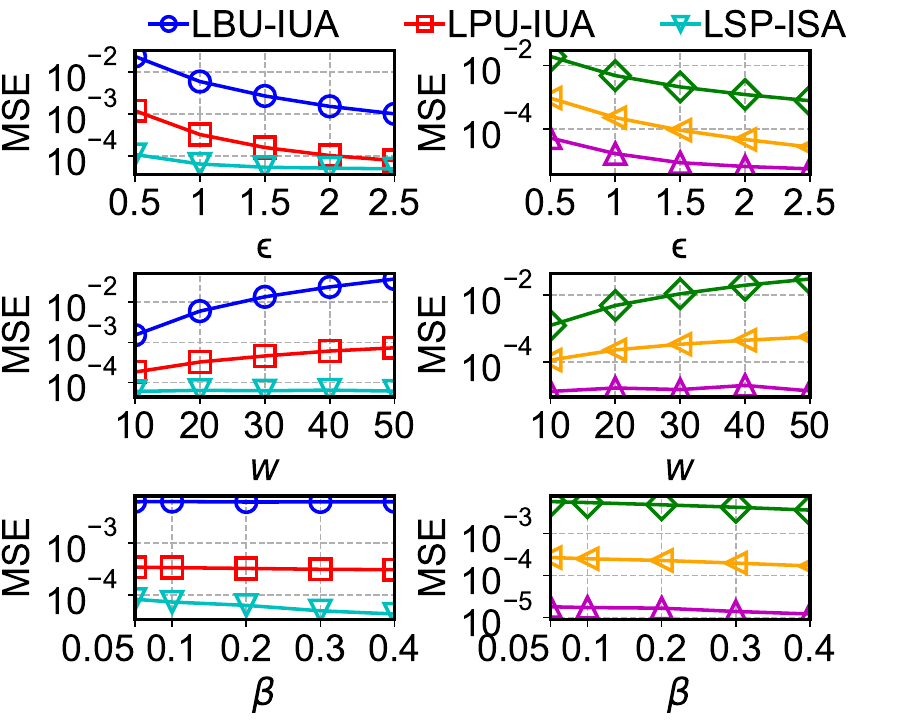}\vspace{-1mm}\hspace{-2mm}
		\includegraphics[width=0.235\textwidth] {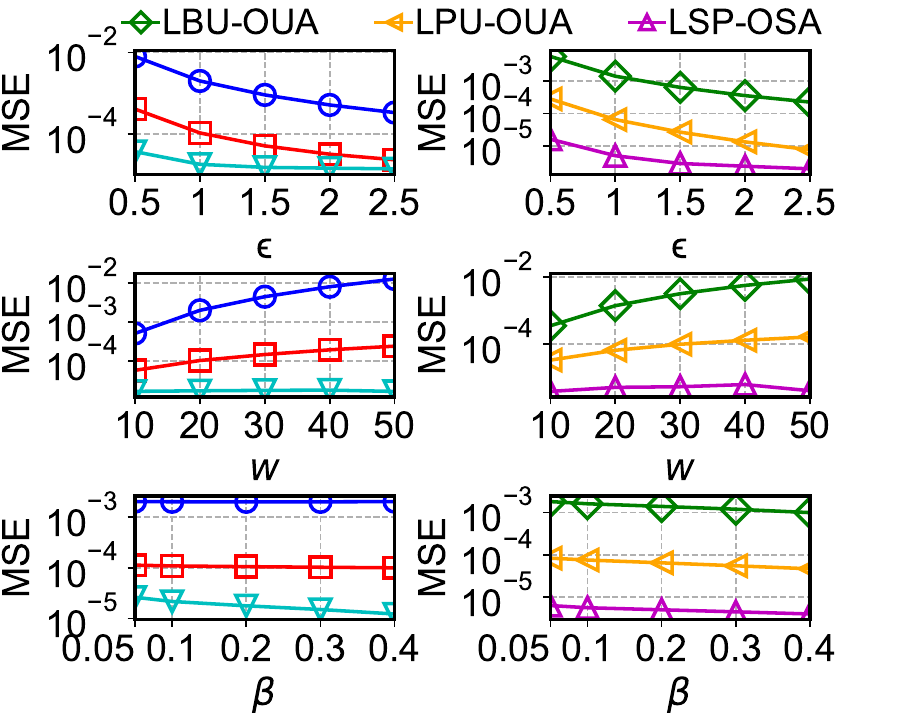}
  \vspace{-1mm}
  \text{\footnotesize (a) Uniform $\mathbf{\tilde{f}}$ \quad\quad\quad\quad\quad\quad\quad\quad (b) Gaussian $\mathbf{\tilde{f}}$} \vspace{-1mm}
	\caption{\small Attack effectiveness on baselines, varying $\epsilon$, $w$ and $\beta$ (\textsf{Foursquare}).}\centering
 \label{fig:attack baseline}
\end{figure}

Fig.~\ref{fig:attack baseline} shows the performance of our attacks against baselines, LBU, LPU and LSP. For LBU and LPU, the Uniform Attack was used. LSP was targeted with the Sampling Attack. The results are much similar to those of attacking adaptive methods, with a larger $\epsilon$ and $\beta$ and a smaller $w$ meaning a more successful attack. More results can be found in Appendix~\ref{appendix:attack baseline}.


%


\subsubsection{Attacking Mean Estimation. }
\begin{figure}[htbp]
	\centering	
	\includegraphics[width=240pt]     {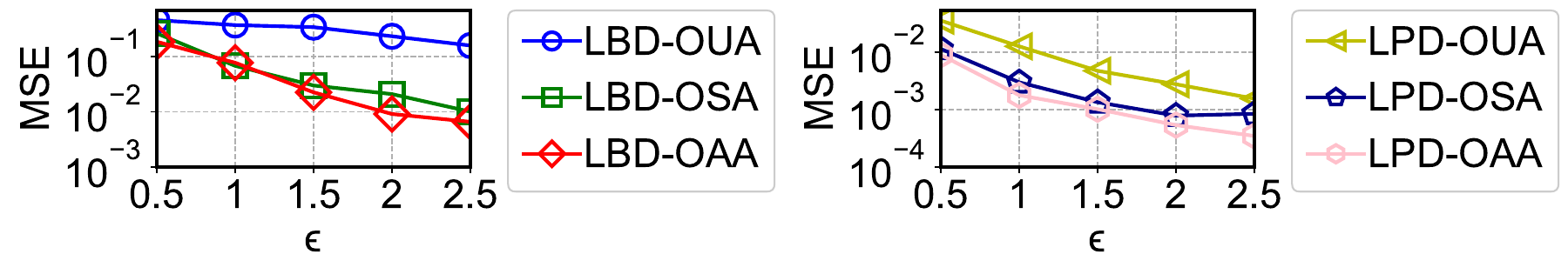}\vspace{-0.5mm}
 \includegraphics[width=240pt]     {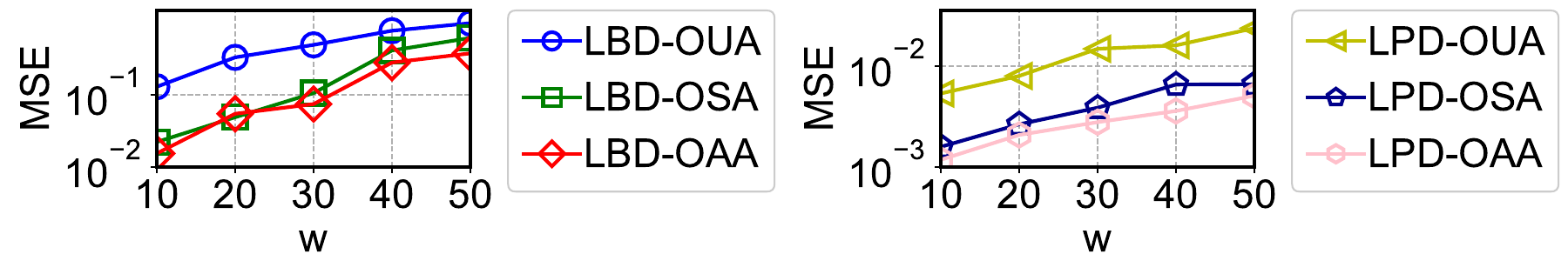}\vspace{-0.5mm}
 \includegraphics[width=240pt]     {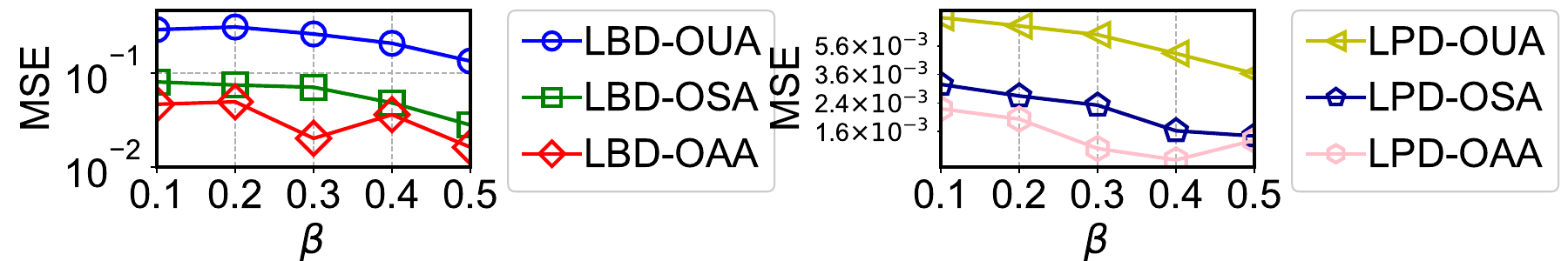}\vspace{-0.5mm}
	\caption{\small Attack effectiveness on HM-based LDP-IDS, varying $\epsilon$, $w$ and $\beta$ (\textsf{Taobao-Price} with Uniform target).}\centering
 \label{fig:attack HM}
\end{figure}

We attacked HM-based mean estimation in the streaming setting. In the experiments, we used \textsf{Taobao-Price} and the Uniform target, and set the parameters as $\epsilon=1$, $w=20$, and $\beta=0.1$. 
Results are shown in Fig.~\ref{fig:attack HM}, demonstrating attack effects across different $\epsilon$, $w$, and $\beta$ values. Still, the Adaptive Attack generally performs the best. Larger $\epsilon$ improves the attack which is align with the analysis in~\cite{Li2023Finegrained}, whereas increasing $w$ reduces effectiveness by decreasing the privacy budget and user population at each timestamp. 
A larger $\beta$ allows more fake user manipulation, enhancing attack effectiveness. More results are shown in Appendix~\ref{appendix:HM}.

\subsubsection{Attacking Other LDP Algorithms. }
\begin{figure}[htbp]
	\centering	
	\includegraphics[width=0.48\textwidth]     {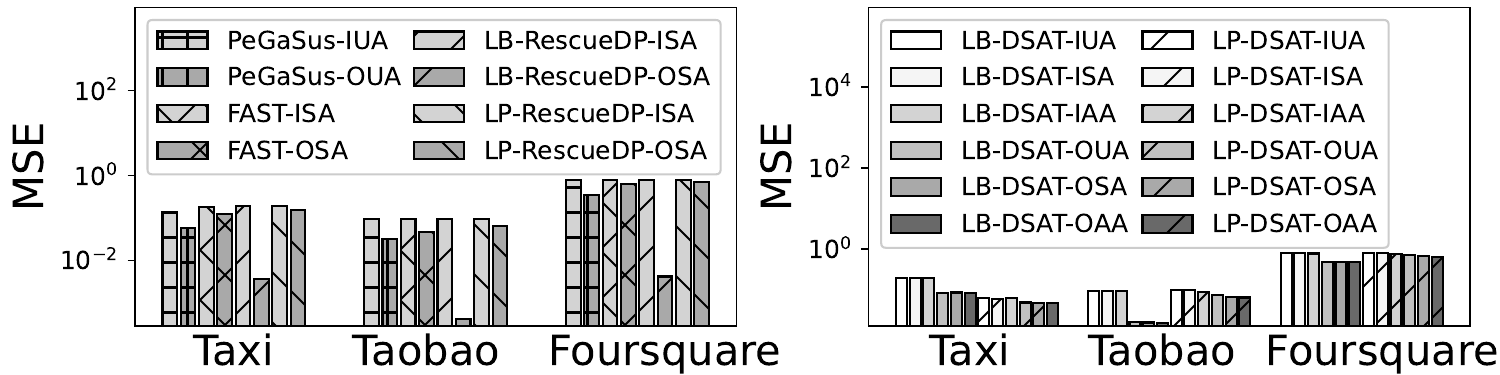} \vspace{-5mm}
	\caption{\small Effectiveness on other stream algs. (Gaussian target).}\centering
 \label{fig:attack other stream algorithms}
\end{figure}

As discussed in Sec.~\ref{sec:Attacking Other LDP Tasks and Protocols}, our attacking methods apply to other streaming LDPs adapted in \cite{LDP_IDS}, where they are categorized into budget-division (LB-*) and population-division (LP-*). Only Uniform Attack work on PeGaSus, and only Sampling Attack on FAST and RescueDP. However, $\text{DSAT}_w$ is susceptible to all our adaptive proposed attacks. Fig.~\ref{fig:attack other stream algorithms} shows the results. For comparison, we binarized the \textsf{Taxi}, \textsf{Taobao}, and \textsf{Foursquare} datasets and set $\beta$ at 0.1. Results show output attacks generally outperform input ones, with Adaptive Attacks notably more effective on $\text{DSAT}_w$.

\begin{figure}[htbp]
	\centering	
	\includegraphics[width=0.48\textwidth]     {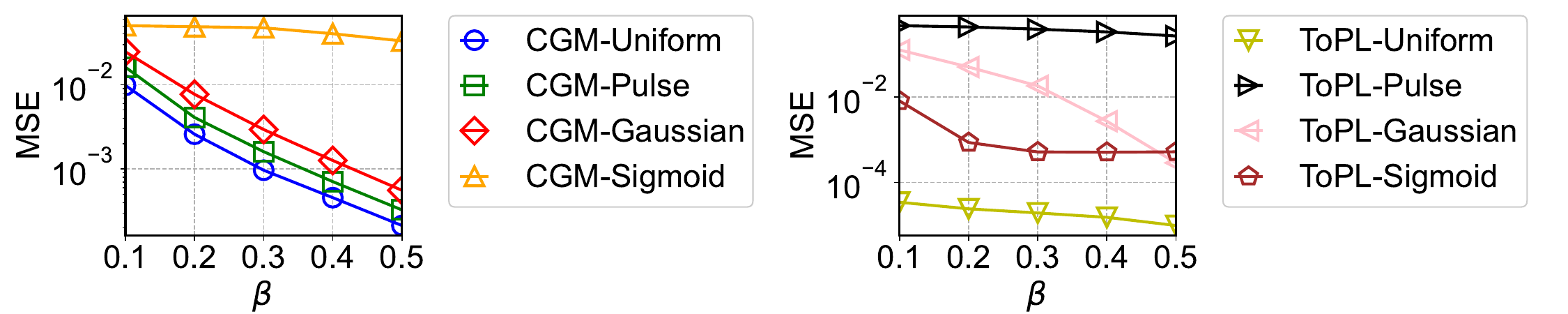}\vspace{-2mm}
	\caption{\small Attack effectiveness against stream LDPs over numerical domain (\textsf{Taxi-Longitude}).}\centering
	\label{fig:attack CGM and topl}
\end{figure}

For other stream LDPs over numerical domain, we normalized Taxi-Longitude and Taobao-Price datasets into $[-\frac{1}{2},\frac{1}{2}]$ for CGM and $[0,1]$ for ToPL. 
We set the default value $\epsilon=1$, $\beta=0.1$, and $\delta=10^{-5}$ (for CGM). Fig.~\ref{fig:attack CGM and topl} shows the attack effectiveness of Output Uniform Attack against CGM with user-level LDP and ToPL with event-level LDP respectively. The results demonstrate that more fake users are beneficial to the attack. More results can be found in Appendix~\ref{appendix:attack numerical}.

\begin{figure}[t]
\vspace{2mm}
	\centering	
 \includegraphics[width=0.48\textwidth]     {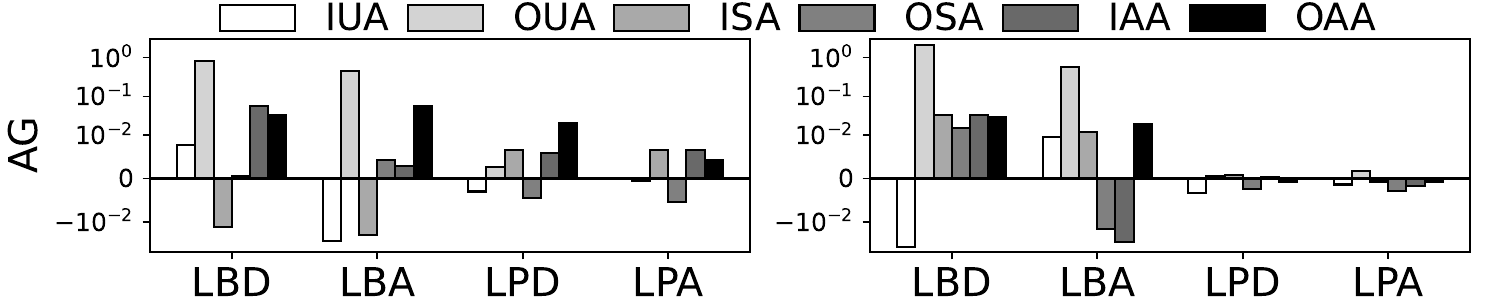}\vspace{0mm}
 \text{\footnotesize \quad \quad\space (a) Sigmoid target.\quad\quad\quad\quad\quad\quad\quad\quad (b) Uniform target.}\vspace{-1mm}
	\caption{\small Accuracy Gain (\textsf{Taxi}, $\epsilon=1$, $r$=0.5).}\centering
 \label{fig:defense} \vspace{-5mm}
\end{figure}

\begin{figure}[t]
	\centering	
 \includegraphics[width=0.48\textwidth]     {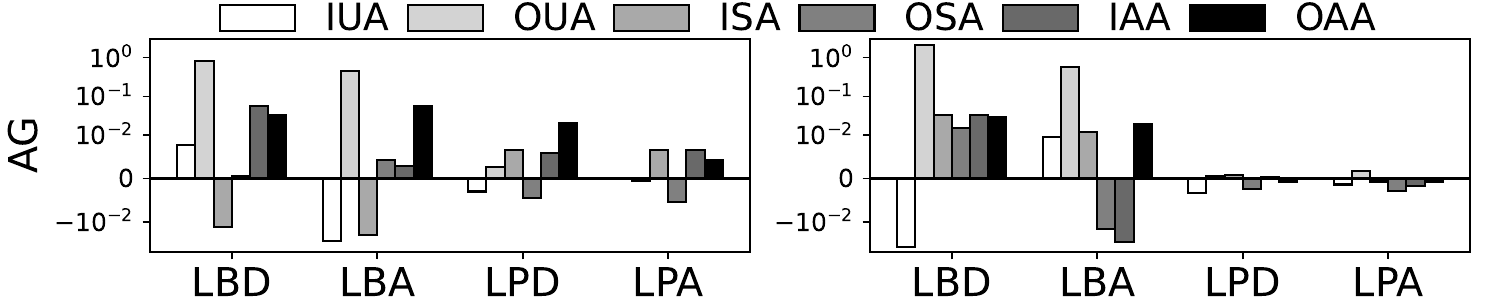}\vspace{0mm}
 \text{\footnotesize \quad \quad\space (a) Sigmoid target.\quad\quad\quad\quad\quad\quad\quad\quad\quad (b) Uniform target.}\vspace{-3mm}
	\caption{\small Accuracy Gain (\textsf{Taxi}, $\epsilon=1$, $r$=0.5).}\centering
 \label{fig:defense} 
\end{figure}

\section{Discussion on Possible Defense}\label{sec:defense}
Inspired by \cite{Cao2021Provably,Li2023Finegrained}, we propose a potential countermeasure.
We first propose our defense for FOs. The first step is sampling. Assuming a total of $m+n$ users, the aggregator samples subsets of received data, each with the size of $r(m+n)$ where $r$ is a fraction between 0 and 1. It then estimates item frequencies for each subset, obtaining different $\mathbf{\hat{f}}$. Most $\mathbf{\hat{f}}$, due to a small proportion of fake users, cluster near the true frequency distribution, but some may appear abnormal with higher fake user concentrations. We then employ Isolation Forest \cite{Liu2008IsolationForest}, an anomaly detection method, to score each $\mathbf{\hat{f}}$ for anomalies. Larger scores indicate less anomalous data, from which the aggregator determines the publication frequency. Based on the defense on FO above, we propose our defense on LDP-IDS. We notice that LDP-IDS invokes FO twice at each timestamp, producing two estimates for the same distribution. However, the proposed attack may break the similarities between two statistics, causing two estimates inconsistent. Thus, the attack can be detected by comparing two estimates. Here we use Kolmogorov-Smirnov test to detect it. Once the attack is detected and LDP-IDS chooses the publication strategy, the defense for FOs above will be invoked.

\textbf{Experiments.} Fig.~\ref{fig:defense} demonstrates the performance of our proposed defense mechanism on \textsf{Taxi}, setting $\epsilon=1$ by default. The defense performance is measured by the Accuracy Gain (AG) of the LDP protocols brought by the defense. Specifically, AG is defined as the reduction in estimation error (measured by MSE, i.e., $MSE_\text{before}(\mathbf{\hat{\textbf{f}}},\mathbf{f})-MSE_\text{after}(\mathbf{\hat{\textbf{f}}},\mathbf{f})$) between the released statistics $\mathbf{\hat{\textbf{f}}}$ from the genuine statistics $\mathbf{f}$, after applying the defense. A positive AG indicates successful mitigation. Results in Fig.~\ref{fig:defense} show our defense mechanism can generally mitigate the attack, more effective for Sigmoid target than Uniform target. It is because Sigmoid target manipulates the dataset from the original more than Uniform target, resulting in more anomalies detected and leaving more space for mitigation. More results are in Appendix~\ref{appendix:defense}.


\section{Related Work}\label{sec:related work}

\textbf{LDP over data streams.} 
Local differential privacy (LDP), a decentralized variant of Differential privacy (DP)~\cite{Dwork2014TheAF,Dwork-405}, has emerged as a popular privacy-preserving paradigm for large-scale data collection and analysis without relying a trusted aggregator.
It has been applied in various analytic tasks like frequency estimation~\cite{arcolezi2021random,Erlingsson2014RAPPORRA,kairouz2016extremal,Wang2017LocallyDPFE,arcolezi2022frequency,arcolezi2022improving,ohrimenko2022randomize}, mean/variance estimation~\cite{Wang2019CollectingAA}, key-value data collection~\cite{ye2019privkv}, frequent itemset mining~\cite{Qin2016HeavyHE,li2022frequent,wang2018privset,wang2018locally}, graph data mining~\cite{Imola2021LDPgraph,Ye2020LDPgraph}, as well as streaming data analysis~\cite{LDP_IDS,xue2022ddrm,li2023locally,li2023privsketch,gao2023privacy,ye2021beyond,wang2020continuous,baocgm}. 

This paper focuses on the LDP analysis over data streams. According to the granularity of protection, related studies can be categorized into event-level DP~\cite{Chen2017PeGaSusDD}, user-level DP~\cite{10.1145/1806689.1806787,Fan2014AnAA}, and $w$-event level DP~\cite{Kellaris2014DifferentiallyPE,schaler2023benchmarking}. Among them, $w$-event level strikes a balance between and can be easily extended to the other two, thus gaining much attention~\cite{schaler2023benchmarking}.  
Wang \emph{et al.} \cite{wang2020continuous} proposed ToPL composed of a threshold optimizer and perturber for outputting streaming data at event-level LDP. It actually publishes at each timestamp and then performs post-processing. 
Bao \emph{et al.} \cite{baocgm} presented CGM that adds correlated noise to ensure approximate user-level LDP. CGM can also be seen as publishing at each timestamp with different budget.
Ren \emph{et al.}~\cite{LDP_IDS} proposed LDP-IDS focusing on $w$-event LDP, which includes two extensible budget-division and population-division frameworks. Based on the frameworks, most existing CDP algorithms on streams can be extended into the LDP setting, and roughly summarized into three phases discussed in Sec.~\ref{subsec:LDP for Infinite Data Streams}.





\textbf{Poisoning attacks to LDP protocols.} Research shows the LDP aggregator's vulnerability to perturbed data, enabling data poisoning attacks. Cheu \emph{et al.}~\cite{cheu2021manipulation} explored untargeted attacks manipulating the $L_p$ norm distance in estimated frequencies pre- and post-attack. Cao \emph{et al.}~\cite{Cao2021poisonFE} and Wu \emph{et al.}~\cite{Wu2022poisonKey} developed targeted attacks on LDP for frequency estimation, heavy-hitter identification, and key-value data, respectively. Li \emph{et al.}~\cite{Li2023Finegrained} examined fine-grained attacks on mean and variance estimation. Zheng \emph{et al.}~\cite{zheng2024data} disrupted LDP-based crowdsensing with bi-level optimization. Tong \emph{et al.}~\cite{tong2024data} presented poisoning attacks against LDP frequent itemset mining protocols. However, all these focus on static settings and cannot directly work for fine-grained manipulation in streaming scenarios.


\section{Conclusion}\label{conclusion}
We conduct a comprehensive study on fine-grained data poisoning attacks to LDP protocols for data streams. By summarizing existing streaming LDP algorithms into three phases with poisoning surfaces, we introduce a unified attack framework with three coordinated attack modules, each with theory-driven designs. Applying the attack framework, we integrate these modules and propose detailed adaptive attacking methods to the state-of-the-art online adaptive LDP algorithms. Furthermore, we explore their potential against other algorithms under different LDP models and analytic tasks. 
Besides theoretical analysis, extensive experiments demonstrate the effectiveness, sufficient conditions, and parameter impacts of the proposed attacks. Finally, we discuss possible countermeasures with experiment validations. Our work highlights the security of streaming LDP protocols and appeals to further research. 


\bibliographystyle{IEEEtran}

\begin{thebibliography}{10}
\providecommand{\url}[1]{#1}
\csname url@samestyle\endcsname
\providecommand{\newblock}{\relax}
\providecommand{\bibinfo}[2]{#2}
\providecommand{\BIBentrySTDinterwordspacing}{\spaceskip=0pt\relax}
\providecommand{\BIBentryALTinterwordstretchfactor}{4}
\providecommand{\BIBentryALTinterwordspacing}{\spaceskip=\fontdimen2\font plus
\BIBentryALTinterwordstretchfactor\fontdimen3\font minus \fontdimen4\font\relax}
\providecommand{\BIBforeignlanguage}[2]{{%
\expandafter\ifx\csname l@#1\endcsname\relax
\typeout{** WARNING: IEEEtran.bst: No hyphenation pattern has been}%
\typeout{** loaded for the language `#1'. Using the pattern for}%
\typeout{** the default language instead.}%
\else
\language=\csname l@#1\endcsname
\fi
#2}}
\providecommand{\BIBdecl}{\relax}
\BIBdecl

\bibitem{duchi2013local}
J.~Duchi, M.~Jordan, and M.~Wainwright, ``Local privacy and statistical minimax rates,'' in \emph{Proc. IEEE FOCS}, 2013, pp. 429--438.

\bibitem{duchi2014privacy}
------, ``Privacy aware learning,'' \emph{Journal of the ACM (JACM)}, vol.~61, no.~6, pp. 1--57, 2014.

\bibitem{kasiviswanathan2011can}
S.~Kasiviswanathan, H.~Lee, N, S.~Raskhodnikova, and A.~Smith, ``What can we learn privately?'' \emph{SIAM J. Computing}, vol.~40, no.~3, pp. 793--826, 2011.

\bibitem{Erlingsson2014RAPPORRA}
{\'U}.~Erlingsson, A.~Korolova, and V.~Pihur, ``Rappor: Randomized aggregatable privacy-preserving ordinal response,'' in \emph{Proc. ACM CCS}, 2014, pp. 1054--1067.

\bibitem{ding2017collecting}
B.~Ding, J.~Kulkarni, and S.~Yekhanin, ``Collecting telemetry data privately,'' in \emph{Proc. NeurIPS}, 2017, pp. 3574--3583.

\bibitem{AppleDP}
\BIBentryALTinterwordspacing
differential privacy team~at Apple, ``Learning with privacy at scale,'' 2017. [Online]. Available: \url{https://machinelearning.apple.com/research/learning-with-privacy-at-scale}
\BIBentrySTDinterwordspacing

\bibitem{arcolezi2021random}
H.~Arcolezi, J.~Couchot, B.~Al~Bouna, and X.~Xiao, ``Random sampling plus fake data: Multidimensional frequency estimates with local differential privacy,'' in \emph{Proc. of ACM CIKM}, 2021, pp. 47--57.

\bibitem{kairouz2016extremal}
P.~Kairouz, S.~Oh, and P.~Viswanath, ``Extremal mechanisms for local differential privacy,'' \emph{J. Mach. Learn. Res.}, vol.~17, no.~1, pp. 492--542, 2016.

\bibitem{Wang2017LocallyDPFE}
T.~Wang, J.~Blocki, N.~Li, and S.~Jha, ``Locally differentially private protocols for frequency estimation,'' in \emph{Proc. USENIX Security}, 2017, pp. 729--745.

\bibitem{arcolezi2022frequency}
H.~H. Arcolezi, C.~Pinz{\'o}n, C.~Palamidessi, and S.~Gambs, ``Frequency estimation of evolving data under local differential privacy,'' \emph{arXiv preprint arXiv:2210.00262}, 2022.

\bibitem{arcolezi2022improving}
H.~H. Arcolezi, J.-F. Couchot, B.~Bouna, and X.~Xiao, ``Improving the utility of locally differentially private protocols for longitudinal and multidimensional frequency estimates. digital communications and networks (2022),'' 2022.

\bibitem{ohrimenko2022randomize}
O.~Ohrimenko, A.~Wirth, and H.~Wu, ``Randomize the future: Asymptotically optimal locally private frequency estimation protocol for longitudinal data,'' in \emph{Proceedings of the 41st ACM SIGMOD-SIGACT-SIGAI Symposium on Principles of Database Systems}, 2022, pp. 237--249.

\bibitem{Wang2019CollectingAA}
N.~Wang, X.~Xiao, Y.~Yang, J.~Zhao, S.~Hui, H.~Shin, J.~Shin, and G.~Yu, ``Collecting and analyzing multidimensional data with local differential privacy,'' in \emph{Proc. IEEE ICDE}, 2019, pp. 638--649.

\bibitem{ye2019privkv}
Q.~Ye, H.~Hu, X.~Meng, and H.~Zheng, ``Privkv: Key-value data collection with local differential privacy,'' in \emph{Proc. IEEE SP}, 2019, pp. 317--331.

\bibitem{Qin2016HeavyHE}
Z.~Qin, Y.~Yang, T.~Yu, I.~Khalil, X.~Xiao, and K.~Ren, ``Heavy hitter estimation over set-valued data with local differential privacy,'' in \emph{Proc. ACM CCS}, 2016, pp. 192--203.

\bibitem{li2022frequent}
J.~Li, W.~Gan, Y.~Gui, Y.~Wu, and P.~S. Yu, ``Frequent itemset mining with local differential privacy,'' in \emph{Proceedings of the 31st ACM International Conference on Information \& Knowledge Management}, 2022, pp. 1146--1155.

\bibitem{wang2018privset}
S.~Wang, L.~Huang, Y.~Nie, P.~Wang, H.~Xu, and W.~Yang, ``Privset: Set-valued data analyses with locale differential privacy,'' in \emph{IEEE INFOCOM 2018-IEEE Conference on Computer Communications}.\hskip 1em plus 0.5em minus 0.4em\relax IEEE, 2018, pp. 1088--1096.

\bibitem{wang2018locally}
T.~Wang, N.~Li, and S.~Jha, ``Locally differentially private frequent itemset mining,'' in \emph{2018 IEEE Symposium on Security and Privacy (SP)}.\hskip 1em plus 0.5em minus 0.4em\relax IEEE, 2018, pp. 127--143.

\bibitem{Imola2021LDPgraph}
J.~Imola, T.~Murakami, and K.~Chaudhuri, ``Locally differentially private analysis of graph statistics,'' in \emph{30th USENIX security symposium (USENIX Security 21)}, 2021, pp. 983--1000.

\bibitem{Ye2020LDPgraph}
Q.~Ye, H.~Hu, M.~H. Au, X.~Meng, and X.~Xiao, ``Towards locally differentially private generic graph metric estimation,'' in \emph{2020 IEEE 36th International Conference on Data Engineering (ICDE)}.\hskip 1em plus 0.5em minus 0.4em\relax IEEE, 2020, pp. 1922--1925.

\bibitem{wang2020continuous}
T.~Wang, J.-Q. Chen, Z.~Zhang, D.~Su, Y.~Cheng, Z.~Li, N.~Li, and S.~Jha, ``Continuous release of data streams under both centralized and local differential privacy,'' in \emph{Proc. ACM CCS}, 2021, pp. 1237--1253.

\bibitem{baocgm}
E.~Bao, Y.~Yang, X.~Xiao, and B.~Ding, ``Cgm: An enhanced mechanism for streaming data collection with local differential privacy,'' \emph{Proc. of VLDB Endow.}, vol.~14, no.~11, pp. 2258--2270, 2021.

\bibitem{LDP_IDS}
X.~Ren, L.~Shi, W.~Yu, S.~Yang, C.~Zhao, and Z.~Xu, ``Ldp-ids: Local differential privacy for infinite data streams,'' in \emph{Proceedings of the 2022 international conference on management of data}, 2022, pp. 1064--1077.

\bibitem{schaler2023benchmarking}
C.~Sch{\"a}ler, T.~H{\"u}tter, and M.~Sch{\"a}ler, ``Benchmarking the utility of w-event differential privacy mechanisms-when baselines become mighty competitors,'' \emph{Proceedings of the VLDB Endowment}, vol.~16, no.~8, pp. 1830--1842, 2023.

\bibitem{li2023privsketch}
Y.~Li, X.~Lee, B.~Peng, T.~Palpanas, and J.~Xue, ``Privsketch: A private sketch-based frequency estimation protocol for data streams,'' \emph{arXiv preprint arXiv:2306.12144}, 2023.

\bibitem{joseph2018local}
M.~Joseph, A.~Roth, J.~Ullman, and B.~Waggoner, ``Local differential privacy for evolving data,'' \emph{Proc. NeurIPS}, vol.~31, pp. 2375--2384, 2018.

\bibitem{Erlingsson2019AmplificationBS}
{\'U}.~Erlingsson, V.~Feldman, I.~Mironov, A.~Raghunathan, K.~Talwar, and A.~Thakurta, ``Amplification by shuffling: From local to central differential privacy via anonymity,'' \emph{ArXiv}, vol. abs/1811.12469, 2019.

\bibitem{10.1145/3639292}
\BIBentryALTinterwordspacing
J.~Zhang, C.~Zhang, G.~Li, and C.~Chai, ``Pace: Poisoning attacks on learned cardinality estimation,'' \emph{Proc. ACM Manag. Data}, vol.~2, no.~1, Mar. 2024. [Online]. Available: \url{https://doi.org/10.1145/3639292}
\BIBentrySTDinterwordspacing

\bibitem{10.1145/3514221.3517867}
\BIBentryALTinterwordspacing
E.~M. Kornaropoulos, S.~Ren, and R.~Tamassia, ``The price of tailoring the index to your data: Poisoning attacks on learned index structures,'' in \emph{Proceedings of the 2022 International Conference on Management of Data}, ser. SIGMOD '22.\hskip 1em plus 0.5em minus 0.4em\relax New York, NY, USA: Association for Computing Machinery, 2022, p. 1331–1344. [Online]. Available: \url{https://doi.org/10.1145/3514221.3517867}
\BIBentrySTDinterwordspacing

\bibitem{Cao2021poisonFE}
X.~Cao, J.~Jia, and N.~Z. Gong, ``Data poisoning attacks to local differential privacy protocols,'' in \emph{30th USENIX Security Symposium (USENIX Security 21)}, 2021, pp. 947--964.

\bibitem{Wu2022poisonKey}
Y.~Wu, X.~Cao, J.~Jia, and N.~Z. Gong, ``Poisoning attacks to local differential privacy protocols for $\{$Key-Value$\}$ data,'' in \emph{31st USENIX Security Symposium (USENIX Security 22)}, 2022, pp. 519--536.

\bibitem{Li2023Finegrained}
X.~Li, N.~Li, W.~Sun, N.~Z. Gong, and H.~Li, ``Fine-grained poisoning attack to local differential privacy protocols for mean and variance estimation,'' in \emph{32nd USENIX Security Symposium (USENIX Security 23)}, 2023, pp. 1739--1756.

\bibitem{Kellaris2014DifferentiallyPE}
G.~Kellaris, S.~Papadopoulos, X.~Xiao, and D.~Papadias, ``Differentially private event sequences over infinite streams,'' \emph{Proc. VLDB Endow.}, vol.~7, pp. 1155--1166, 2014.

\bibitem{ZhangWLHC18}
Z.~Zhang, T.~Wang, N.~Li, S.~He, and J.~Chen, ``Calm: Consistent adaptive local marginal for marginal release under local differential privacy,'' in \emph{Proc. ACM CCS}, 2018, pp. 212--229.

\bibitem{Duchi2016MinimaxOP}
J.~Duchi, M.~Wainwright, and M.~Jordan, ``Minimax optimal procedures for locally private estimation,'' \emph{J. Am. Stat. Assoc}, vol. 113, pp. 182 -- 201, 2016.

\bibitem{Wang2016RescueDPRS}
Q.~Wang, Y.~Zhang, X.~Lu, Z.~Wang, Z.~Qin, and K.~Ren, ``Rescuedp: Real-time spatio-temporal crowd-sourced data publishing with differential privacy,'' \emph{Proc. IEEE INFOCOM}, pp. 1--9, 2016.

\bibitem{10.1145/2806416.2806441}
H.~Li, L.~Xiong, X.~Jiang, and J.~Liu, ``Differentially private histogram publication for dynamic datasets: An adaptive sampling approach,'' in \emph{Proc. of ACM CIKM}, 2015, p. 1001–1010.

\bibitem{Chen2017PeGaSusDD}
Y.~Chen, A.~Machanavajjhala, M.~Hay, and G.~Miklau, ``Pegasus: Data-adaptive differentially private stream processing,'' \emph{Proc. ACM CCS}, pp. 1375--1388, 2017.

\bibitem{Fan2014AnAA}
L.~Fan and L.~Xiong, ``An adaptive approach to real-time aggregate monitoring with differential privacy,'' \emph{IEEE Trans. on Knowl. Data Eng.}, vol.~26, pp. 2094--2106, 2014.

\bibitem{thomas2013trafficking}
K.~Thomas, D.~McCoy, C.~Grier, A.~Kolcz, and V.~Paxson, ``$\{$Trafficking$\}$ fraudulent accounts: The role of the underground market in twitter spam and abuse,'' in \emph{22nd USENIX Security Symposium (USENIX Security 13)}, 2013, pp. 195--210.

\bibitem{tong2024data}
W.~Tong, H.~Chen, J.~Niu, and S.~Zhong, ``Data poisoning attacks to locally differentially private frequent itemset mining protocols,'' in \emph{Proc. ACM CCS}, 2024.

\bibitem{huangfu2018parallelizing}
Q.~Huangfu and J.~J. Hall, ``Parallelizing the dual revised simplex method,'' \emph{Mathematical Programming Computation}, vol.~10, no.~1, pp. 119--142, 2018.

\bibitem{Cao2021Provably}
X.~Cao, J.~Jia, and N.~Z. Gong, ``Provably secure federated learning against malicious clients,'' in \emph{Proceedings of the AAAI conference on artificial intelligence}, vol.~35, no.~8, 2021, pp. 6885--6893.

\bibitem{Liu2008IsolationForest}
F.~T. Liu, K.~M. Ting, and Z.-H. Zhou, ``Isolation forest,'' in \emph{2008 eighth ieee international conference on data mining}.\hskip 1em plus 0.5em minus 0.4em\relax IEEE, 2008, pp. 413--422.

\bibitem{Dwork2014TheAF}
C.~Dwork and A.~Roth, ``The algorithmic foundations of differential privacy,'' \emph{Found. Trends Theor. Comput. Sci.}, vol.~9, pp. 211--407, 2014.

\bibitem{Dwork-405}
C.~Dwork, ``Differential privacy,'' in \emph{Proc. ICALP}, 2006, pp. 1--12.

\bibitem{xue2022ddrm}
Q.~Xue, Q.~Ye, H.~Hu, Y.~Zhu, and J.~Wang, ``Ddrm: A continual frequency estimation mechanism with local differential privacy,'' \emph{IEEE Transactions on Knowledge and Data Engineering}, 2022.

\bibitem{li2023locally}
X.~Li, Y.~Cao, and M.~Yoshikawa, ``Locally private streaming data release with shuffling and subsampling,'' in \emph{2023 IEEE 39th International Conference on Data Engineering Workshops (ICDEW)}.\hskip 1em plus 0.5em minus 0.4em\relax IEEE, 2023, pp. 125--131.

\bibitem{gao2023privacy}
W.~Gao and S.~Zhou, ``Privacy-preserving for dynamic real-time published data streams based on local differential privacy,'' \emph{IEEE Internet of Things Journal}, 2023.

\bibitem{ye2021beyond}
Q.~Ye, H.~Hu, N.~Li, X.~Meng, H.~Zheng, and H.~Yan, ``Beyond value perturbation: Local differential privacy in the temporal setting,'' in \emph{IEEE INFOCOM 2021-IEEE Conference on Computer Communications}.\hskip 1em plus 0.5em minus 0.4em\relax IEEE, 2021, pp. 1--10.

\bibitem{10.1145/1806689.1806787}
C.~Dwork, M.~Naor, T.~Pitassi, and G.~N. Rothblum, ``Differential privacy under continual observation,'' in \emph{Proc. ACM STOC}, 2010, pp. 715--724.

\bibitem{cheu2021manipulation}
A.~Cheu, A.~Smith, and J.~Ullman, ``Manipulation attacks in local differential privacy,'' in \emph{2021 IEEE Symposium on Security and Privacy (SP)}.\hskip 1em plus 0.5em minus 0.4em\relax IEEE, 2021, pp. 883--900.

\bibitem{zheng2024data}
Z.~Zheng, Z.~Li, C.~Huang, S.~Long, M.~Li, and X.~Shen, ``Data poisoning attacks and defenses to ldp-based privacy-preserving crowdsensing,'' \emph{IEEE Transactions on Dependable and Secure Computing}, 2024.

\bibitem{li2020estimating}
Z.~Li, T.~Wang, M.~Lopuha{\"a}-Zwakenberg, N.~Li, and B.~{\v{S}}koric, ``Estimating numerical distributions under local differential privacy,'' in \emph{Proc. ACM SIGMOD}, 2020, pp. 621--635.

\end{thebibliography}

\clearpage
\normalsize
\appendix





\section{Proofs of Theorems and Corollaries}

\subsection{Proof of Theorem~\ref{Theorem:G_{I,t}}}\label{Proof G_{I,t}}

We consider to manipulate an FO with the privacy budget of $\epsilon$. Suppose that the solution to the above optimization problem is $\left(m^\ast\left[1\right],m^\ast\left[2\right],\ldots,m^\ast\left[d\right]\right)$. Then, 
the expected manipulation gap for IPMA $G_{I,t}\left(m,n,\epsilon\right)$ can be calculated as follows:
\begin{small}
    \begin{align}
\nonumber
&G_{I,t}\left(m,n,\epsilon\right)=\mathbb{E}\left(\frac{1}{d}\sum\nolimits_{k=1}^{d}\left(\mathbf{\hat{f}}_t\left[k\right]-\mathbf{\tilde{f}}_t\left[k\right]\right)^2\right) \\ \nonumber
&=\frac{1}{d}\sum\nolimits_{k=1}^{d}\left(\text{Var}\left(\mathbf{\hat{f}}_t\left[k\right]\right)+\left(\mathbb{E}\left(\mathbf{\hat{f}}_t\left[k\right]\right)-\mathbf{\tilde{f}}_t\left[k\right]\right)^2\right) \\ 
&=\frac{1}{d}\sum\nolimits_{k=1}^{d}\left(\frac{m^\ast\left[k\right]+n\cdot\mathbf{f}_t\left[k\right]}{n+m}-\mathbf{\tilde{f}}_t\left[k\right]\right)^2+\text{Var}\left(n+m,\epsilon\right)
\end{align}
\end{small}where $\text{Var}\left(n+m,\epsilon\right)=\frac{1}{d}\sum_{k=1}^d \text{Var}\left(\mathbf{\hat{f}}_t\left[k\right]\right)$. The solution to the optimization problem is independent of $\epsilon$, affecting only the term $\text{Var}(n+m,\epsilon)$ in terms of attack impact. 

\subsection{Proof of Corollary~\ref{Theorem:IPMA_epsilon}} \label{Proof IPMA_epsilon}
The optimization problem is unrelated to $\epsilon$, so $m^\ast[k]$ is independent of $\epsilon$. In $G_{I,t}$, only $\text{Var}(n+m, \epsilon)$ varies with $\epsilon$: a larger $\epsilon$ reduces $\text{Var}$, decreasing $G_{I,t}(m,n,\epsilon)$.

\subsection{Proof of Corollary~\ref{Theorem:IPMA_mn}}\label{sec: Proof of Theorem 5.2}
Suppose $m$ and $n$ become $\alpha>1$ times larger, and the solution to the optimization problem before expanding $m$ and $n$ (denoted as problem A) is $\left(m^\ast\left[1\right],m^\ast\left[2\right],\ldots,m^\ast\left[d\right]\right)$. 
For the optimization problem after expanding $m$ and $n$ (called problem B), we can replace $m\left[k\right]$ with $m^\prime\left[k\right]$, where $m^\prime\left[k\right]=m\left[k\right]/\alpha $. Thus, problem B can be transformed into a problem similar to problem A. Solving $m^\prime\left[k\right]$ of problem B is equivalent to solving $m\left[k\right]$ of problem A. So, considering that the solution to problem A is $\left(m^\ast\left[1\right],\ldots,m^\ast\left[d\right]\right)$, $\left(m^\prime\left[1\right], \ldots,m^\prime\left[d\right]\right)$ can be solved as $\left(m^\ast\left[1\right],\ldots,m^\ast\left[d\right]\right)$.
Thus, the solution to problem B is $\left(\alpha m^\ast\left[1\right],{\alpha m}^\ast\left[2\right],\ldots,{\alpha m}^\ast\left[d\right]\right)$. The first term of $G_{I,t}\left(m,n,\epsilon\right)$ after expanding $m$ and $n$ would be equal to the first term of $G_{I,t}\left(m,n,\epsilon\right)$ before expanding $m$ and $n$. However, $\text{Var}$ becomes smaller after expanding $m$ and $n$. 
Hence, if $n+m$ is larger, $G_{I,t}\left(m,n,\epsilon\right)$ will be smaller.

\subsection{Proof of Corollary~\ref{Theorem:condition_IPMA}}\label{Proof condition_IPMA}
When $m$ satisfies Eq.~(\ref{condition_IPMA}), it is easy to prove that one feasible solution of Eq.~(\ref{IPMA_optimization}) is $m[k]=(m+n)\mathbf{\tilde{f}}_t[k]-n\mathbf{f}_t[k]$, i.e., $m[k]$ satisfies $\sum\nolimits_{k=1}^{d}m\left[k\right]=m$ and $0\le m\left[k\right]\le m$. Also, when $m[k]=(m+n)\mathbf{\tilde{f}}_t[k]-n\mathbf{f}_t[k]$,
\begin{small}
    \begin{align}
    \nonumber
&\frac{1}{d}\sum\nolimits_{k=1}^{d}\left(\mathbf{f}_t\left[k\right]-\mathbf{\tilde{f}}_t\left[k\right]\right)^2\\
=&\frac{1}{d}\sum\nolimits_{k=1}^{d}\left(\frac{m\left[k\right]+n\cdot\mathbf{f}_t\left[k\right]}{n+m}-\mathbf{\tilde{f}}_t\left[k\right]\right)^2=0,
\end{align}    \end{small}
which reaches the minimum of quadratic sum. Considering that Eq.~(\ref{IPMA_optimization}) is a convex optimization problem which has only one global optimal solution and $\frac{1}{d}\sum\nolimits_{k=1}^{d}\left(\mathbf{f}_t\left[k\right]-\mathbf{\tilde{f}}_t\left[k\right]\right)^2$ reaches the minimum when $m[k]=(m+n)\mathbf{\tilde{f}}_t[k]-n\mathbf{f}_t[k]$, $m[k]=(m+n)\mathbf{\tilde{f}}_t[k]-n\mathbf{f}_t[k]$ is the optimal solution of Eq.~(\ref{IPMA_optimization}). Substitute $m^\ast[k]=(m+n)\mathbf{\tilde{f}}_t[k]-n\mathbf{f}_t[k]$ into Eq.~(\ref{G_{I,t}}), $G_{I,t}$ can be calculated as $G_{I,t}^\ast\left(m,n,\epsilon\right)=\text{Var}\left(n+m,\epsilon\right)$.

\subsection{Proof of Theorem~\ref{Theorem:G_{O,t}}}\label{Proof G_{O,t}}
Suppose the attacker targets a publication strategy FO whose privacy budget is $\epsilon$ and the solution to the above optimization problem is $\left(m^\ast\left[1\right],m^\ast\left[2\right],\ldots,m^\ast\left[d\right]\right)$. Thus, the manipulation gap for OPMA $G_{O,t}\left(m,n,\epsilon\right)$ can be calculated as follows:
\begin{small}
    \begin{align}
\nonumber
&G_{O,t}\left(m,n,\epsilon\right)=\mathbb{E}\left(\frac{1}{d}\sum_{k=1}^{d}\left(\mathbf{\hat{f}}_t\left[k\right]-\mathbf{\tilde{f}}_t\left[k\right]\right)^2\right) \\ \nonumber
&=\frac{1}{d}\sum_{k=1}^{d}\left(\frac{n\mathbf{f}_t\left[k\right]\left(p-q\right)+m^\ast\left[k\right]-mq}{\left(m+n\right)\left(p-q\right)}-\mathbf{\tilde{f}}_t\left[k\right]\right)^2 \\
&+\frac{1}{d}\sum_{k=1}^{d}\text{Var}\left(\frac{\sum_{j=1}^{n}\mathbb{I}_{S\left(y_j\right)}^{\left(k\right)}+\sum_{j=1}^{m}\mathbb{I}_{S\left(z_j\right)}^{\left(k\right)}-q\left(m+n\right)}{\left(m+n\right)\left(p-q\right)}\right). 
\end{align}
\end{small}
For the second term, $\sum_{j=1}^{m}\mathbb{I}_{S\left(z_j\right)}^{\left(k\right)}$ is treated as a constant as the optimization problem is solved as $\left(m^\ast\left[1\right],m^\ast\left[2\right],\ldots,m^\ast\left[d\right]\right)$. So,
\begin{small}
    \begin{align}
\nonumber
G_{O,t}\left(m,n,\epsilon\right)&=\frac{1}{d}\sum_{k=1}^{d}{\left(\frac{n\mathbf{f}_t\left[k\right]\left(p-q\right)+m^\ast\left[k\right]-mq}{\left(m+n\right)\left(p-q\right)}-\mathbf{\tilde{f}}_t\left[k\right]\right)^2} \\
&+\left(\frac{n}{m+n}\right)^2\text{Var}\left(n,\epsilon\right). 
\end{align}
\end{small}

\subsection{Proof of Corollary~\ref{Theorem:condition_OPMA}}\label{Proof condition_OPMA}
When $m$ satisfies Eq.~(\ref{condition_OPMA}), it is easy to prove that one feasible solution of $\min~\frac{1}{d}\sum_{k=1}^{d}\left|\mathbb{E}\left(\mathbf{\hat{f}}_t\left[k\right]\right)-\mathbf{\tilde{f}}_t\left[k\right]\right|$ is \\ $m[k]=\left(p-q\right)\left[\left(m+n\right)\mathbf{\tilde{f}}_t[k]-n\mathbf{f}_t[k]\right]+mq$, i.e., $m[k]$ satisfies $\sum\nolimits_{k=1}^{d}m\left[k\right]=m$ and $0\le m\left[k\right]\le m$ when attacking kRR or $0\le m\left[k\right]\le m$ when attacking OUE. Also, when \\ $m[k]=\left(p-q\right)\left[\left(m+n\right)\mathbf{\tilde{f}}_t[k]-n\mathbf{f}_t[k]\right]+mq$,
\begin{small}
    \begin{align}
    \nonumber
&\frac{1}{d}\sum\nolimits_{k=1}^{d}\left|\mathbb{E}\left(\mathbf{\hat{f}}_t\left[k\right]\right)-\mathbf{\tilde{f}}_t\left[k\right]\right|\\
=&\frac{1}{d}\sum_{k=1}^{d}{\left(\frac{n\mathbf{f}_t\left[k\right]\left(p-q\right)+m\left[k\right]-mq}{\left(m+n\right)\left(p-q\right)}-\mathbf{\tilde{f}}_t\left[k\right]\right)^2}=0,
\end{align}\end{small}
which reaches the minimum of sum of absolute values. Note that, the optimization problem (i.e., $\min~\frac{1}{d}\sum_{k=1}^{d}\left|\mathbb{E}\left(\mathbf{\hat{f}}_t\left[k\right]\right)-\mathbf{\tilde{f}}_t\left[k\right]\right|$) is convex with the only optimal solution \\ $m[k]=\left(p-q\right)\left[\left(m+n\right)\mathbf{\tilde{f}}_t[k]-n\mathbf{f}_t[k]\right]+mq$. So, we substitute $m^\ast[k]=\left(p-q\right)\left[\left(m+n\right)\mathbf{\tilde{f}}_t[k]-n\mathbf{f}_t[k]\right]+mq$ into Eq.~(\ref{G_{O,t}}), we can obtain that $G_{O,t}^\ast\left(m,n,\epsilon\right)=\left(\frac{n}{m+n}\right)^2\text{Var}\left(n,\epsilon\right)=\frac{n}{m+n}\text{Var}\left(n+m,\epsilon\right)$.    

\subsection{Proof of Corollary~\ref{Theorem:OPMA_epsilon}}\label{Proof OPMA_epsilon}
According to Eq.~(\ref{G^*_O}), when $\epsilon$ is larger, $\text{Var}$ becomes smaller and $G_{O,t}^\ast\left(m,n,\epsilon\right)$ becomes smaller.

\subsection{Proof of Corollary~\ref{Theorem:OPMA_mn}}\label{Proof OPMA_mn}
The proof of Corollary~\ref{Theorem:OPMA_mn} is similar to that of Corollary~\ref{Theorem:IPMA_mn}.

\subsection{Proof of Theorem~\ref{Theorem:IDMA}}\label{Proof IDMA}
We define $n[k]=n^e\mathbf{f}^e_t[k]$ and $b_k=n[k]/(m+n^e)-\mathbf{\hat{f}}^{t-1}[k]$ with $k^\ast=\arg\max_{k\in[1,..,d]}(b_k)$, aiming to maximize the formula.
\begin{small}
    \begin{align}
\nonumber
&\frac{1}{d}\sum_{k=1}^{d}\left(\mathbf{f}\left[k\right]-\mathbf{\hat{f}}_{t-1}\left[k\right]\right)^2=\frac{1}{d}\sum_{k=1}^{d}\left(\frac{m\left[k\right]}{n^e+m}+b_k\right)^2 \\ 
&=\frac{1}{d\left(n^e+m\right)^2}\sum_{k=1}^{d}{m\left[k\right]}^2+\frac{2}{d\left(n^e+m\right)}\sum_{k=1}^{d}{m\left[k\right]b_k}+\frac{1}{d}\sum_{k=1}^{d}b_k^2
\end{align}
\end{small}

For the first term, the maximum value is taken if one of $m\left[k\right]$ ($m\left[i\right]$) is $m$ and the other $m\left[k\right]$ (e.g., $k\neq\ i$) is 0. For the second term, the maximum value is taken if $m\left[k^\ast\right]=m$ and other $m\left[k\right]\ (k\neq\ k^\ast)$ are 0. And the third term is a constant. So the solution to this problem is $m\left[k^\ast\right]=m$ and $m\left[k\right]\ =0\ (k\neq\ k^\ast)$.

\subsection{Proof of Theorem~\ref{Theorem:ODMA}}\label{Proof ODMA}
    It's assumed that the attacker targets a FO with a privacy budget of $\epsilon$. Real users have outputs $(y_1, y_2, \ldots, y_n)$, and fake users have outputs $(z_1, z_2, \ldots, z_m)$. Let $m[k]=\mathbb{E}(\sum_{j=1}^{m}\mathbb{I}_{S(z_j)}^{(k)})$.
 So for $\forall k\in\left[1,...,d\right]$,
\begin{small}
    \begin{align}
\nonumber
&\mathbb{E}\left(\left(\mathbf{\bar{f}}_t\left[k\right]-\mathbf{\hat{f}}_{t-1}\left[k\right]\right)^2\right)=\left(\frac{n\mathbf{f}_t\left[k\right]\left(p-q\right)+m\left[k\right]-mq}{\left(m+n\right)\left(p-q\right)}-\mathbf{\hat{f}}_{t-1}\left[k\right]\right)^2 \\
&+\text{Var}\left(\frac{\sum_{j=1}^{n}\mathbb{I}_{S\left(y_j\right)}^{\left(k\right)}+\sum_{j=1}^{m}\mathbb{I}_{S\left(z_j\right)}^{\left(k\right)}-q\left(m+n\right)}{\left(m+n\right)\left(p-q\right)}\right)
\end{align}
\end{small}
Here, for the second term, $\left(z_1,z_2,\ldots,z_m\right)$ are random variables, whose variance are hard to determine, while $\left(y_1,y_2,\ldots,y_n\right)$ are outputs of local perturbation, whose variance has been shown in \cite{Wang2017LocallyDPFE}. Thus, Let $Y_k=\frac{\sum_{j=1}^{n}\mathbb{I}_{S\left(y_j\right)}^{\left(k\right)}}{\left(m+n\right)\left(p-q\right)}$ and $Z_k=\frac{\sum_{j=1}^{m}\mathbb{I}_{S\left(z_j\right)}^{\left(k\right)}}{\left(m+n\right)\left(p-q\right)}$,
\begin{small}
\begin{align}
\nonumber
&\text{Var}\left(\frac{\sum_{j=1}^{n}\mathbb{I}_{S\left(y_j\right)}^{\left(k\right)}+\sum_{j=1}^{m}\mathbb{I}_{S\left(z_j\right)}^{\left(k\right)}-q\left(m+n\right)}{\left(m+n\right)\left(p-q\right)}\right) \\ \nonumber
&=\text{Var}(Y_k)+\text{Var}(Z_k)+2Cov(Y_k,Z_k) \\
&=\text{Var}\left(Y_k\right)+E\left({Z_k}^2\right)-\left(E\left(Z_k\right)\right)^2+2(E(Y_kZ_k)-E(Y_k)E(Z_k)), 
\end{align}
\end{small}where $\mathbb{E}\left({Z_k}^2\right)\geq0$ and $\mathbb{E}(Y_kZ_k)\geq0$. Thus, we have
\begin{small}
    \begin{align}
\nonumber
&\mathbb{E}\left(\left(\mathbf{\bar{f}}_t\left[k\right]-\mathbf{\hat{f}}_{t-1}\left[k\right]\right)^2\right) \\ \nonumber
&\geq\left(\frac{n\mathbf{f}_t\left[k\right]\left(p-q\right)+m\left[k\right]-mq}{\left(m+n\right)\left(p-q\right)}-\mathbf{\hat{f}}_{t-1}\left[k\right]\right)^2 \\
&+\text{Var}\left(Y_k\right)-\left(E\left(Z_k\right)\right)^2-2E\left(Y_k\right)E\left(Z_k\right).
\end{align}
\end{small}where $\mathbb{E}\left(Z_k\right)=\frac{m\left[k\right]}{\left(m+n\right)\left(p-q\right)}$, $\mathbb{E}\left(Y
_k\right)=\frac{\sum_{j=1}^{n}\mathbb{E}\left(\mathbb{I}_{S\left(y_j\right)}^{\left(k\right)}\right)}{\left(m+n\right)\left(p-q\right)}=\frac{A_k+\left(m+n\right)q}{\left(m+n\right)\left(p-q\right)}$, $A_k=n\mathbf{f}_t\left[k\right]\left(p-q\right)-mq$. Thus, 
\begin{small}
    \begin{align}
\nonumber
&\mathbb{E}\left(\left(\mathbf{\bar{f}}_t\left[k\right]-\mathbf{\hat{f}}_{t-1}\left[k\right]\right)^2\right) \\ \nonumber
&\geq\left(\frac{A_k+m\left[k\right]}{\left(m+n\right)\left(p-q\right)}-\mathbf{\hat{f}}_{t-1}\left[k\right]\right)^2+\frac{n^2}{\left(m+n\right)^2}\text{Var}\left(\frac{\sum_{j=1}^{n}\mathbb{I}_{S\left(y_j\right)}^{\left(k\right)}}{n\left(p-q\right)}\right) \\ \nonumber
&-\left(\frac{m\left[k\right]}{\left(m+n\right)\left(p-q\right)}\right)^2-2\frac{m\left[k\right]\left(A_k+\left(m+n\right)q\right)}{\left(m+n\right)^2\left(p-q\right)^2} \\ \nonumber
&=\frac{A_k^2-2m\left[k\right]\left(nq+mq\right)}{\left(p-q\right)^2\left(m+n\right)^2}+\frac{n^2}{\left(m+n\right)^2}\text{Var}\left(\frac{\sum_{j=1}^{n}\mathbb{I}_{S\left(y_j\right)}^{\left(k\right)}-nq}{n\left(p-q\right)}\right)\\ 
&-2\frac{A_k+m\left[k\right]}{\left(m+n\right)\left(p-q\right)}\mathbf{\hat{f}}_{t-1}\left[k\right]+\left(\mathbf{\hat{f}}_{t-1}\left[k\right]\right)^2
\end{align}
\end{small}where $\sum_{k=1}^{d}\text{Var}\left(\frac{\sum_{j=1}^{n}\mathbb{I}_{S\left(y_j\right)}^{\left(k\right)}-nq}{n\left(p-q\right)}\right)=d\ \text{Var}\left(n,\epsilon\right)$. After summing the inequality from $k=1$ to $d$, disregarding constants like $d\ \text{Var}(n,\epsilon)$ and focusing on $m[k]$ terms, maximizing $\overline{dis}$ becomes an optimization problem to maximize its lower bound.
\begin{small}
    \begin{align}
&\max~~~\sum_{k=1}^{d}{\frac{-2\mathbf{\hat{f}}_{t-1}\left[k\right]\left(m+n\right)\left(p-q\right)-2\left(m+n\right)q}{\left(p-q\right)^2\left(m+n\right)^2}m\left[k\right]}
\end{align}
\end{small}
Also, $0\le m\left[k\right]\le m$ and for kRR, $\sum_{k=1}^{d}m\left[k\right]=m$. To solve this problem, we only need calculate the coefficient in front of $m\left[k\right]$ (i.e.find the minimum $\mathbf{\hat{f}}_{t-1}\left[k\right]$). Assume the index of the maximum coefficient (the minimum $\mathbf{\hat{f}}_{t-1}\left[k\right]$) is $k^\ast\in\left[d\right]$. Then, the solution is $m\left[k^\ast\right]=m$, $m\left[k\right]=0\left(k\neq k^\ast\right)$.

\section{Experimental Results}\label{appendix:experiment results}

\subsection{Results on Real-World Datasets}
\label{sec: Experiment Results on Real-World Datasets}

\begin{figure*}[htbp]
	\centering	
	\subfigure[\textsf{Taxi} dataset, Uniform $\mathbf{\tilde{f}}$]{
		\includegraphics[width=0.24\textwidth]{./Figures/Attack/Taxi/Taxi_Uniform-m.pdf}}\vspace{-0.06cm}\hspace{-4mm}
	\subfigure[\textsf{Taxi} dataset, Pulse $\mathbf{\tilde{f}}$]{
		\includegraphics[width=0.24\textwidth]{./Figures/Attack/Taxi/Taxi_Pulse-m.pdf}}\vspace{-0.06cm}\hspace{-4mm}
	\subfigure[\textsf{Taxi} dataset, Gaussian $\mathbf{\tilde{f}}$]{
		\includegraphics[width=0.24\textwidth]{./Figures/Attack/Taxi/Taxi_Brown-m.pdf}}\vspace{-0.06cm}\hspace{-4mm}
	\subfigure[\textsf{Taxi} dataset, Sigmoid $\mathbf{\tilde{f}}$]{
		\includegraphics[width=0.24\textwidth]{./Figures/Attack/Taxi/Taxi_Sigmoid-m.pdf}}\vspace{-0.06cm}\hspace{-4mm}
	\subfigure[\scriptsize\textsf{Foursquare} dataset, Uniform $\mathbf{\tilde{f}}$]{
		\includegraphics[width=0.24\textwidth]{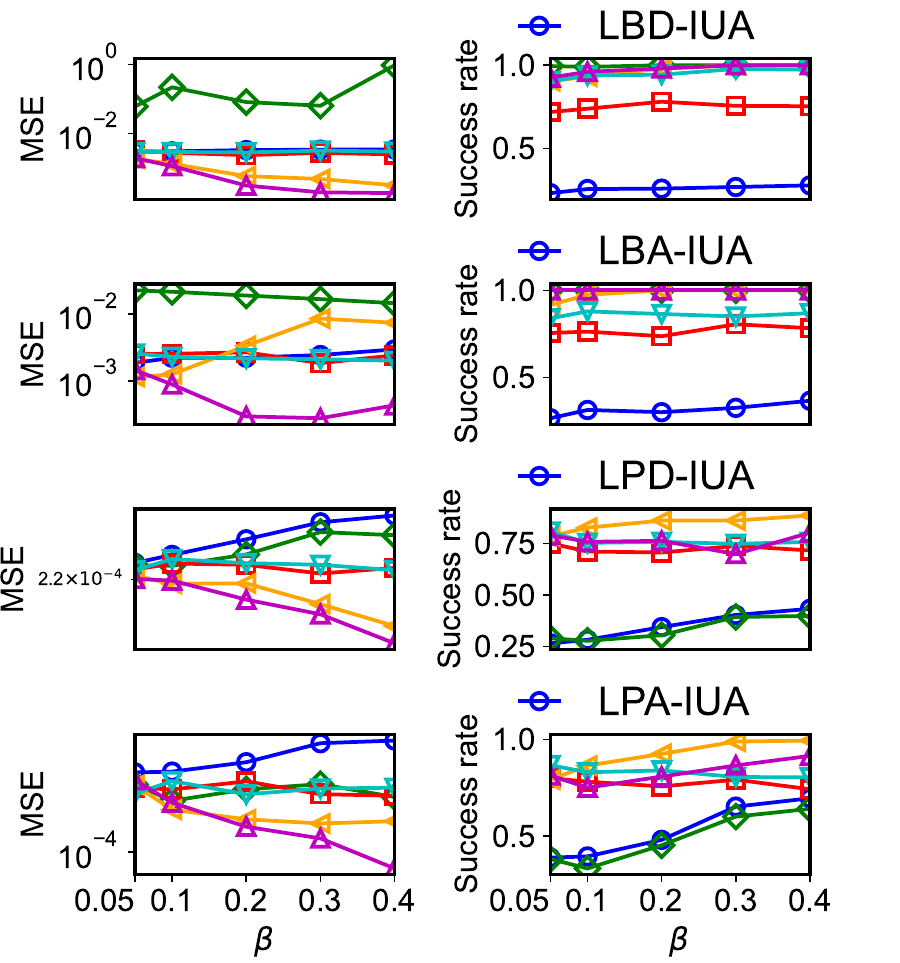}}\vspace{-0.05cm}\hspace{-4mm}
	\subfigure[\scriptsize\textsf{Foursquare} dataset, Pulse $\mathbf{\tilde{f}}$]{
		\includegraphics[width=0.24\textwidth]{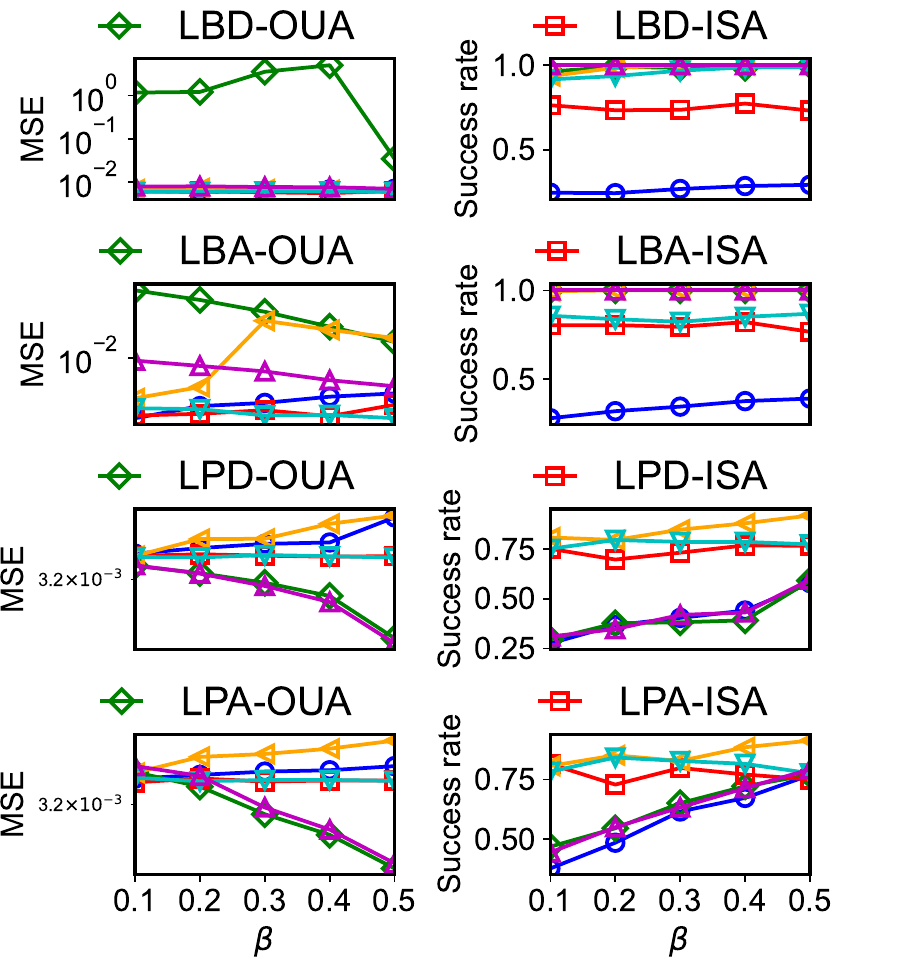}}\vspace{-0.06cm}\hspace{-4mm}
	\subfigure[\scriptsize\textsf{Foursquare} dataset, Gaussian $\mathbf{\tilde{f}}$]{
		\includegraphics[width=0.24\textwidth]{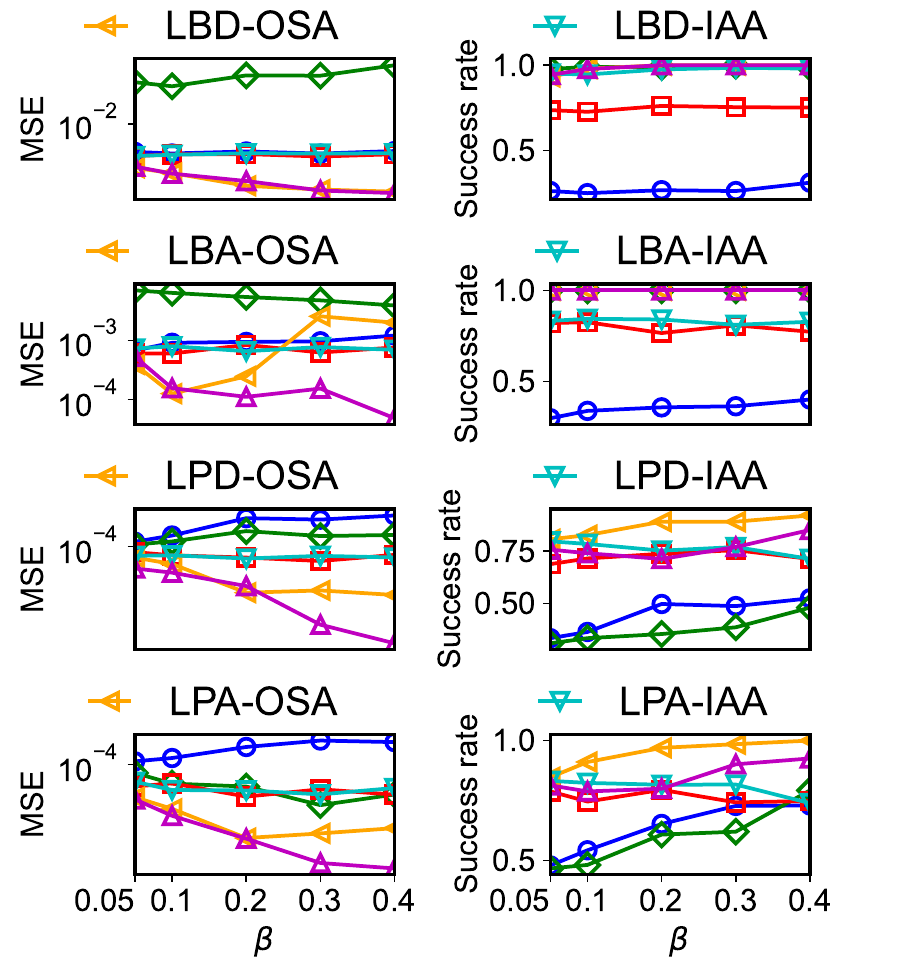}}\vspace{-0.06cm}\hspace{-4mm}
	\subfigure[\scriptsize\textsf{Foursquare} dataset, Sigmoid $\mathbf{\tilde{f}}$]{
		\includegraphics[width=0.24\textwidth]{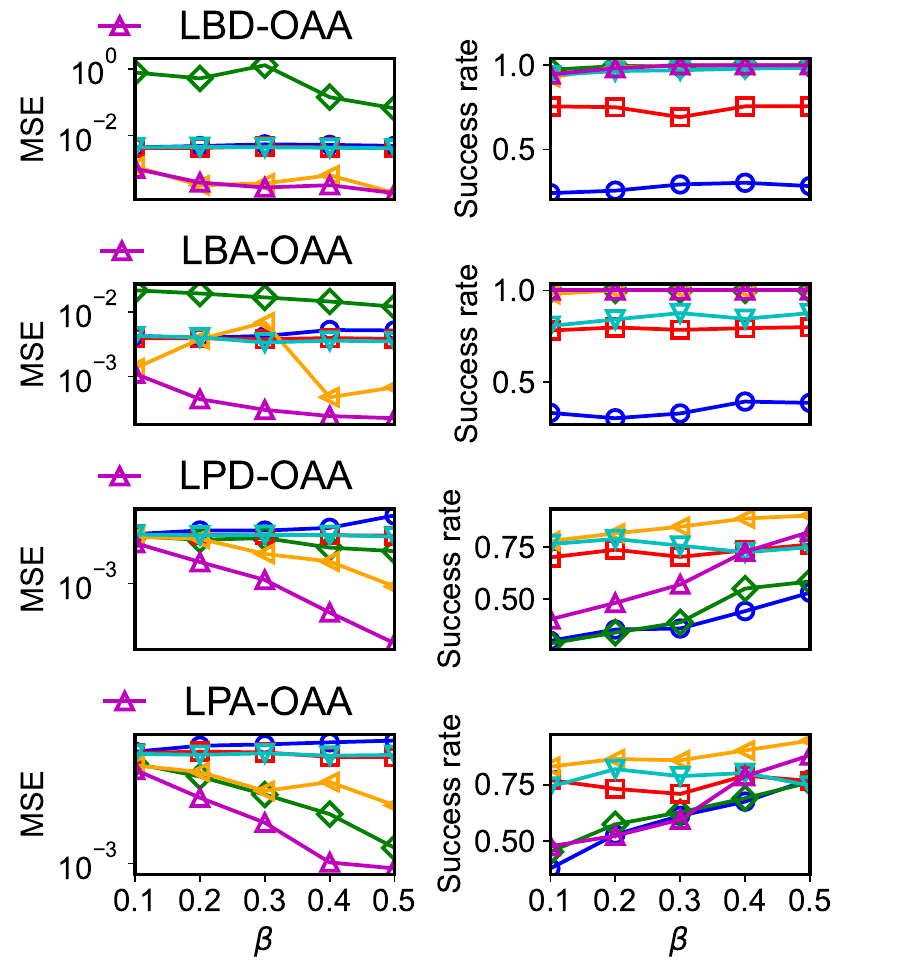}}\vspace{-0.06cm}\hspace{-4mm}
	\subfigure[\textsf{Taobao} dataset, Uniform $\mathbf{\tilde{f}}$]{
		\includegraphics[width=0.24\textwidth]{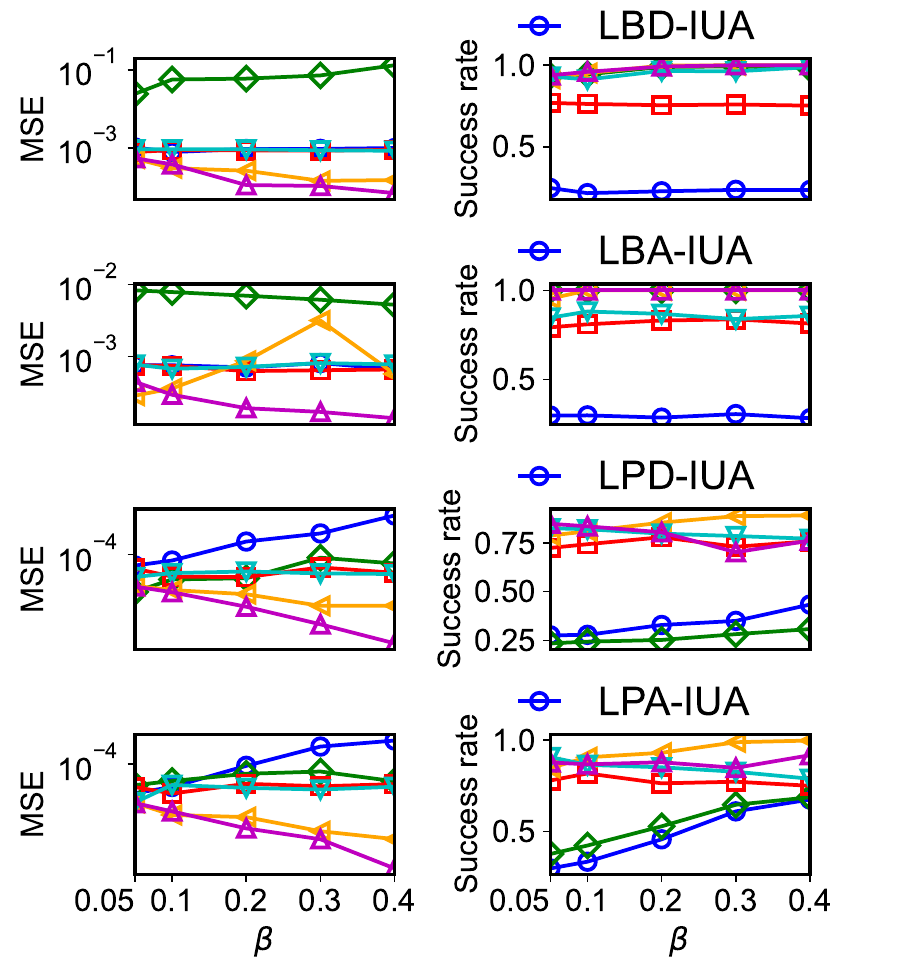}}\vspace{-0.13cm}\hspace{-4mm}
	\subfigure[\textsf{Taobao} dataset, Pulse $\mathbf{\tilde{f}}$]{
		\includegraphics[width=0.24\textwidth]{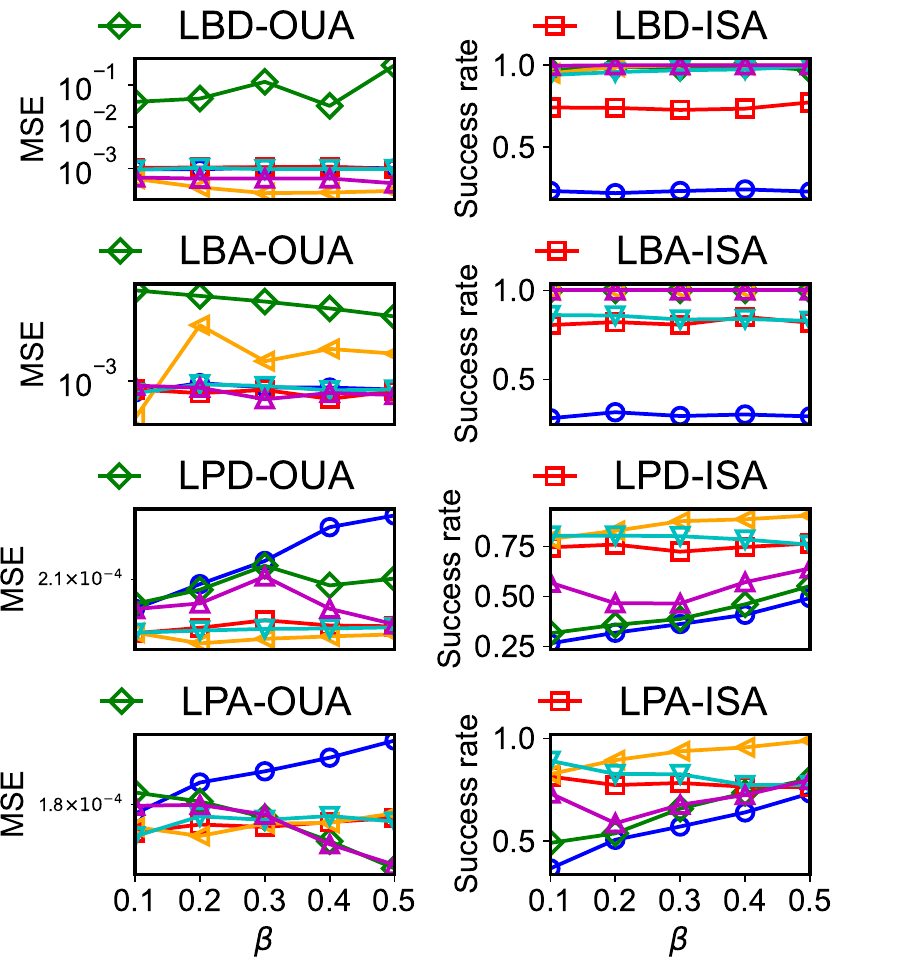}}\vspace{-0.13cm}\hspace{-4mm}
	\subfigure[\textsf{Taobao} dataset, Gaussian $\mathbf{\tilde{f}}$]{
		\includegraphics[width=0.24\textwidth]{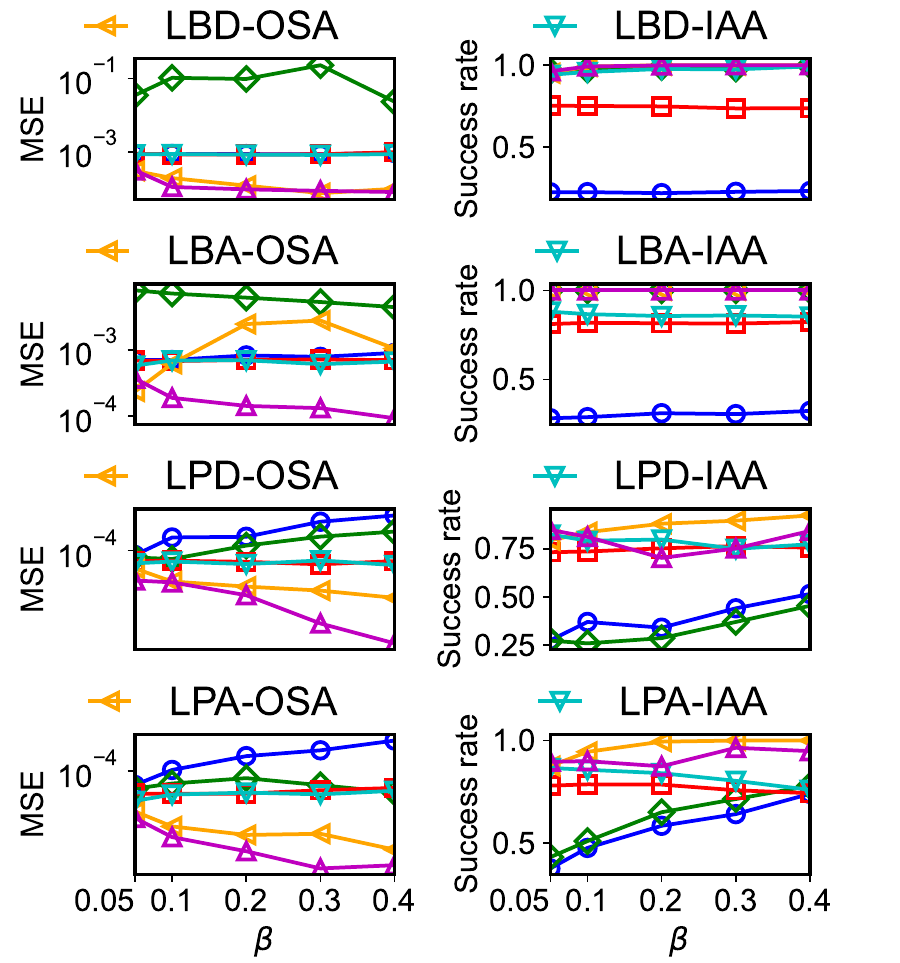}}\vspace{-0.13cm}\hspace{-4mm}
	\subfigure[\textsf{Taobao} dataset, Sigmoid $\mathbf{\tilde{f}}$]{
		\includegraphics[width=0.24\textwidth]{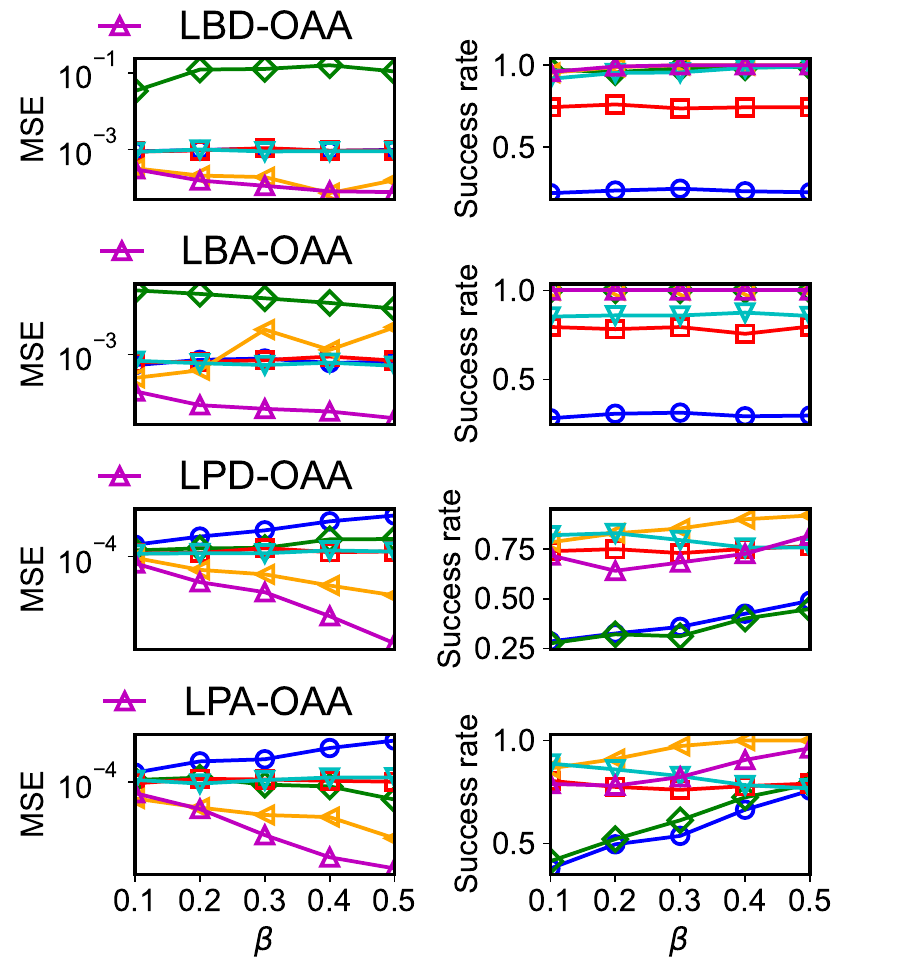}}
	\caption{\small Attacking effectiveness for real-world datasets with varying $\beta$. 
	}\centering
	\label{fig:beta} 
	\vspace{-0.5cm}
\end{figure*}

\begin{figure*}[htbp]
	\centering	
	\subfigure[\textsf{Taxi} dataset, Uniform $\mathbf{\tilde{f}}$]{
		\includegraphics[width=0.24\textwidth]{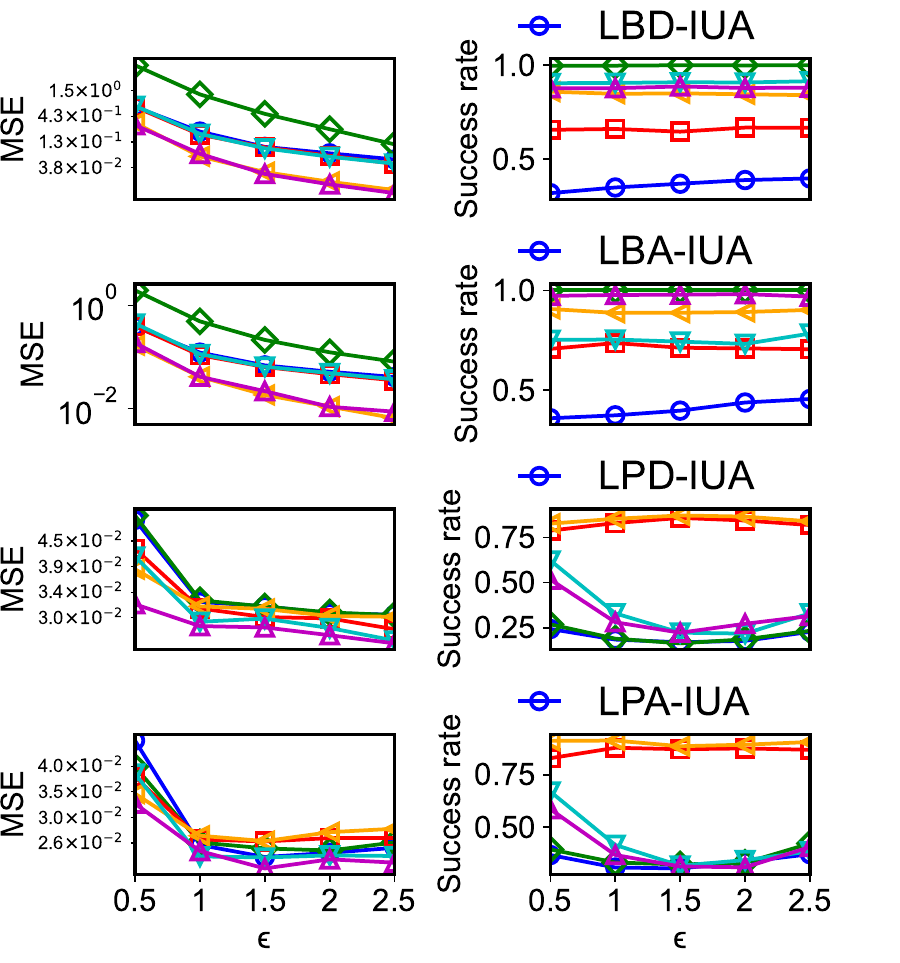}}\vspace{-0.06cm}\hspace{-4mm}
	\subfigure[\textsf{Taxi} dataset, Pulse $\mathbf{\tilde{f}}$]{
		\includegraphics[width=0.24\textwidth]{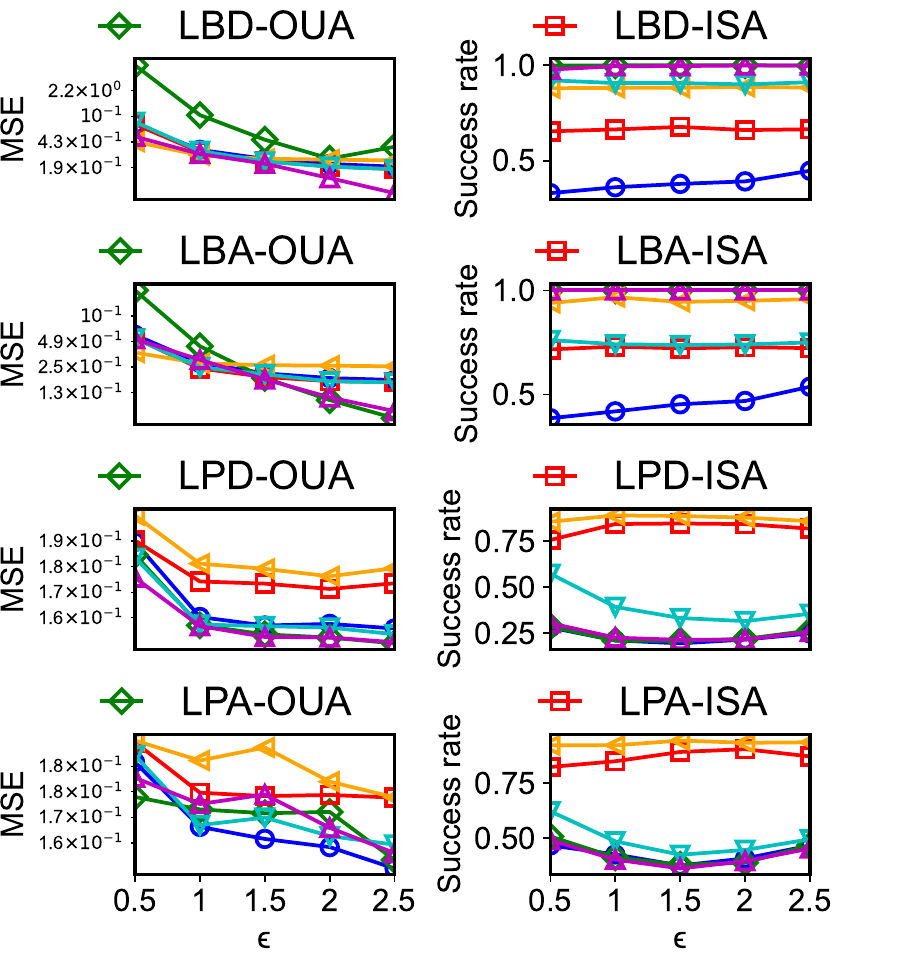}}\vspace{-0.06cm}\hspace{-4mm}
	\subfigure[\textsf{Taxi} dataset, Gaussian $\mathbf{\tilde{f}}$]{
		\includegraphics[width=0.24\textwidth]{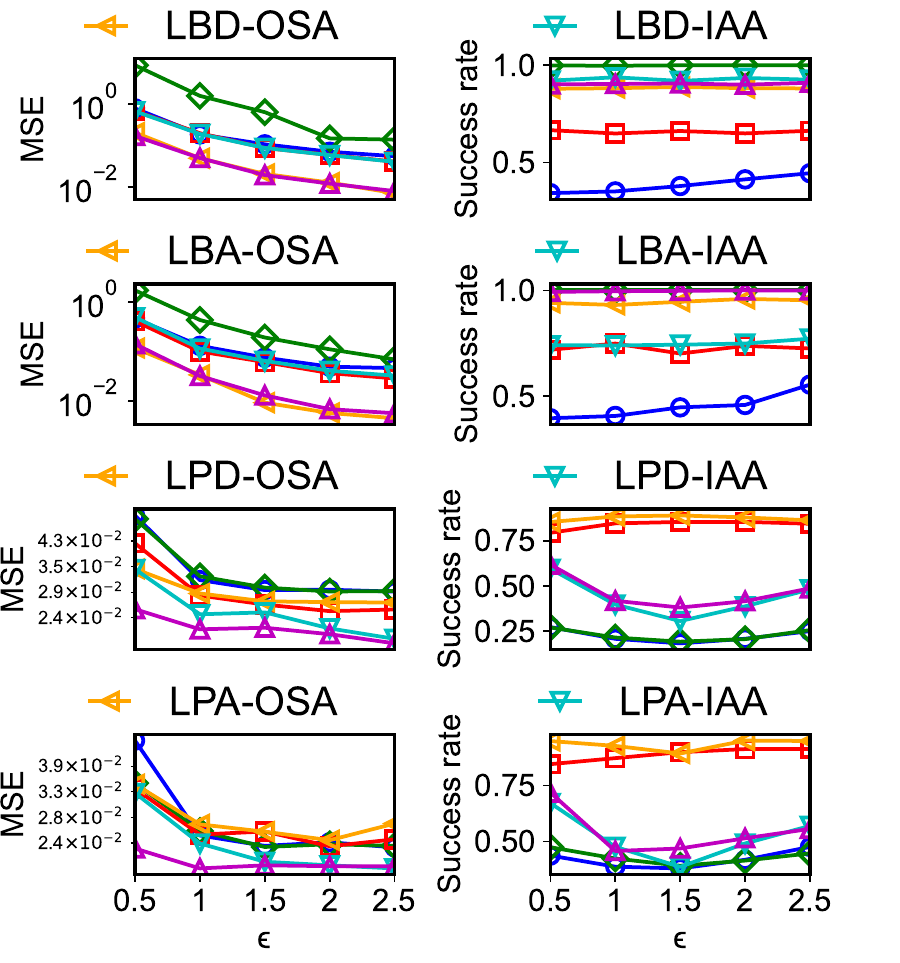}}\vspace{-0.06cm}\hspace{-4mm}
	\subfigure[\textsf{Taxi} dataset, Sigmoid $\mathbf{\tilde{f}}$]{
		\includegraphics[width=0.24\textwidth]{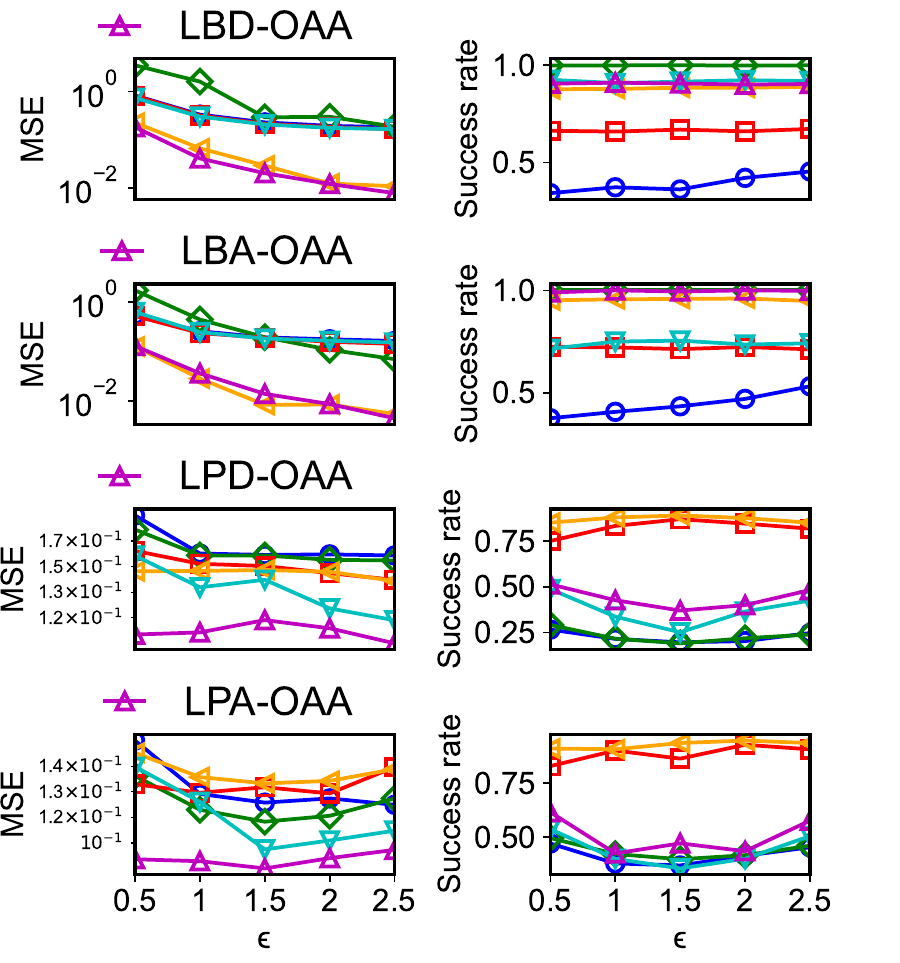}}\vspace{-0.06cm}\hspace{-4mm}
	\subfigure[\scriptsize\textsf{Foursquare} dataset, Uniform $\mathbf{\tilde{f}}$]{
		\includegraphics[width=0.24\textwidth]{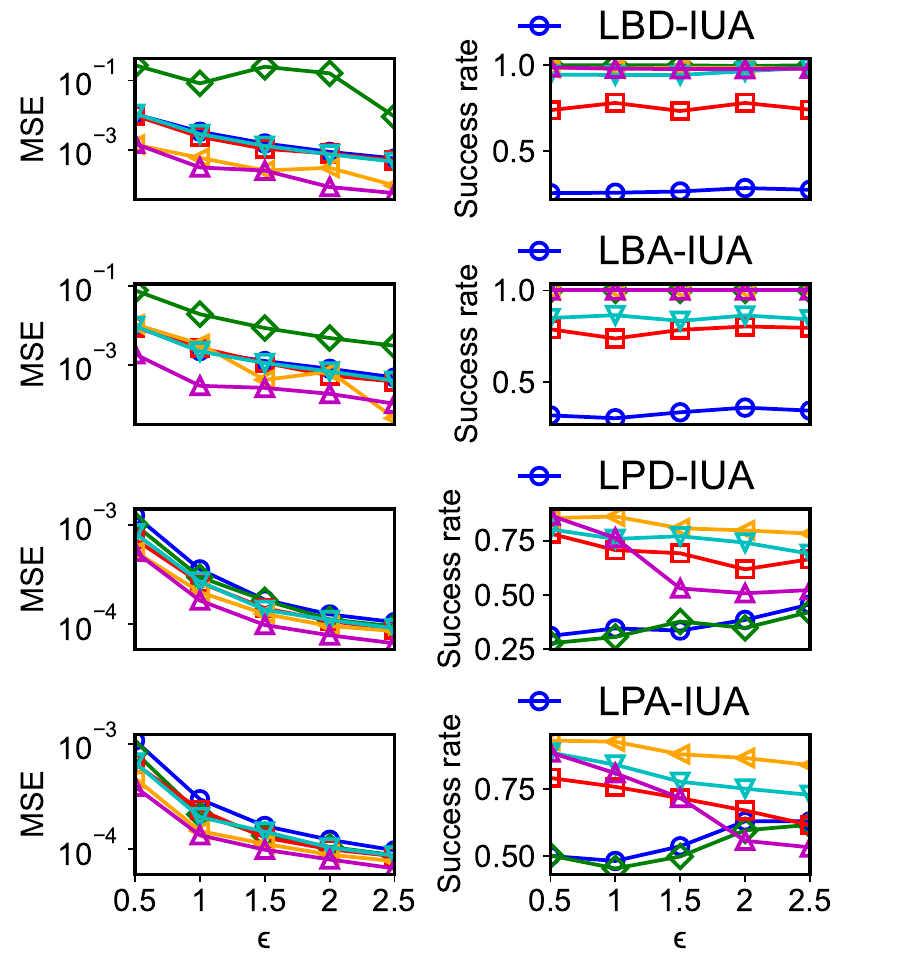}}\vspace{-0.05cm}\hspace{-4mm}
	\subfigure[\scriptsize\textsf{Foursquare} dataset, Pulse $\mathbf{\tilde{f}}$]{
		\includegraphics[width=0.24\textwidth]{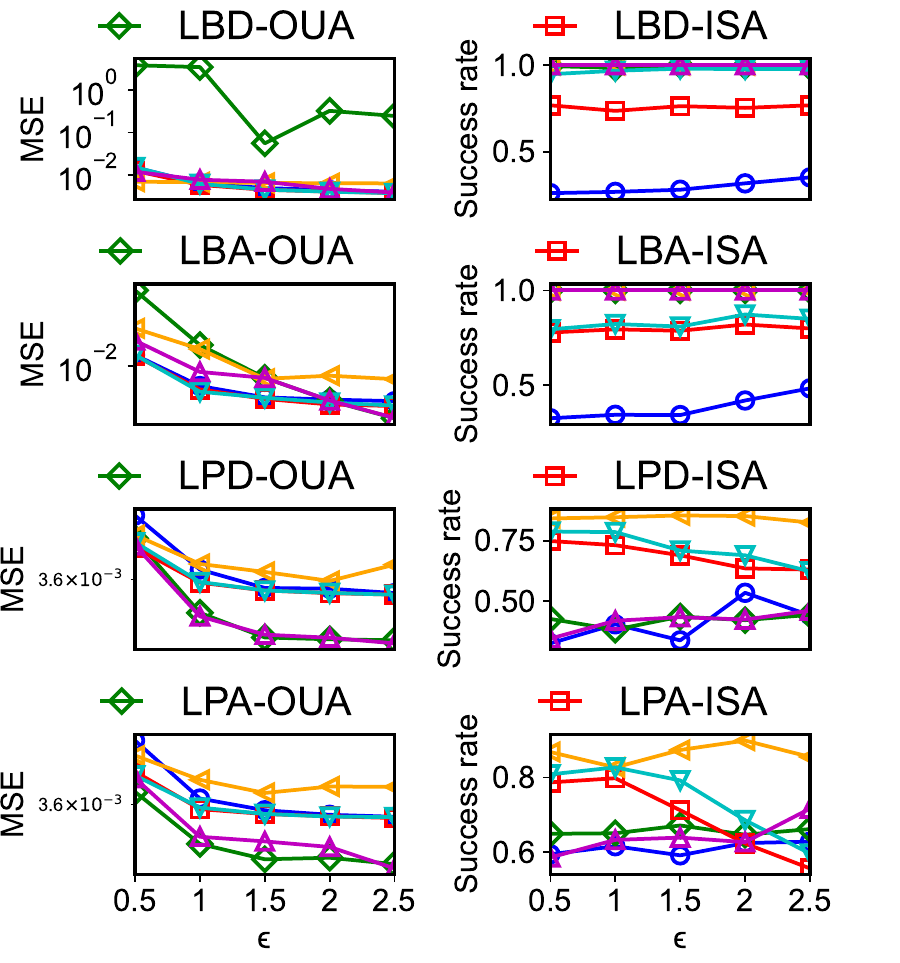}}\vspace{-0.06cm}\hspace{-4mm}
	\subfigure[\scriptsize\textsf{Foursquare} dataset, Gaussian $\mathbf{\tilde{f}}$]{
		\includegraphics[width=0.24\textwidth]{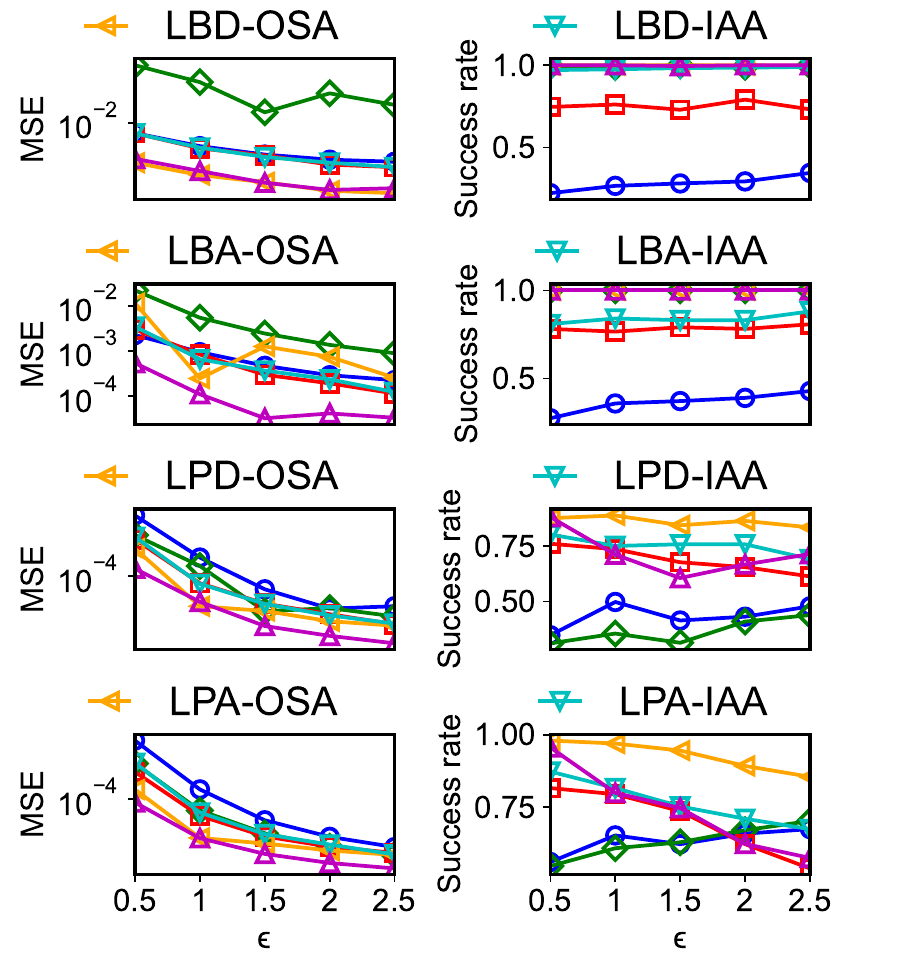}}\vspace{-0.06cm}\hspace{-4mm}
	\subfigure[\scriptsize\textsf{Foursquare} dataset, Sigmoid $\mathbf{\tilde{f}}$]{
		\includegraphics[width=0.24\textwidth]{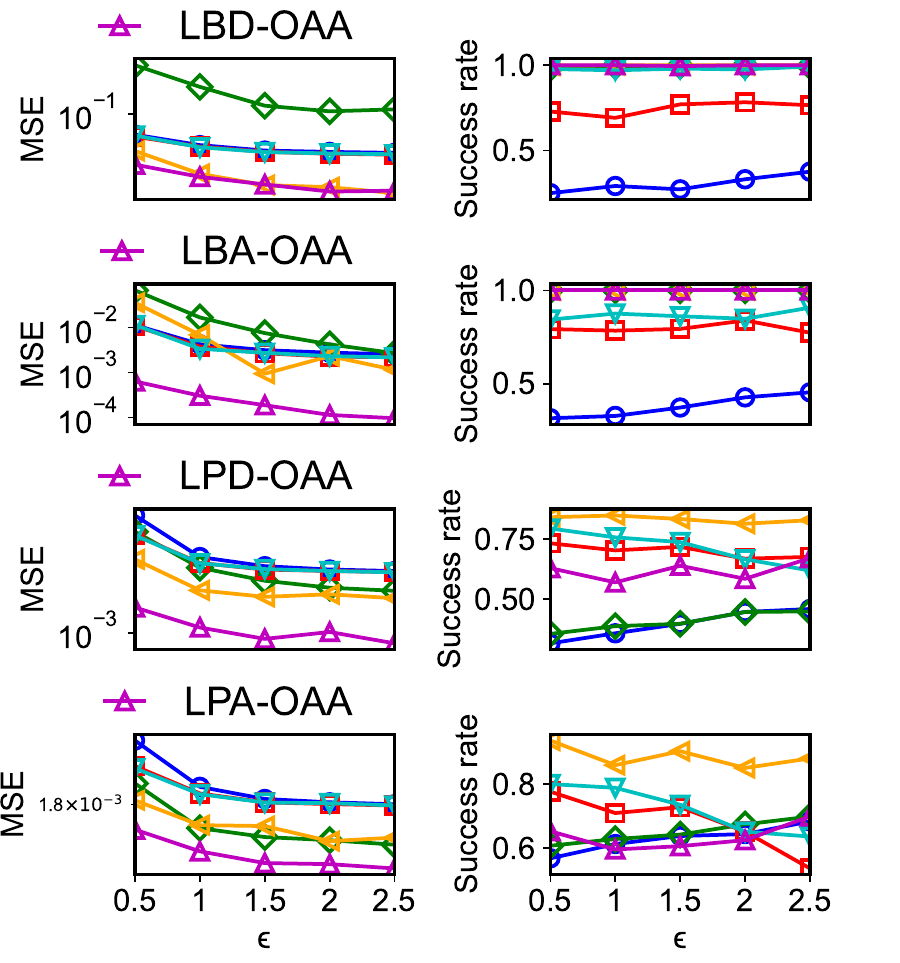}}\vspace{-0.06cm}\hspace{-4mm}
	\subfigure[\textsf{Taobao} dataset, Uniform $\mathbf{\tilde{f}}$]{
		\includegraphics[width=0.24\textwidth]{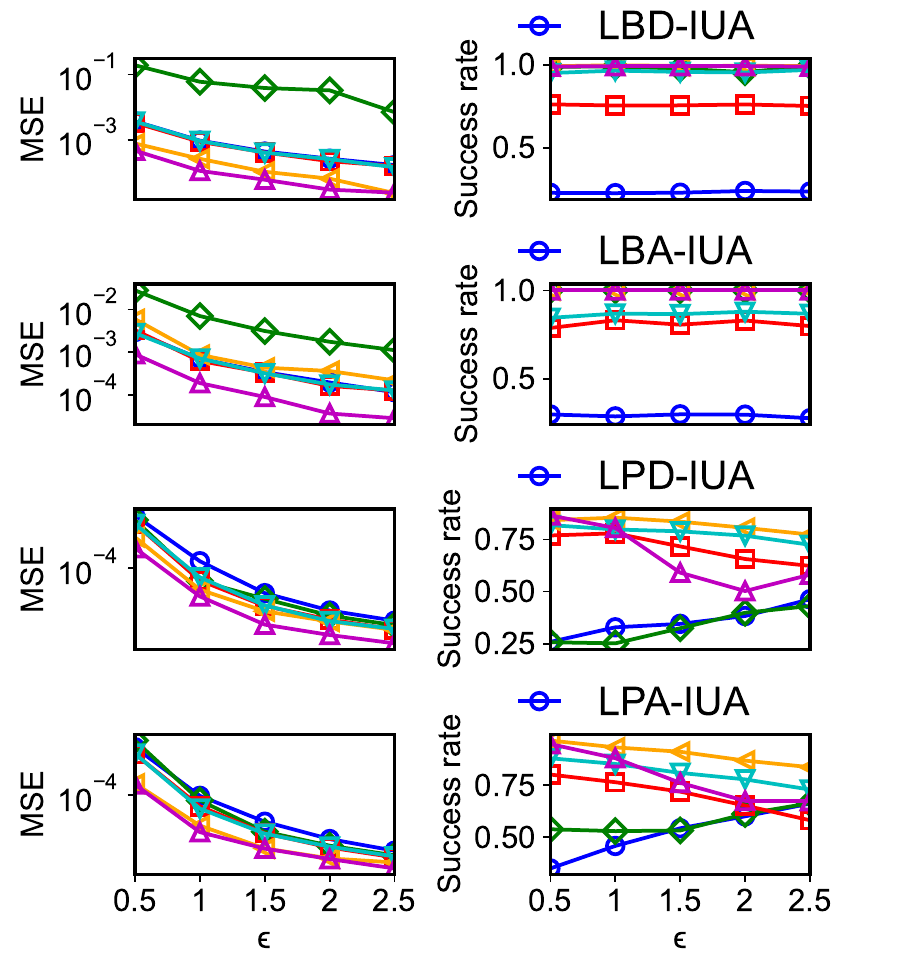}}\vspace{-0.13cm}\hspace{-4mm}
	\subfigure[\textsf{Taobao} dataset, Pulse $\mathbf{\tilde{f}}$]{
		\includegraphics[width=0.24\textwidth]{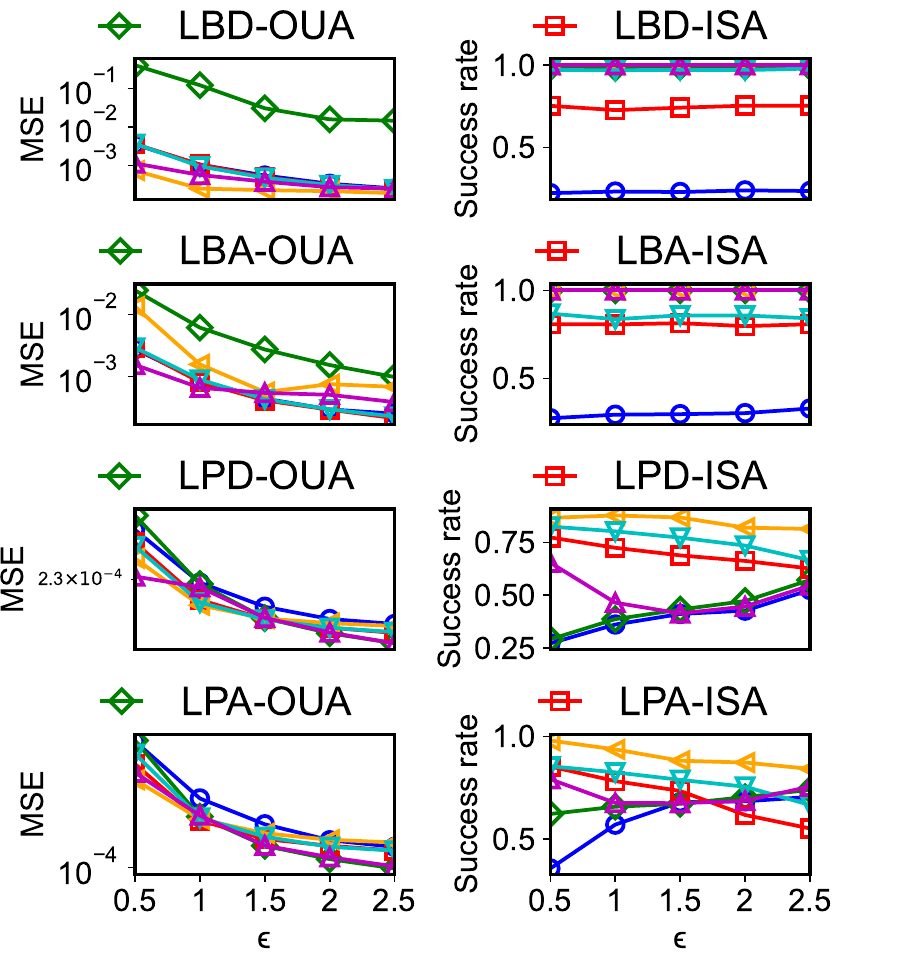}}\vspace{-0.13cm}\hspace{-4mm}
	\subfigure[\textsf{Taobao} dataset, Gaussian $\mathbf{\tilde{f}}$]{
		\includegraphics[width=0.24\textwidth]{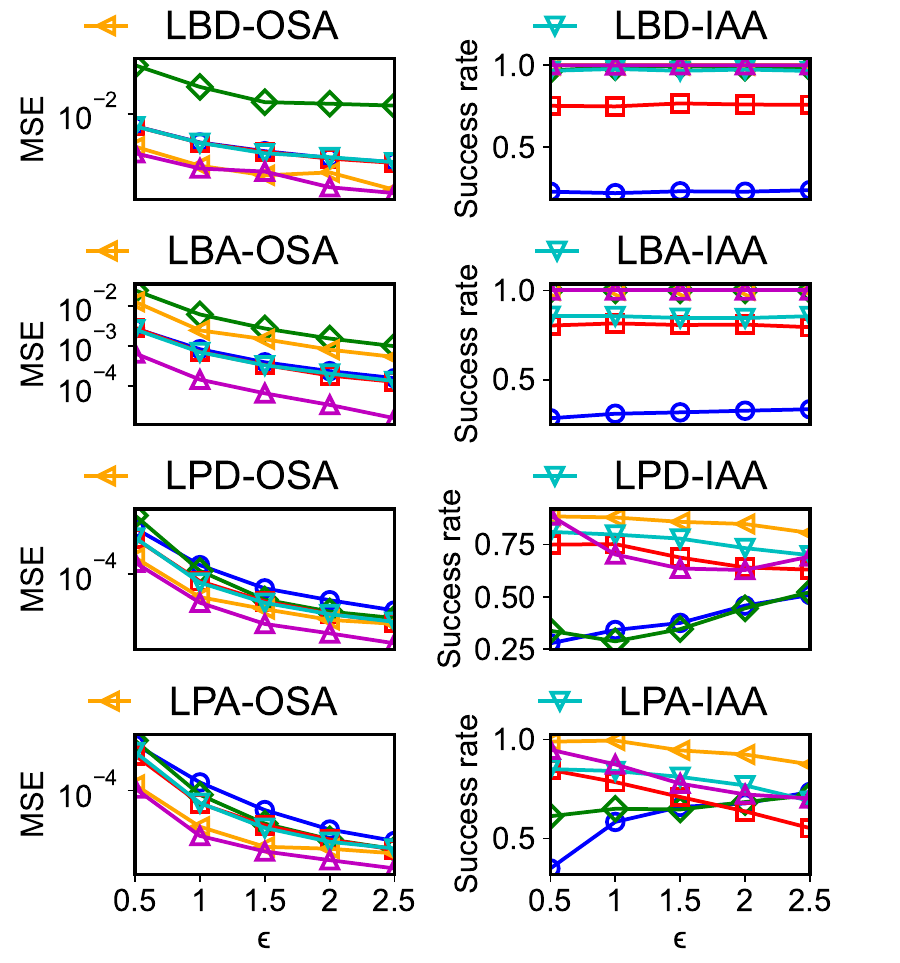}}\vspace{-0.13cm}\hspace{-4mm}
	\subfigure[\textsf{Taobao} dataset, Sigmoid $\mathbf{\tilde{f}}$]{
		\includegraphics[width=0.24\textwidth]{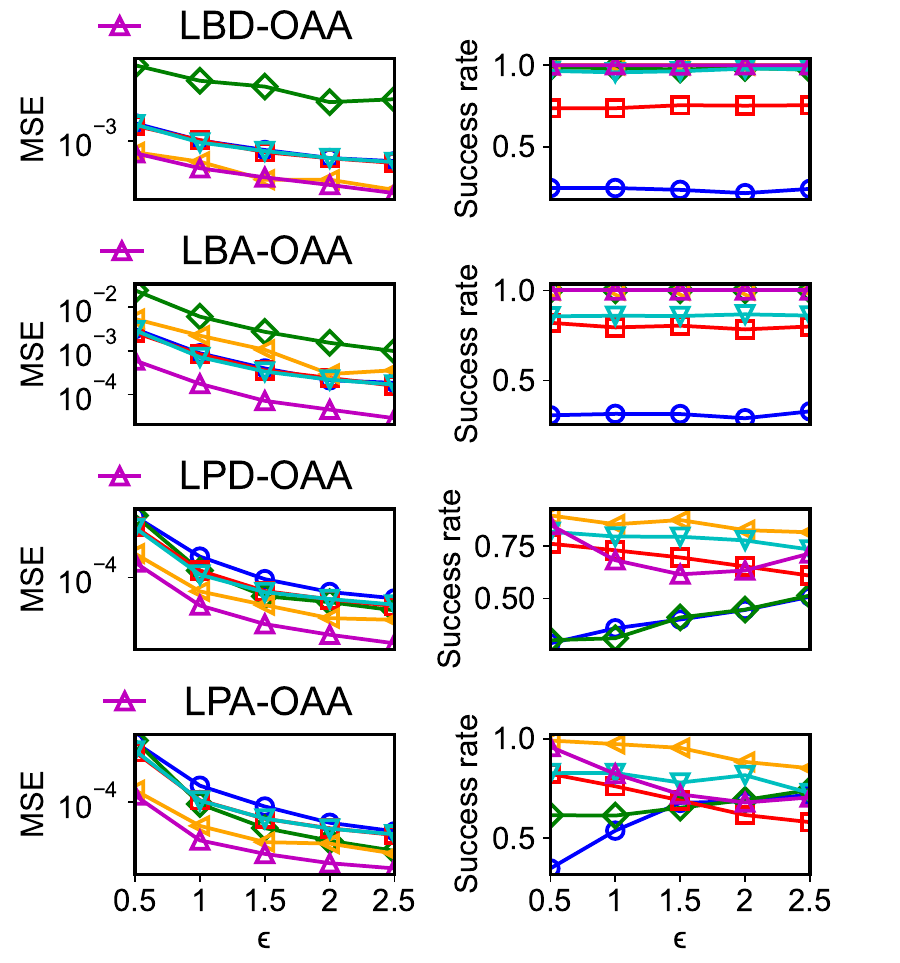}}
	\caption{\small Attack effectiveness for real-world datasets with varying $\epsilon$. }\centering
	\label{fig:epsilon} 
	\vspace{-0.5cm}
\end{figure*}

\begin{figure*}[htbp]
	\centering	
	\subfigure[\textsf{Taxi} dataset, Uniform $\mathbf{\tilde{f}}$]{
		\includegraphics[width=0.24\textwidth]{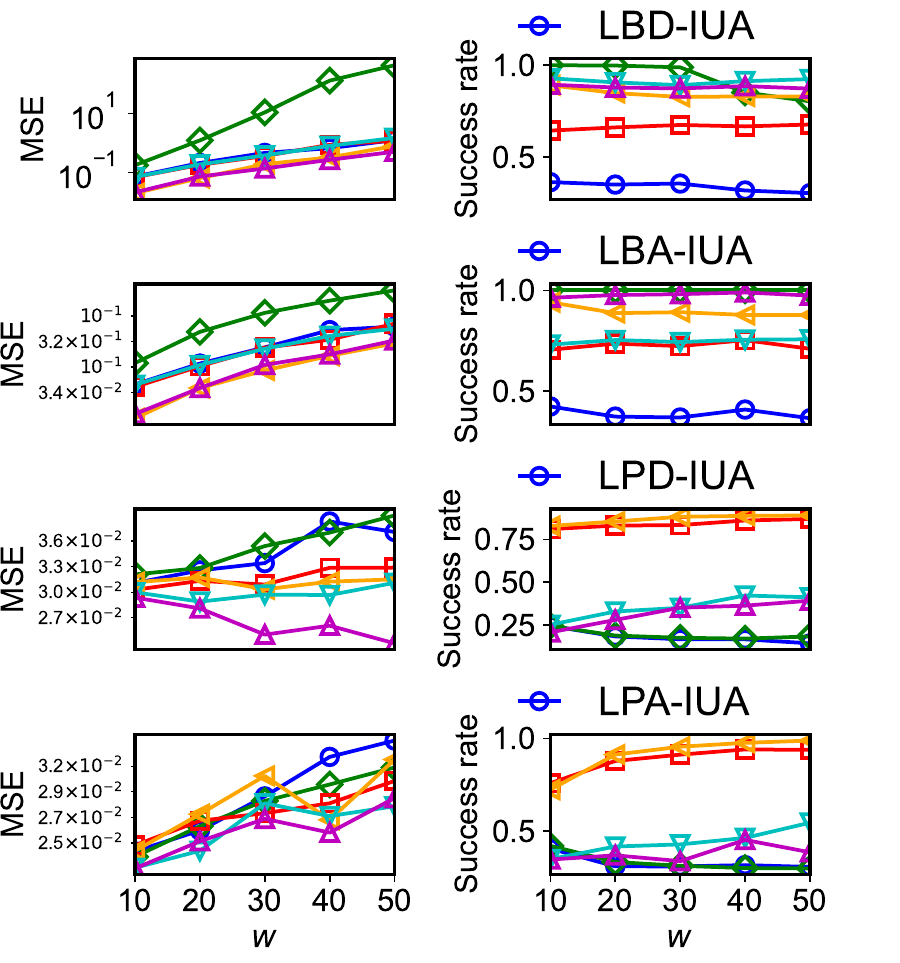}}\vspace{-0.06cm}\hspace{-4mm}
	\subfigure[\textsf{Taxi} dataset, Pulse $\mathbf{\tilde{f}}$]{
		\includegraphics[width=0.24\textwidth]{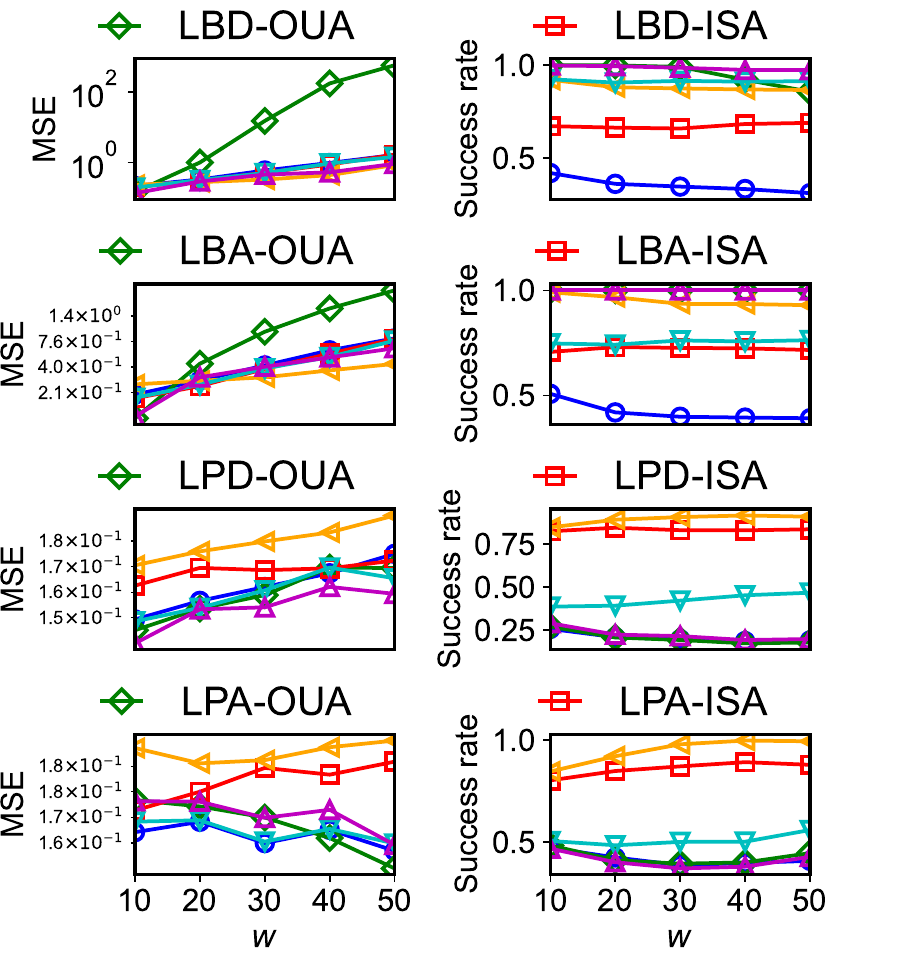}}\vspace{-0.06cm}\hspace{-4mm}
	\subfigure[\textsf{Taxi} dataset, Gaussian $\mathbf{\tilde{f}}$]{
		\includegraphics[width=0.24\textwidth]{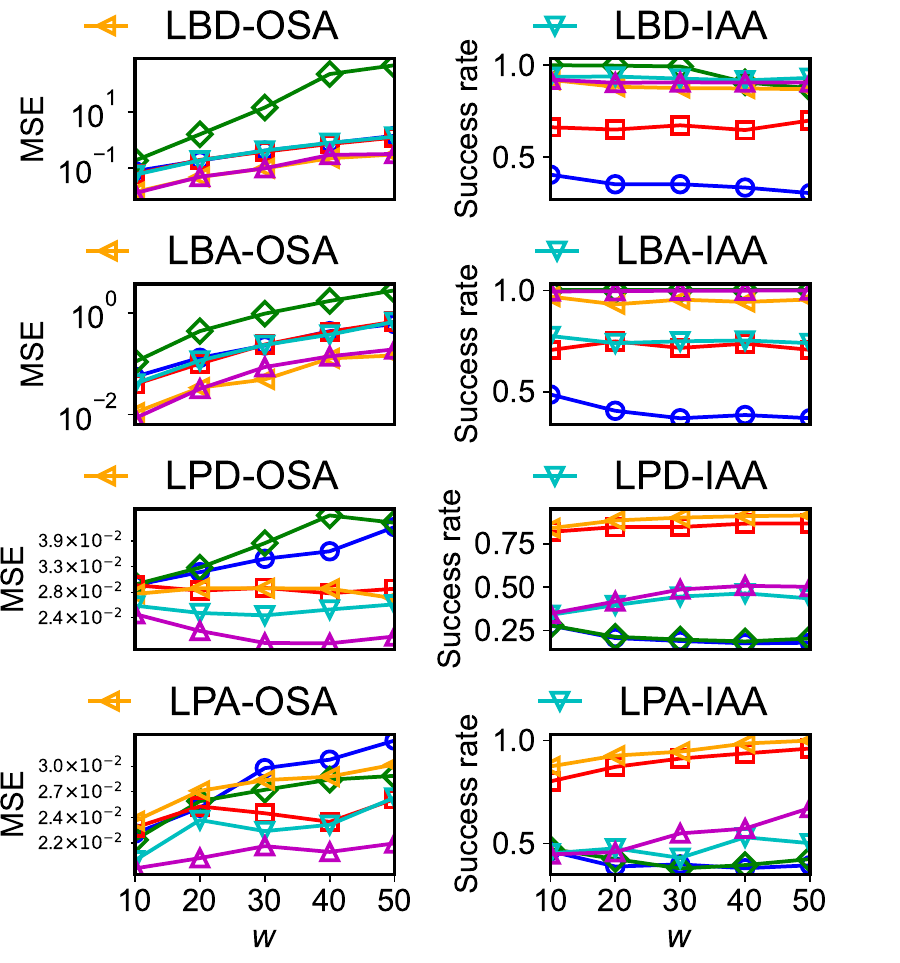}}\vspace{-0.06cm}\hspace{-4mm}
	\subfigure[\textsf{Taxi} dataset, Sigmoid $\mathbf{\tilde{f}}$]{
		\includegraphics[width=0.24\textwidth]{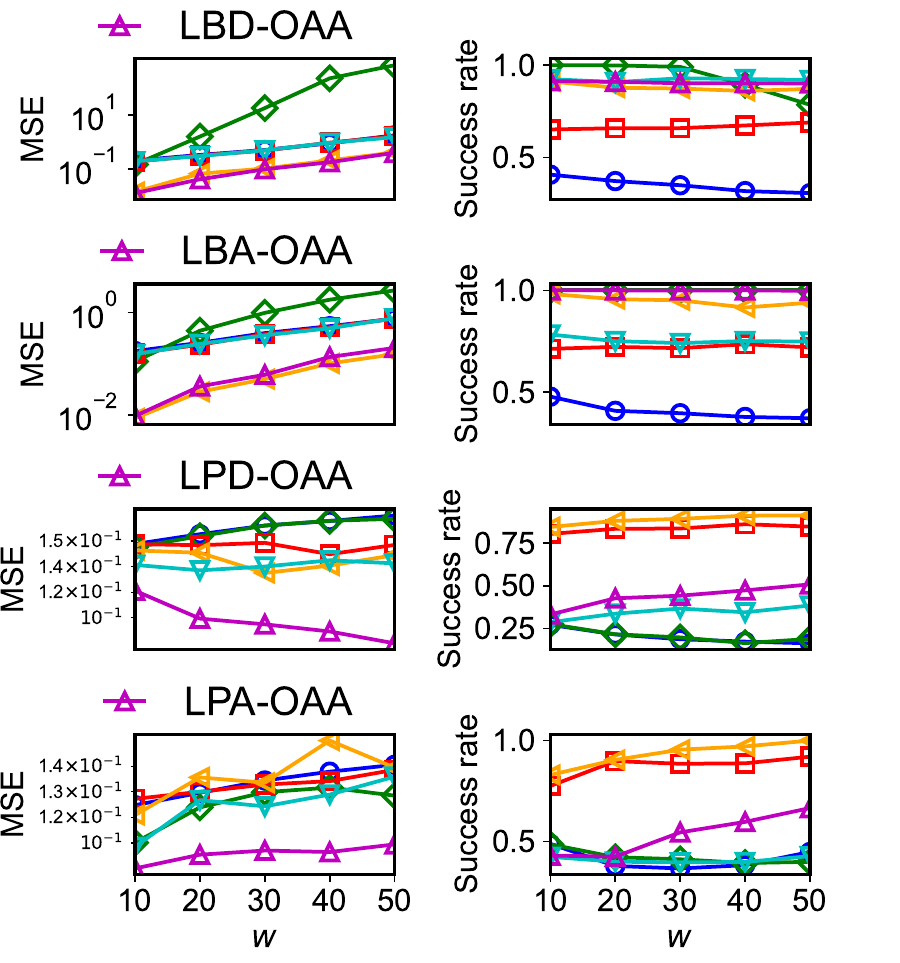}}\vspace{-0.06cm}\hspace{-4mm}
	\subfigure[\scriptsize\textsf{Foursquare} dataset, Uniform $\mathbf{\tilde{f}}$]{
		\includegraphics[width=0.24\textwidth]{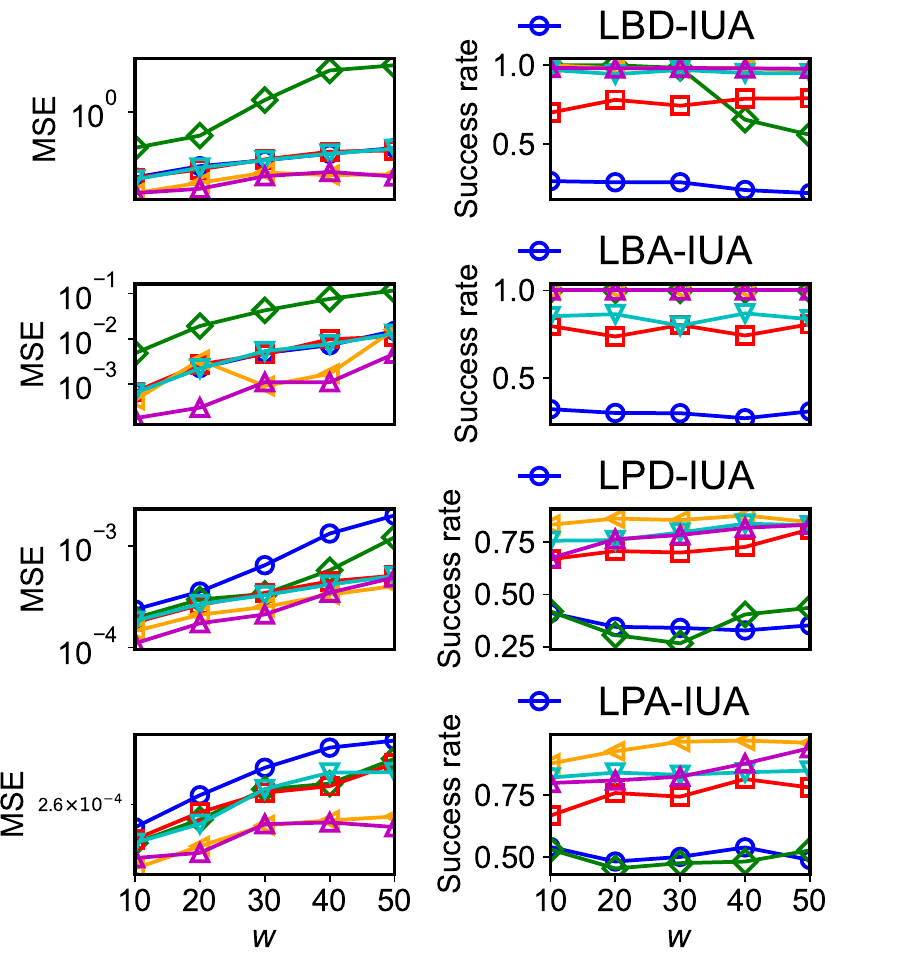}}\vspace{-0.05cm}\hspace{-4mm}
	\subfigure[\scriptsize\textsf{Foursquare} dataset, Pulse $\mathbf{\tilde{f}}$]{
		\includegraphics[width=0.24\textwidth]{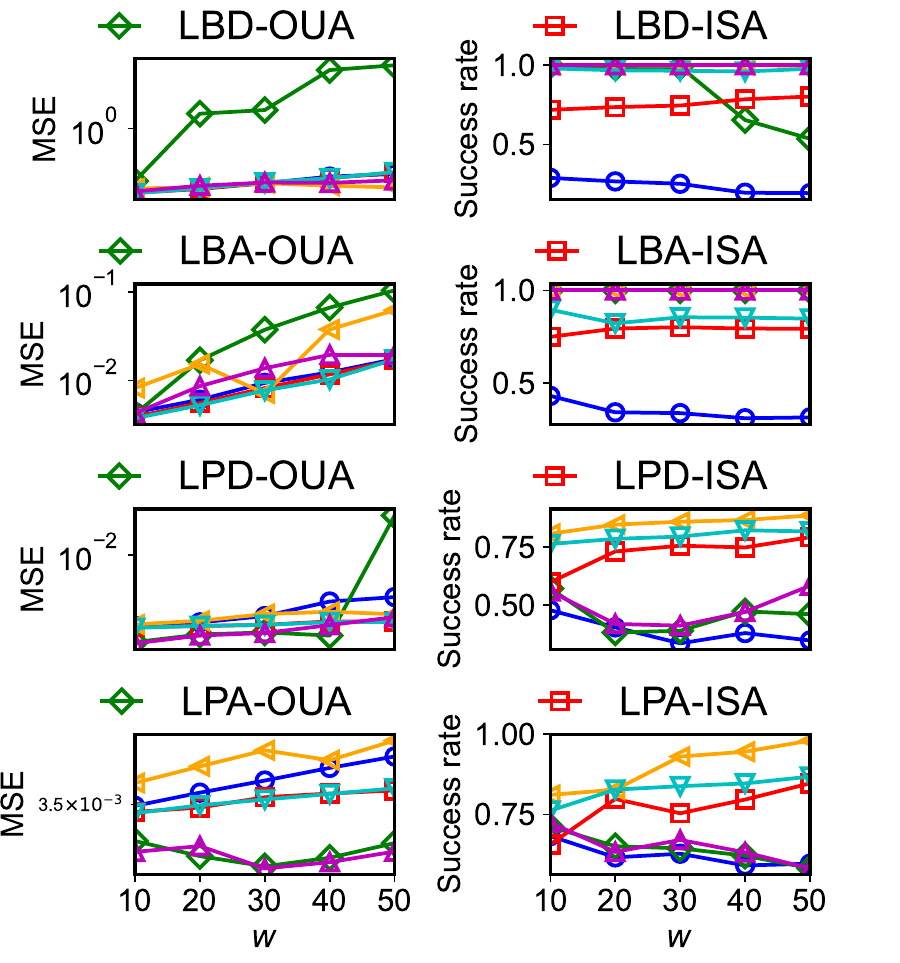}}\vspace{-0.06cm}\hspace{-4mm}
	\subfigure[\scriptsize\textsf{Foursquare} dataset, Gaussian $\mathbf{\tilde{f}}$]{
		\includegraphics[width=0.24\textwidth]{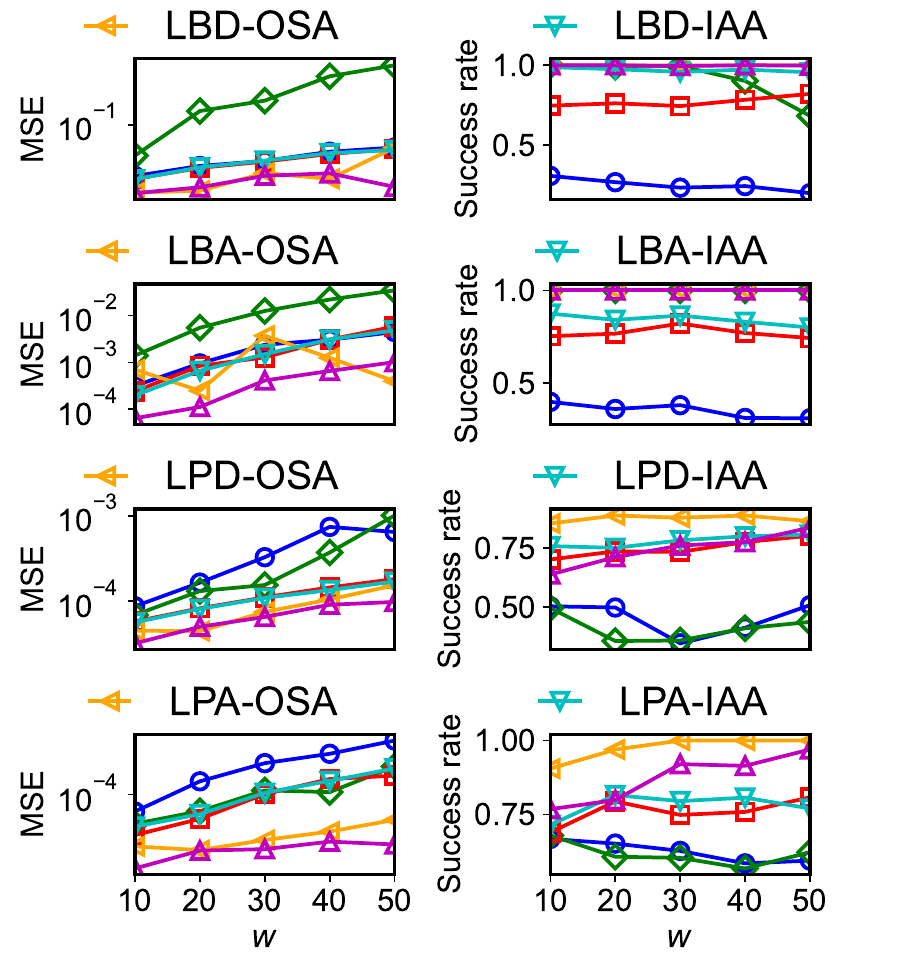}}\vspace{-0.06cm}\hspace{-4mm}
	\subfigure[\scriptsize\textsf{Foursquare} dataset, Sigmoid $\mathbf{\tilde{f}}$]{
		\includegraphics[width=0.24\textwidth]{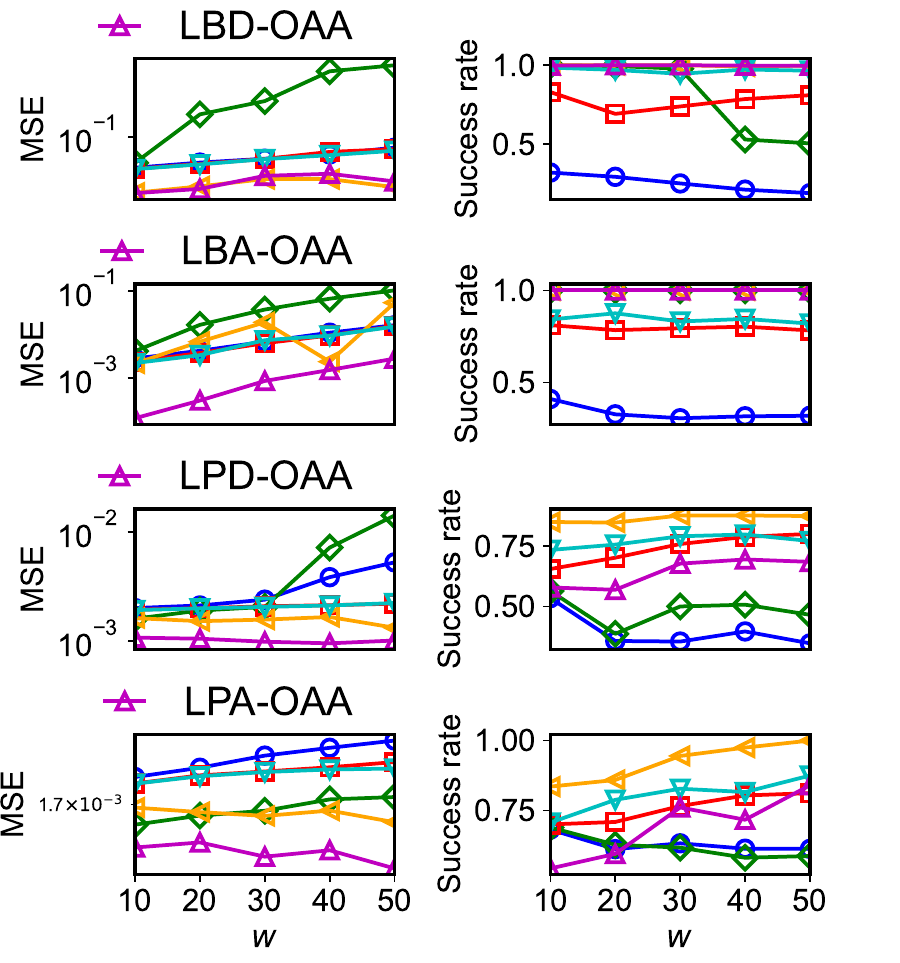}}\vspace{-0.06cm}\hspace{-4mm}
	\subfigure[\textsf{Taobao} dataset, Uniform $\mathbf{\tilde{f}}$]{
		\includegraphics[width=0.24\textwidth]{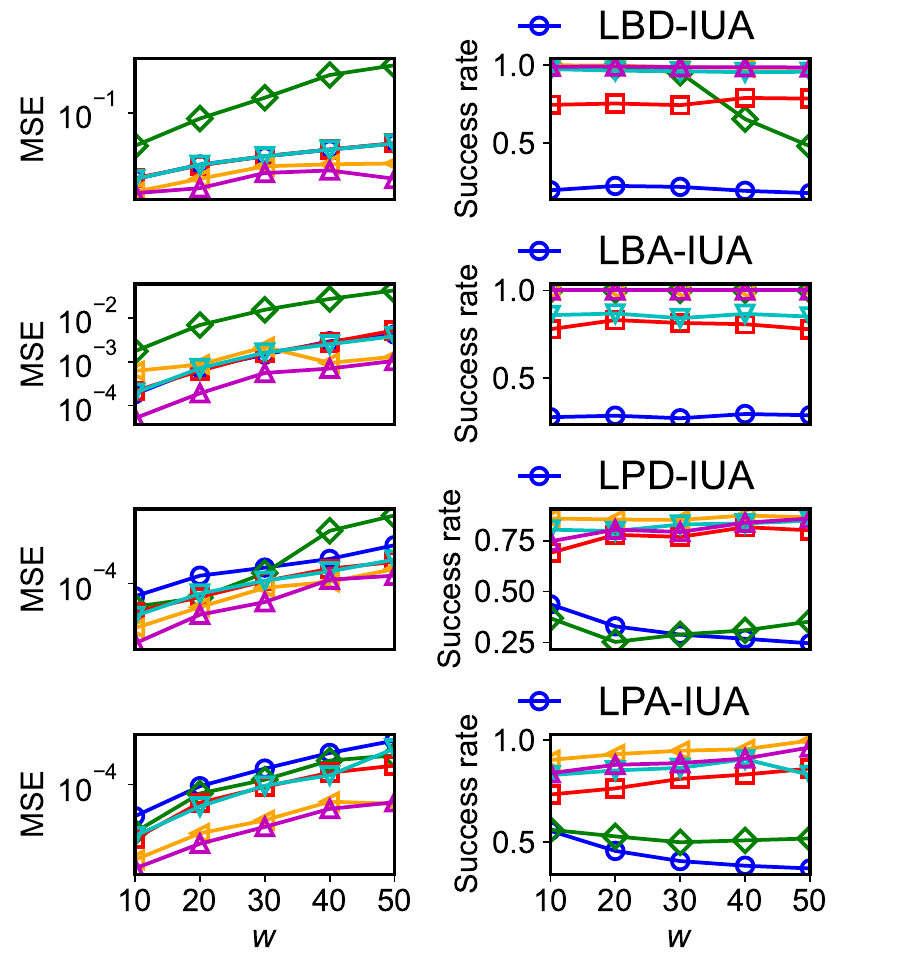}}\vspace{-0.13cm}\hspace{-4mm}
	\subfigure[\textsf{Taobao} dataset, Pulse $\mathbf{\tilde{f}}$]{
		\includegraphics[width=0.24\textwidth]{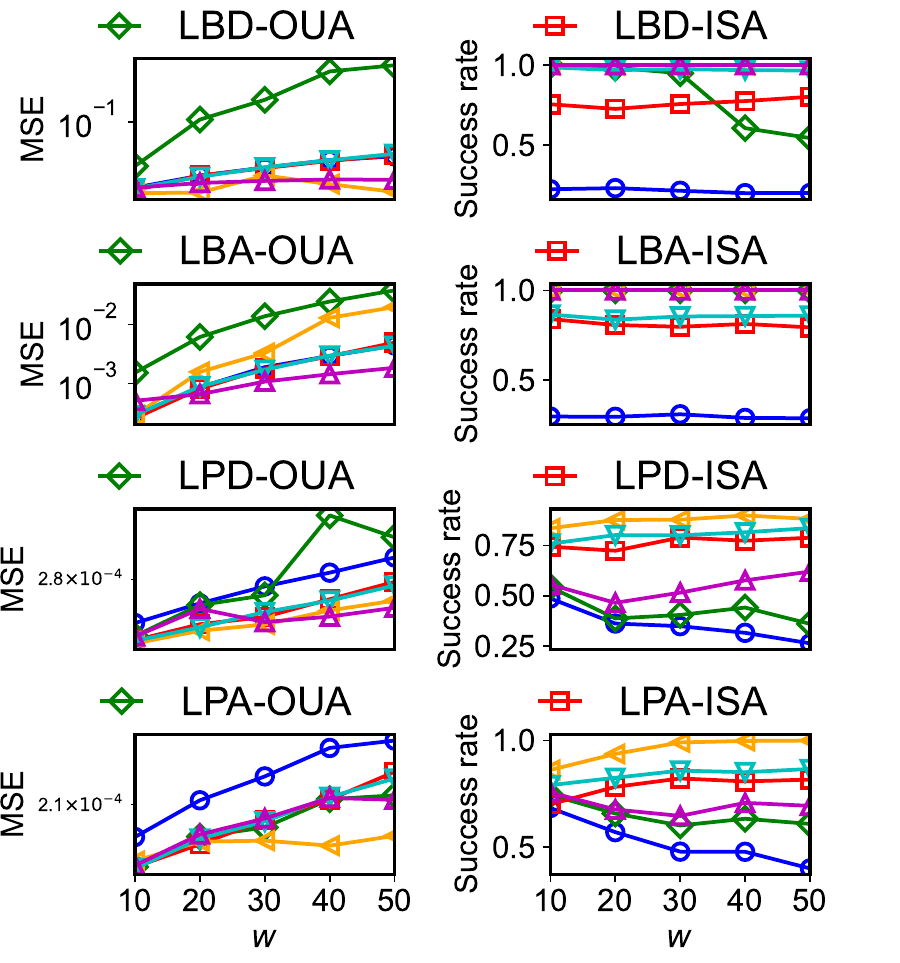}}\vspace{-0.13cm}\hspace{-4mm}
	\subfigure[\textsf{Taobao} dataset, Gaussian $\mathbf{\tilde{f}}$]{
		\includegraphics[width=0.24\textwidth]{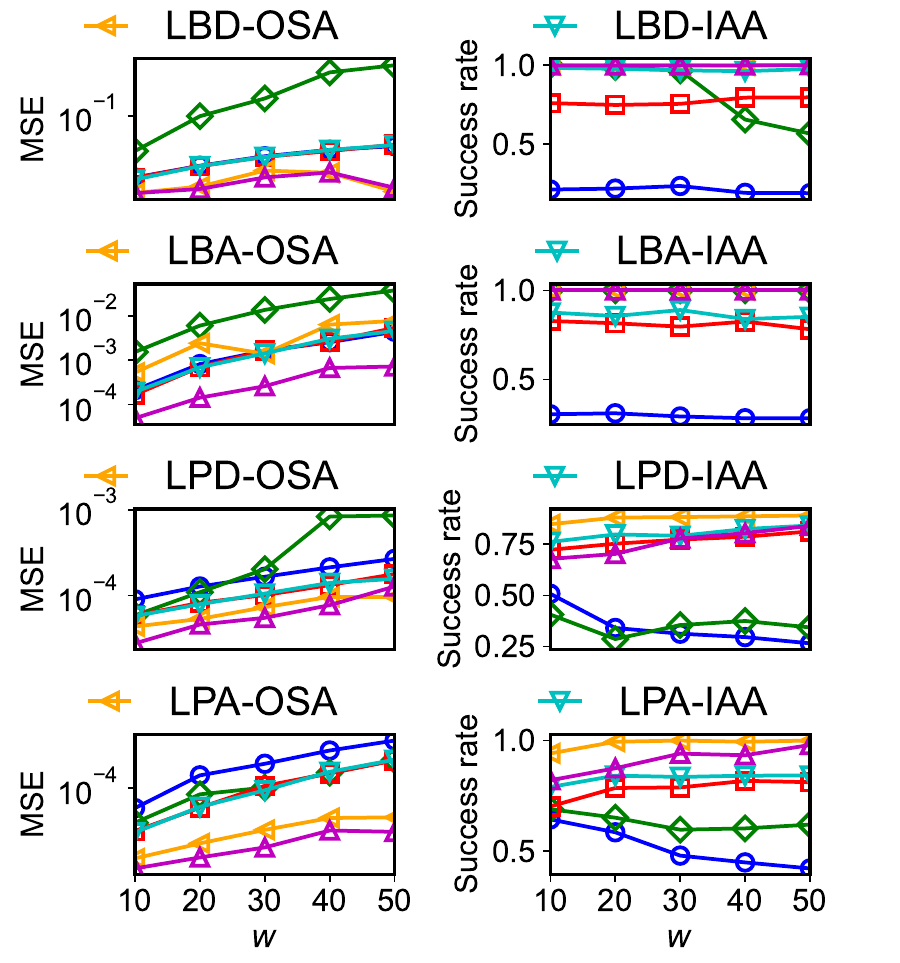}}\vspace{-0.13cm}\hspace{-4mm}
	\subfigure[\textsf{Taobao} dataset, Sigmoid $\mathbf{\tilde{f}}$]{
		\includegraphics[width=0.24\textwidth]{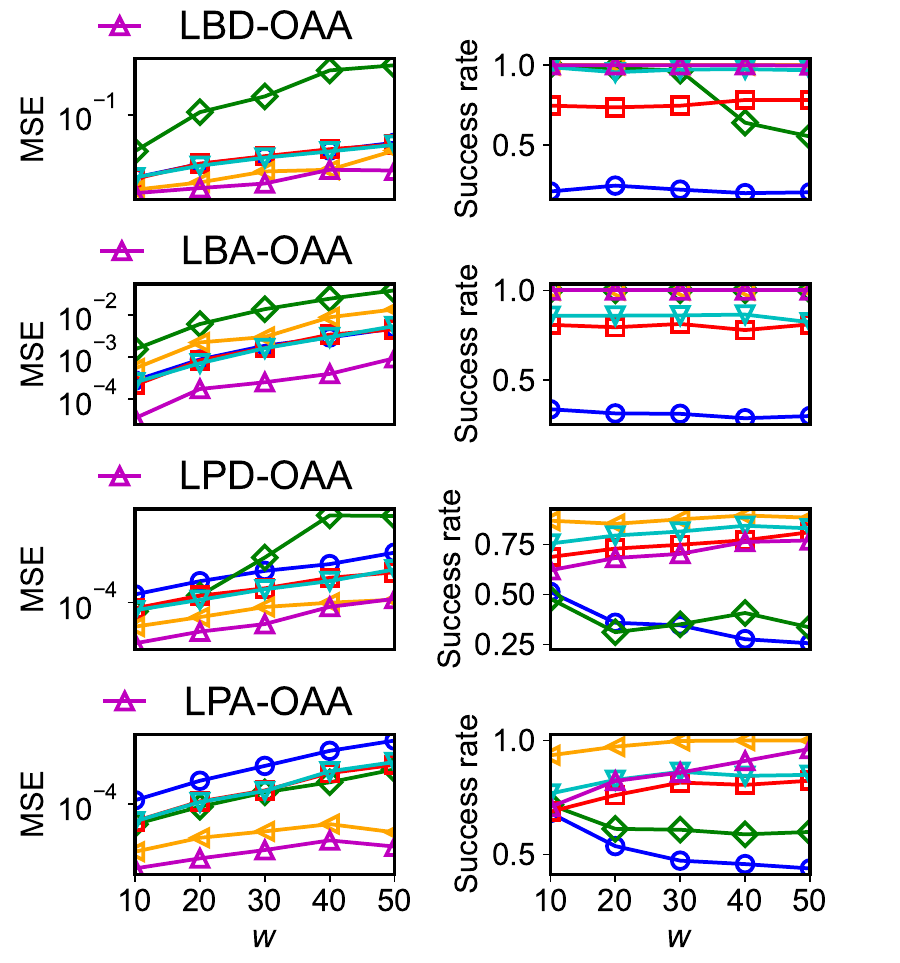}}
	\caption{\small Attacking effectiveness for real-world datasets with varying $w$.
	}\centering
	\label{fig:w} 
	\vspace{-0.5cm}
\end{figure*}

\begin{figure*}[htbp]
	\centering	
	\subfigure[\textsf{Taxi} dataset, Uniform $\mathbf{\tilde{f}}$]{
		\includegraphics[width=0.24\textwidth]{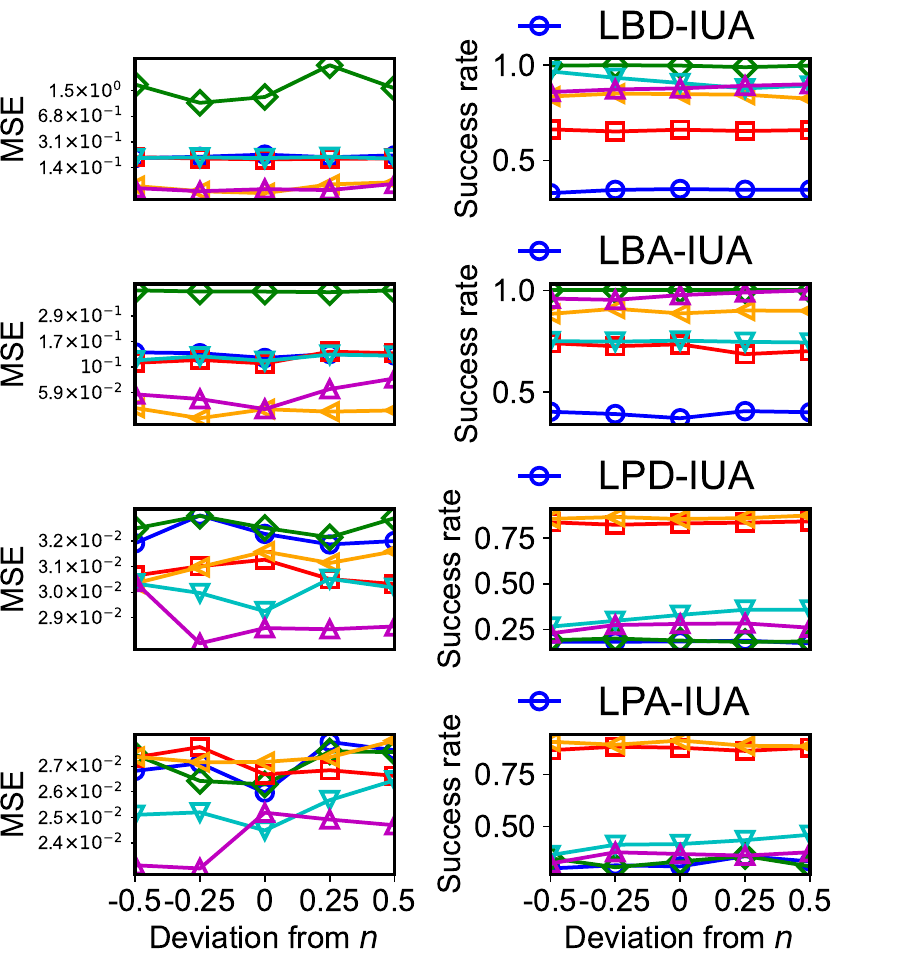}}\vspace{-0.06cm}\hspace{-4mm}
	\subfigure[\textsf{Taxi} dataset, Pulse $\mathbf{\tilde{f}}$]{
		\includegraphics[width=0.24\textwidth]{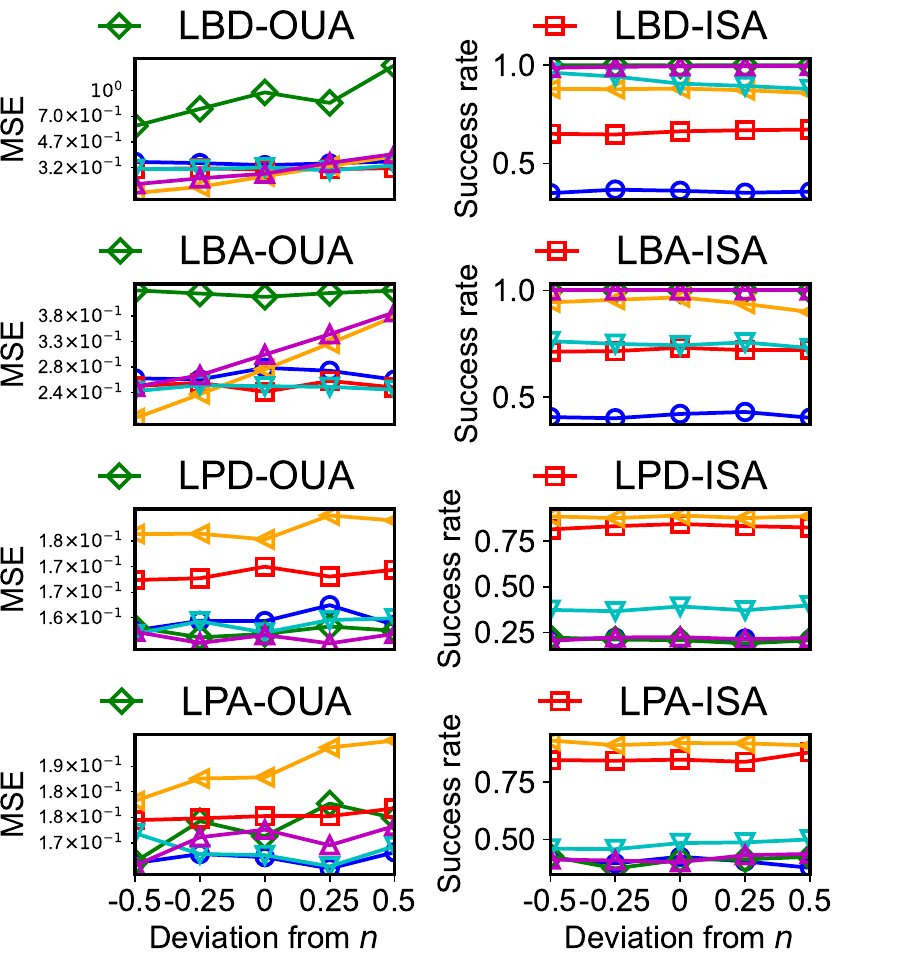}}\vspace{-0.06cm}\hspace{-4mm}
	\subfigure[\textsf{Taxi} dataset, Gaussian $\mathbf{\tilde{f}}$]{
		\includegraphics[width=0.24\textwidth]{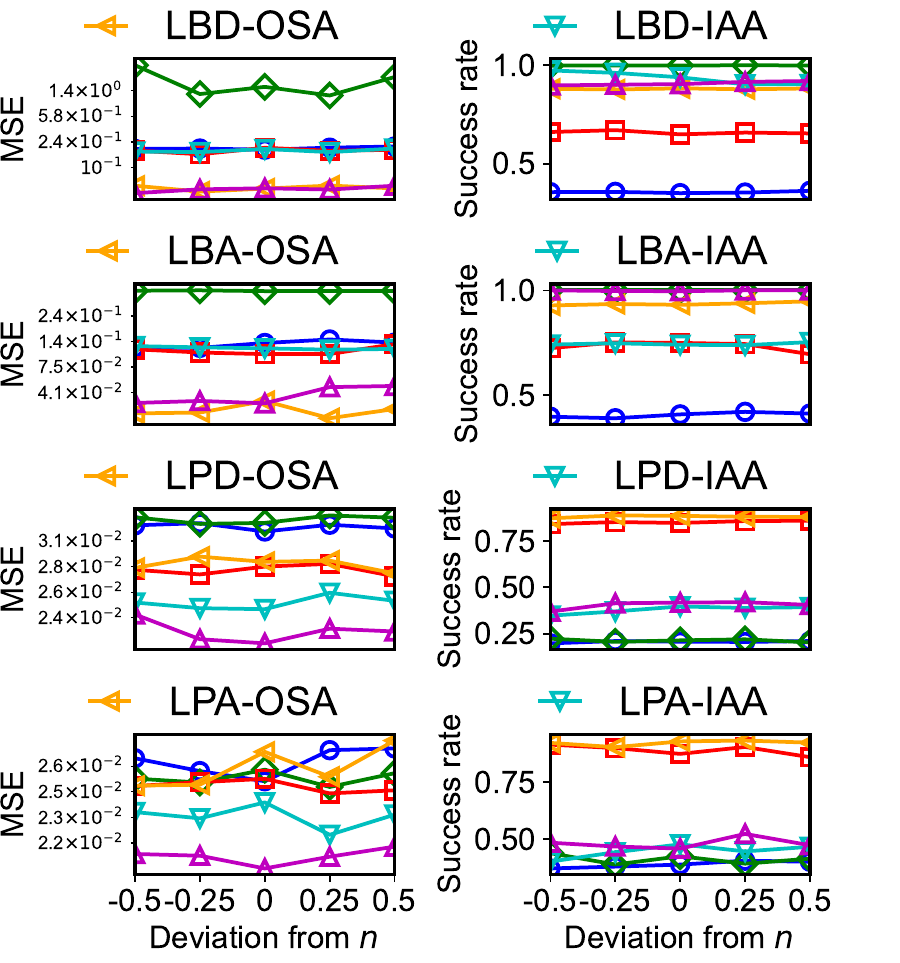}}\vspace{-0.06cm}\hspace{-4mm}
	\subfigure[\textsf{Taxi} dataset, Sigmoid $\mathbf{\tilde{f}}$]{
		\includegraphics[width=0.24\textwidth]{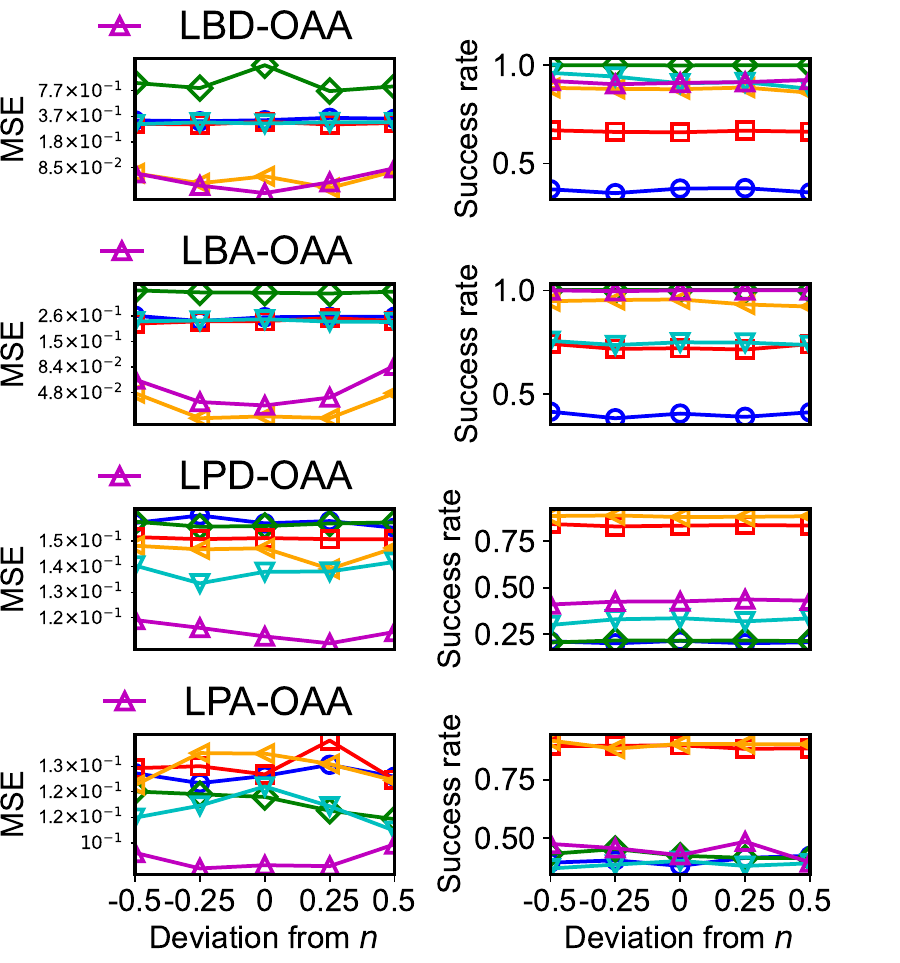}}\vspace{-0.06cm}\hspace{-4mm}
	\subfigure[\scriptsize\textsf{Foursquare} dataset, Uniform $\mathbf{\tilde{f}}$]{
		\includegraphics[width=0.24\textwidth]{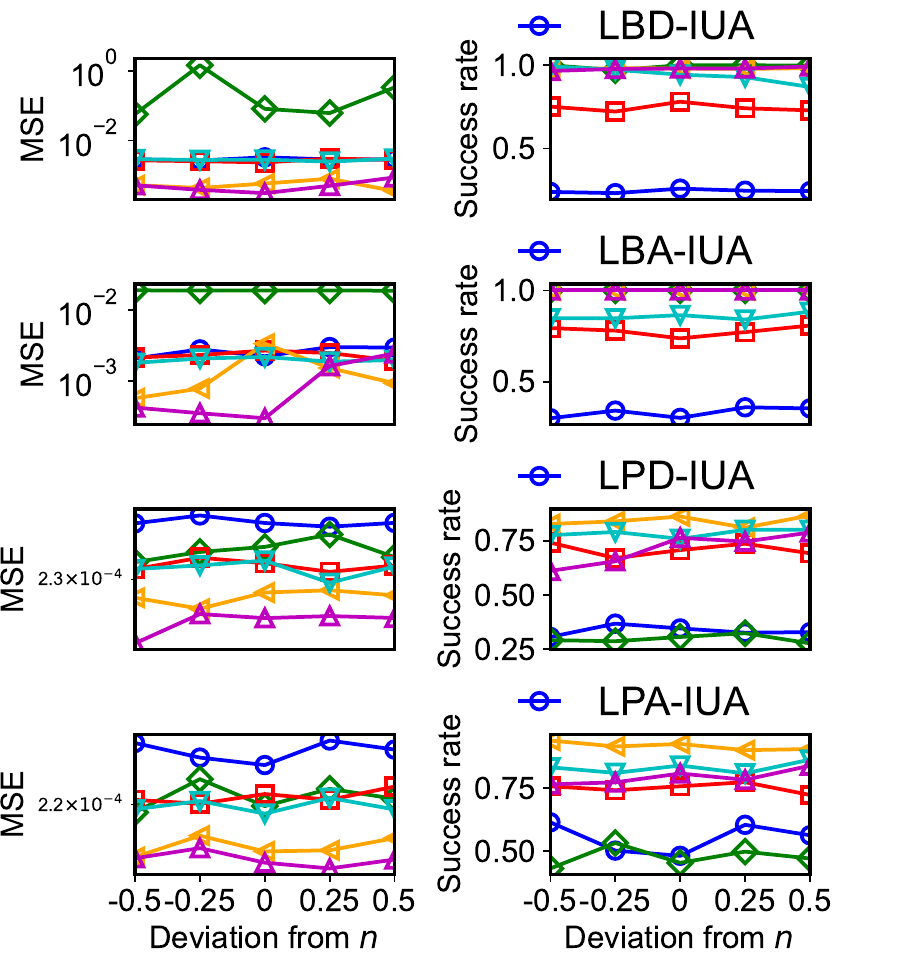}}\vspace{-0.05cm}\hspace{-4mm}
	\subfigure[\scriptsize\textsf{Foursquare} dataset, Pulse $\mathbf{\tilde{f}}$]{
		\includegraphics[width=0.24\textwidth]{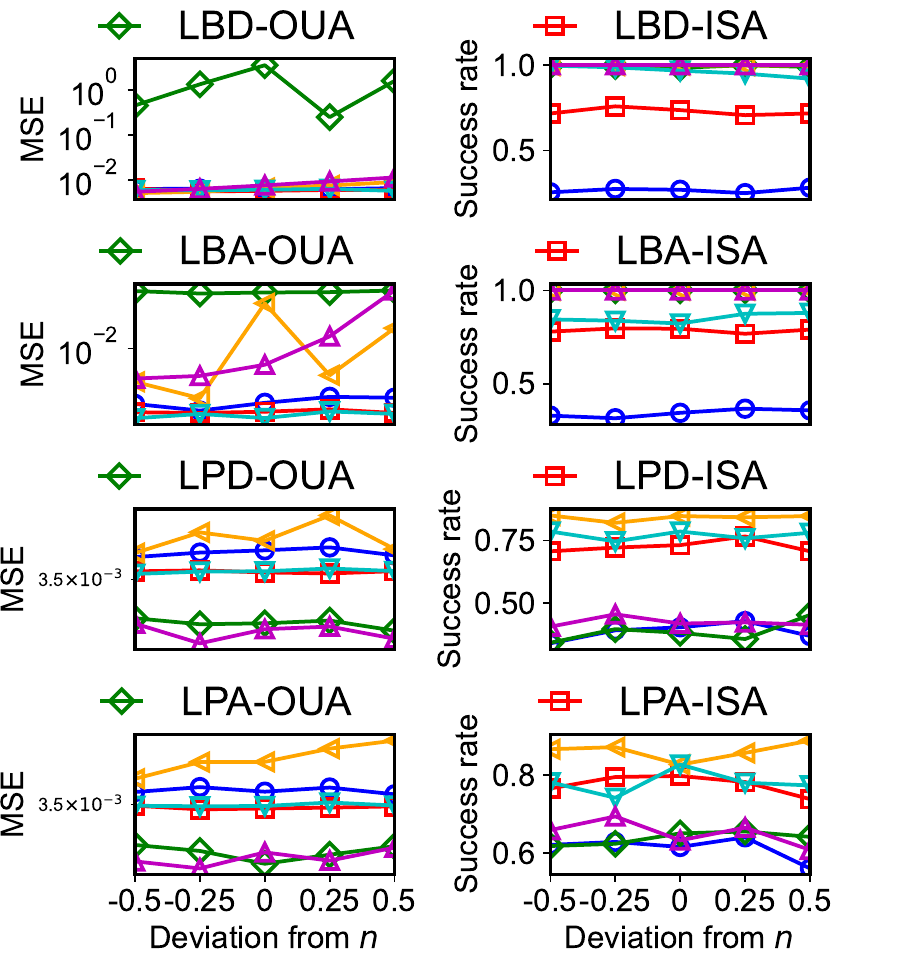}}\vspace{-0.06cm}\hspace{-4mm}
	\subfigure[\scriptsize\textsf{Foursquare} dataset, Gaussian $\mathbf{\tilde{f}}$]{
		\includegraphics[width=0.24\textwidth]{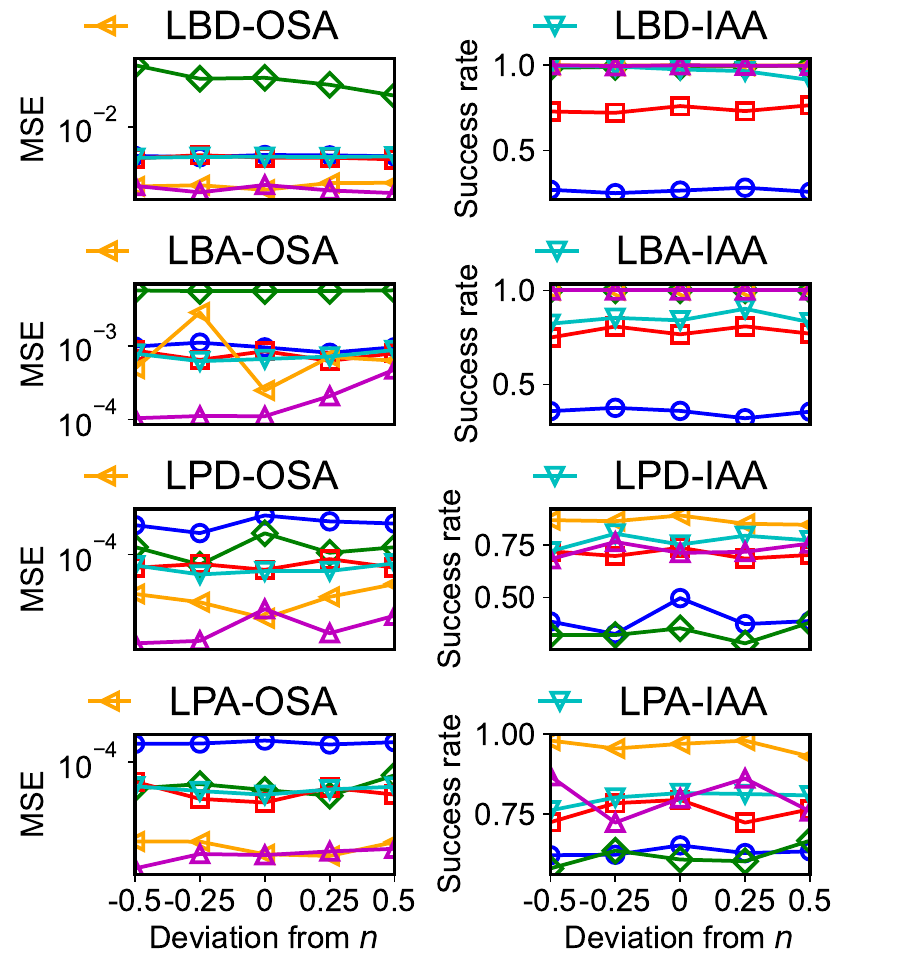}}\vspace{-0.06cm}\hspace{-4mm}
	\subfigure[\scriptsize\textsf{Foursquare} dataset, Sigmoid $\mathbf{\tilde{f}}$]{
		\includegraphics[width=0.24\textwidth]{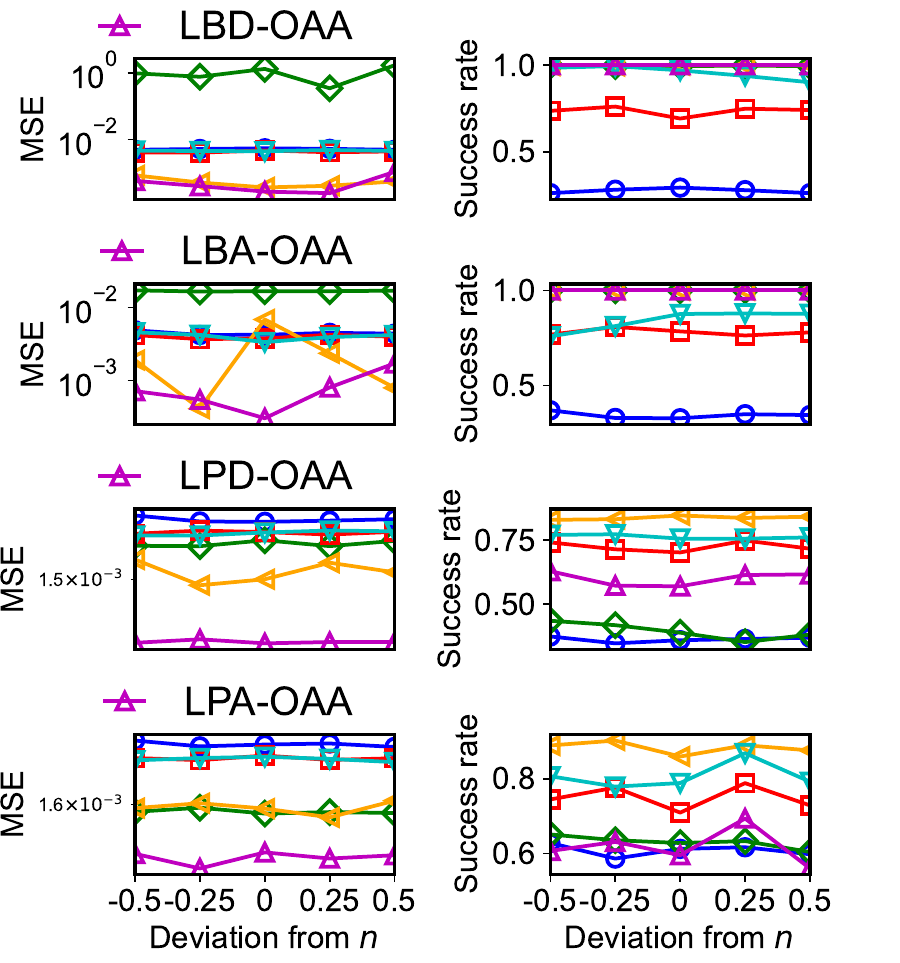}}\vspace{-0.06cm}\hspace{-4mm}
	\subfigure[\textsf{Taobao} dataset, Uniform $\mathbf{\tilde{f}}$]{
		\includegraphics[width=0.24\textwidth]{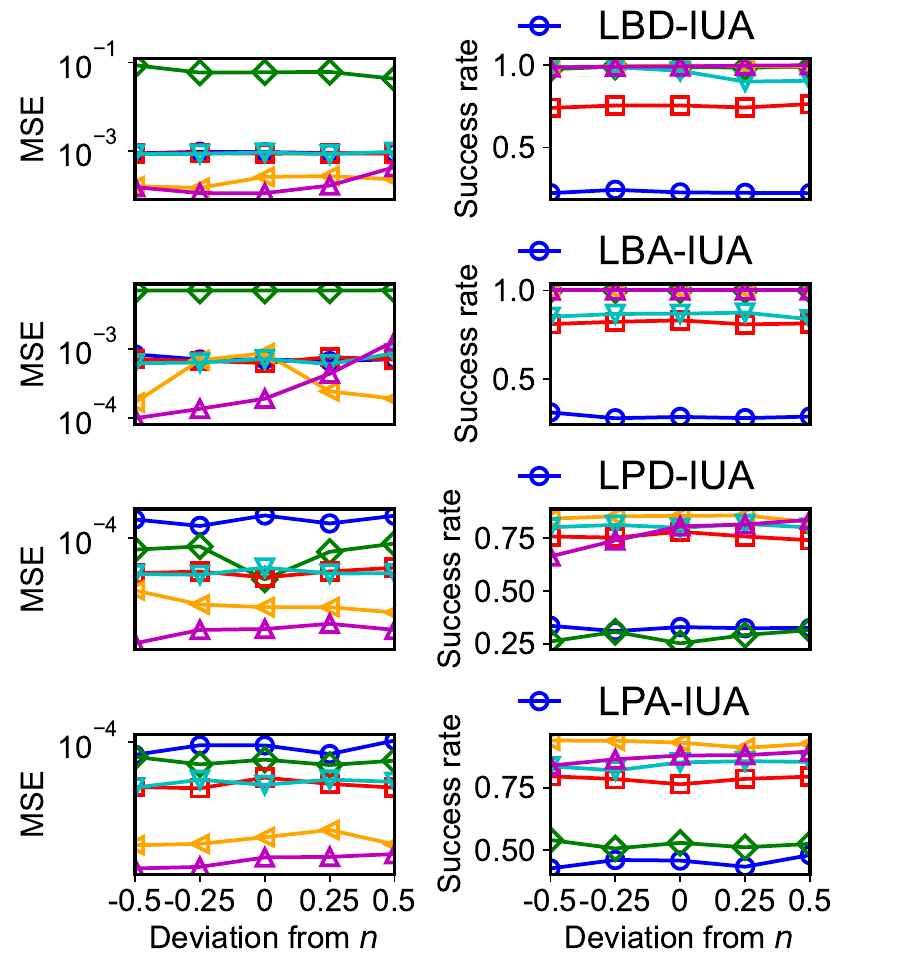}}\vspace{-0.13cm}\hspace{-4mm}
	\subfigure[\textsf{Taobao} dataset, Pulse $\mathbf{\tilde{f}}$]{
		\includegraphics[width=0.24\textwidth]{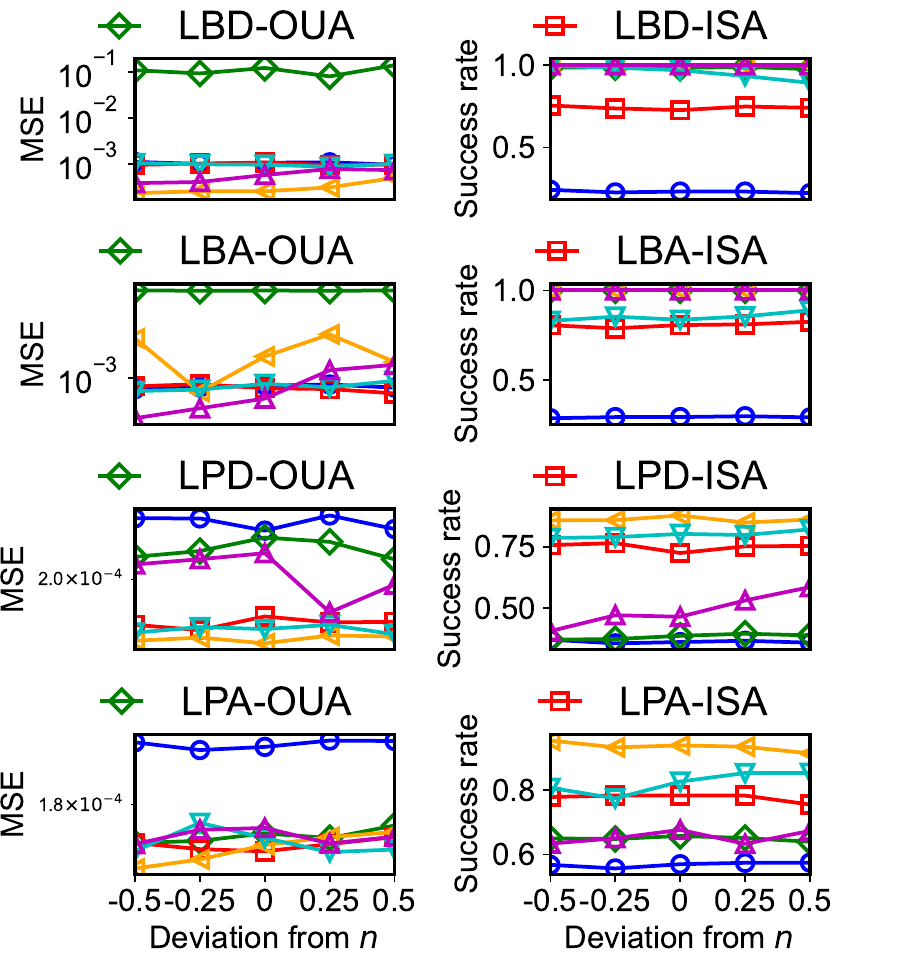}}\vspace{-0.13cm}\hspace{-4mm}
	\subfigure[\textsf{Taobao} dataset, Gaussian $\mathbf{\tilde{f}}$]{
		\includegraphics[width=0.24\textwidth]{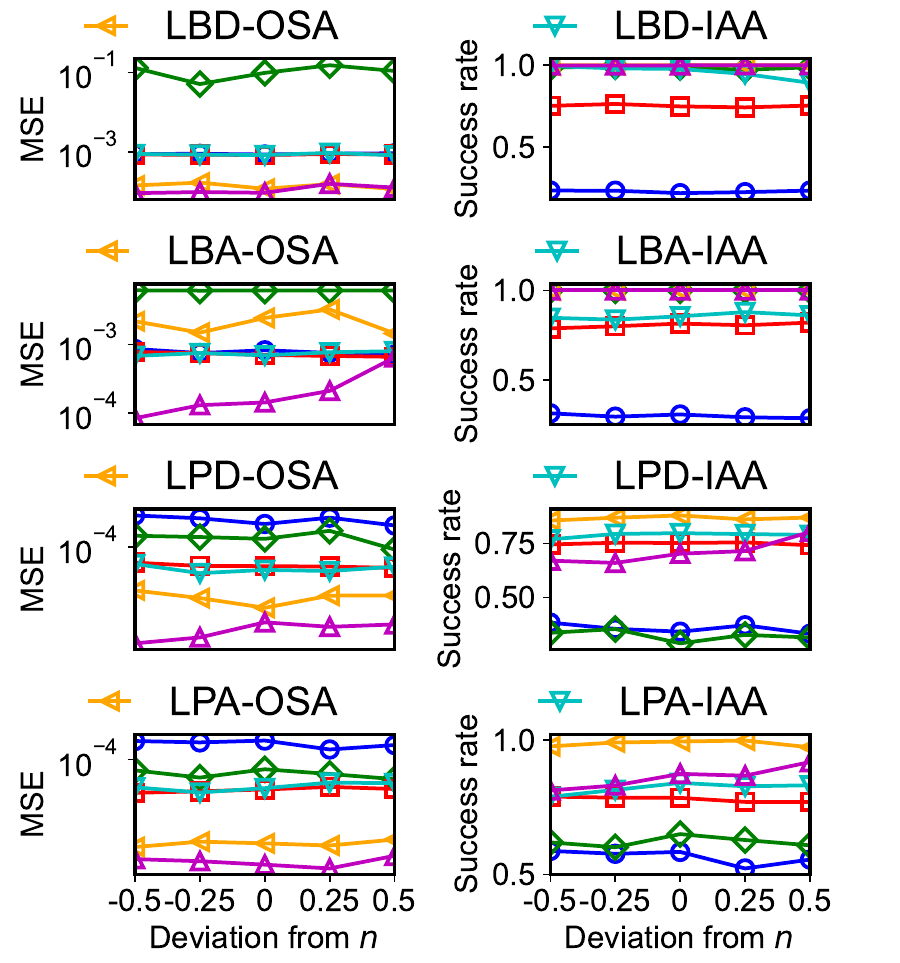}}\vspace{-0.13cm}\hspace{-4mm}
	\subfigure[\textsf{Taobao} dataset, Sigmoid $\mathbf{\tilde{f}}$]{
		\includegraphics[width=0.24\textwidth]{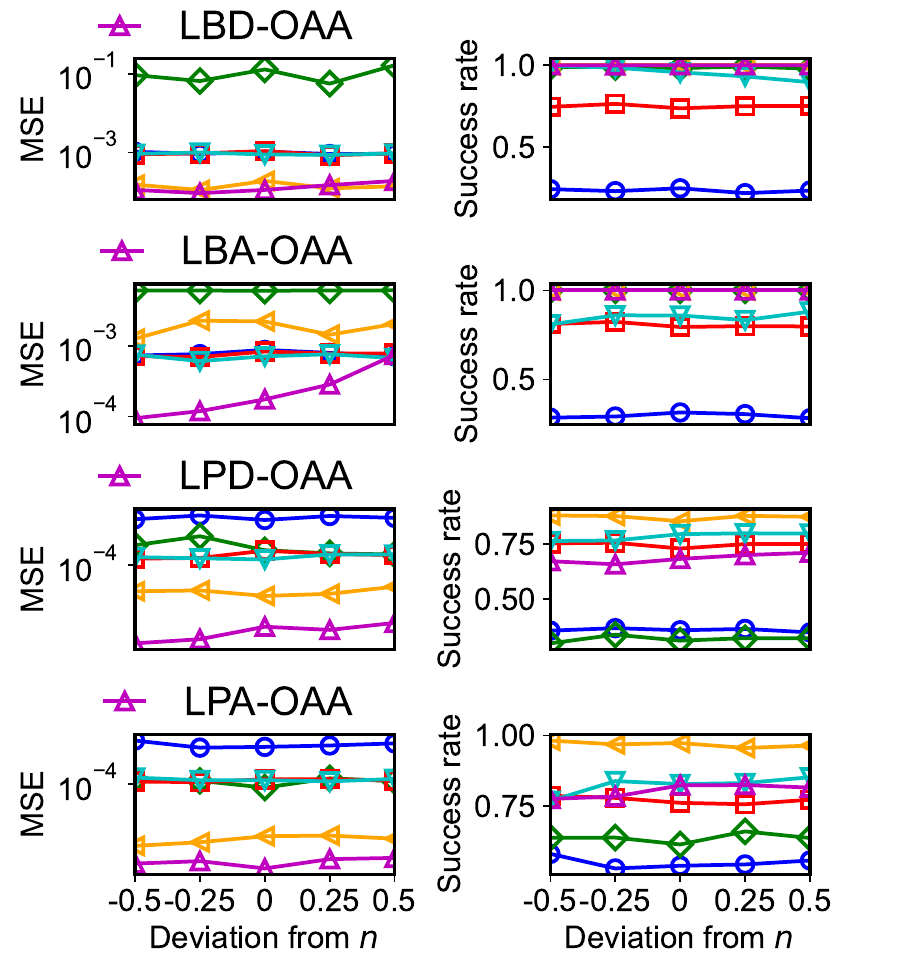}}
	\caption{\small Attacking effectiveness for real-world datasets with varying $n^e$.
	}\centering
	\label{fig:n^e} 
	\vspace{-0.5cm}
\end{figure*}

\begin{figure*}[htbp]
	\centering	
	\subfigure[\textsf{Taxi} dataset, Uniform $\mathbf{\tilde{f}}$]{
		\includegraphics[width=0.24\textwidth]{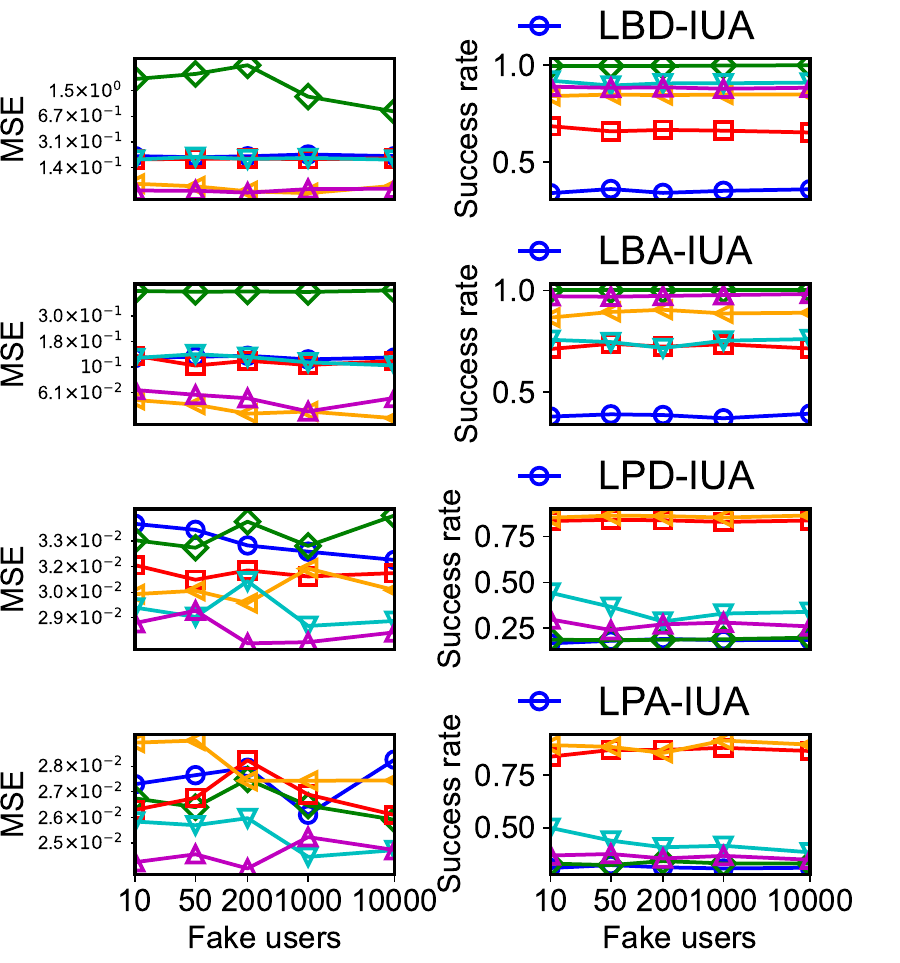}}\vspace{-0.06cm}\hspace{-4mm}
	\subfigure[\textsf{Taxi} dataset, Pulse $\mathbf{\tilde{f}}$]{
		\includegraphics[width=0.24\textwidth]{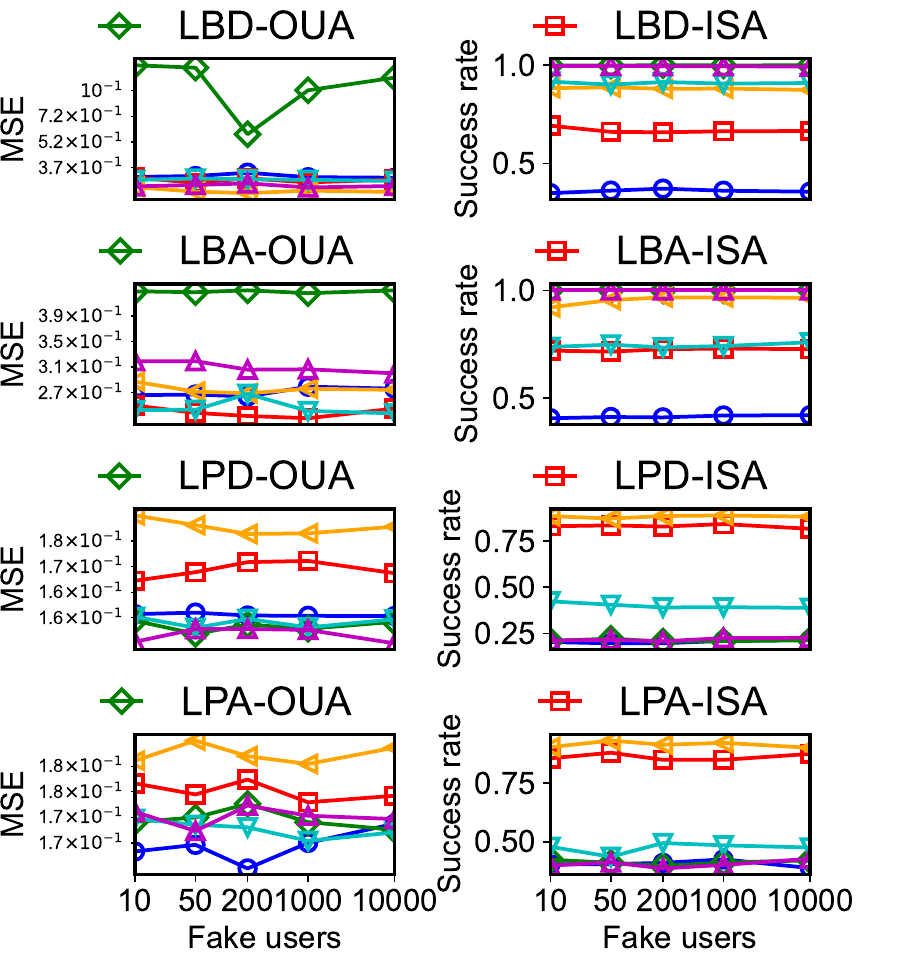}}\vspace{-0.06cm}\hspace{-4mm}
	\subfigure[\textsf{Taxi} dataset, Gaussian $\mathbf{\tilde{f}}$]{
		\includegraphics[width=0.24\textwidth]{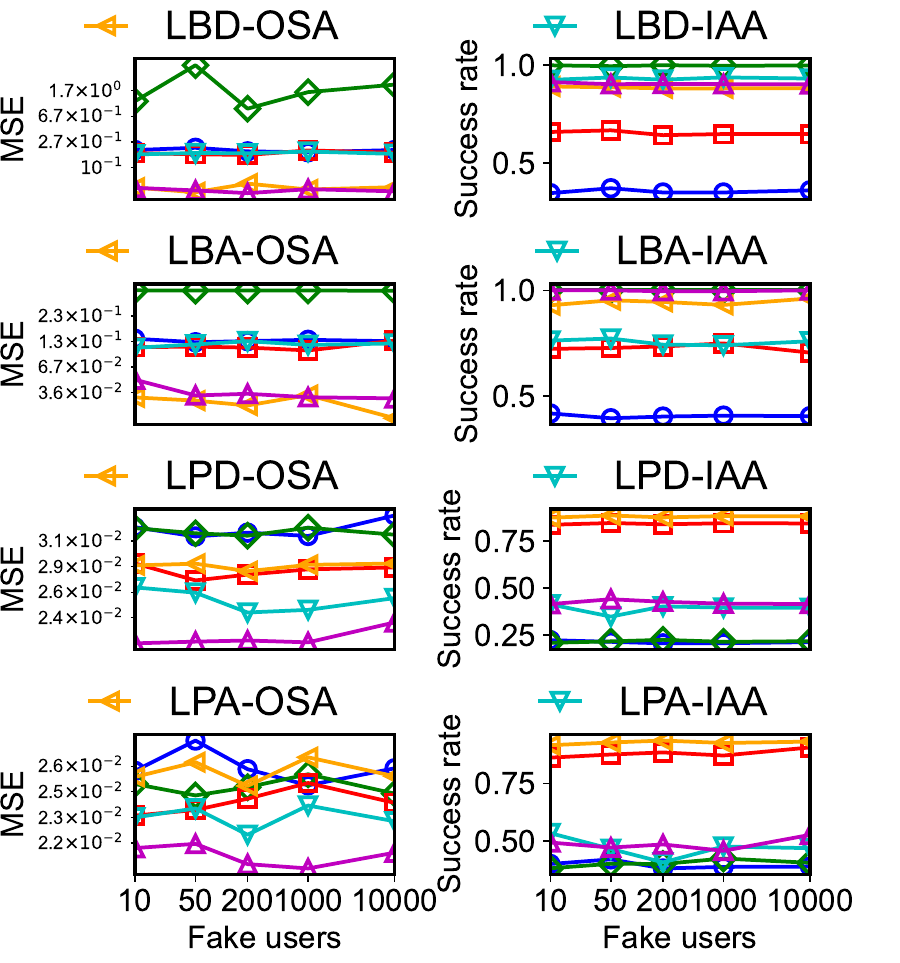}}\vspace{-0.06cm}\hspace{-4mm}
	\subfigure[\textsf{Taxi} dataset, Sigmoid $\mathbf{\tilde{f}}$]{
		\includegraphics[width=0.24\textwidth]{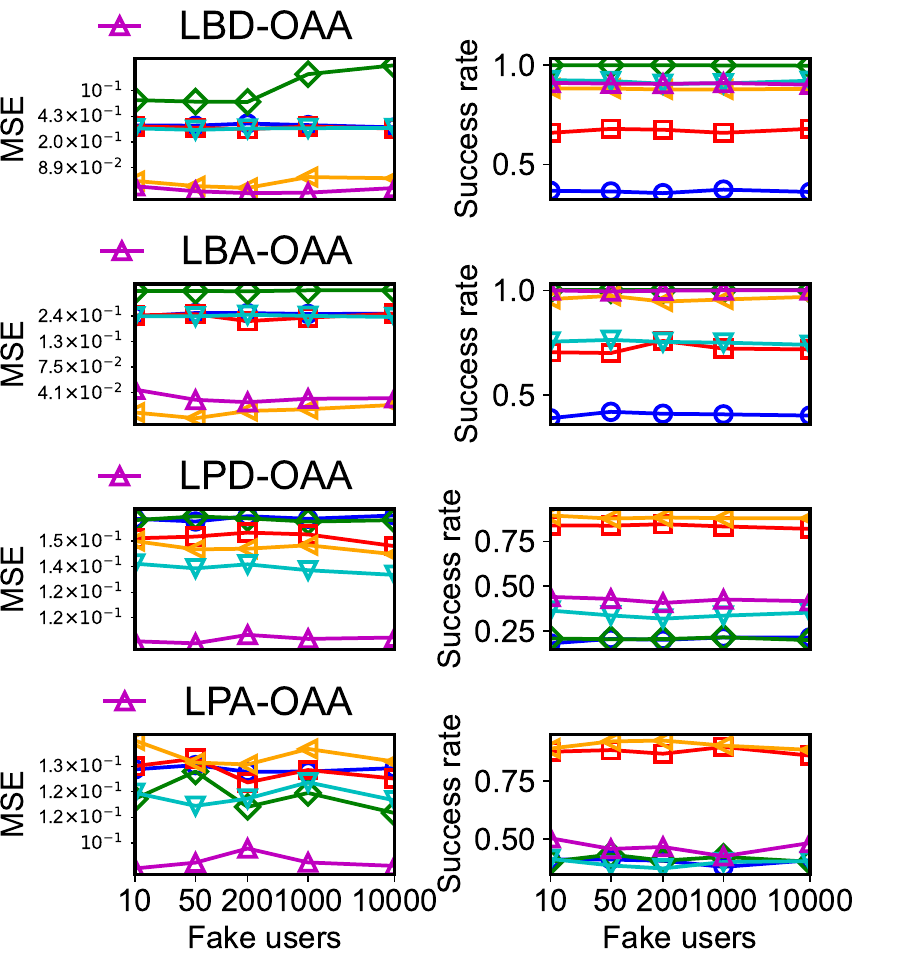}}\vspace{-0.06cm}\hspace{-4mm}
	\subfigure[\scriptsize\textsf{Foursquare} dataset, Uniform $\mathbf{\tilde{f}}$]{
		\includegraphics[width=0.24\textwidth]{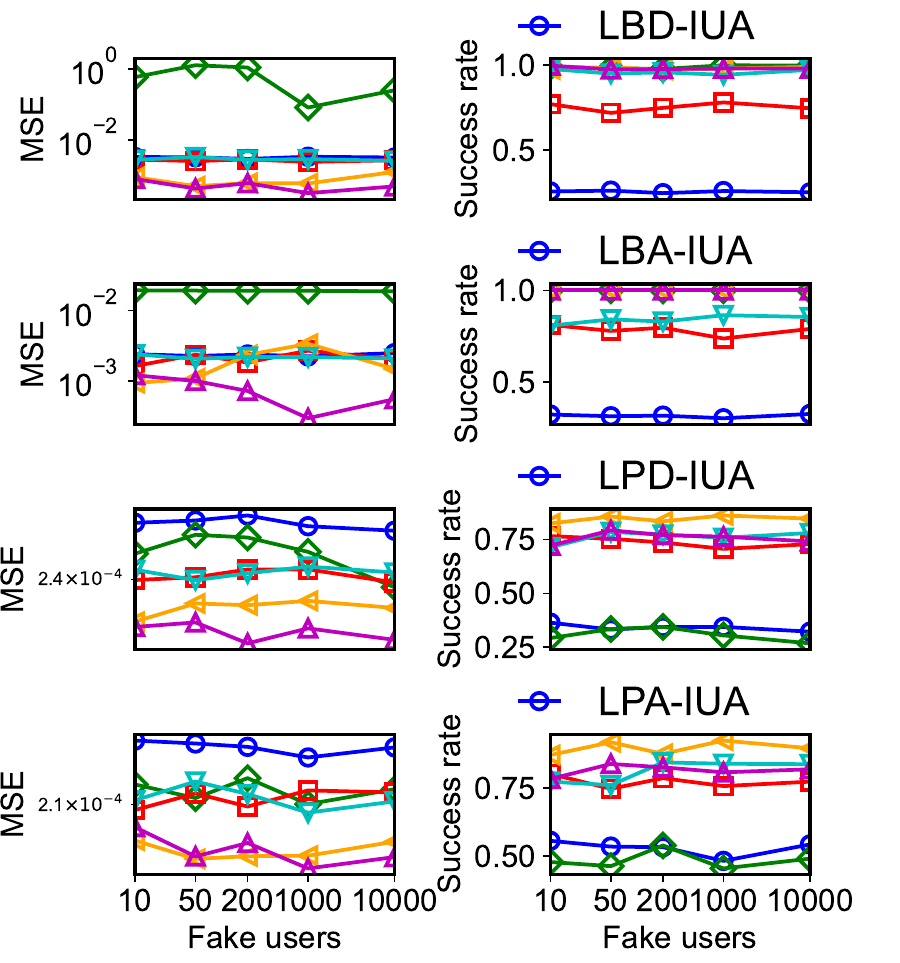}}\vspace{-0.05cm}\hspace{-4mm}
	\subfigure[\scriptsize\textsf{Foursquare} dataset, Pulse $\mathbf{\tilde{f}}$]{
		\includegraphics[width=0.24\textwidth]{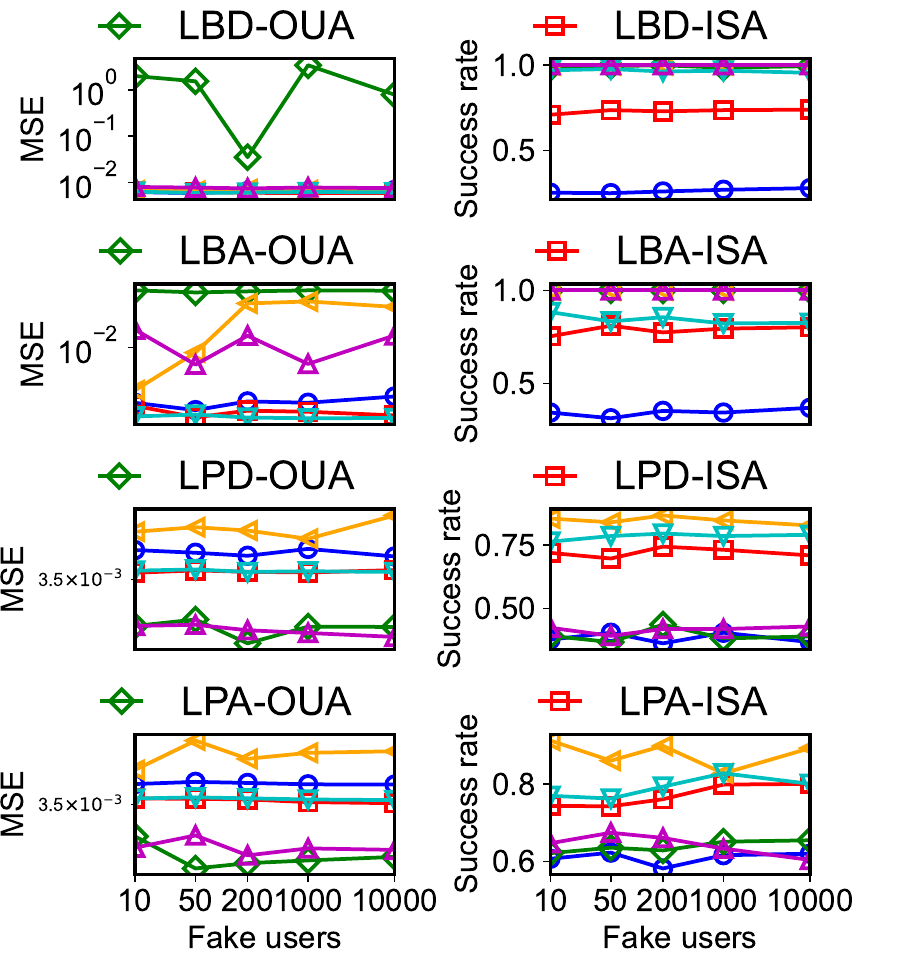}}\vspace{-0.06cm}\hspace{-4mm}
	\subfigure[\scriptsize\textsf{Foursquare} dataset, Gaussian $\mathbf{\tilde{f}}$]{
		\includegraphics[width=0.24\textwidth]{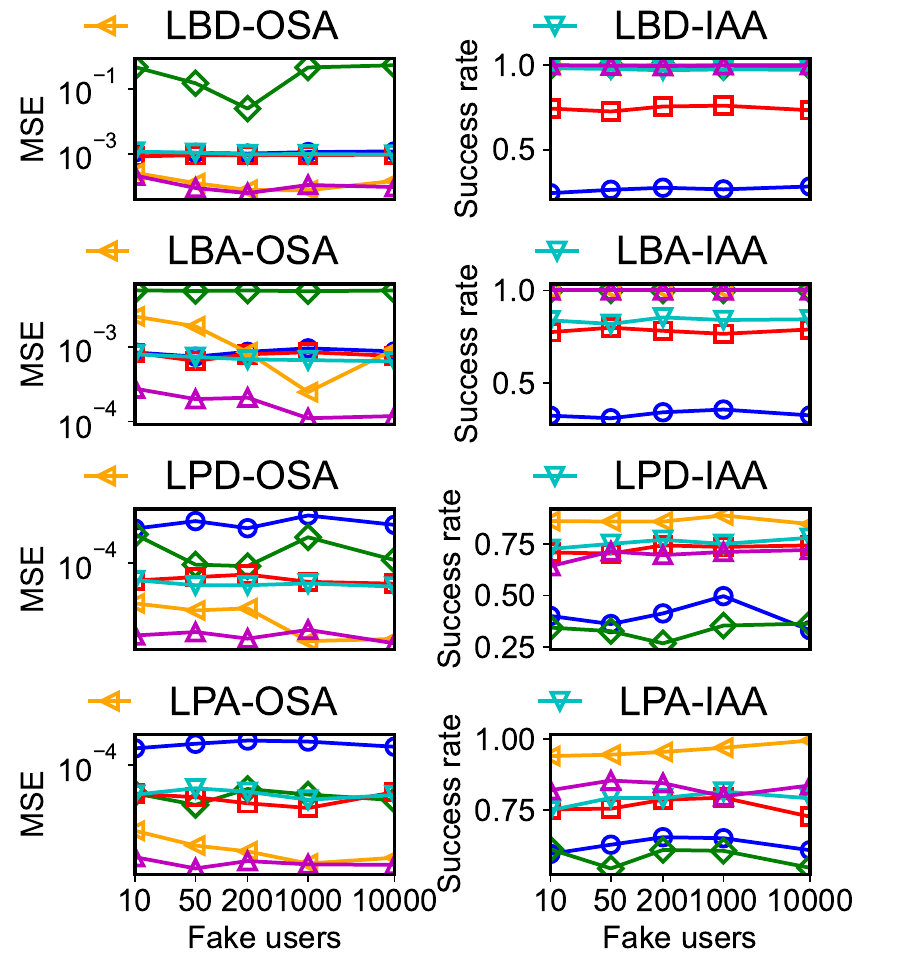}}\vspace{-0.06cm}\hspace{-4mm}
	\subfigure[\scriptsize\textsf{Foursquare} dataset, Sigmoid $\mathbf{\tilde{f}}$]{
		\includegraphics[width=0.24\textwidth]{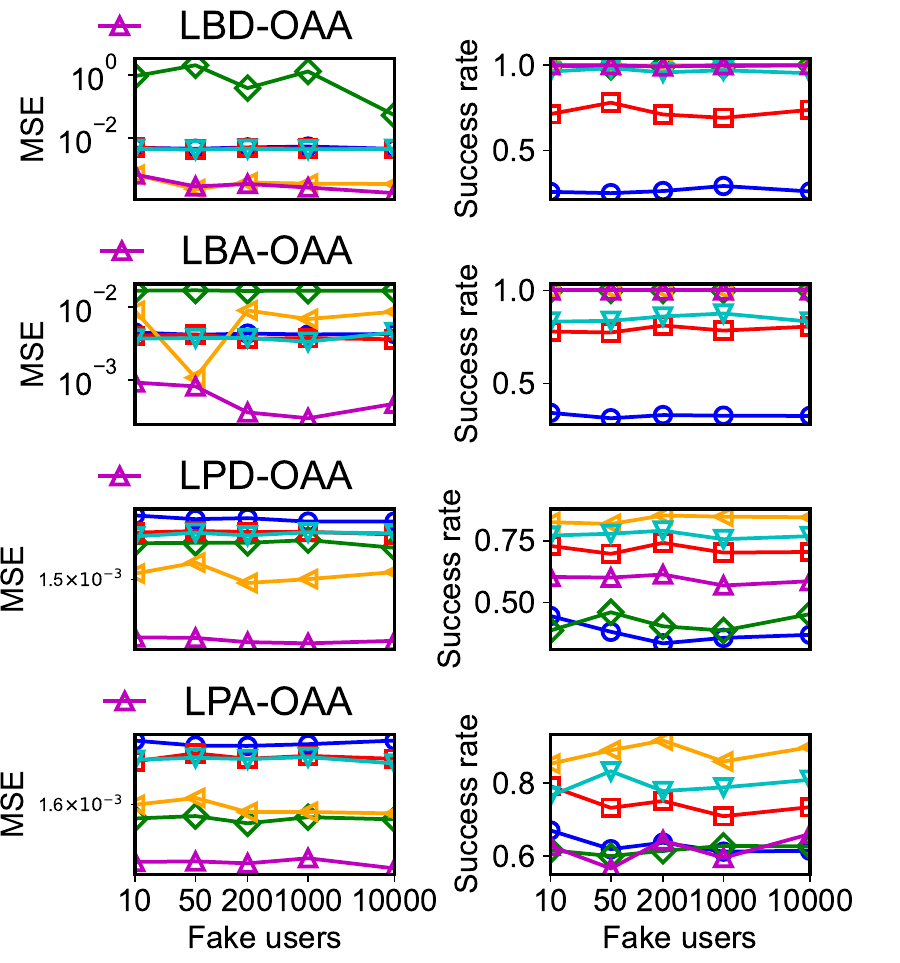}}\vspace{-0.06cm}\hspace{-4mm}
	\subfigure[\textsf{Taobao} dataset, Uniform $\mathbf{\tilde{f}}$]{
		\includegraphics[width=0.24\textwidth]{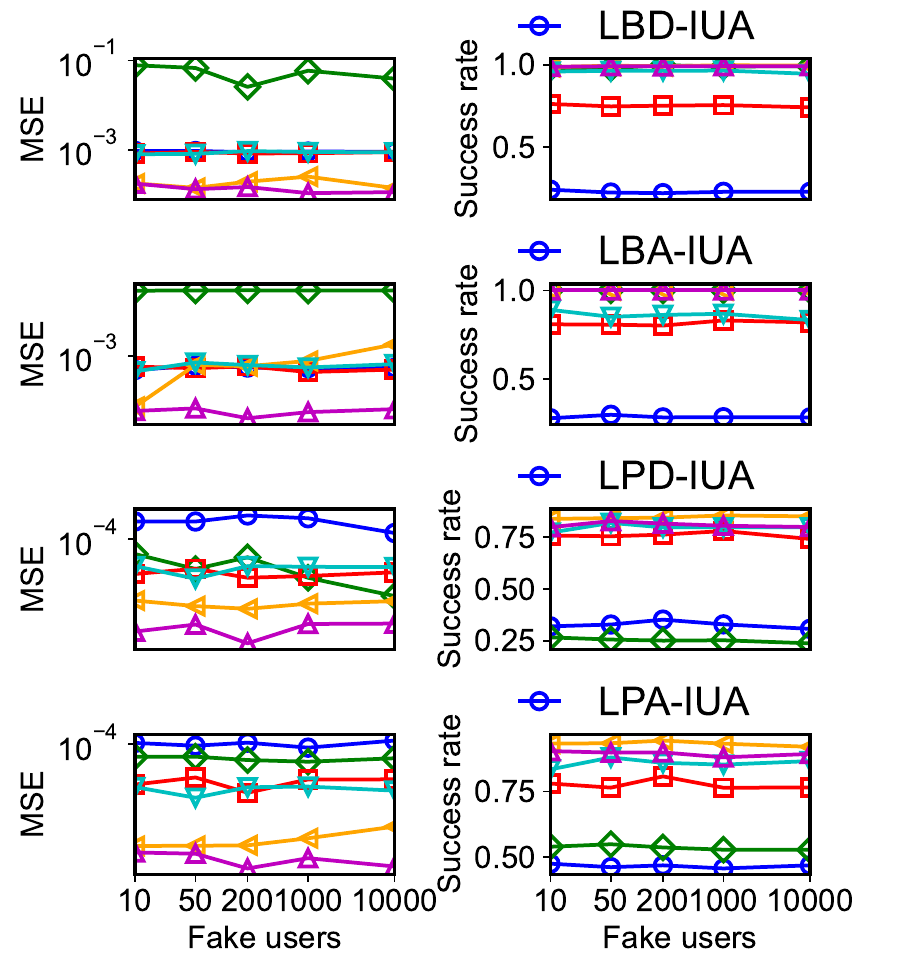}}\vspace{-0.13cm}\hspace{-4mm}
	\subfigure[\textsf{Taobao} dataset, Pulse $\mathbf{\tilde{f}}$]{
		\includegraphics[width=0.24\textwidth]{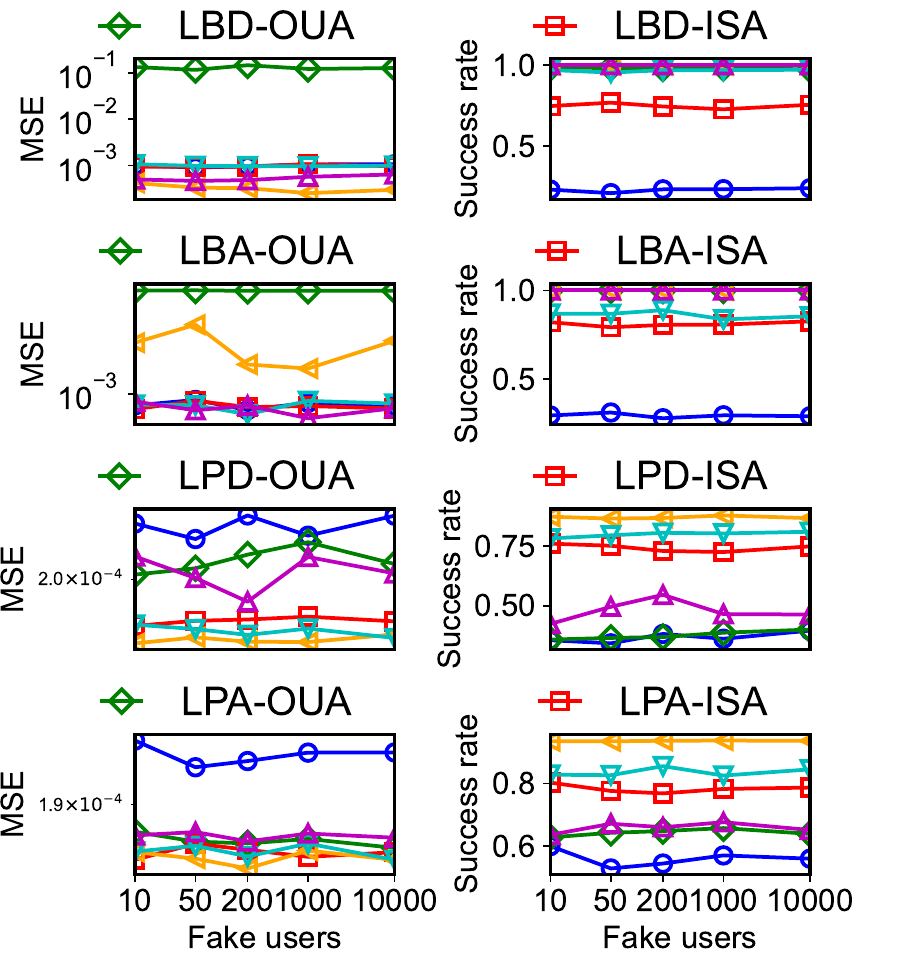}}\vspace{-0.13cm}\hspace{-4mm}
	\subfigure[\textsf{Taobao} dataset, Gaussian $\mathbf{\tilde{f}}$]{
		\includegraphics[width=0.24\textwidth]{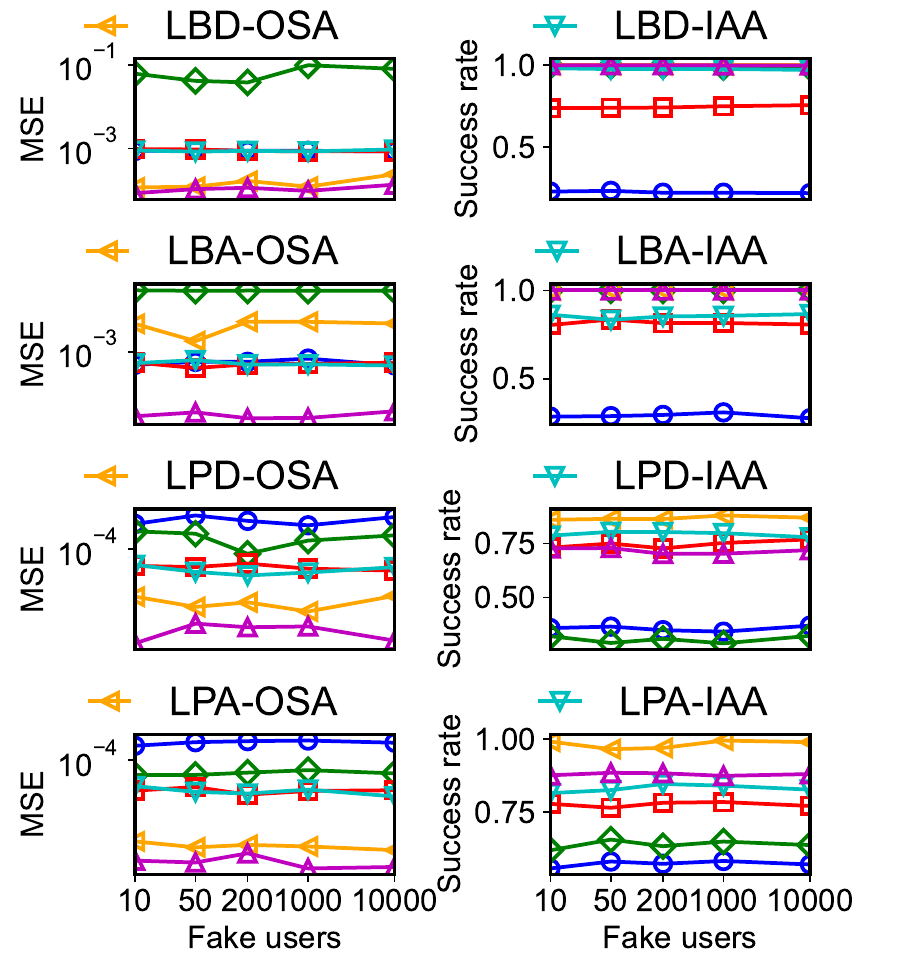}}\vspace{-0.13cm}\hspace{-4mm}
	\subfigure[\textsf{Taobao} dataset, Sigmoid $\mathbf{\tilde{f}}$]{
		\includegraphics[width=0.24\textwidth]{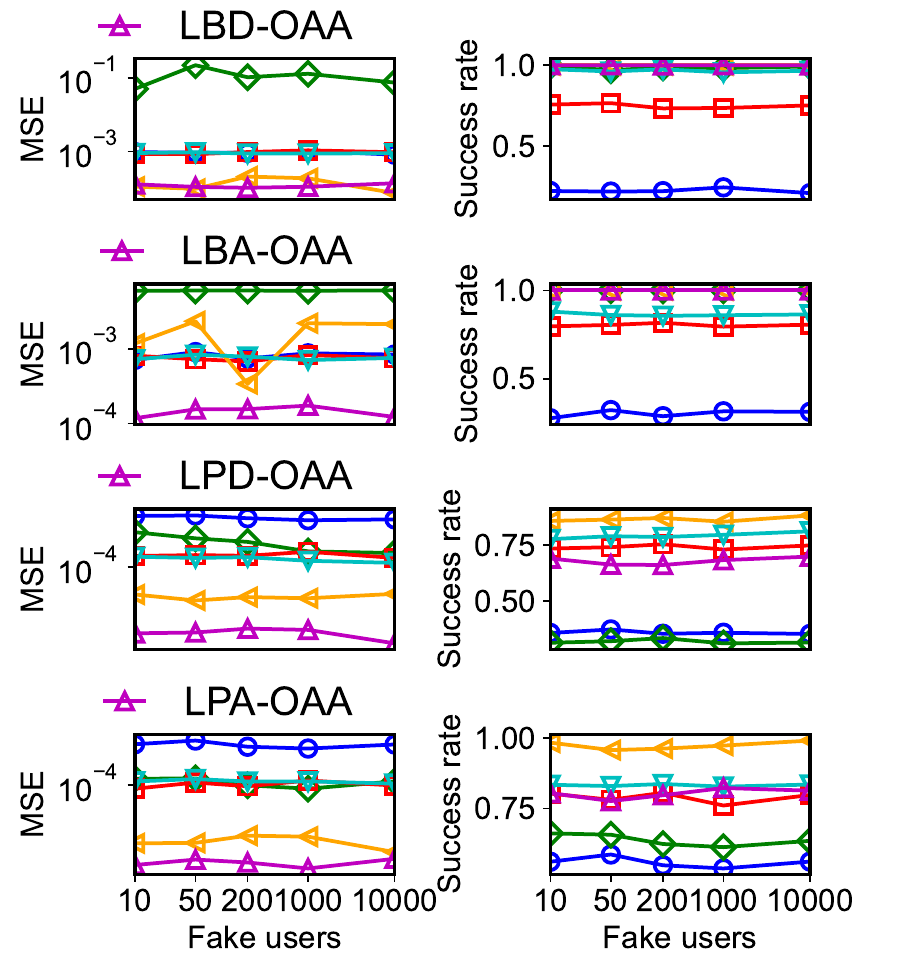}}
	\caption{\small Attacking effectiveness for real-world datasets with varying the number of fake users for $\mathbf{f}^e$ calculation.
	}\centering
	\label{fig:f_e} 
	\vspace{-0.5cm}
\end{figure*}

The average manipulation gap against adaptive methods for real-world datasets with varying $\beta$, $\epsilon$, $w$, $n^e$, and $\mathbf{f}^e$ are shown in Fig.~\ref{fig:beta}, Fig.~\ref{fig:epsilon}, Fig.~\ref{fig:w}, Fig.~\ref{fig:n^e}, and Fig.~\ref{fig:f_e}, respectively.

\subsection{Result on Different FOs, Varying $d$}\label{appendix:different FOs}
\begin{figure}[h]
	\centering 
	\includegraphics[width=240pt,height=80pt]    {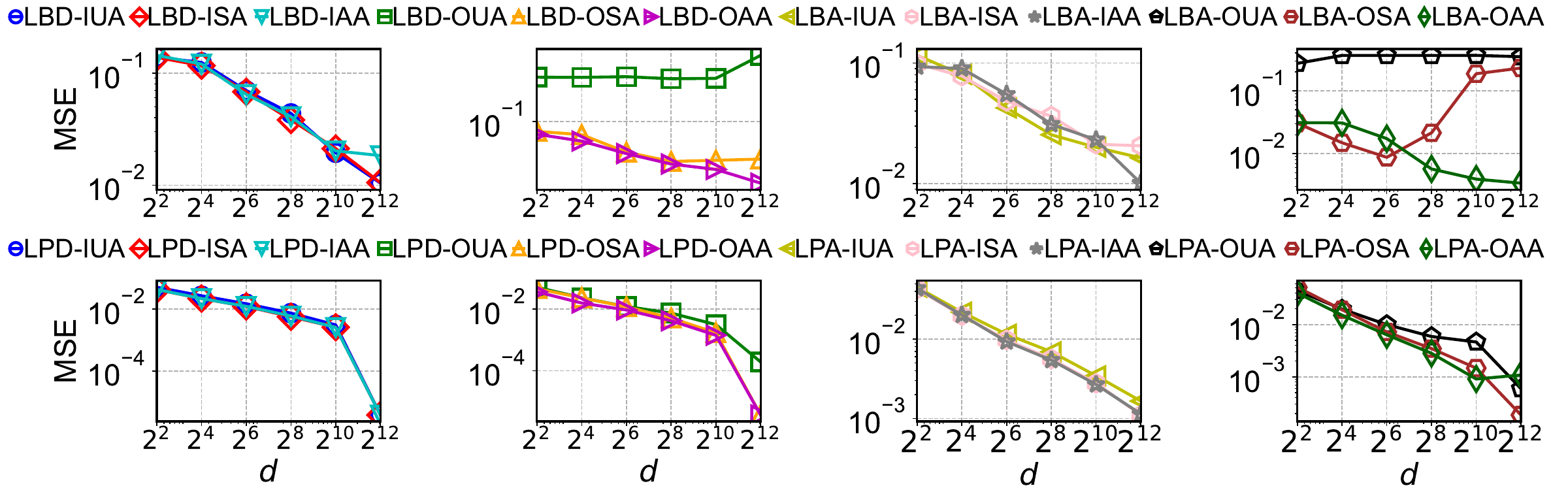}\vspace{-1.5mm}
	\text{\footnotesize (a) Attack effectiveness when using Ada.}
	
	\vspace{2mm}
	\includegraphics[width=240pt,height=80pt]     {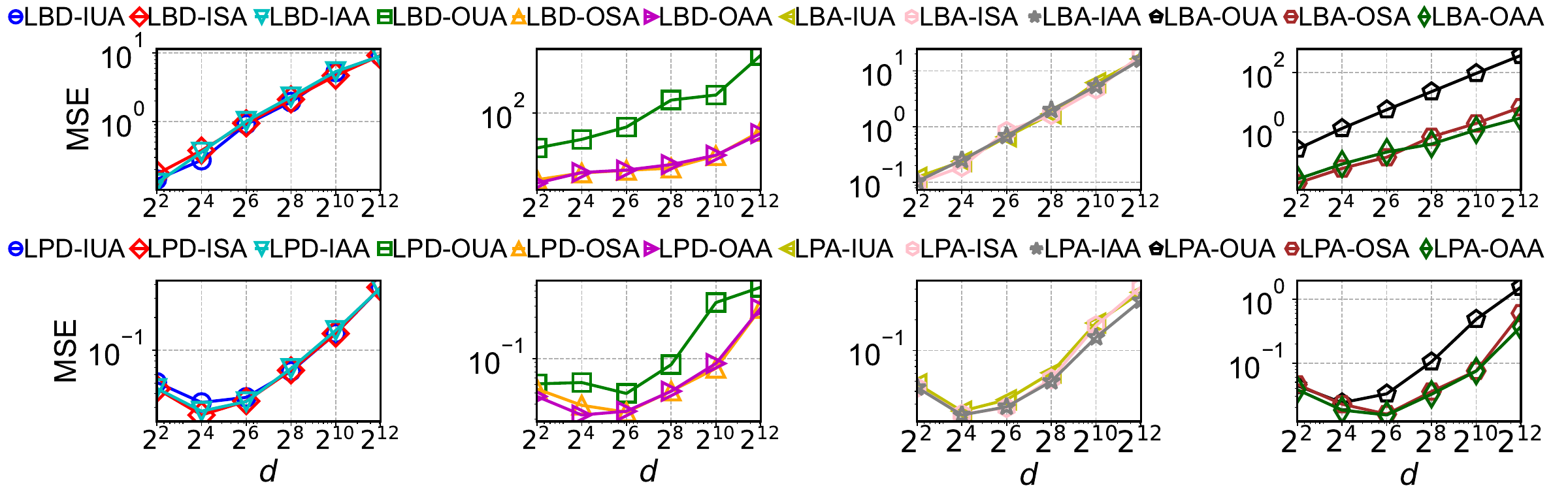}\vspace{-1.5mm}
	\text{\footnotesize (b) Attack effectiveness when using kRR.}
	
	\vspace{2mm}
	\includegraphics[width=240pt,height=80pt]     {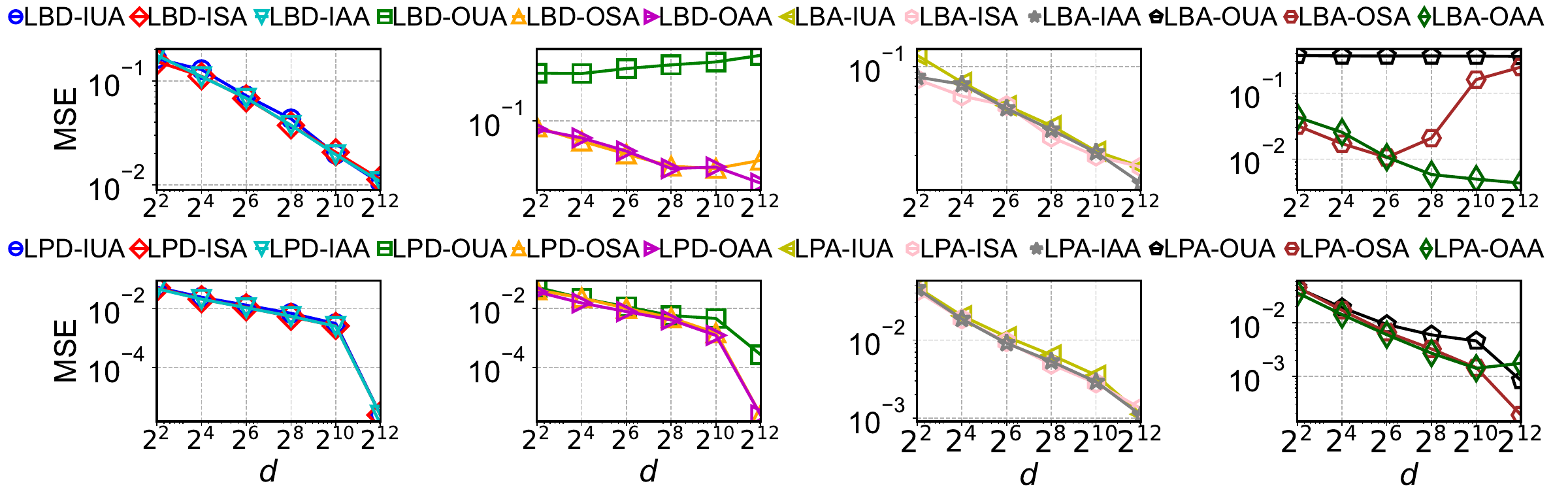}\vspace{-1.5mm}
	\text{\footnotesize (c) Attack effectiveness when using OUE.}
	\vspace{-0.25cm}
	\caption{\small Attack effectiveness, varying $d$ (\textsf{Taobao} with Gaussian target).}\centering
	\label{fig:attack different FO appendix}
\end{figure}

Fig.~\ref{fig:attack different FO appendix} shows attack effectiveness against Ada, kRR and OUE on \textsf{Taobao} dataset with Gaussian target, varying $d$.

\subsection{Result on Different Levels of Knowledge}\label{appendix:different knowledge}
\begin{figure}[htbp]
	\centering	
	\includegraphics[width=240pt,height=80pt] {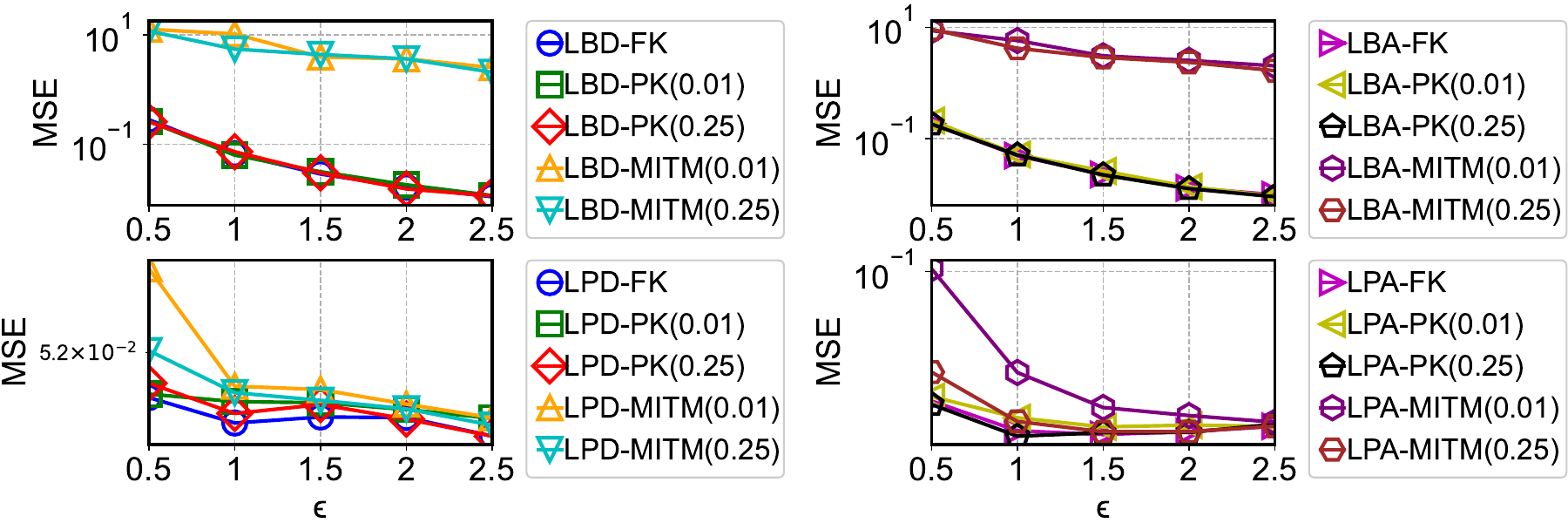}\vspace{-2mm}
	\text{\footnotesize (a) Uniform target.}\vspace{1mm}
	\includegraphics[width=240pt,height=80pt] {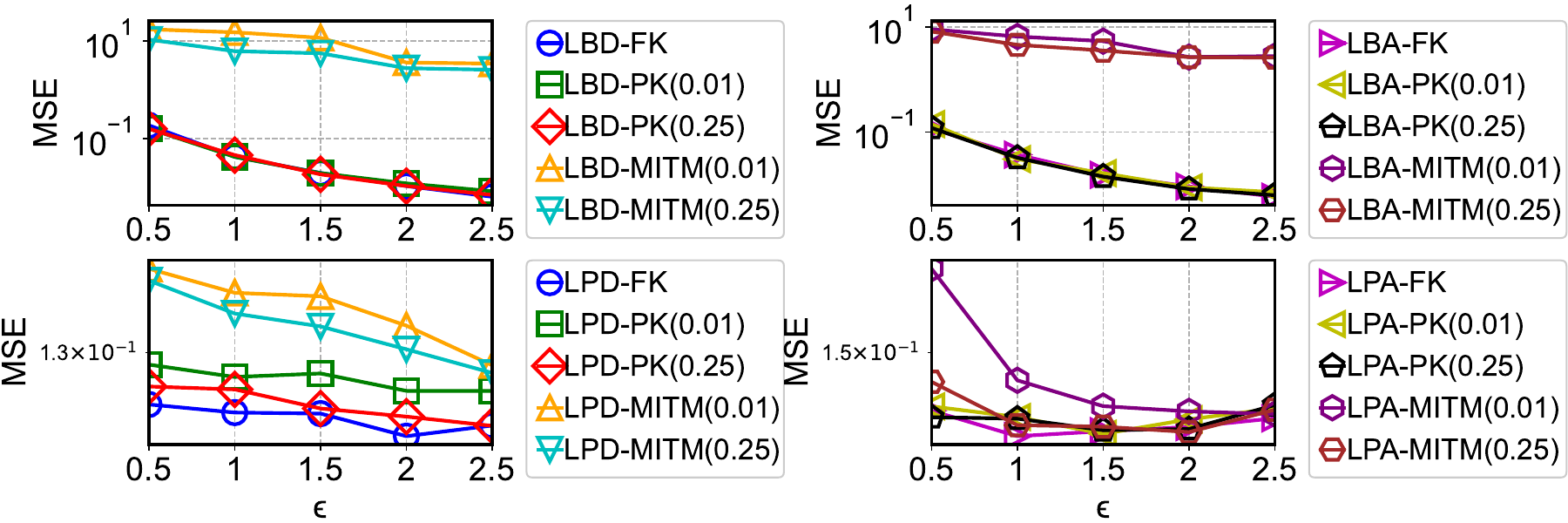}\vspace{-2mm}
	\text{\footnotesize (b) Sigmoid target.}
	\vspace{-0.3cm}
	\caption{\small Attack effectiveness using OAA with different levels of knowledge, varying $\epsilon$ (\textsf{Taxi}).}\centering
	\label{fig:different level knowledge appendix}
\end{figure}

Fig.~\ref{fig:different level knowledge appendix} shows attack effectiveness using OAA on \textsf{Taxi} dataset with different levels of knowledge, varying $\epsilon$.

\subsection{Result on Baseline Methods}\label{appendix:attack baseline}
\begin{figure*}[htbp]
	\centering	
	\subfigure[\textsf{Taxi} dataset, Uniform $\mathbf{\tilde{f}}$]{
		\includegraphics[width=0.24\textwidth]{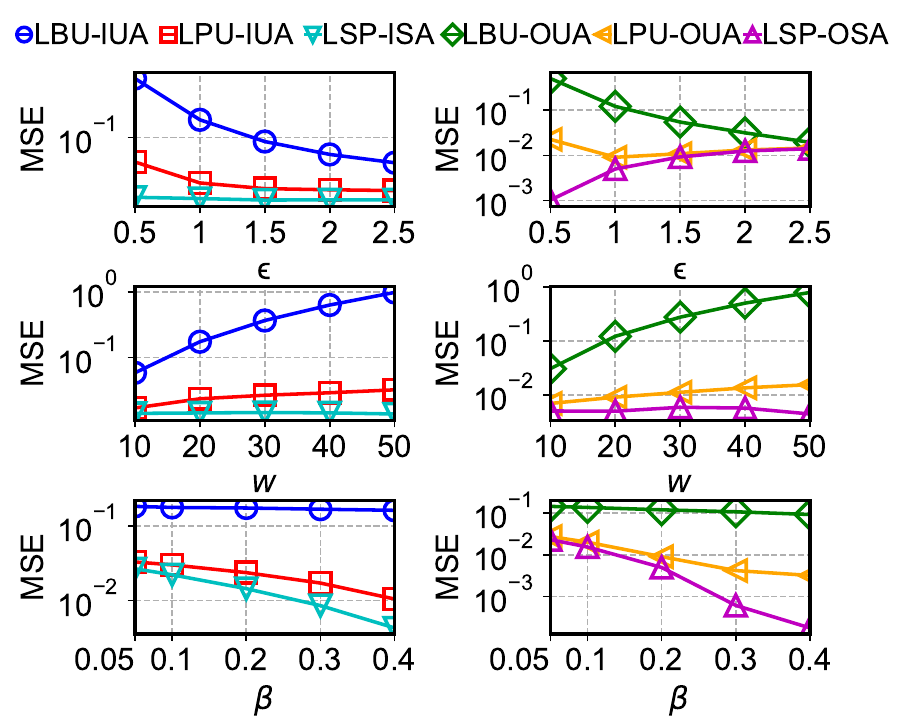}}\vspace{-0.06cm}
	\subfigure[\textsf{Taxi} dataset, Pulse $\mathbf{\tilde{f}}$]{
		\includegraphics[width=0.24\textwidth]{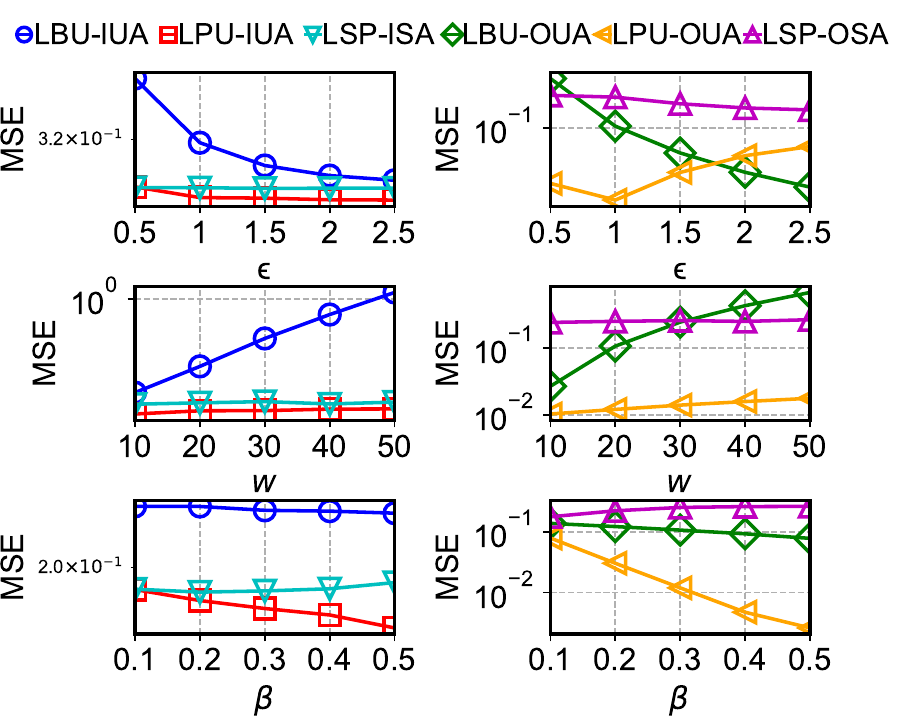}}\vspace{-0.06cm}
	\subfigure[\textsf{Taxi} dataset, Gaussian $\mathbf{\tilde{f}}$]{
		\includegraphics[width=0.24\textwidth]{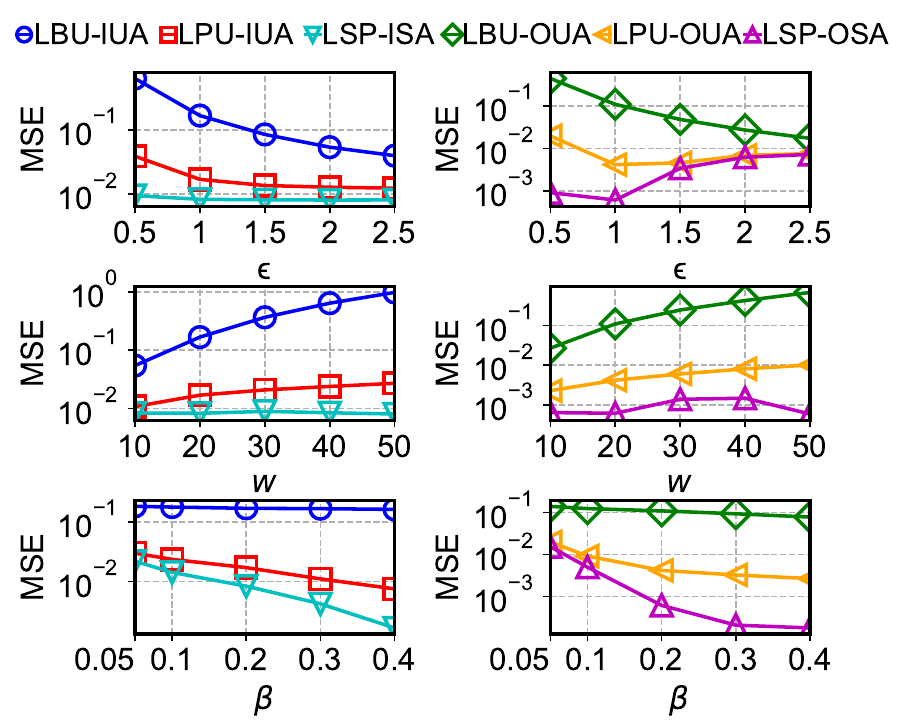}}\vspace{-0.06cm}
	\subfigure[\textsf{Taxi} dataset, Sigmoid $\mathbf{\tilde{f}}$]{
		\includegraphics[width=0.24\textwidth]{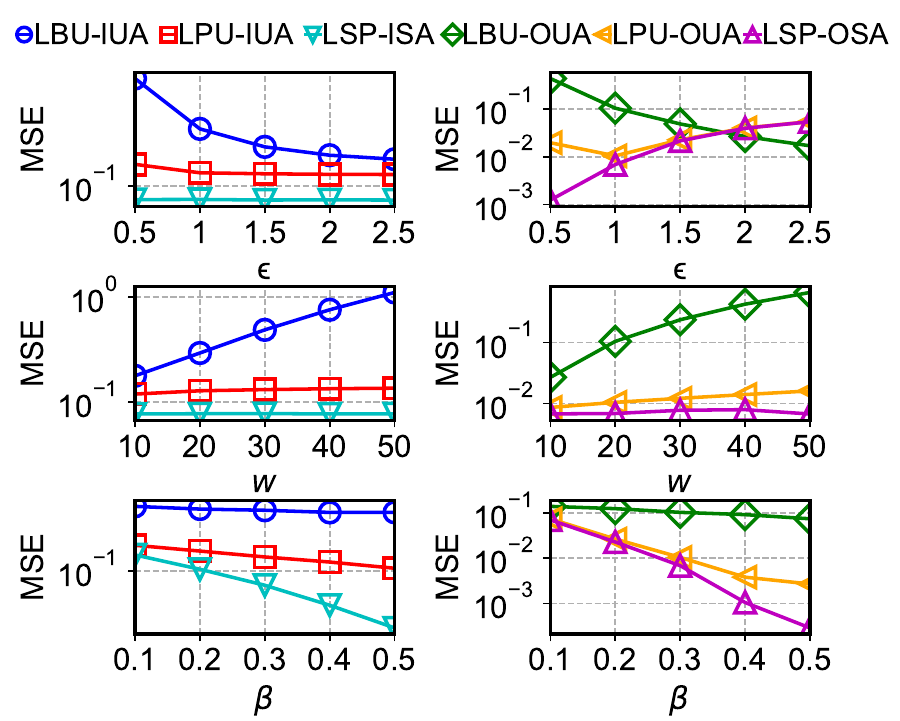}}\vspace{-0.06cm}
	\subfigure[\scriptsize\textsf{Foursquare} dataset, Uniform $\mathbf{\tilde{f}}$]{
		\includegraphics[width=0.24\textwidth]{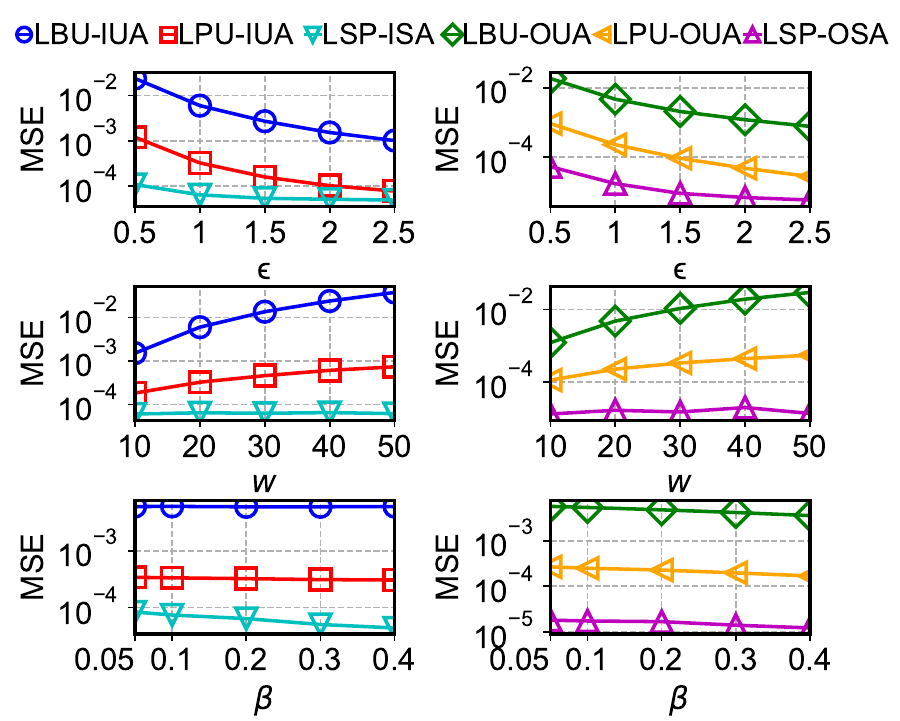}}\vspace{-0.05cm}
	\subfigure[\scriptsize\textsf{Foursquare} dataset, Pulse $\mathbf{\tilde{f}}$]{
		\includegraphics[width=0.24\textwidth]{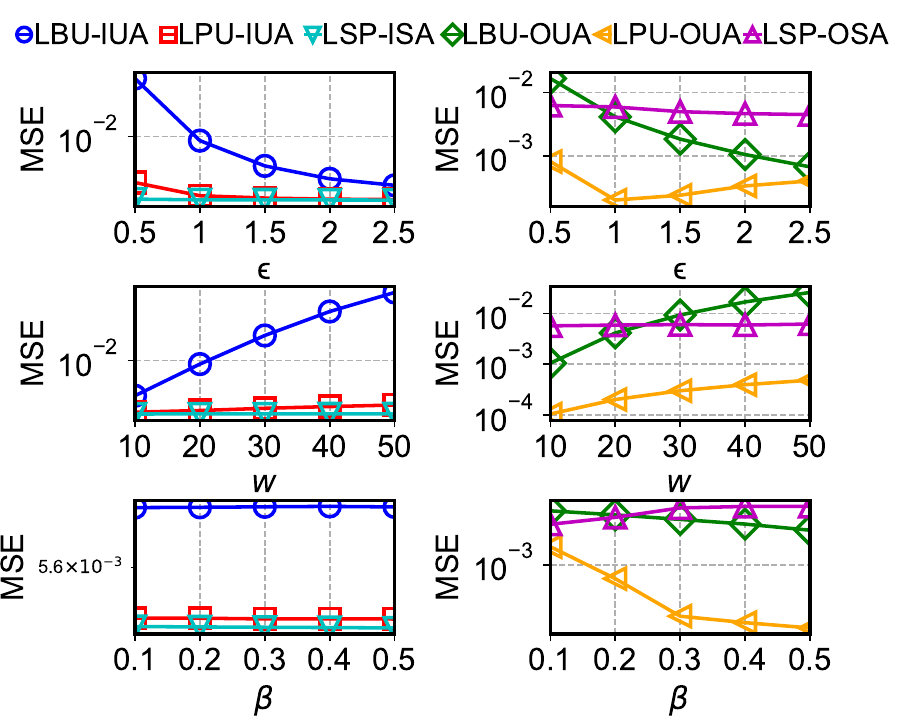}}\vspace{-0.06cm}
	\subfigure[\scriptsize\textsf{Foursquare} dataset, Gaussian $\mathbf{\tilde{f}}$]{
		\includegraphics[width=0.24\textwidth]{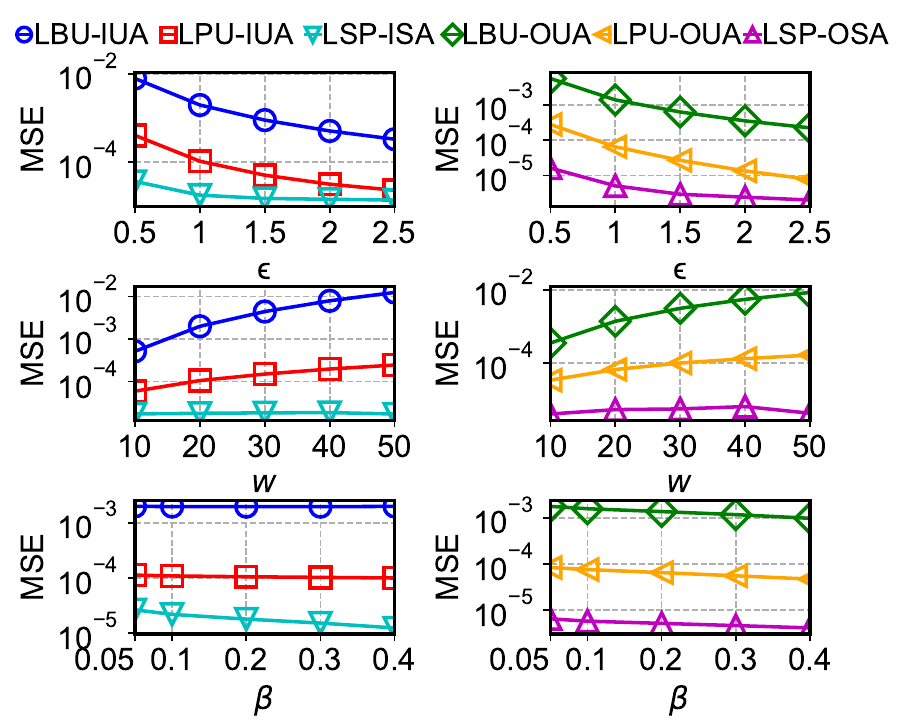}}\vspace{-0.06cm}
	\subfigure[\scriptsize\textsf{Foursquare} dataset, Sigmoid $\mathbf{\tilde{f}}$]{
		\includegraphics[width=0.24\textwidth]{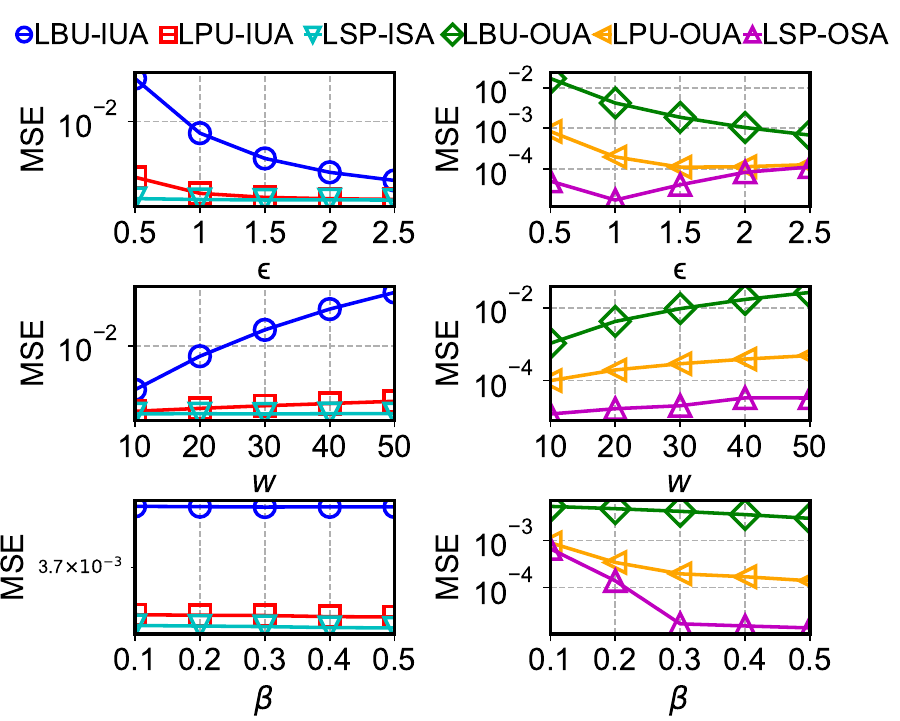}}\vspace{-0.06cm}
	\subfigure[\textsf{Taobao} dataset, Uniform $\mathbf{\tilde{f}}$]{
		\includegraphics[width=0.24\textwidth]{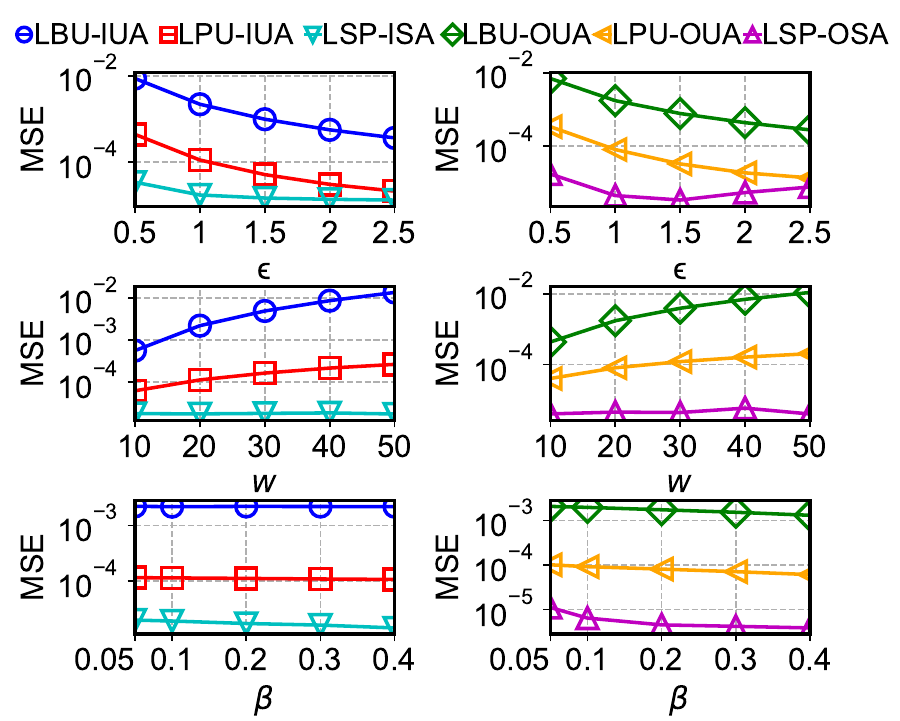}}\vspace{-0.13cm}
	\subfigure[\textsf{Taobao} dataset, Pulse $\mathbf{\tilde{f}}$]{
		\includegraphics[width=0.24\textwidth]{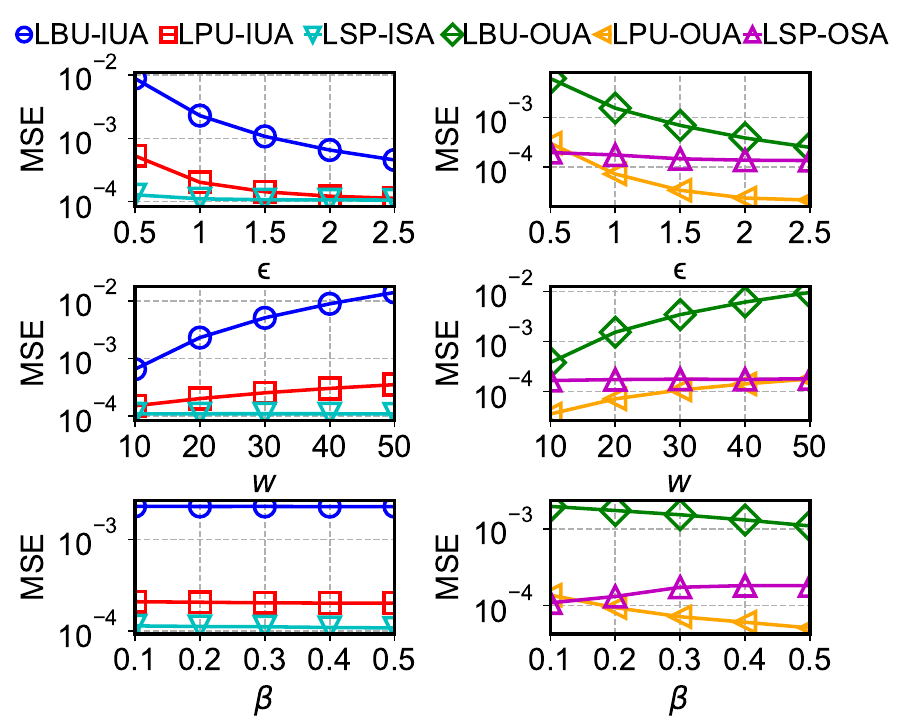}}\vspace{-0.13cm}
	\subfigure[\textsf{Taobao} dataset, Gaussian $\mathbf{\tilde{f}}$]{
		\includegraphics[width=0.24\textwidth]{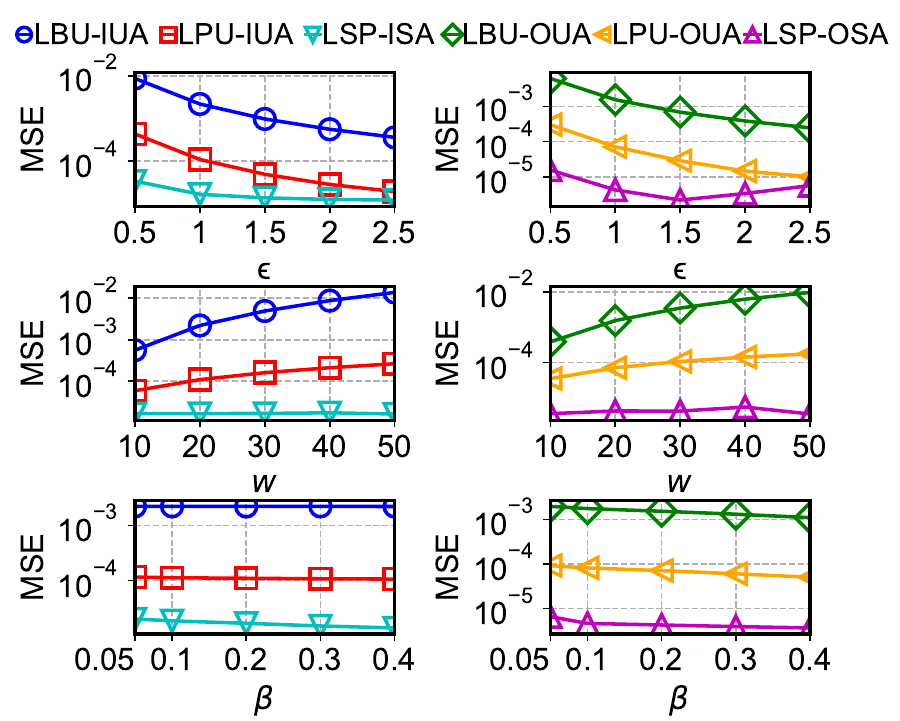}}\vspace{-0.13cm}
	\subfigure[\textsf{Taobao} dataset, Sigmoid $\mathbf{\tilde{f}}$]{
		\includegraphics[width=0.24\textwidth]{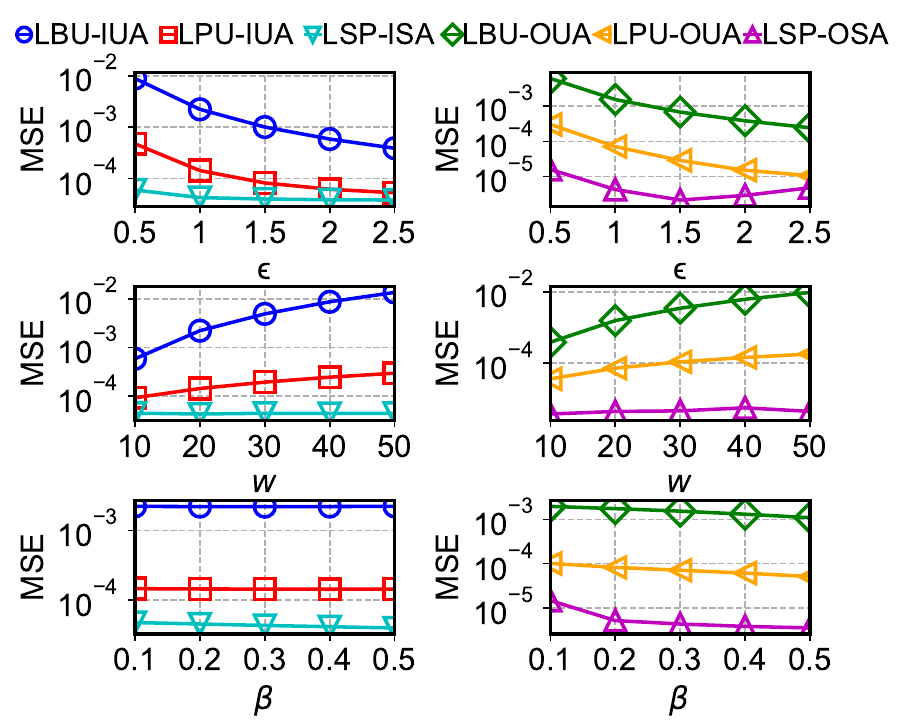}}
	\caption{\small Attack effectiveness for real-world datasets on baseline methods, varying $\epsilon$, $w$ and $\beta$. }\centering
	\label{fig:baseline appendix} 
\end{figure*}

Results for baseline methods are shown in Fig.~\ref{fig:baseline appendix}.

\subsection{Results on Mean Estimation}
\label{appendix:HM}
\begin{figure}[htbp]
	\centering	
	\includegraphics[width=240pt,height=48pt]     {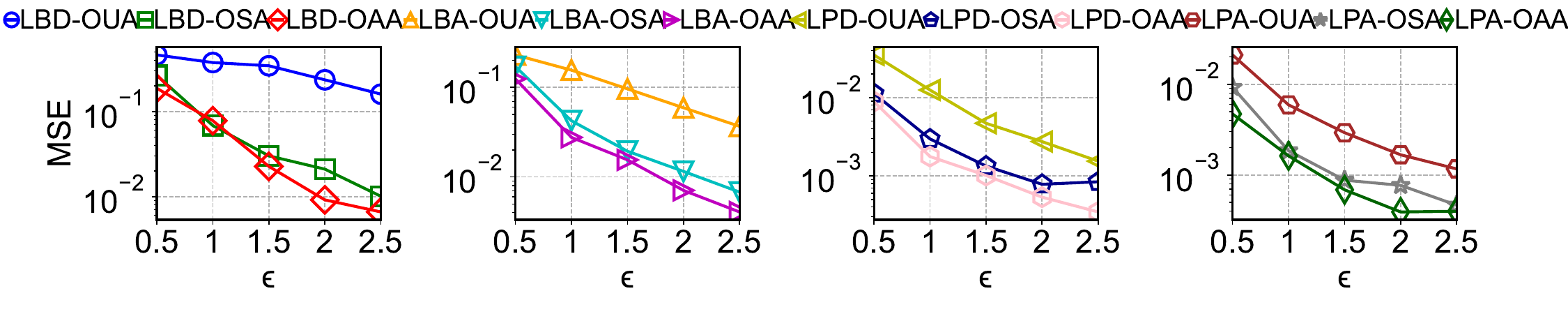}\vspace{-1mm}
	\includegraphics[width=240pt,height=48pt]     {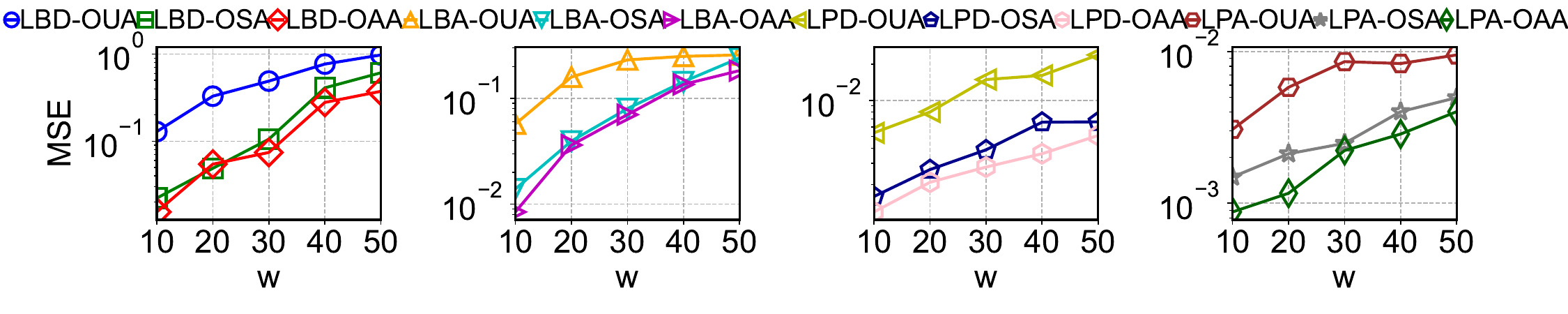}\vspace{-1mm}
	\includegraphics[width=240pt,height=48pt]     {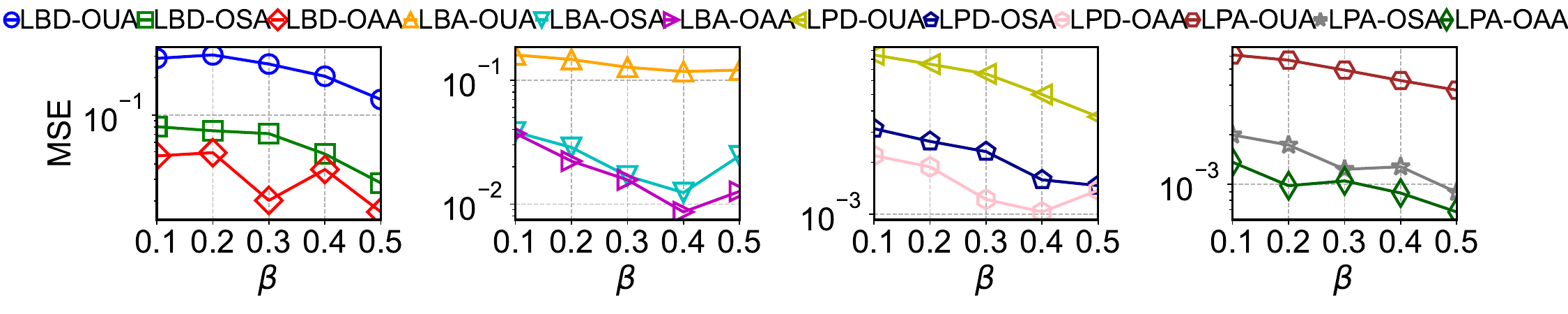}\vspace{-1mm}
	
	\caption{\small Attack effectiveness on HM-based LDP-IDS, varying $\epsilon$, $w$ and $\beta$ (\textsf{Taobao-Price} with Uniform target).}
	\label{fig:attack taobao HM appendix}
\end{figure}

\begin{figure}[htbp]
	\centering	
	\includegraphics[width=240pt,height=48pt]     {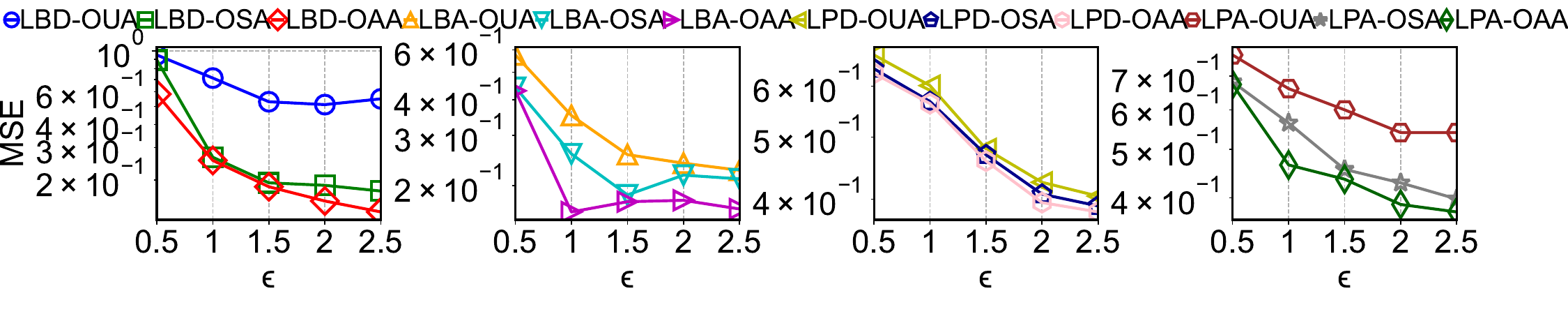}\vspace{-1mm}
	\includegraphics[width=240pt,height=48pt]     {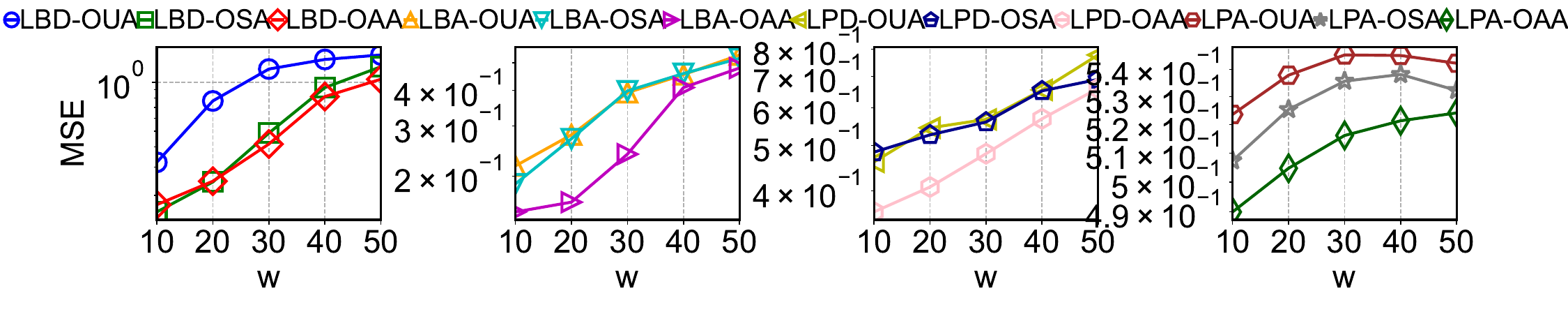}\vspace{-1mm}
	\includegraphics[width=240pt,height=48pt]     {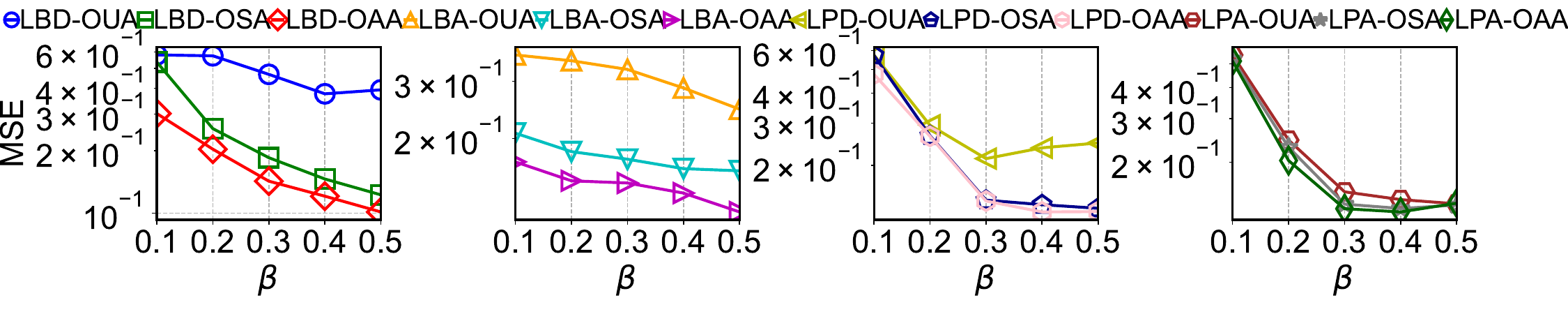}\vspace{-1mm}
	
	\caption{\small Attack effectiveness on HM-based LDP-IDS, varying $\epsilon$, $w$ and $\beta$ (\textsf{Taobao-Price} with Uniform target).}
	\label{fig:attack taxi HM appendix}
\end{figure}

Figs.~\ref{fig:attack taobao HM appendix} and~\ref{fig:attack taxi HM appendix} shows attack effectiveness on HM-based LDP-IDS, \textsf{Taobao-Price} and \textsf{Taxi-Longitude} datasets, varying $\epsilon$, $w$ and $\beta$.

\subsection{Results on Stream LDPs over Numerical domain}\label{appendix:attack numerical}
\begin{figure}[htbp]
	\centering	
	\includegraphics[width=240pt,height=48pt]     {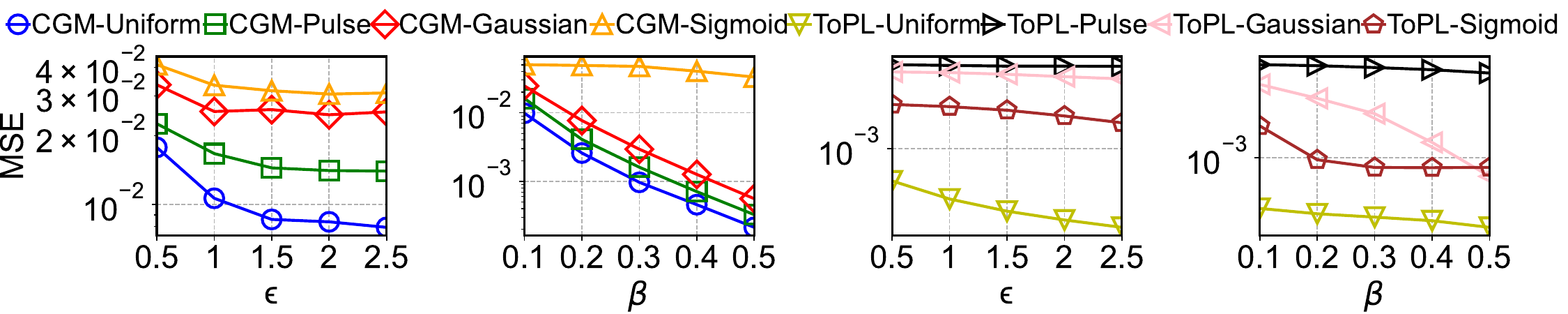}\vspace{-1mm}
	\text{\footnotesize (a) \textsf{Taxi-Longitude} dataset.}\vspace{1mm}
	\includegraphics[width=240pt,height=48pt]     {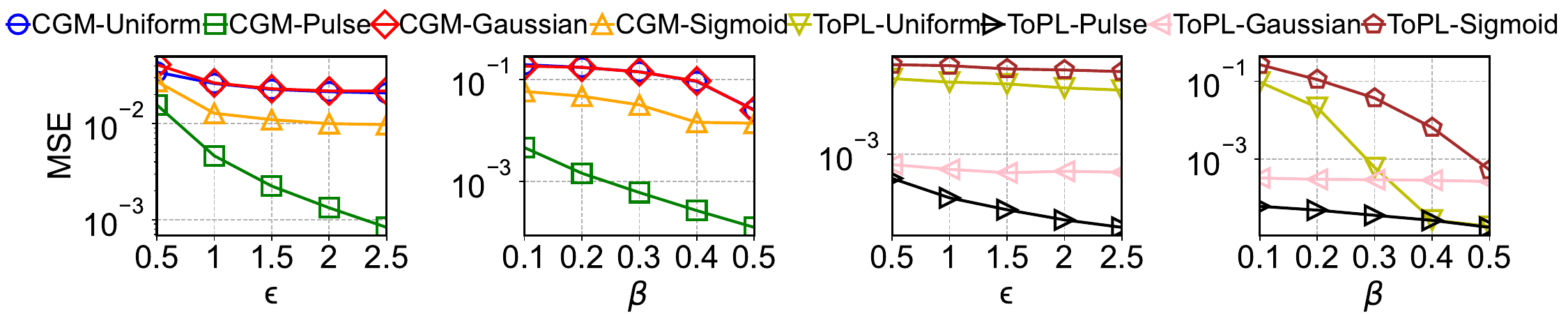}\vspace{-1mm}
	\text{\footnotesize (b) \textsf{Taobao-Price} dataset.}\vspace{-3.5mm}
	\caption{\small Attack effectiveness on stream LDPs over numerical domain.}\centering
	\label{fig:attack CGM and topl appendix}
\end{figure}

Fig.~\ref{fig:attack CGM and topl appendix} presents attack effectiveness on stream LDPs over numerical domain on \textsf{Taxi-Longitude} and \textsf{Taobao-Price} datasets.

\subsection{Results on Defense}\label{appendix:defense}
\begin{figure}[htbp]
	\centering	
 \includegraphics[width=0.48\textwidth]     {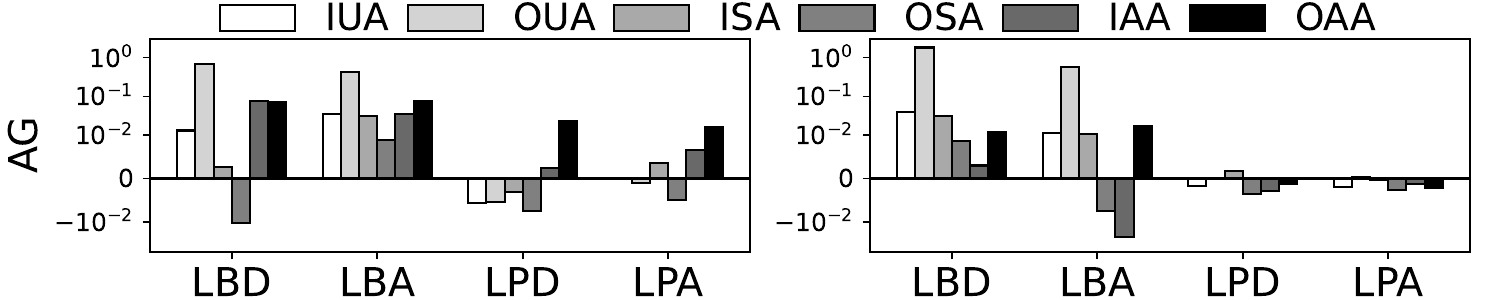}\vspace{0mm}
 \text{\footnotesize \quad \quad\space (a) Sigmoid target, $r=0.3$. \quad\quad\quad\quad\quad (b) Uniform target, $r=0.3$.}\vspace{1mm}
 \includegraphics[width=0.48\textwidth]     {./Figures/Stream_Defense/Taxi_r_0.5.pdf}\vspace{0mm}
 \text{\footnotesize \quad \quad\space (c) Sigmoid target, $r=0.5$. \quad\quad\quad\quad\quad (d) Uniform target, $r=0.5$.}\vspace{1mm}
  \includegraphics[width=0.48\textwidth]     {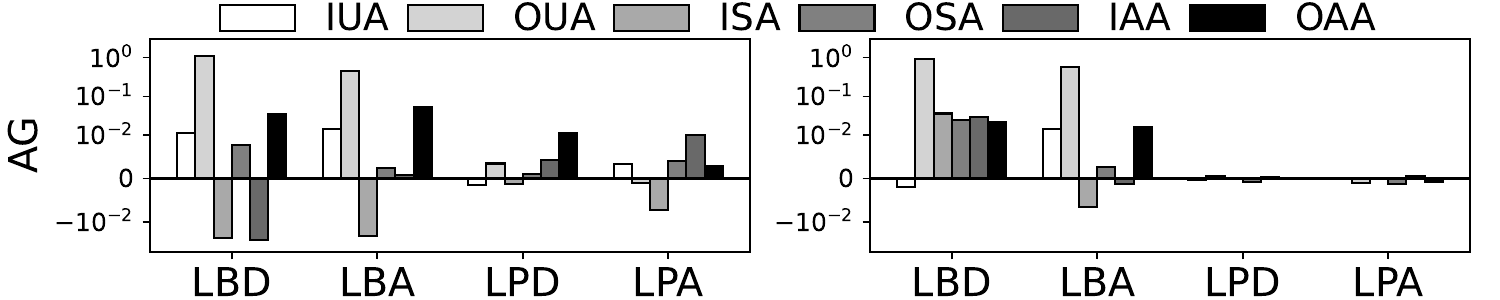}\vspace{0mm}
 \text{\footnotesize \quad \quad\space (e) Sigmoid target, $r=0.7$. \quad\quad\quad\quad\quad (f) Uniform target, $r=0.7$.}\vspace{-3mm}
	\caption{\small Accuracy Gain (\textsf{Taxi}, $\epsilon=1$).}\centering
 \label{fig:appendix defense}\vspace{-0mm}
\end{figure}

Fig.~\ref{fig:appendix defense} shows accuracy gain on different $r$.

\subsection{Results on Synthetic Datasets}
\label{sec: Experiment Results on Synthetic Datasets}
We created binary streaming datasets using various sequence models. Starting with a probability process model $p_t=f(t)$, time $T$, and user count $N$, we generated a sequence $\left(p_1,p_2,\ldots,p_T\right)$. At each timestamp $t$, we randomly assigned $p_t$ fraction of $N$ users a value of 1, with others set to 0. $p_t$ is defined accordingly.

\begin{itemize}
    \item \textbf{\textsf{LNS}} is a linear process $p_t=p_{t-1}+\mathcal{N}(0,Q_\sigma)$, where $p_0=0.5$ and $\mathcal{N}(0,Q_\sigma)$ is Gaussian noise with the standard variance $\sqrt{Q_\sigma}=0.025$.
    \item \textbf{\textsf{Sin}} is a sequence composed by a sine curve $p_t=A sin(bt)+h$ with $A=0.05$, $b=0.01$ and $h=0.5$.
    \item \textbf{\textsf{Log}} is a series with the logistic model $p_t=A/(1+e^{-bt})$ where $A=0.75$ and $b=0.01$.
    \item \textbf{\textsf{Pulse}} is an extreme case that randomly sets $p_t$ as 0 or 1.
\end{itemize}

Using default parameters and models, we generated synthetic binary streams from 100,000 users across 800 timestamps.

\begin{figure*}[htbp]
 	\centering	
         \subfigure[\textsf{LNS} dataset, Uniform $\mathbf{\tilde{f}}$]{
 		\includegraphics[width=0.24\textwidth]{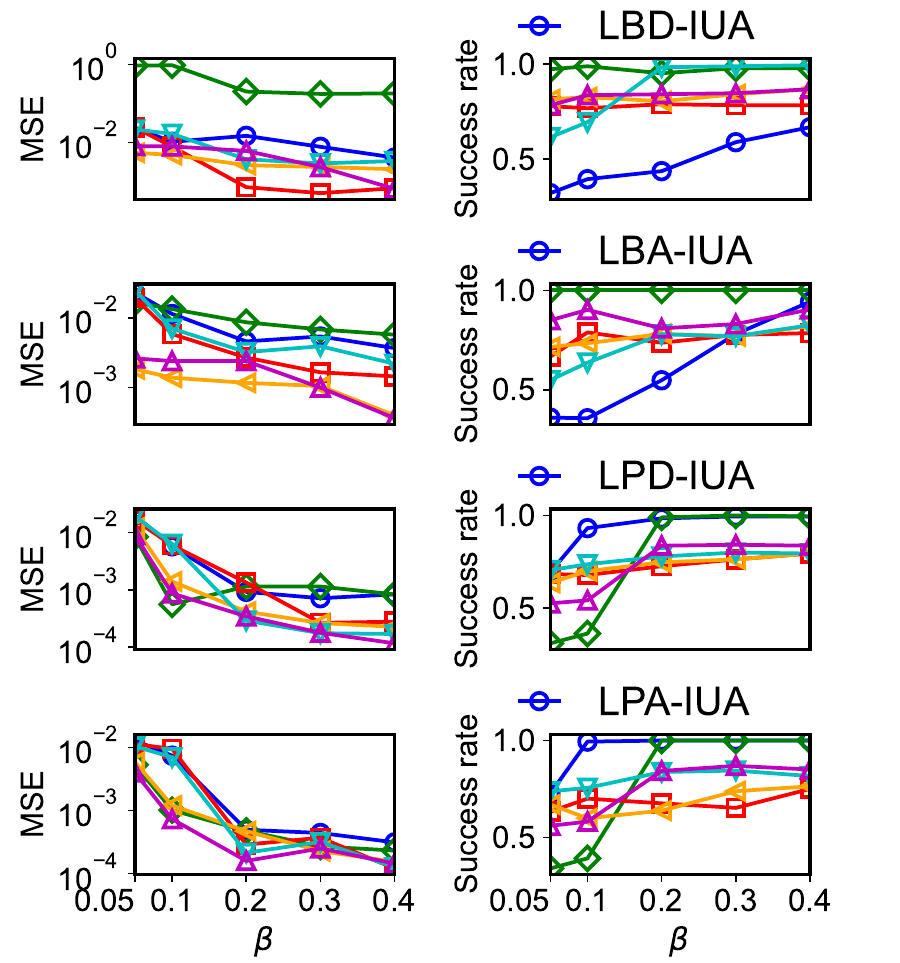}}\vspace{-0.06cm}\hspace{-4mm}
 	\subfigure[\textsf{LNS} dataset, Pulse $\mathbf{\tilde{f}}$]{
 		\includegraphics[width=0.24\textwidth]{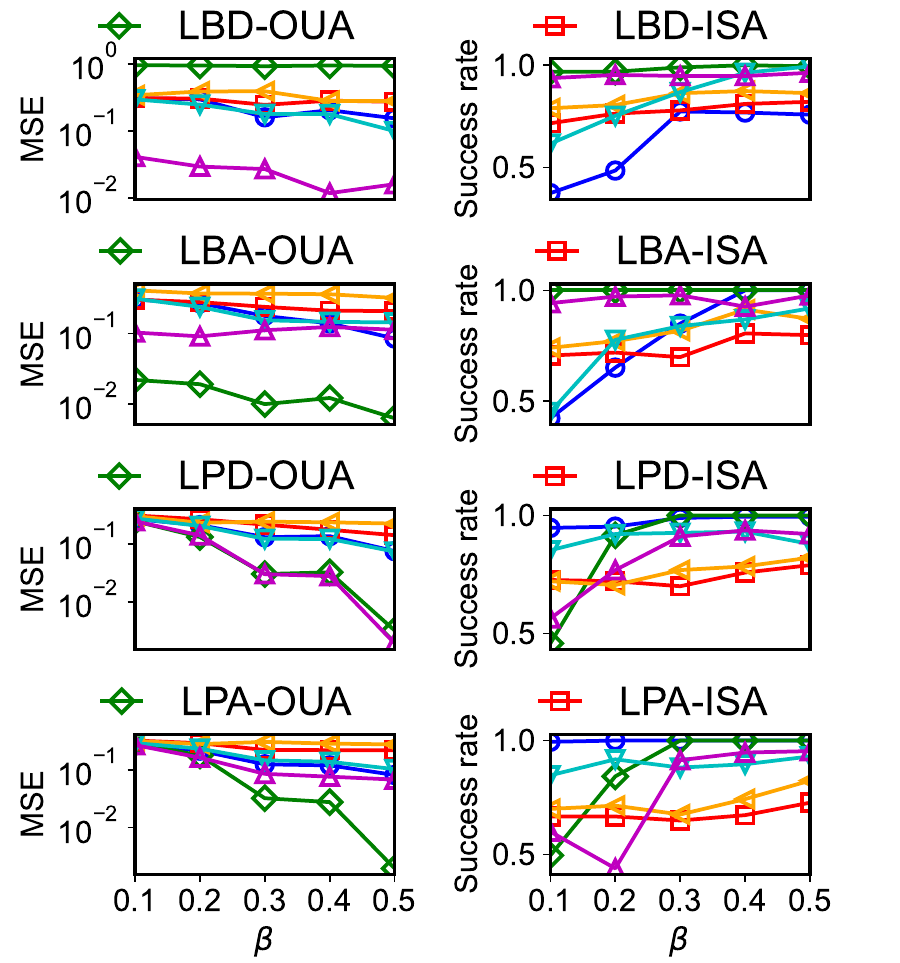}}\vspace{-0.06cm}\hspace{-4mm}
 	\subfigure[\textsf{LNS} dataset, Gaussian $\mathbf{\tilde{f}}$]{
 		\includegraphics[width=0.24\textwidth]{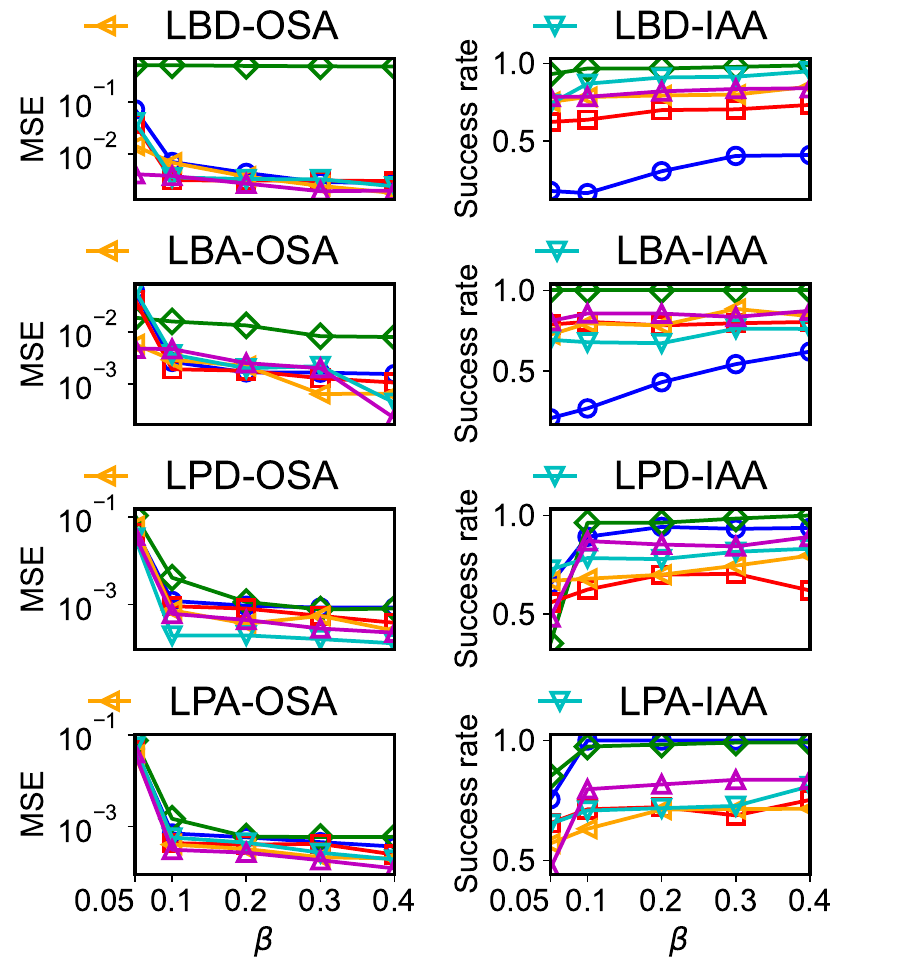}}\vspace{-0.06cm}\hspace{-4mm}
 	\subfigure[\textsf{LNS} dataset, Sigmoid $\mathbf{\tilde{f}}$]{
 		\includegraphics[width=0.24\textwidth]{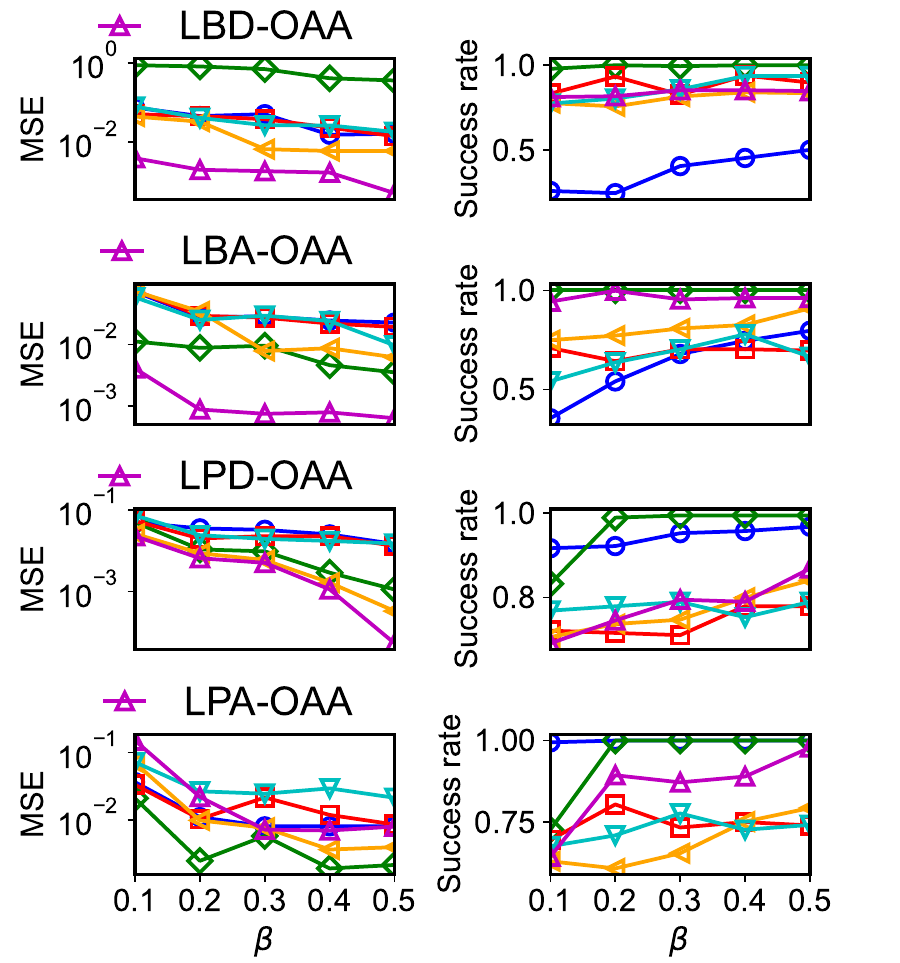}}\vspace{-0.06cm}\hspace{-4mm}
 	\\
 
         \subfigure[\textsf{Sin} dataset, Uniform $\mathbf{\tilde{f}}$]{
 		\includegraphics[width=0.24\textwidth]{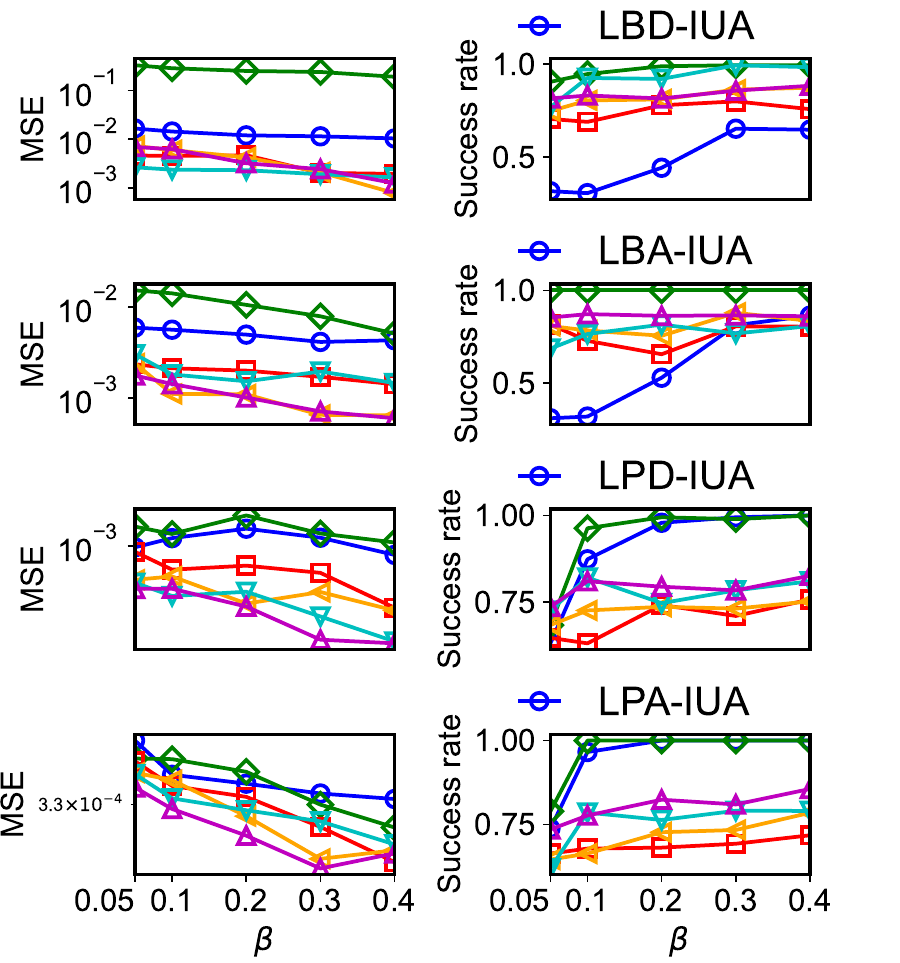}}\vspace{-0.06cm}\hspace{-4mm}
 	\subfigure[\textsf{Sin} dataset, Pulse $\mathbf{\tilde{f}}$]{
 		\includegraphics[width=0.24\textwidth]{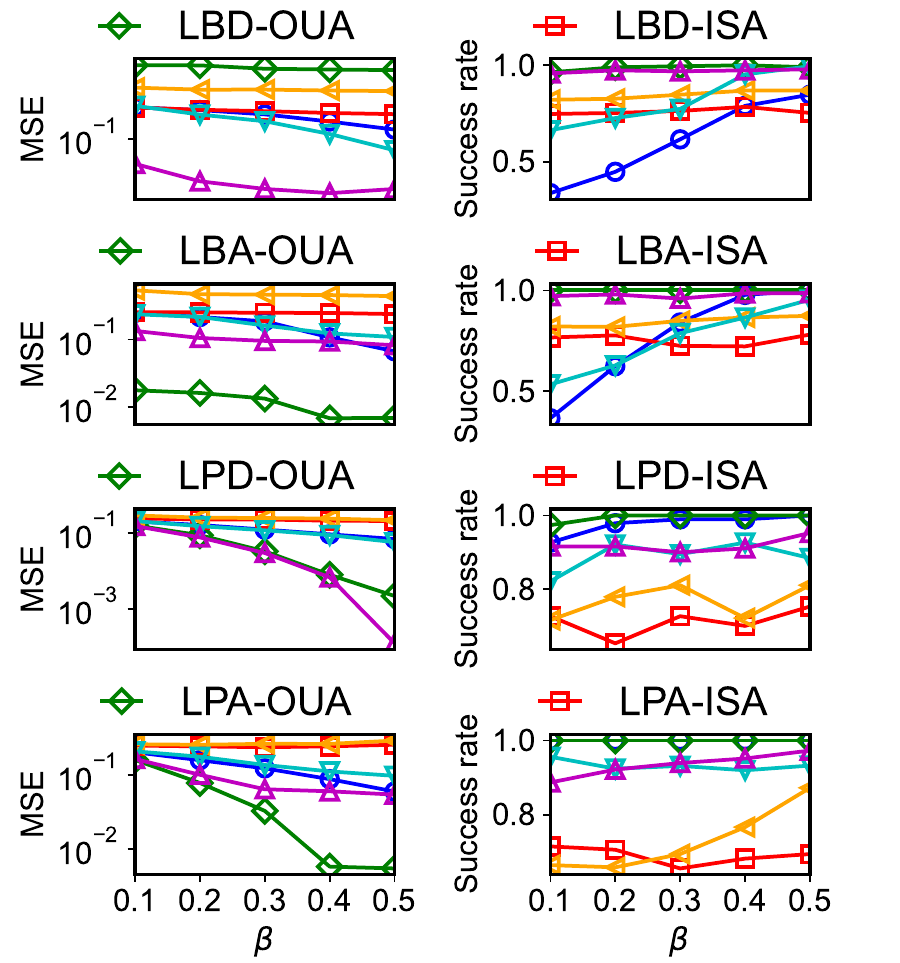}}\vspace{-0.06cm}\hspace{-4mm}
 	\subfigure[\textsf{Sin} dataset, Gaussian $\mathbf{\tilde{f}}$]{
 		\includegraphics[width=0.24\textwidth]{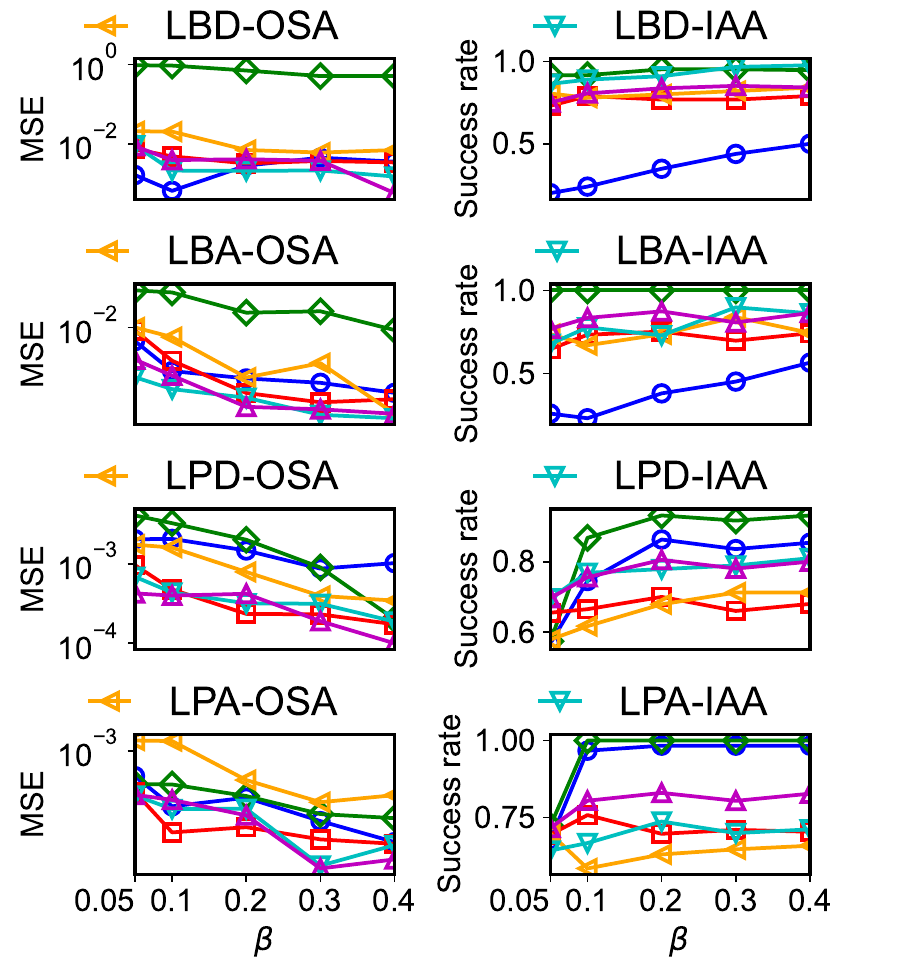}}\vspace{-0.06cm}\hspace{-4mm}
 	\subfigure[\textsf{Sin} dataset, Sigmoid $\mathbf{\tilde{f}}$]{
 		\includegraphics[width=0.24\textwidth]{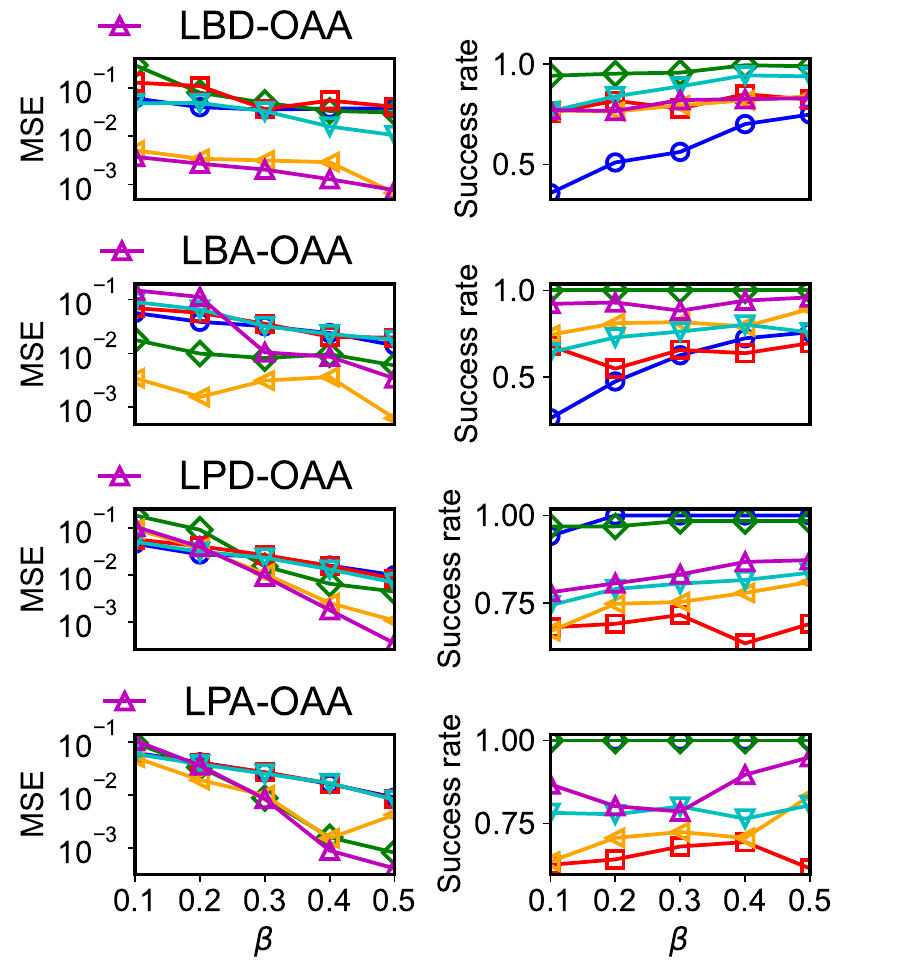}}\vspace{-0.06cm}\hspace{-4mm}
         \\
        
         \subfigure[\textsf{Log} dataset, Uniform $\mathbf{\tilde{f}}$]{
 		\includegraphics[width=0.24\textwidth]{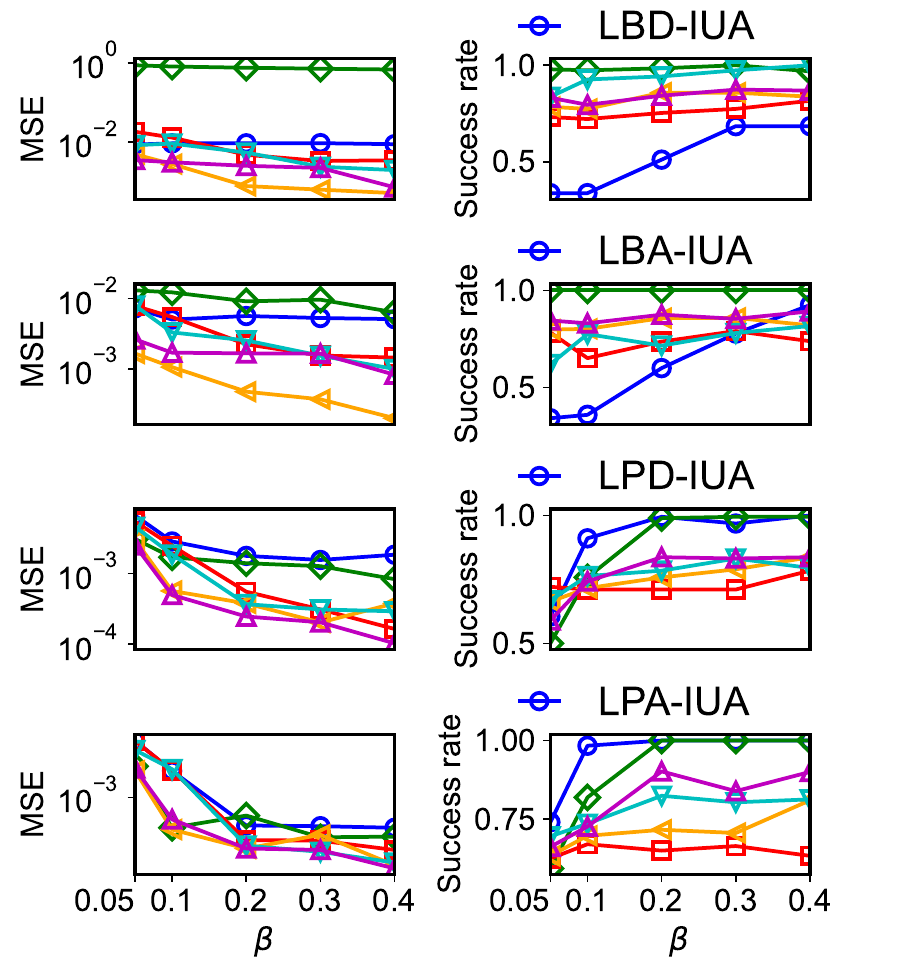}}\vspace{-0.05cm}\hspace{-4mm}
 	\subfigure[\textsf{Log} dataset, Pulse $\mathbf{\tilde{f}}$]{
 		\includegraphics[width=0.24\textwidth]{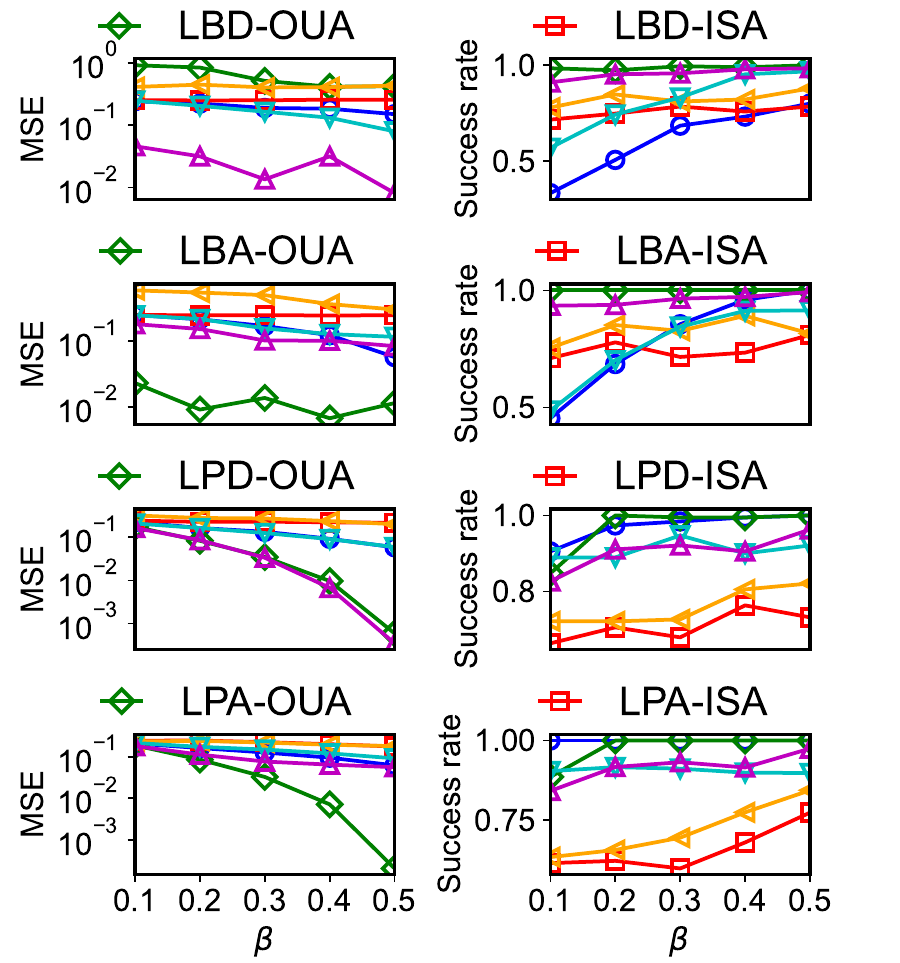}}\vspace{-0.06cm}\hspace{-4mm}
 	\subfigure[\textsf{Log} dataset, Gaussian $\mathbf{\tilde{f}}$]{
 		\includegraphics[width=0.24\textwidth]{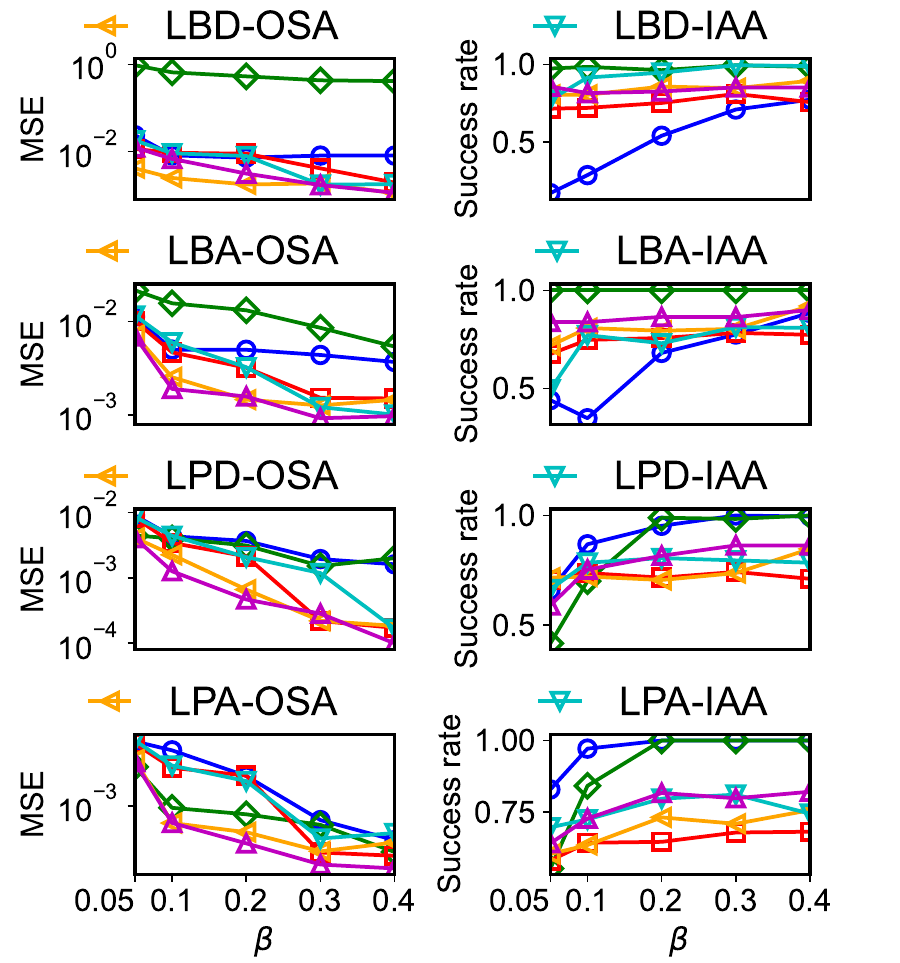}}\vspace{-0.06cm}\hspace{-4mm}
 	\subfigure[\textsf{Log} dataset, Sigmoid $\mathbf{\tilde{f}}$]{
 		\includegraphics[width=0.24\textwidth]{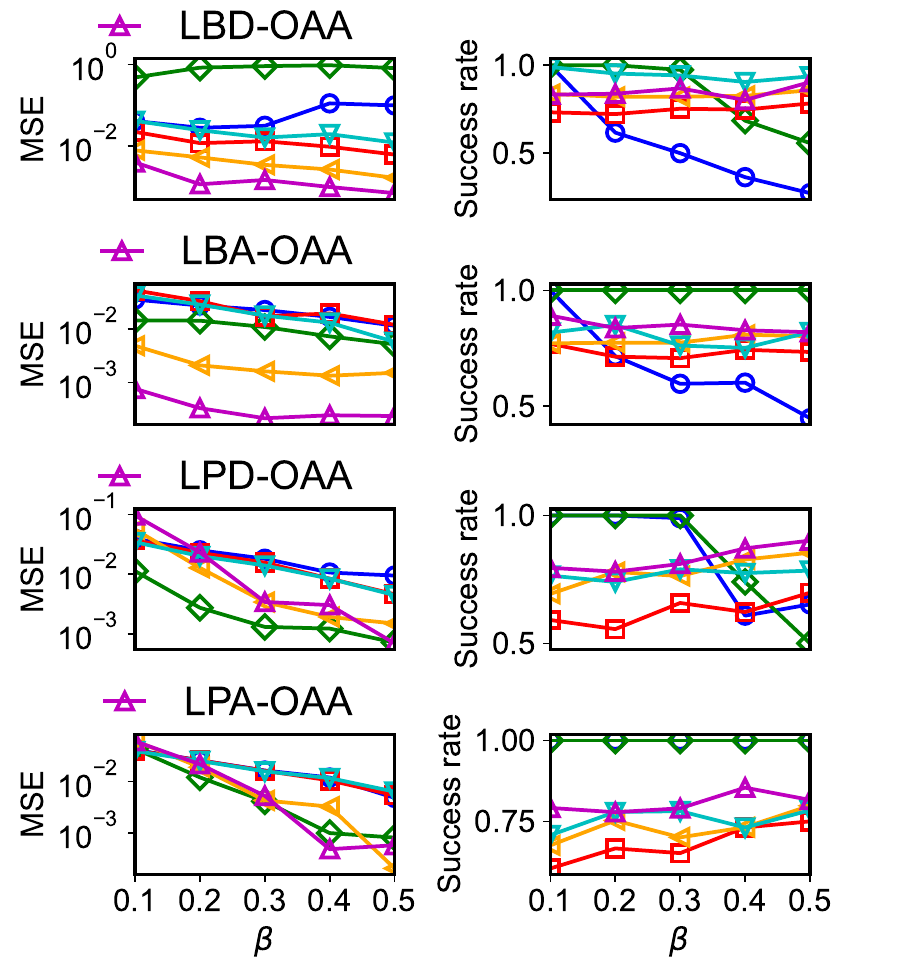}}\vspace{-0.06cm}\hspace{-4mm}
         \\
        
         \subfigure[\textsf{Pulse} dataset, Uniform $\mathbf{\tilde{f}}$]{
 		\includegraphics[width=0.24\textwidth]{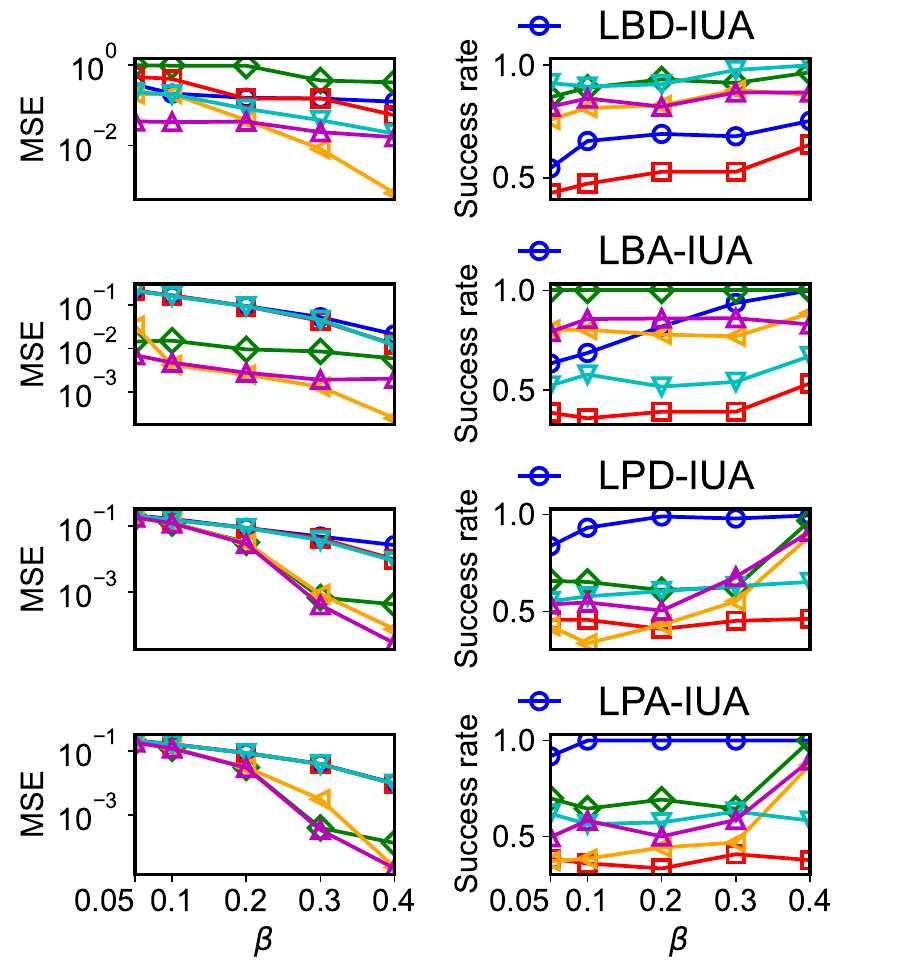}}\vspace{-0.13cm}\hspace{-4mm}
 	\subfigure[\textsf{Pulse} dataset, Pulse $\mathbf{\tilde{f}}$]{
 		\includegraphics[width=0.24\textwidth]{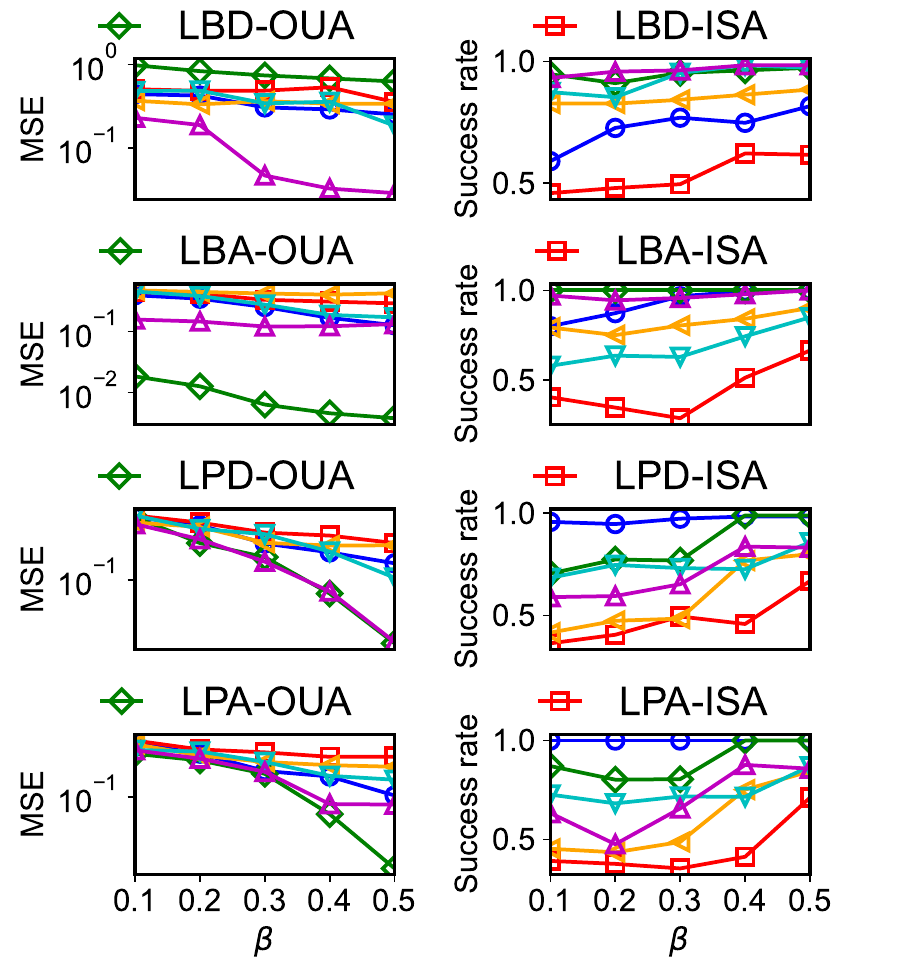}}\vspace{-0.13cm}\hspace{-4mm}
 	\subfigure[\textsf{Pulse} dataset, Gaussian $\mathbf{\tilde{f}}$]{
 		\includegraphics[width=0.24\textwidth]{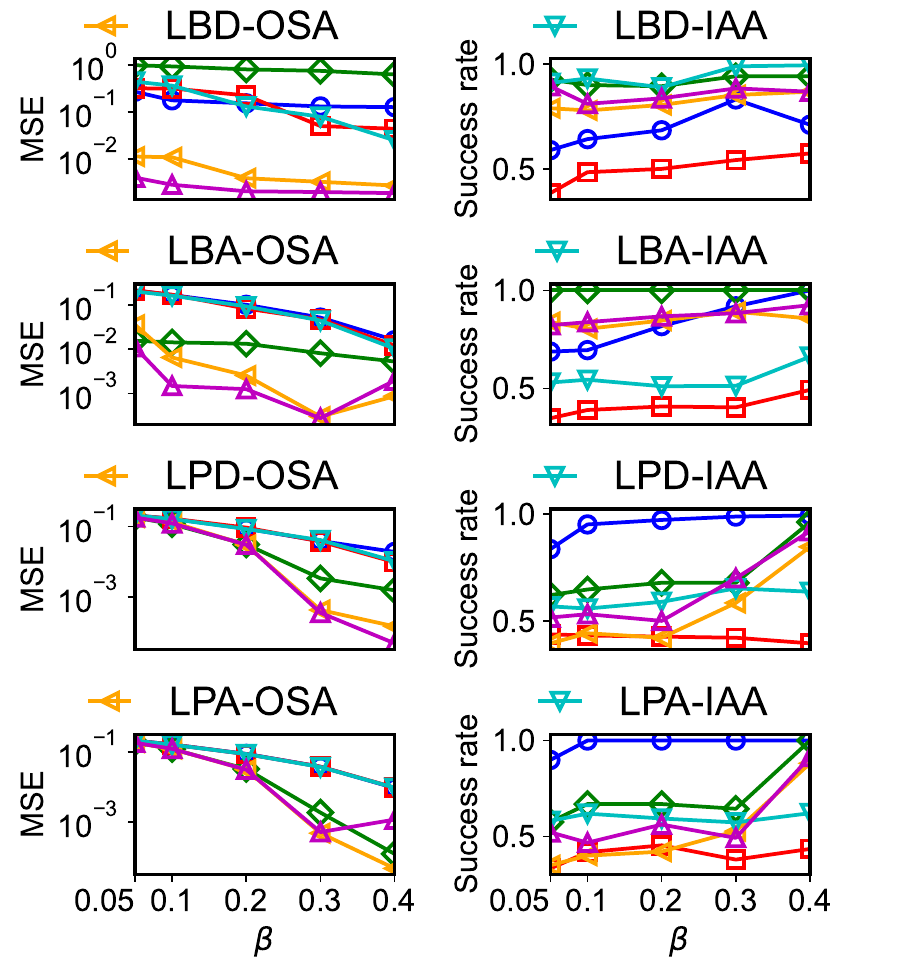}}\vspace{-0.13cm}\hspace{-4mm}
 	\subfigure[\textsf{Pulse} dataset, Sigmoid $\mathbf{\tilde{f}}$]{
 		\includegraphics[width=0.24\textwidth]{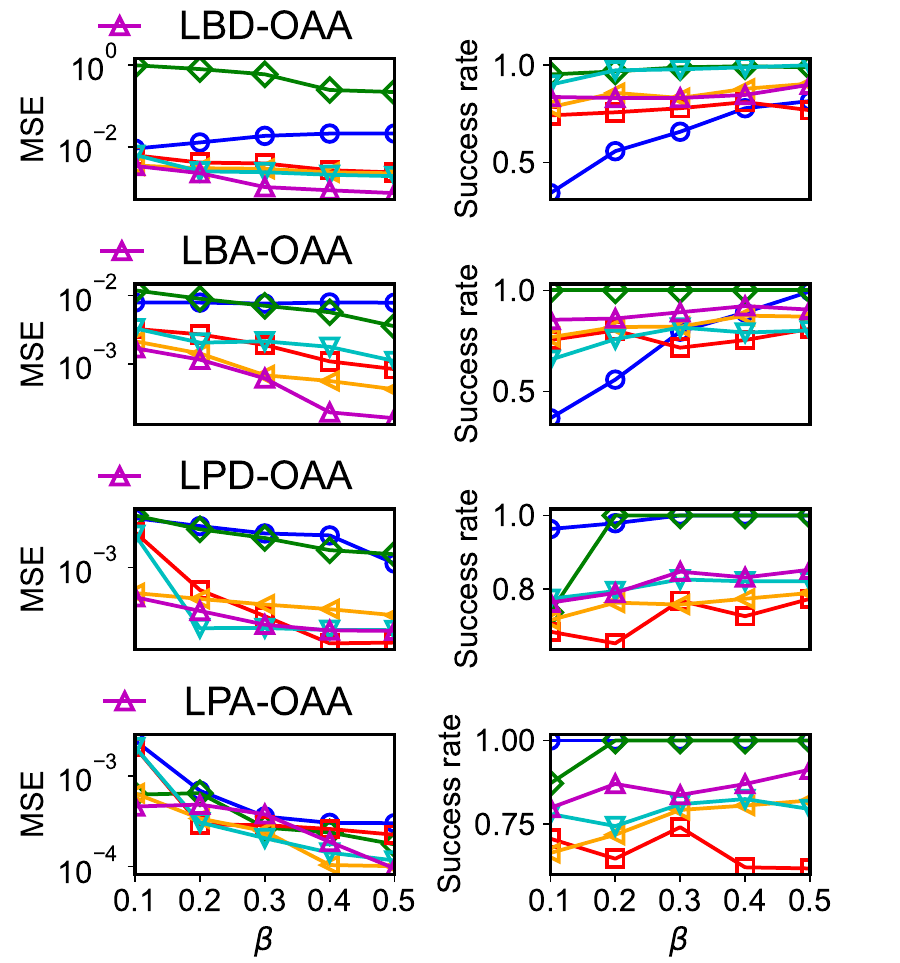}}
 	\caption{\small Attack effectiveness for Synthetic datasets, varying $\beta$.
  }\centering
 	\label{fig:beta synthetic data} 
 	\vspace{-0.5cm}
 \end{figure*}

\begin{figure*}[htbp]
 	\centering	
         \subfigure[\textsf{LNS} dataset, Uniform $\mathbf{\tilde{f}}$]{
 		\includegraphics[width=0.24\textwidth]{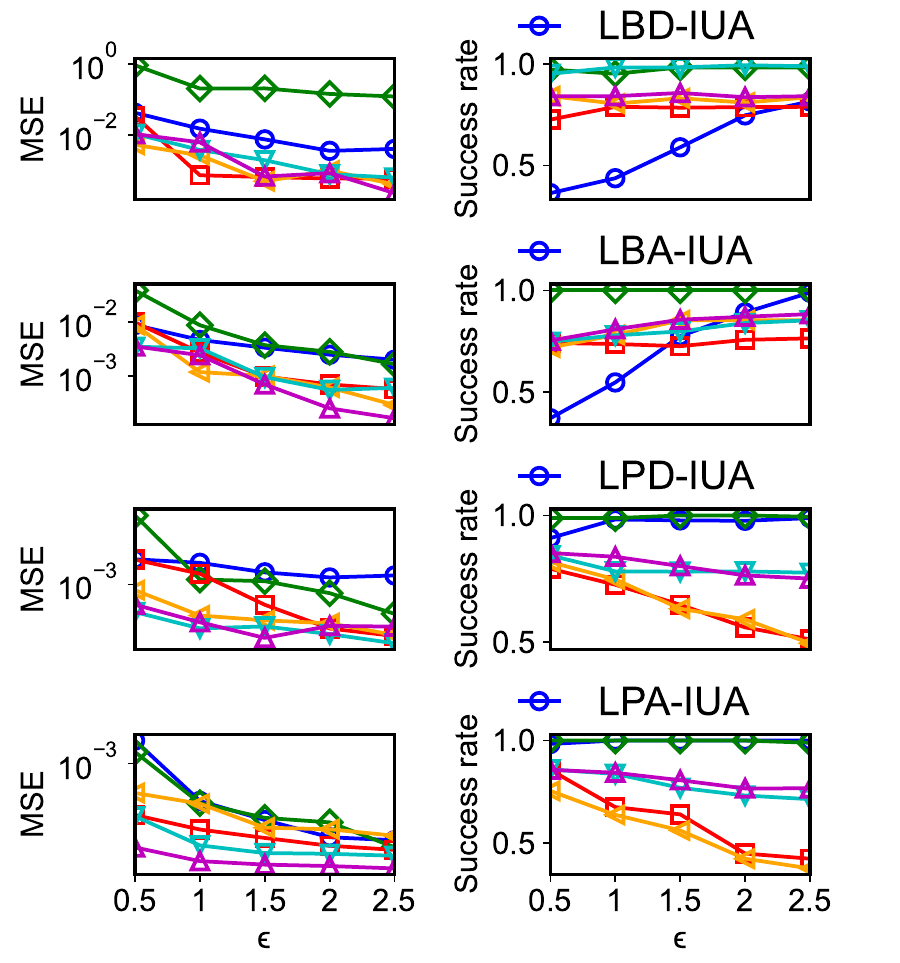}}\vspace{-0.06cm}\hspace{-4mm}
 	\subfigure[\textsf{LNS} dataset, Pulse $\mathbf{\tilde{f}}$]{
 		\includegraphics[width=0.24\textwidth]{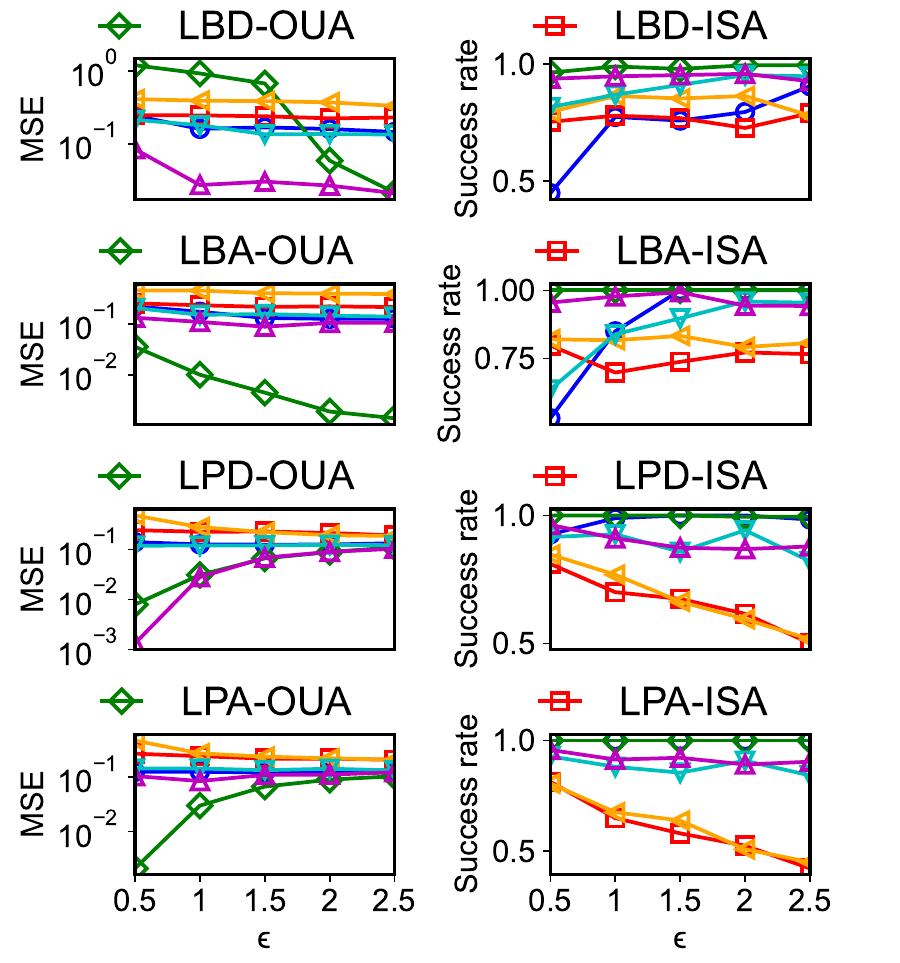}}\vspace{-0.06cm}\hspace{-4mm}
 	\subfigure[\textsf{LNS} dataset, Gaussian $\mathbf{\tilde{f}}$]{
 		\includegraphics[width=0.24\textwidth]{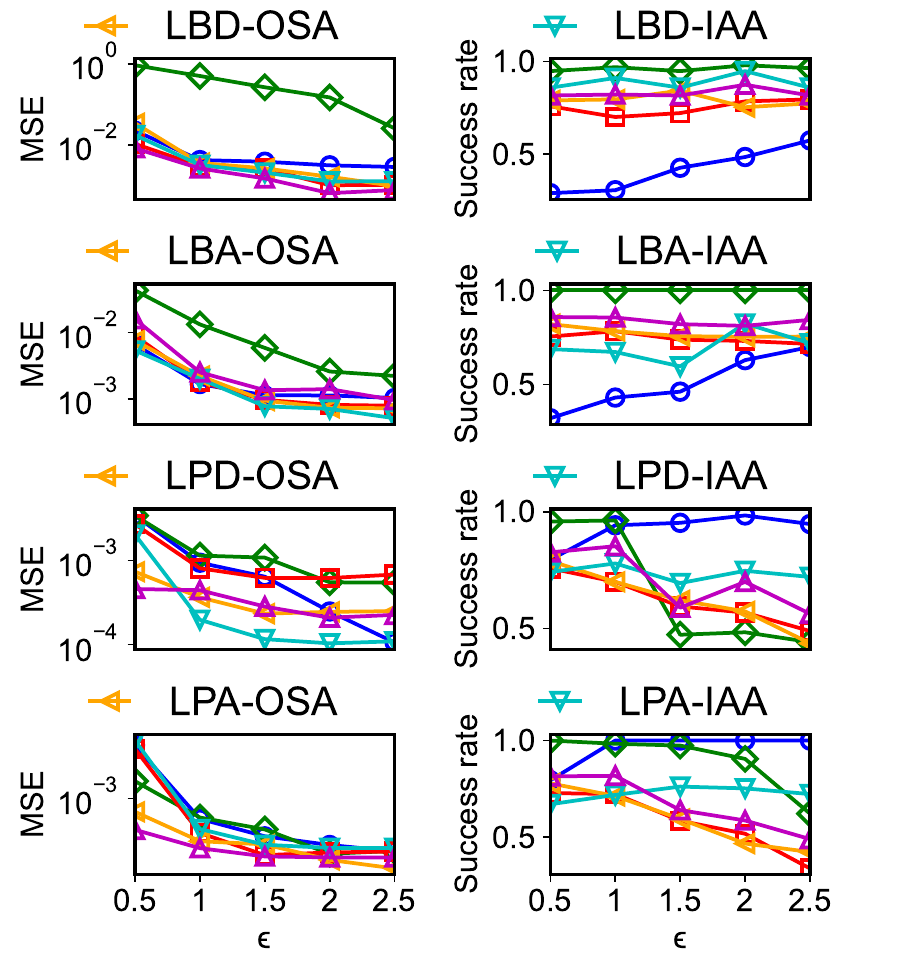}}\vspace{-0.06cm}\hspace{-4mm}
 	\subfigure[\textsf{LNS} dataset, Sigmoid $\mathbf{\tilde{f}}$]{
 		\includegraphics[width=0.24\textwidth]{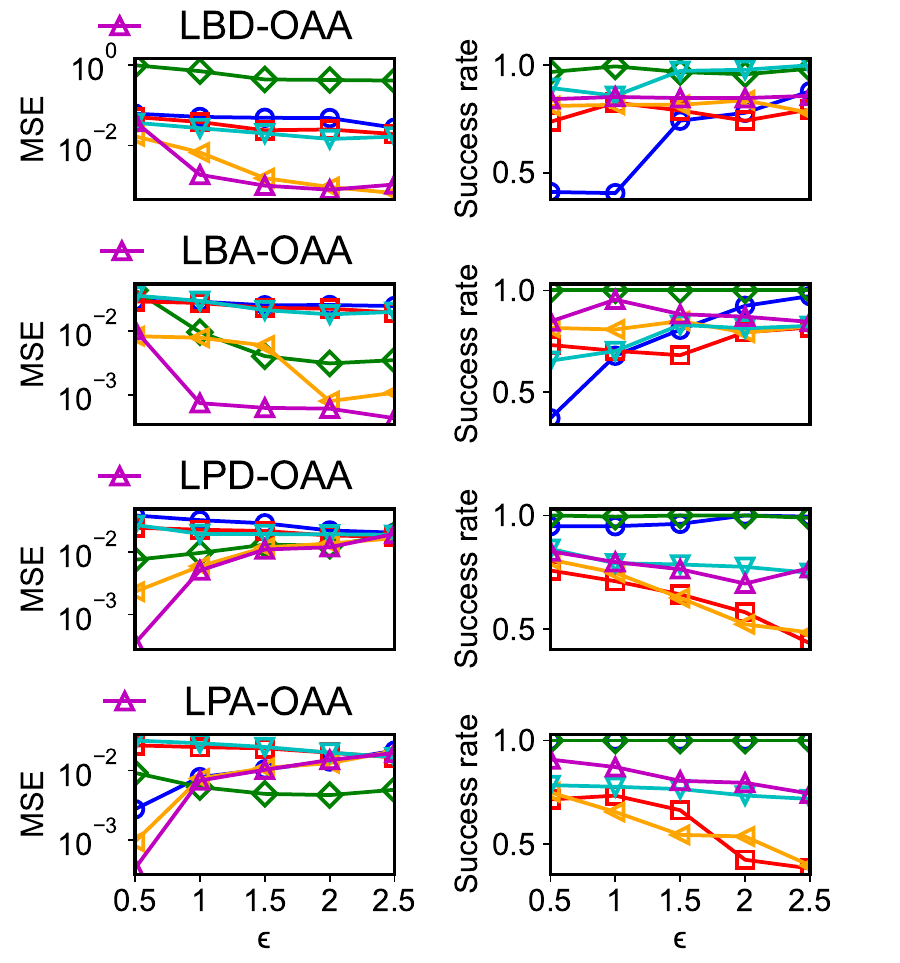}}\vspace{-0.06cm}\hspace{-4mm}
         \\
  
 	\subfigure[\textsf{Sin} dataset, Uniform $\mathbf{\tilde{f}}$]{
 		\includegraphics[width=0.24\textwidth]{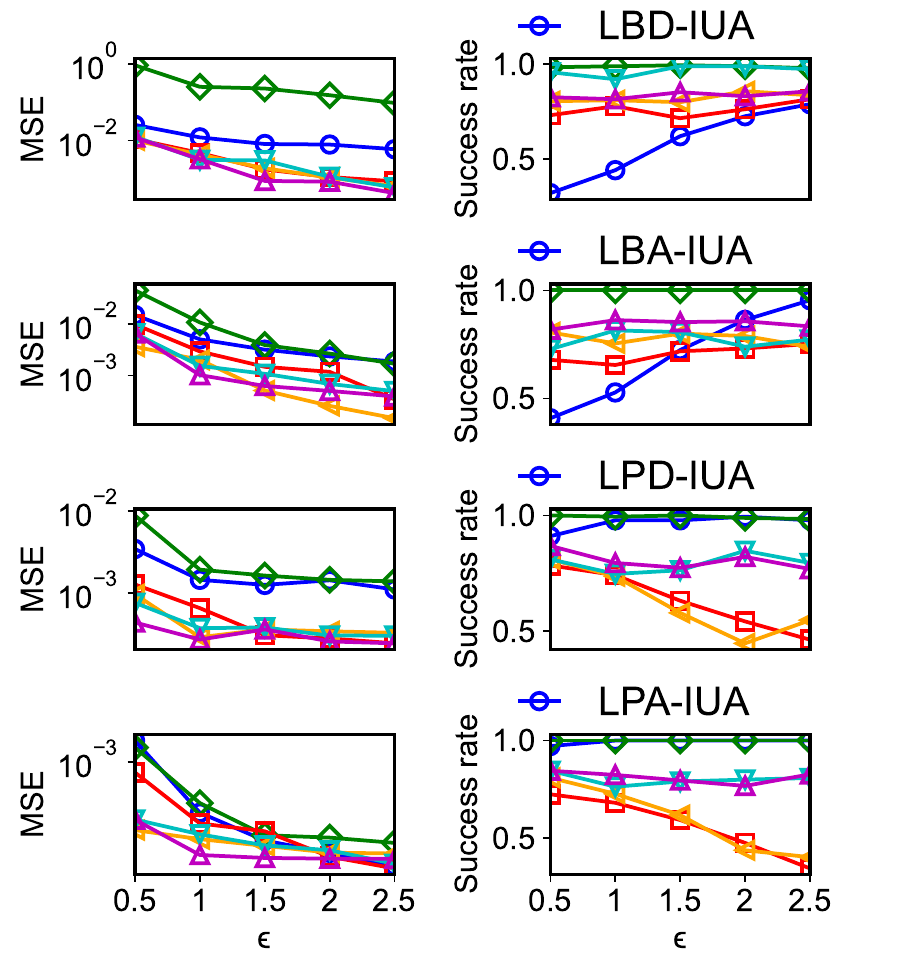}}\vspace{-0.06cm}\hspace{-4mm}
 	\subfigure[\textsf{Sin} dataset, Pulse $\mathbf{\tilde{f}}$]{
 		\includegraphics[width=0.24\textwidth]{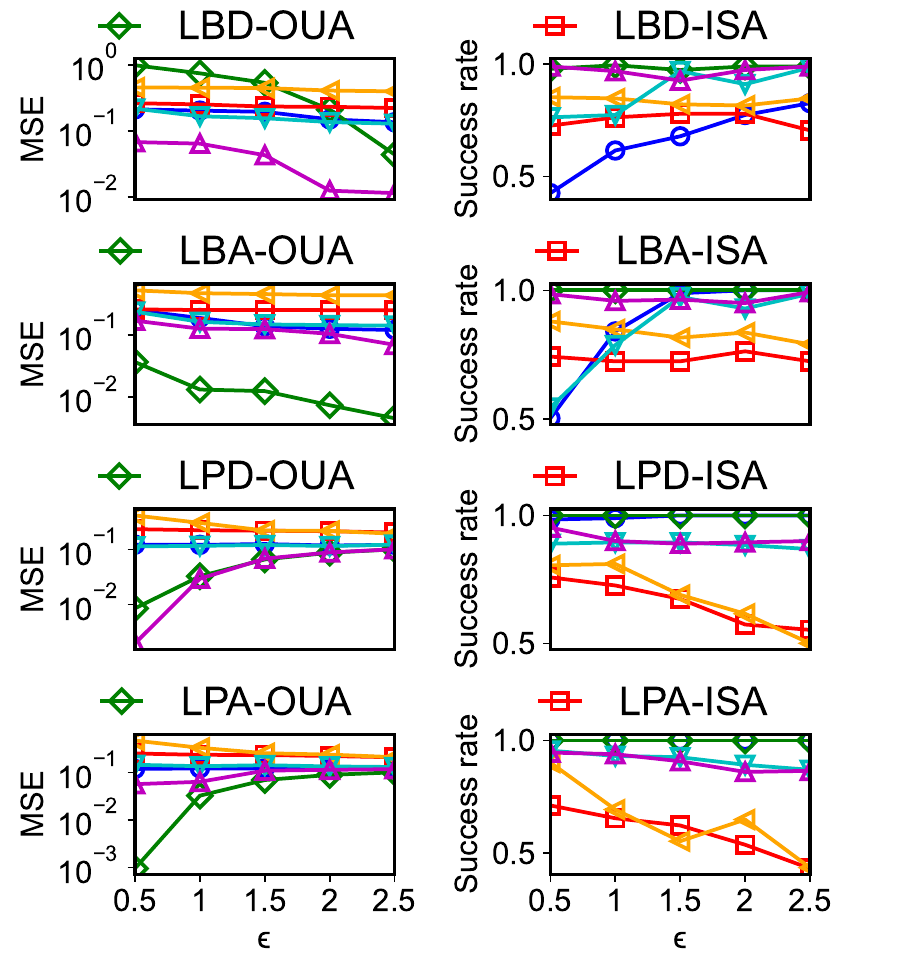}}\vspace{-0.06cm}\hspace{-4mm}
 	\subfigure[\textsf{Sin} dataset, Gaussian $\mathbf{\tilde{f}}$]{
 		\includegraphics[width=0.24\textwidth]{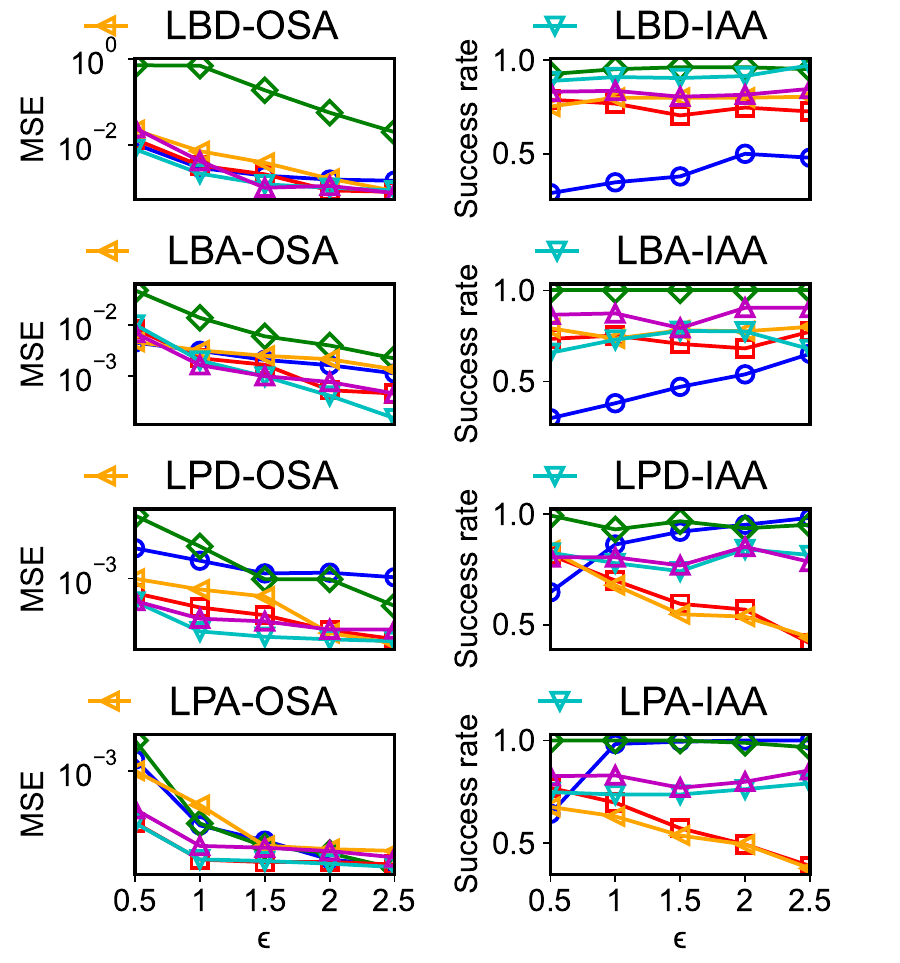}}\vspace{-0.06cm}\hspace{-4mm}
 	\subfigure[\textsf{Sin} dataset, Sigmoid $\mathbf{\tilde{f}}$]{
 		\includegraphics[width=0.24\textwidth]{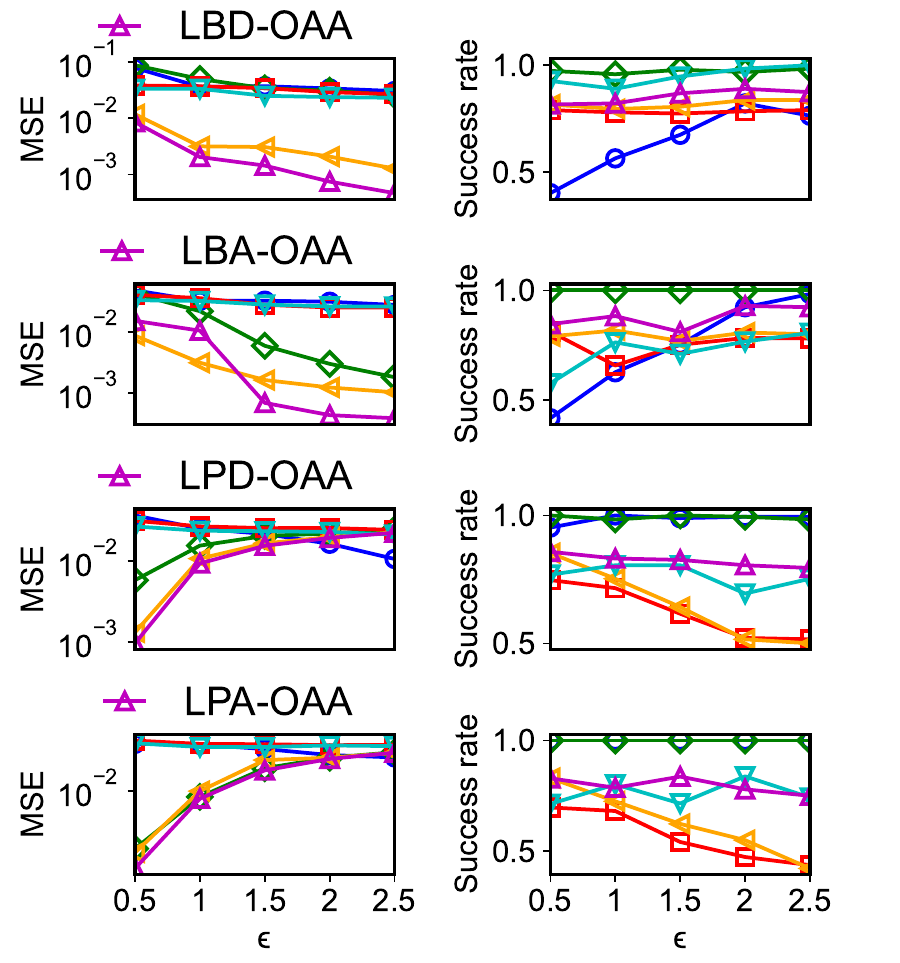}}\vspace{-0.06cm}\hspace{-4mm}
         \\
        
         \subfigure[\textsf{Log} dataset, Uniform $\mathbf{\tilde{f}}$]{
 		\includegraphics[width=0.24\textwidth]{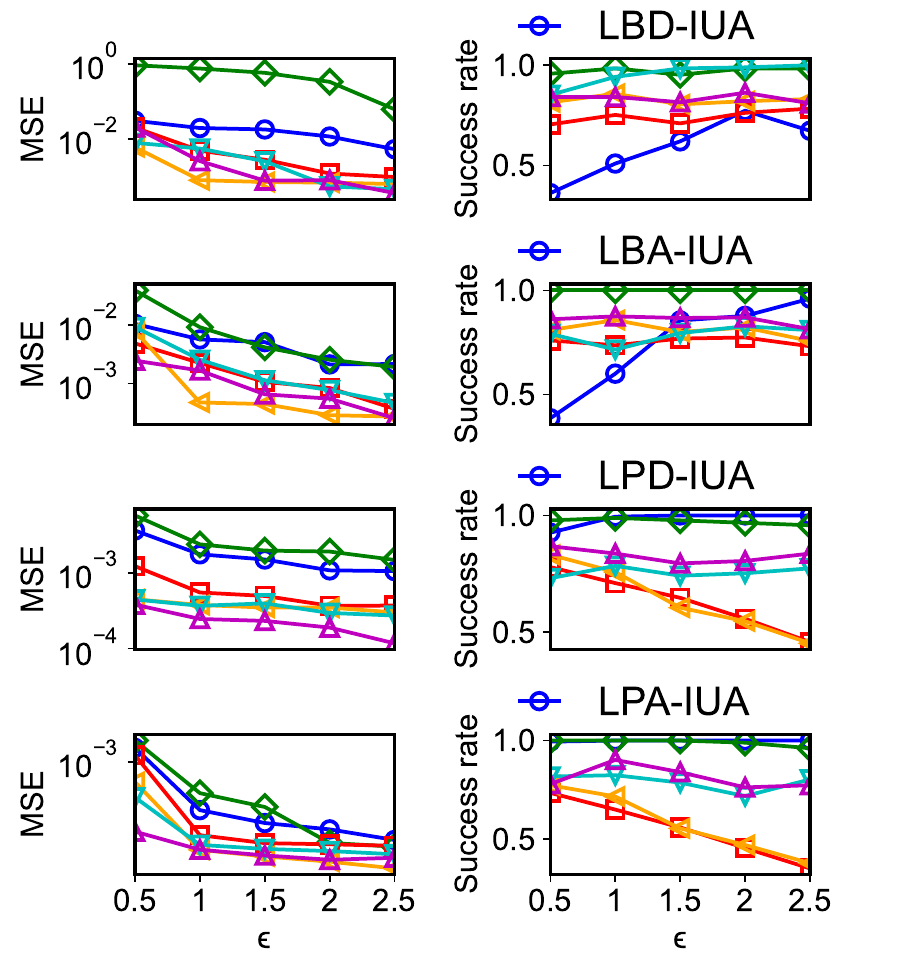}}\vspace{-0.05cm}\hspace{-4mm}
 	\subfigure[\textsf{Log} dataset, Pulse $\mathbf{\tilde{f}}$]{
 		\includegraphics[width=0.24\textwidth]{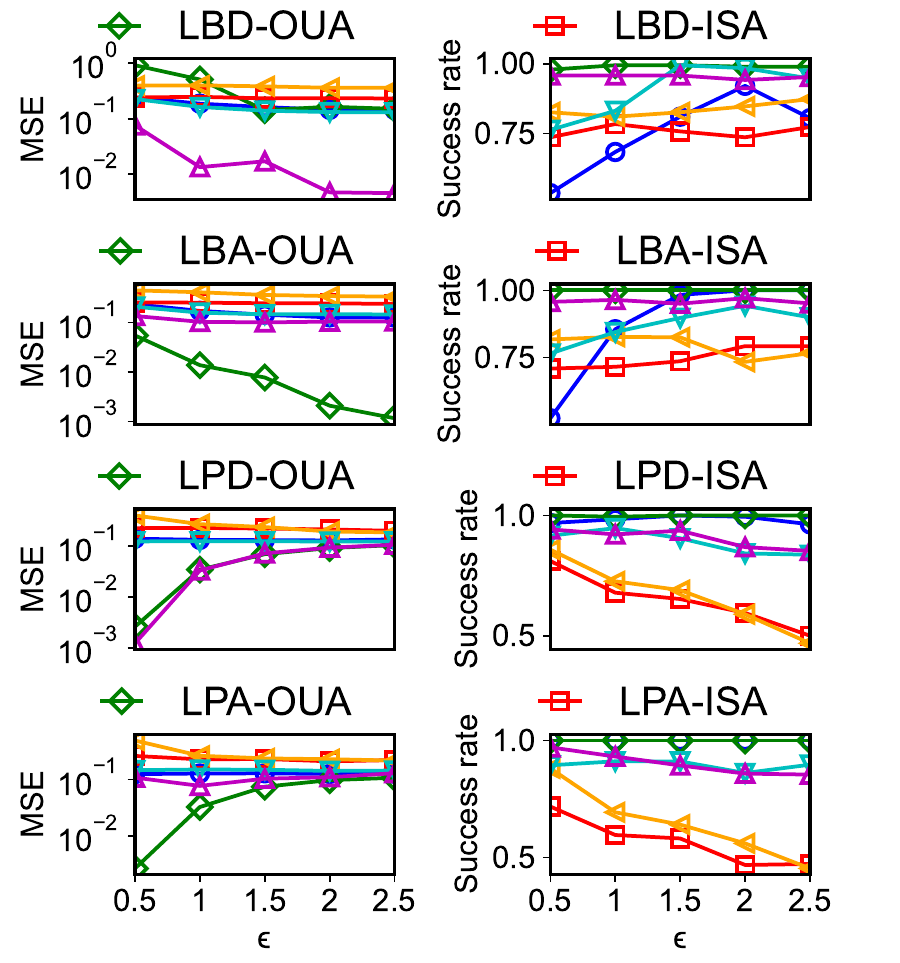}}\vspace{-0.06cm}\hspace{-4mm}
 	\subfigure[\textsf{Log} dataset, Gaussian $\mathbf{\tilde{f}}$]{
 		\includegraphics[width=0.24\textwidth]{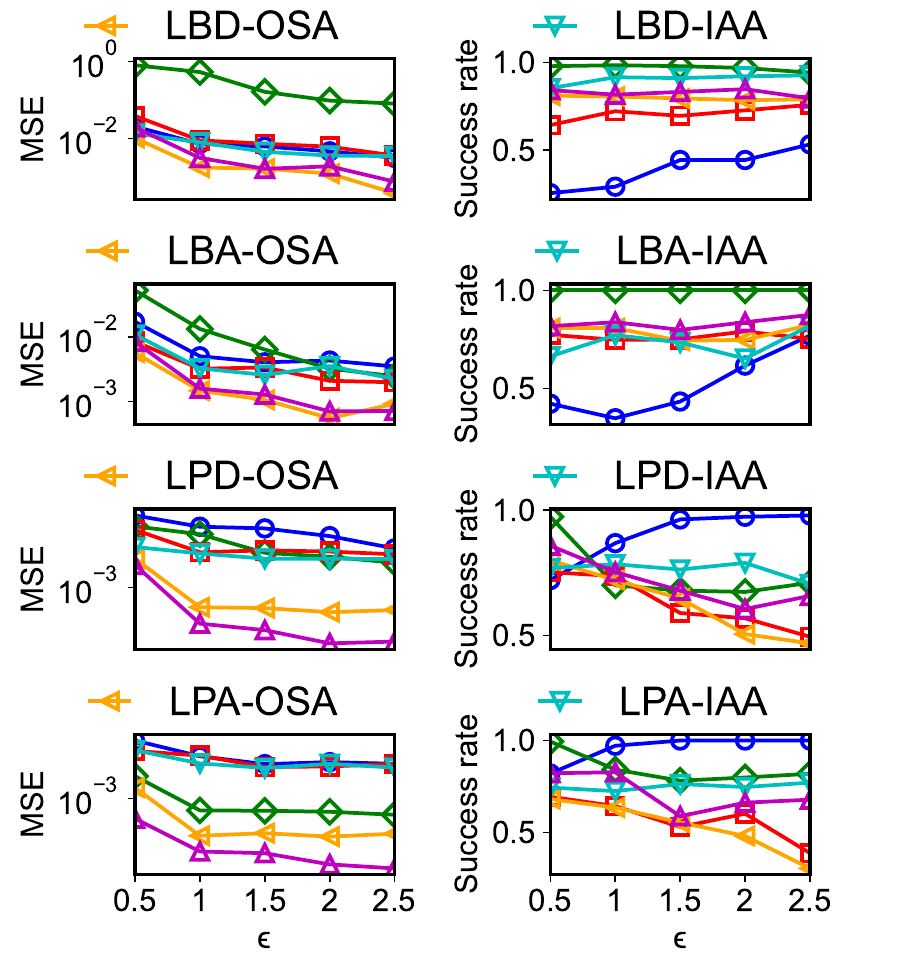}}\vspace{-0.06cm}\hspace{-4mm}
 	\subfigure[\textsf{Log} dataset, Sigmoid $\mathbf{\tilde{f}}$]{
 		\includegraphics[width=0.24\textwidth]{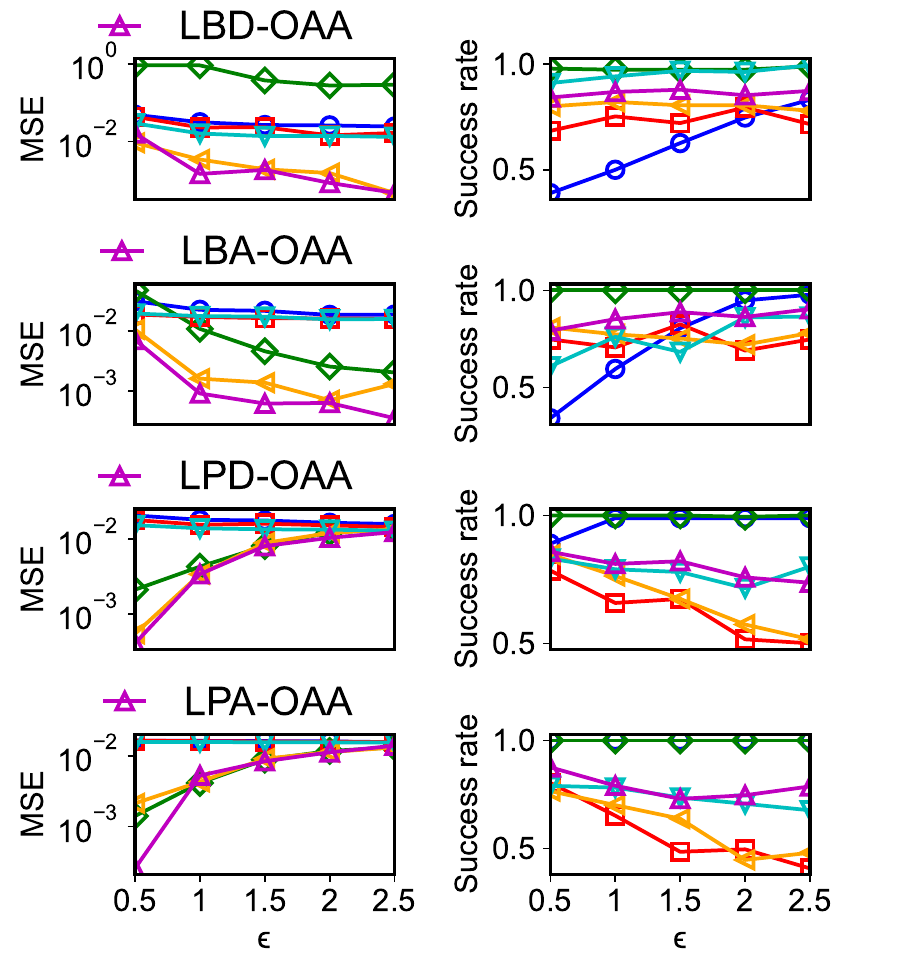}}\vspace{-0.06cm}\hspace{-4mm}
         \\
        
         \subfigure[\textsf{Pulse} dataset, Uniform $\mathbf{\tilde{f}}$]{
 		\includegraphics[width=0.24\textwidth]{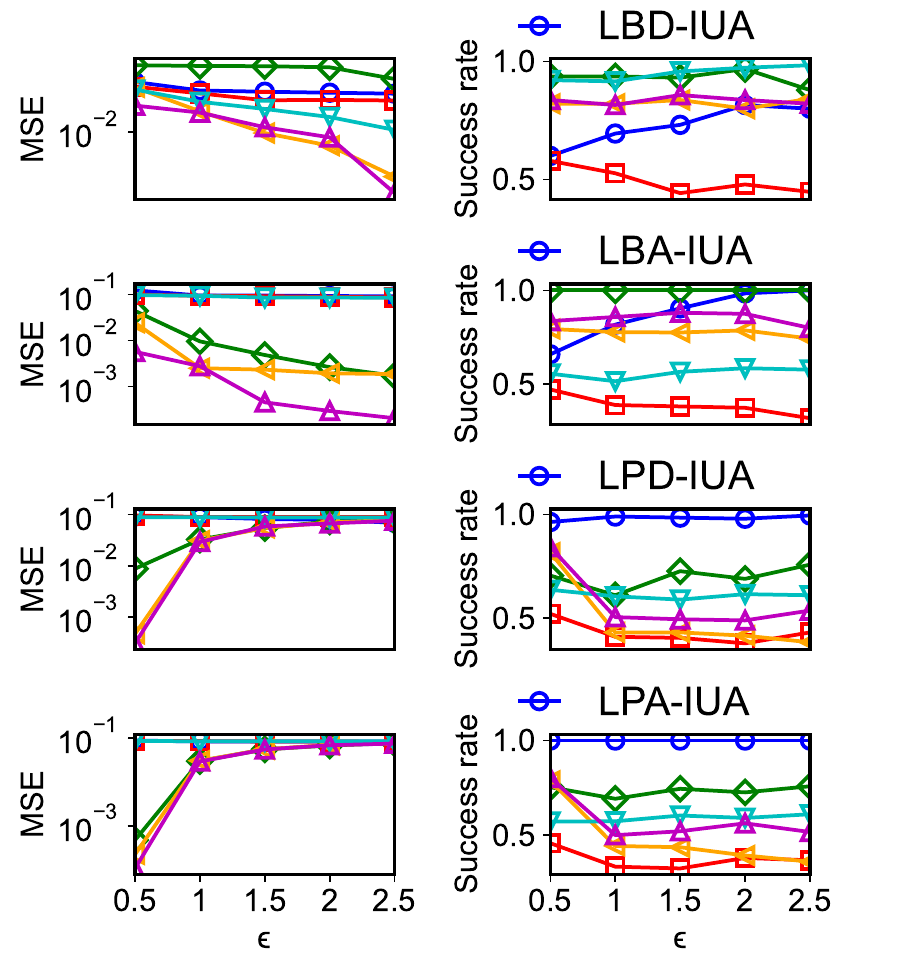}}\vspace{-0.13cm}\hspace{-4mm}
 	\subfigure[\textsf{Pulse} dataset, Pulse $\mathbf{\tilde{f}}$]{
 		\includegraphics[width=0.24\textwidth]{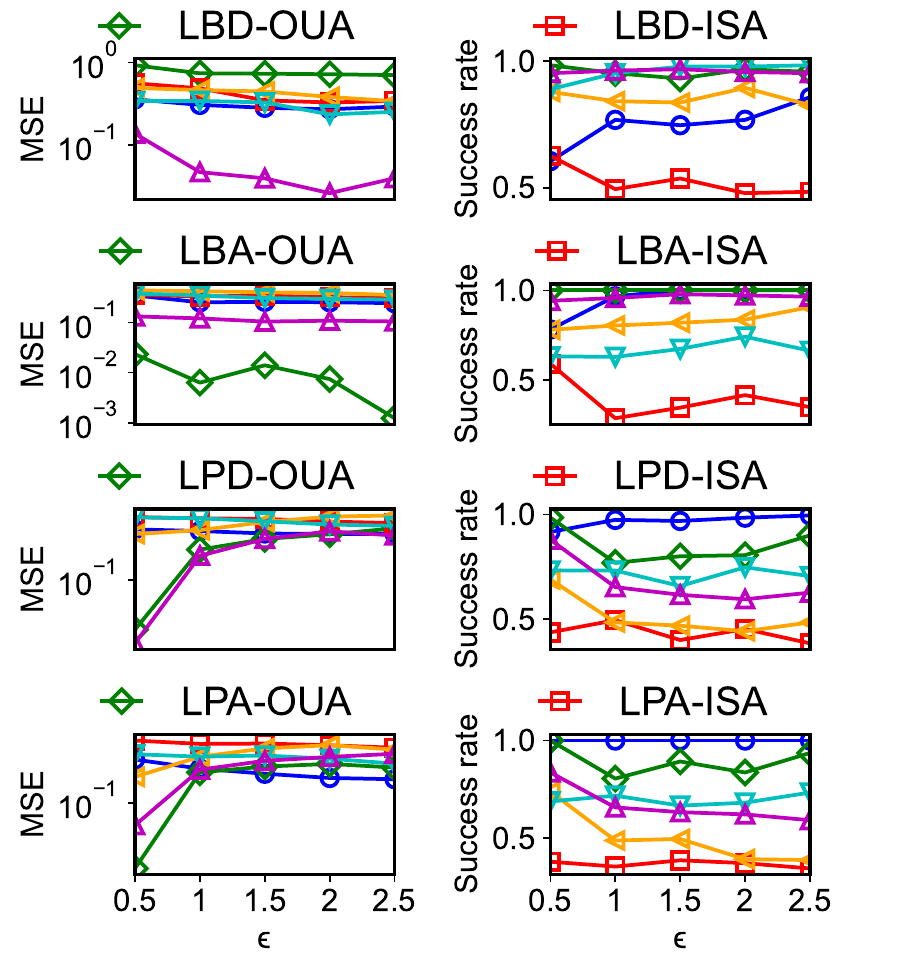}}\vspace{-0.13cm}\hspace{-4mm}
 	\subfigure[\textsf{Pulse} dataset, Gaussian $\mathbf{\tilde{f}}$]{
 		\includegraphics[width=0.24\textwidth]{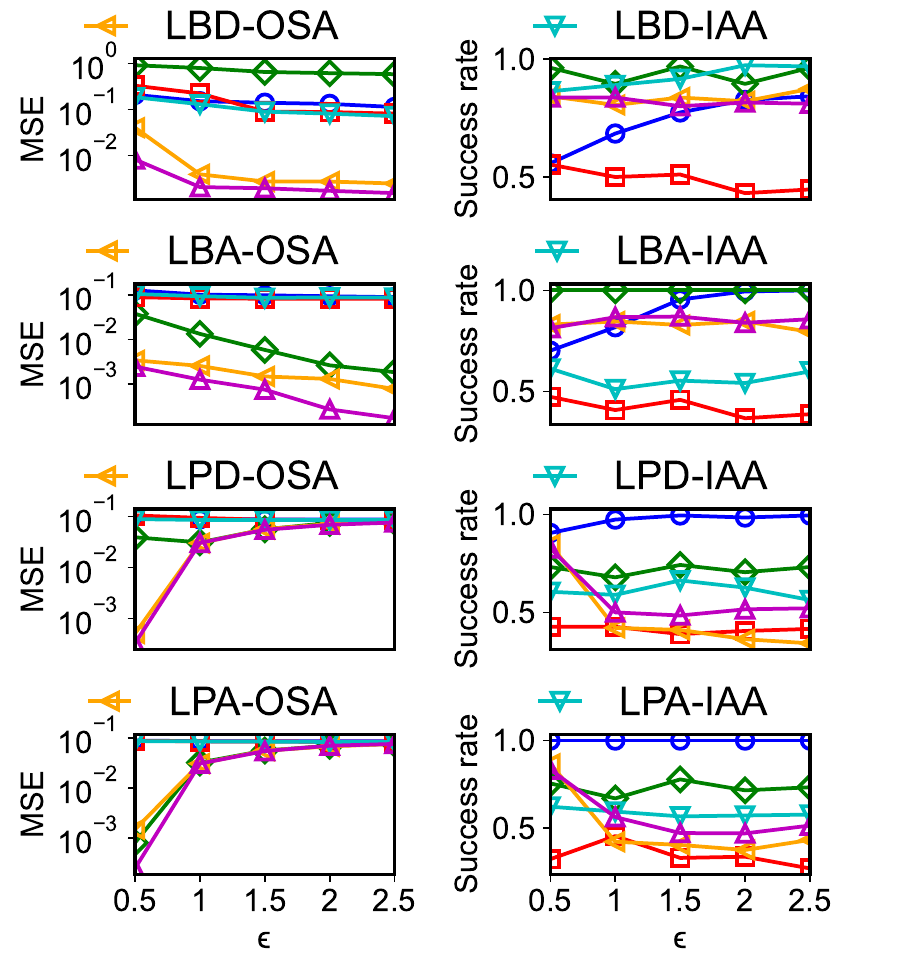}}\vspace{-0.13cm}\hspace{-4mm}
 	\subfigure[\textsf{Pulse} dataset, Sigmoid $\mathbf{\tilde{f}}$]{
 		\includegraphics[width=0.24\textwidth]{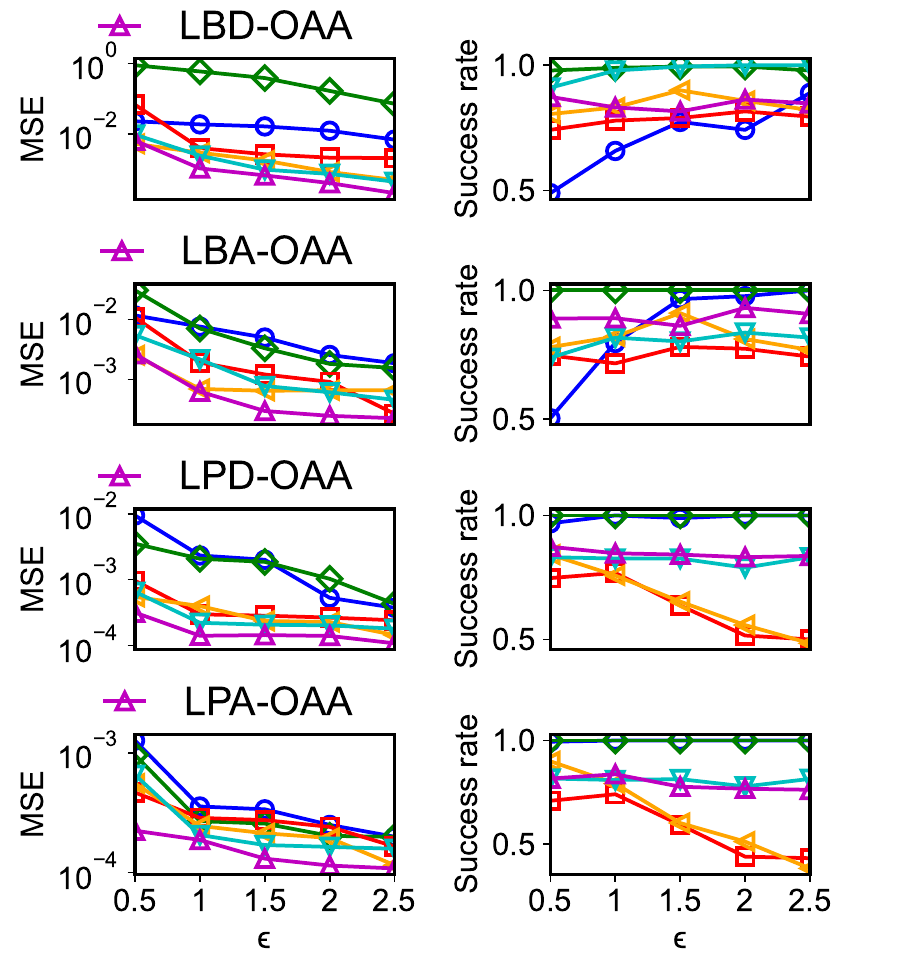}}
 	\caption{\small Attack effectiveness for Synthetic datasets, varying $\epsilon$.
  }\centering
 	\label{fig:epsilon synthetic data} 
 	\vspace{-0.5cm}
 \end{figure*}

\begin{figure*}[htbp]
 	\centering	
         \subfigure[\textsf{LNS} dataset, Uniform $\mathbf{\tilde{f}}$]{
 		\includegraphics[width=0.24\textwidth]{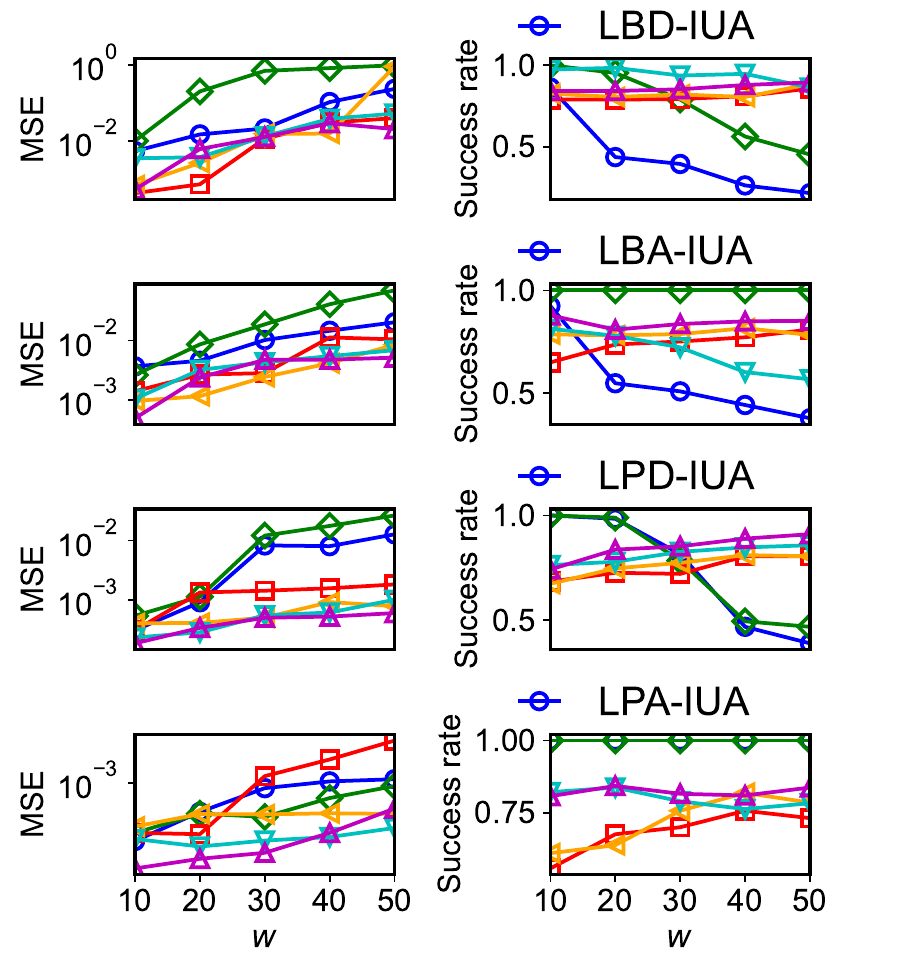}}\vspace{-0.06cm}\hspace{-4mm}
 	\subfigure[\textsf{LNS} dataset, Pulse $\mathbf{\tilde{f}}$]{
 		\includegraphics[width=0.24\textwidth]{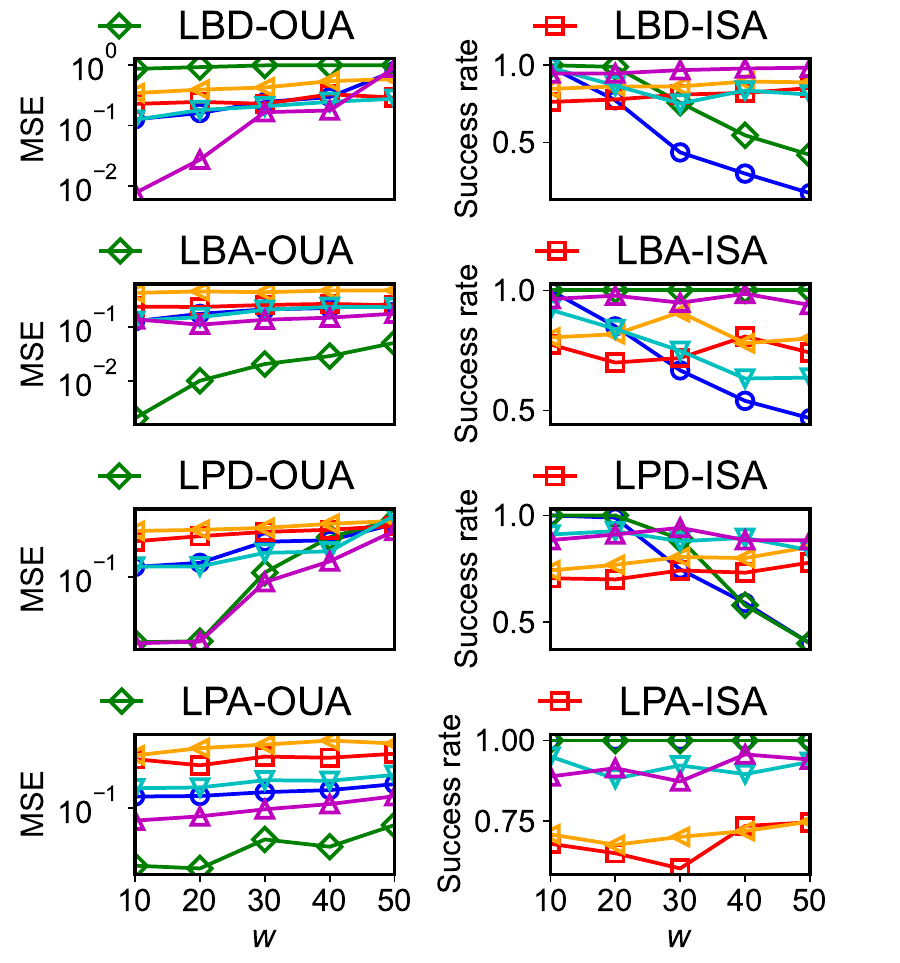}}\vspace{-0.06cm}\hspace{-4mm}
 	\subfigure[\textsf{LNS} dataset, Gaussian $\mathbf{\tilde{f}}$]{
 		\includegraphics[width=0.24\textwidth]{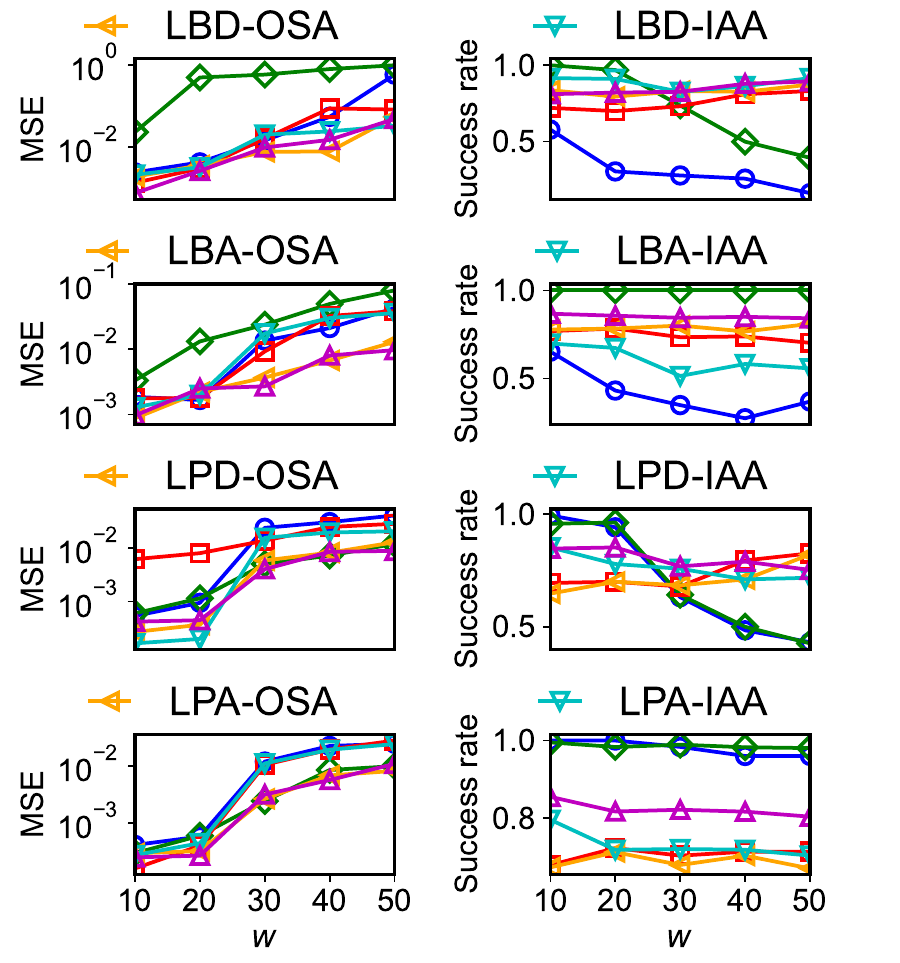}}\vspace{-0.06cm}\hspace{-4mm}
 	\subfigure[\textsf{LNS} dataset, Sigmoid $\mathbf{\tilde{f}}$]{
 		\includegraphics[width=0.24\textwidth]{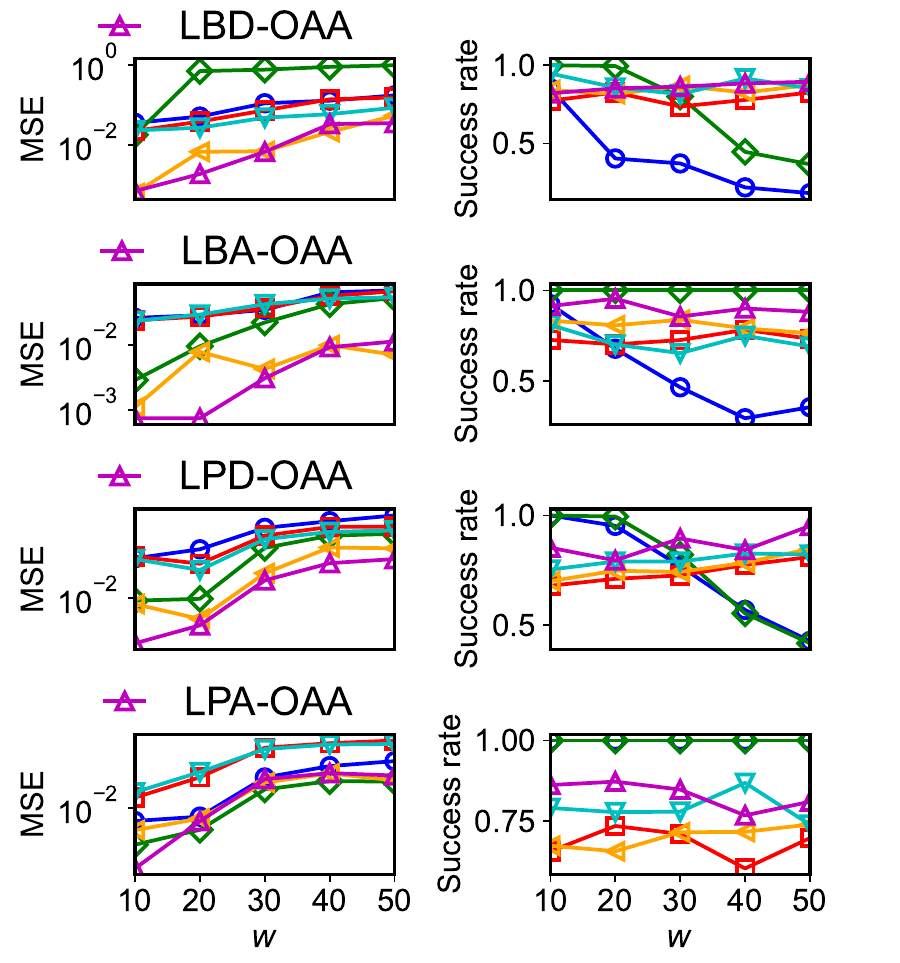}}\vspace{-0.06cm}\hspace{-4mm}
 	\\
 
         \subfigure[\textsf{Sin} dataset, Uniform $\mathbf{\tilde{f}}$]{
 		\includegraphics[width=0.24\textwidth]{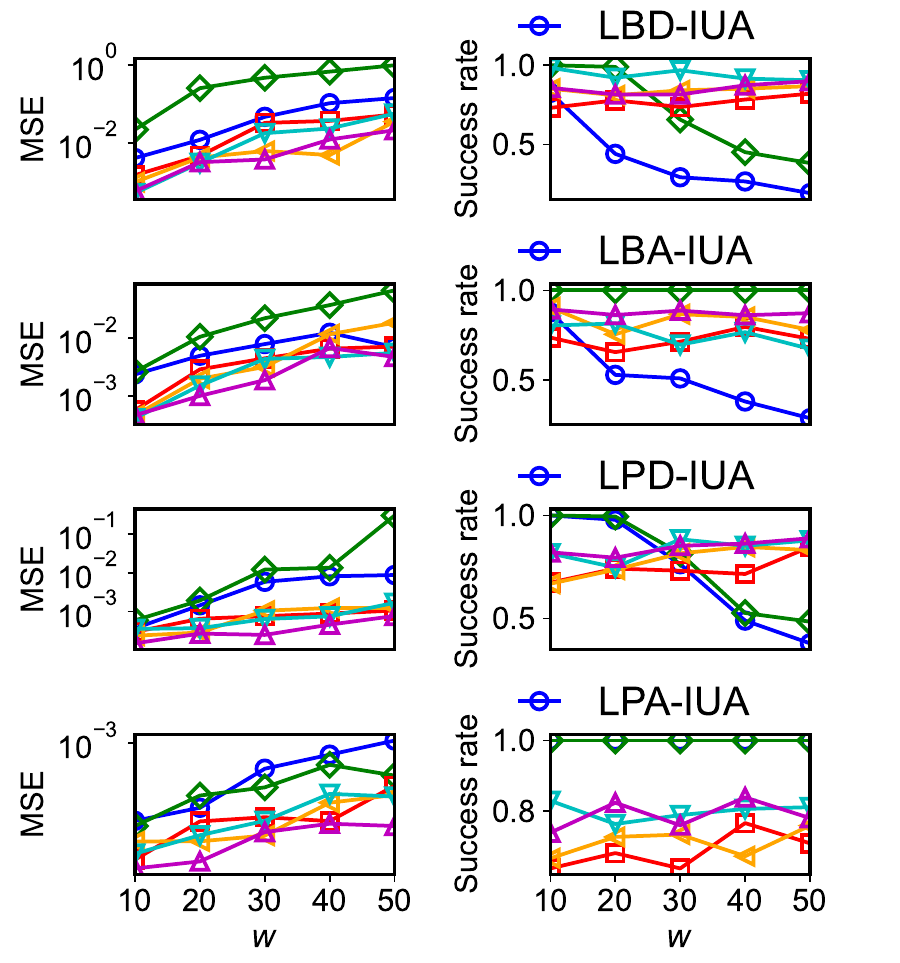}}\vspace{-0.06cm}\hspace{-4mm}
 	\subfigure[\textsf{Sin} dataset, Pulse $\mathbf{\tilde{f}}$]{
 		\includegraphics[width=0.24\textwidth]{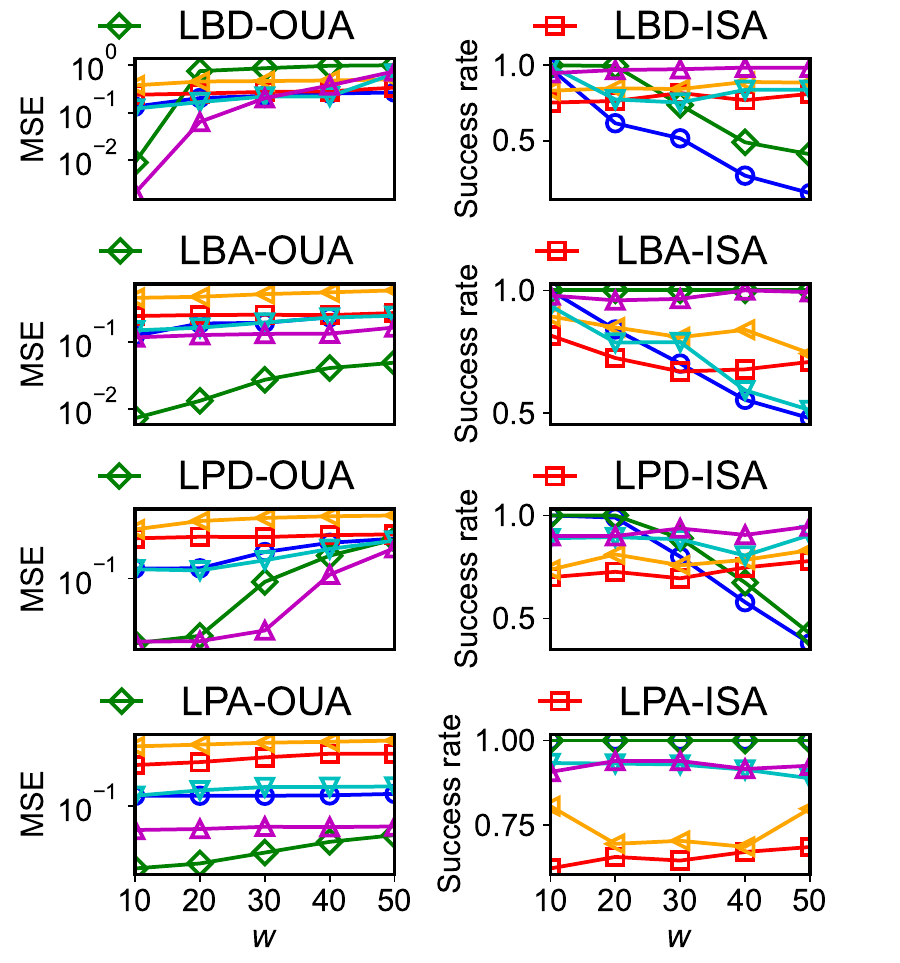}}\vspace{-0.06cm}\hspace{-4mm}
 	\subfigure[\textsf{Sin} dataset, Gaussian $\mathbf{\tilde{f}}$]{
 		\includegraphics[width=0.24\textwidth]{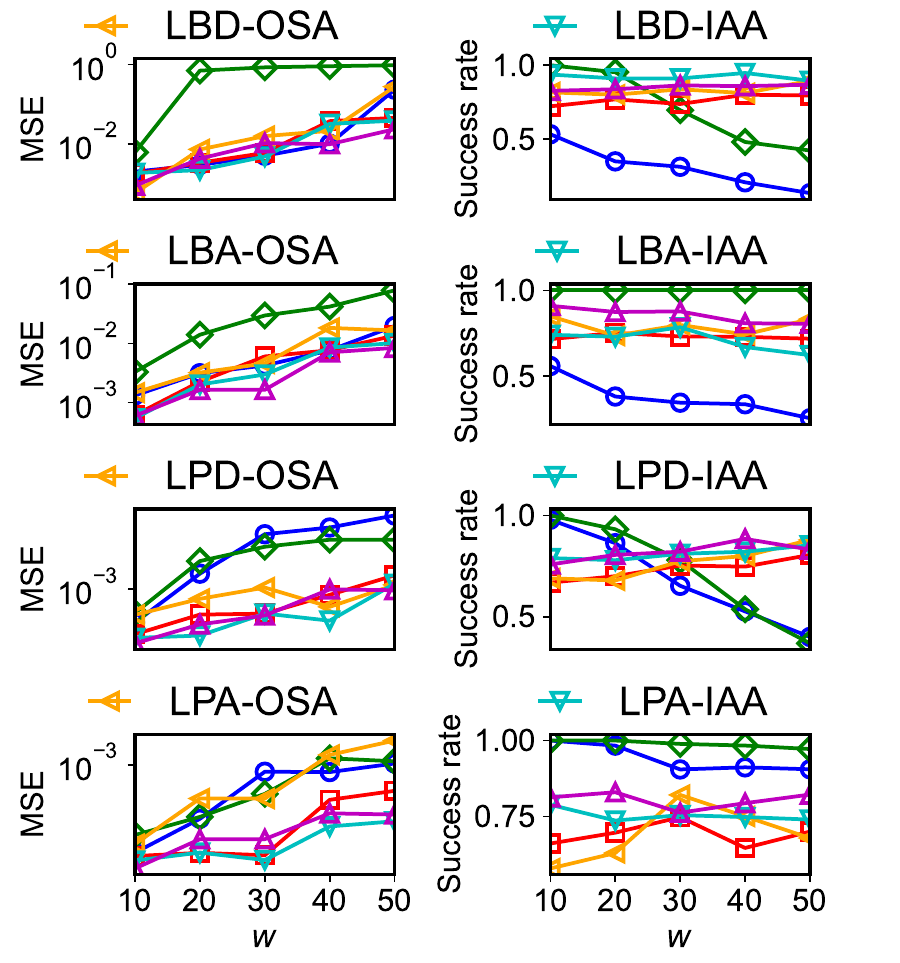}}\vspace{-0.06cm}\hspace{-4mm}
 	\subfigure[\textsf{Sin} dataset, Sigmoid $\mathbf{\tilde{f}}$]{
 		\includegraphics[width=0.24\textwidth]{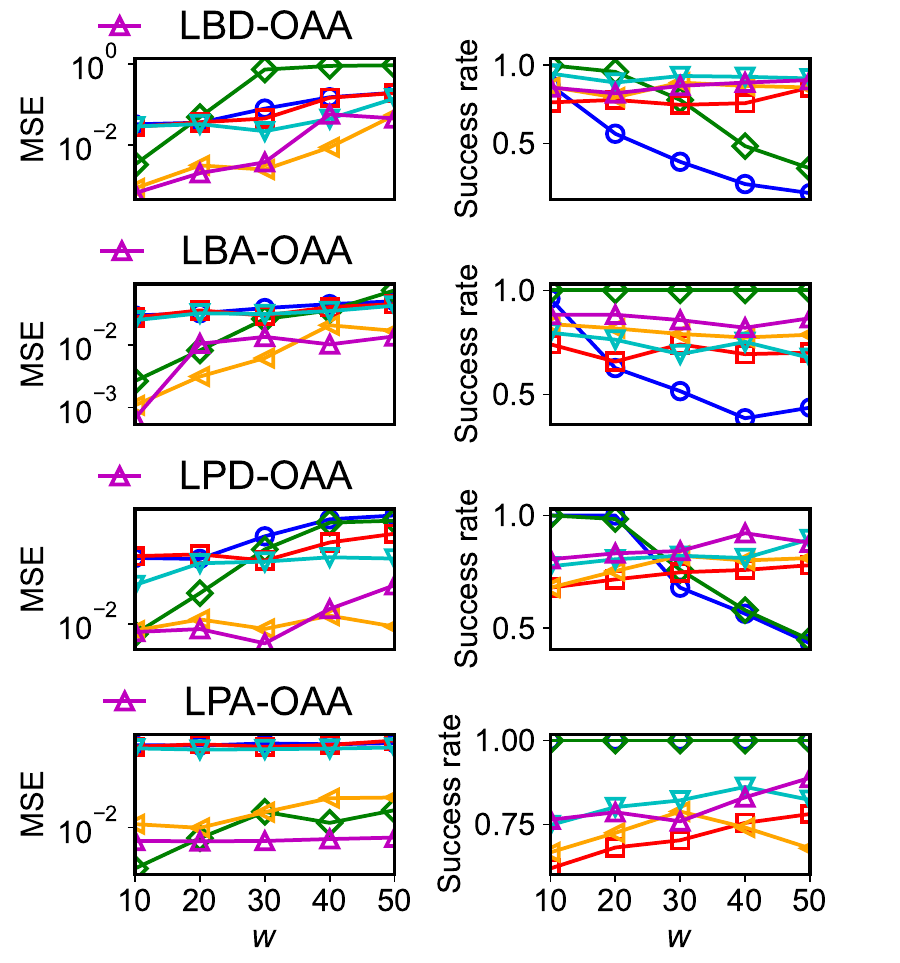}}\vspace{-0.06cm}\hspace{-4mm}
         \\
        
         \subfigure[\textsf{Log} dataset, Uniform $\mathbf{\tilde{f}}$]{
 		\includegraphics[width=0.24\textwidth]{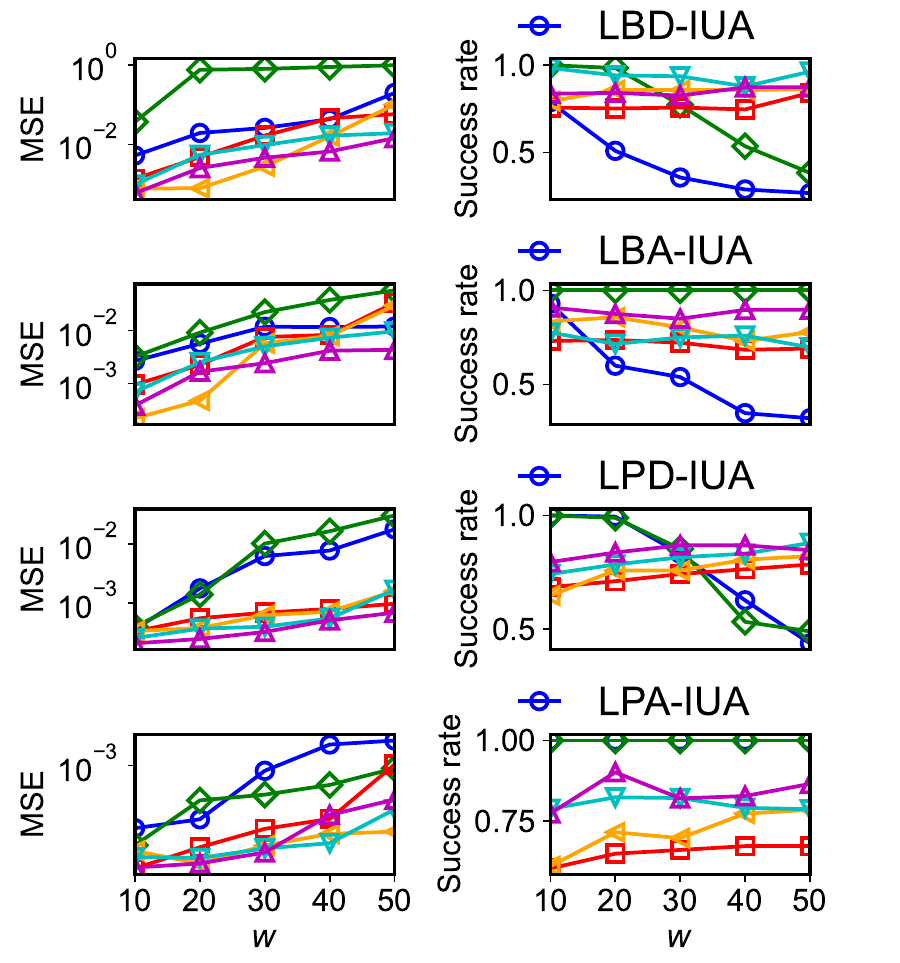}}\vspace{-0.05cm}\hspace{-4mm}
 	\subfigure[\textsf{Log} dataset, Pulse $\mathbf{\tilde{f}}$]{
 		\includegraphics[width=0.24\textwidth]{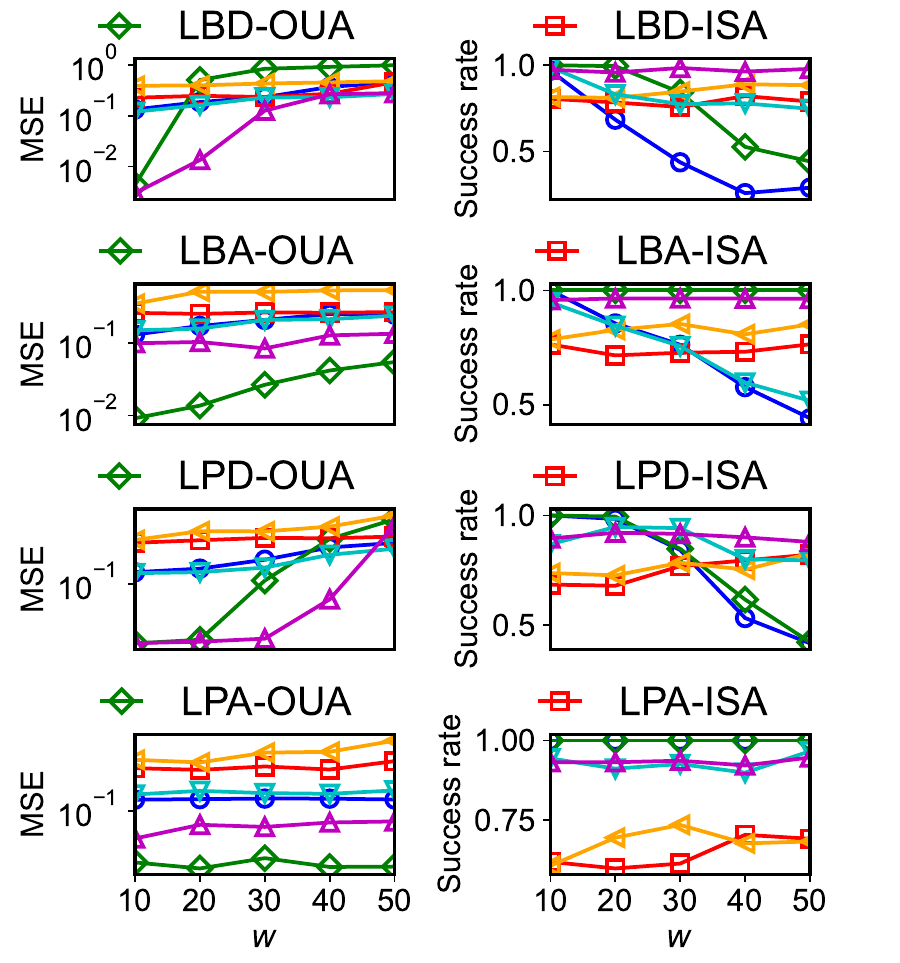}}\vspace{-0.06cm}\hspace{-4mm}
 	\subfigure[\textsf{Log} dataset, Gaussian $\mathbf{\tilde{f}}$]{
 		\includegraphics[width=0.24\textwidth]{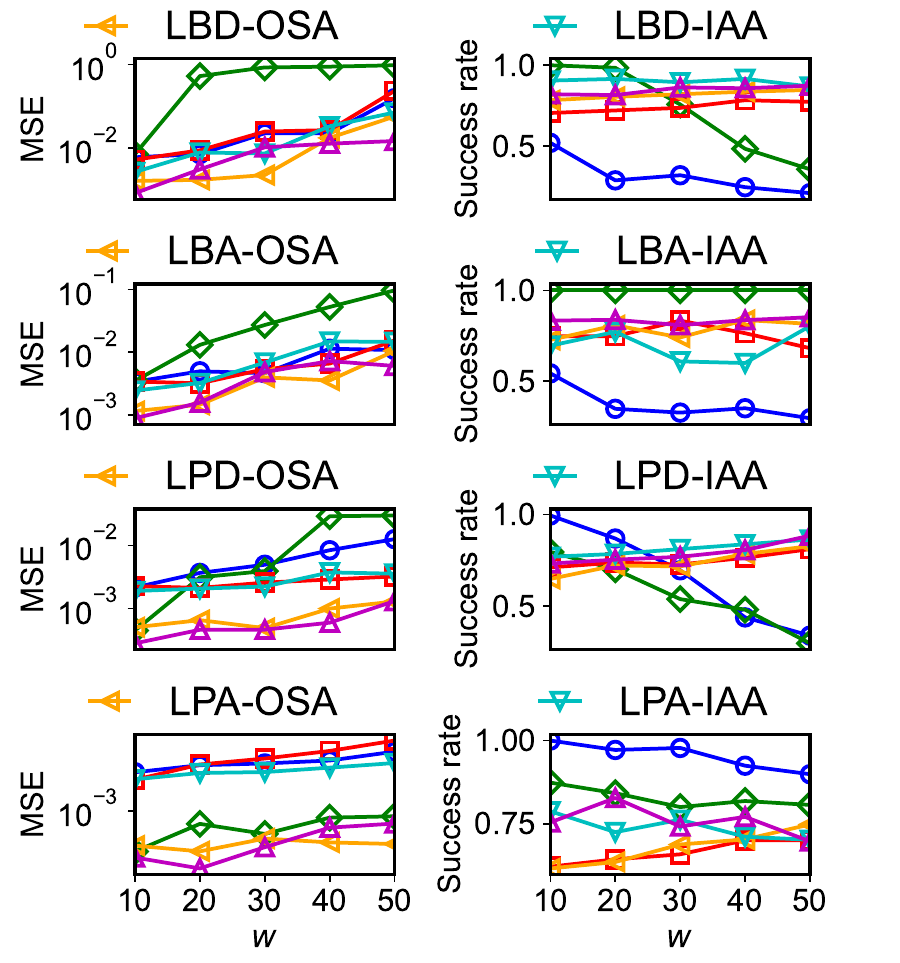}}\vspace{-0.06cm}\hspace{-4mm}
 	\subfigure[\textsf{Log} dataset, Sigmoid $\mathbf{\tilde{f}}$]{
 		\includegraphics[width=0.24\textwidth]{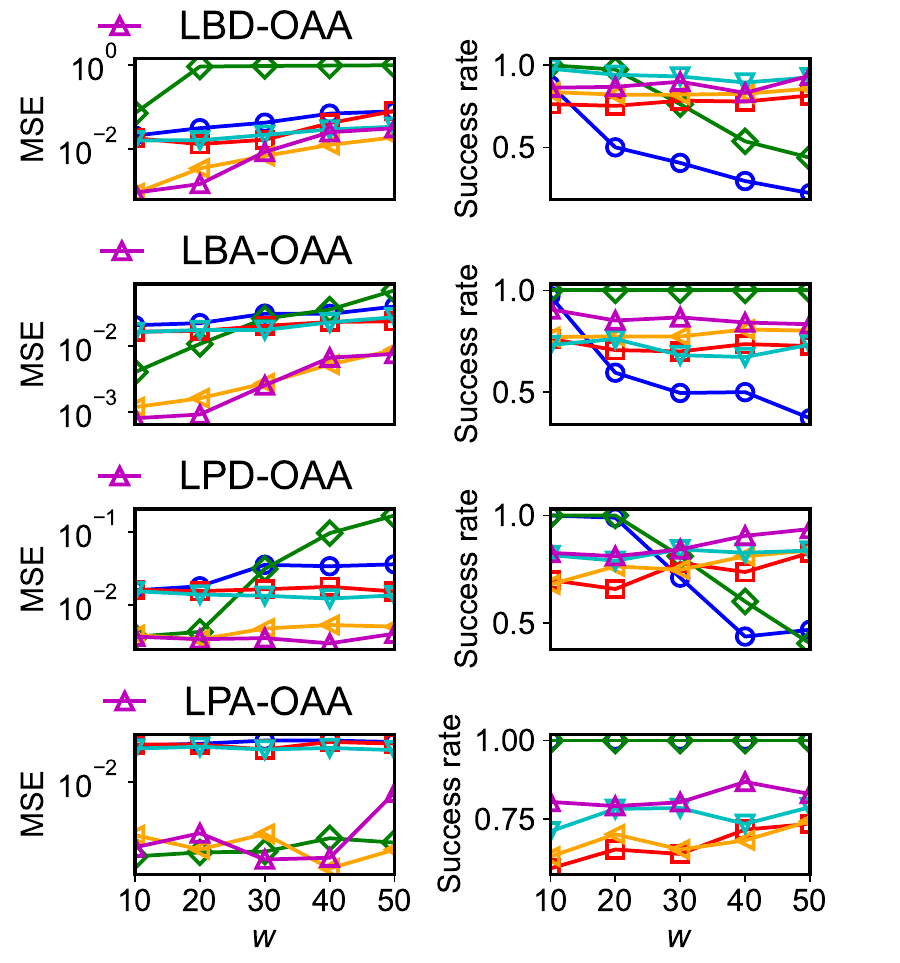}}\vspace{-0.06cm}\hspace{-4mm}
         \\
        
         \subfigure[\textsf{Pulse} dataset, Uniform $\mathbf{\tilde{f}}$]{
 		\includegraphics[width=0.24\textwidth]{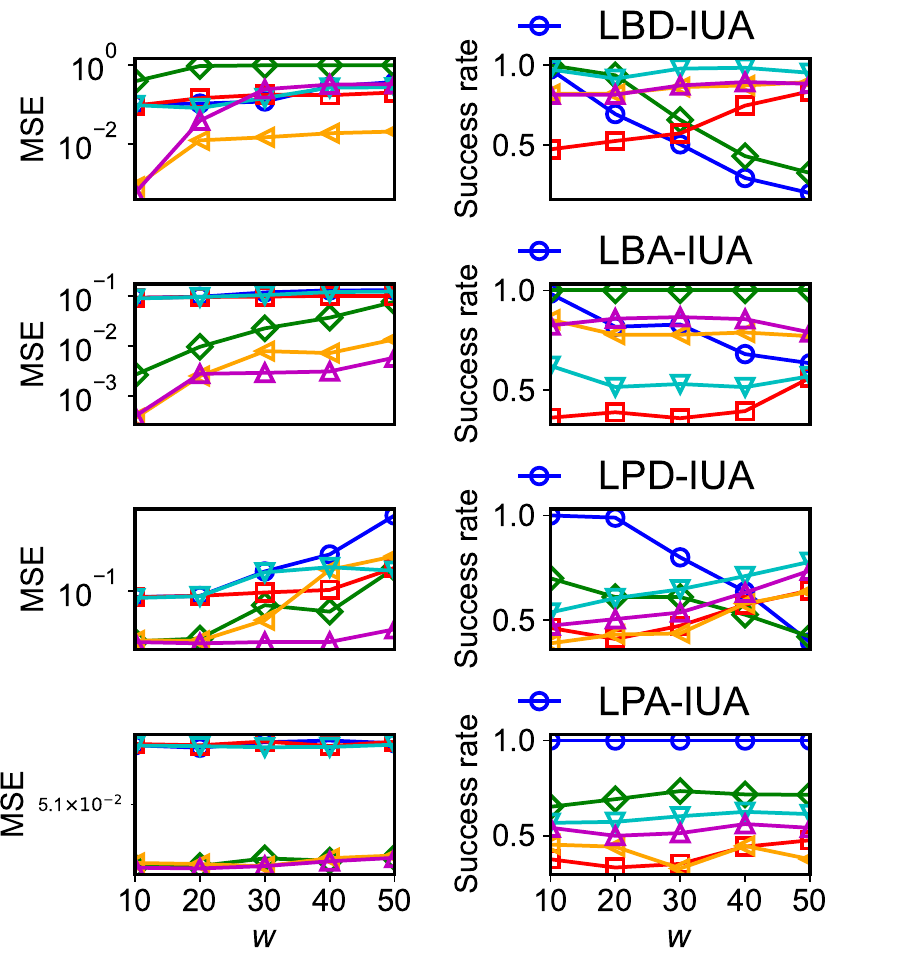}}\vspace{-0.13cm}\hspace{-4mm}
 	\subfigure[\textsf{Pulse} dataset, Pulse $\mathbf{\tilde{f}}$]{
 		\includegraphics[width=0.24\textwidth]{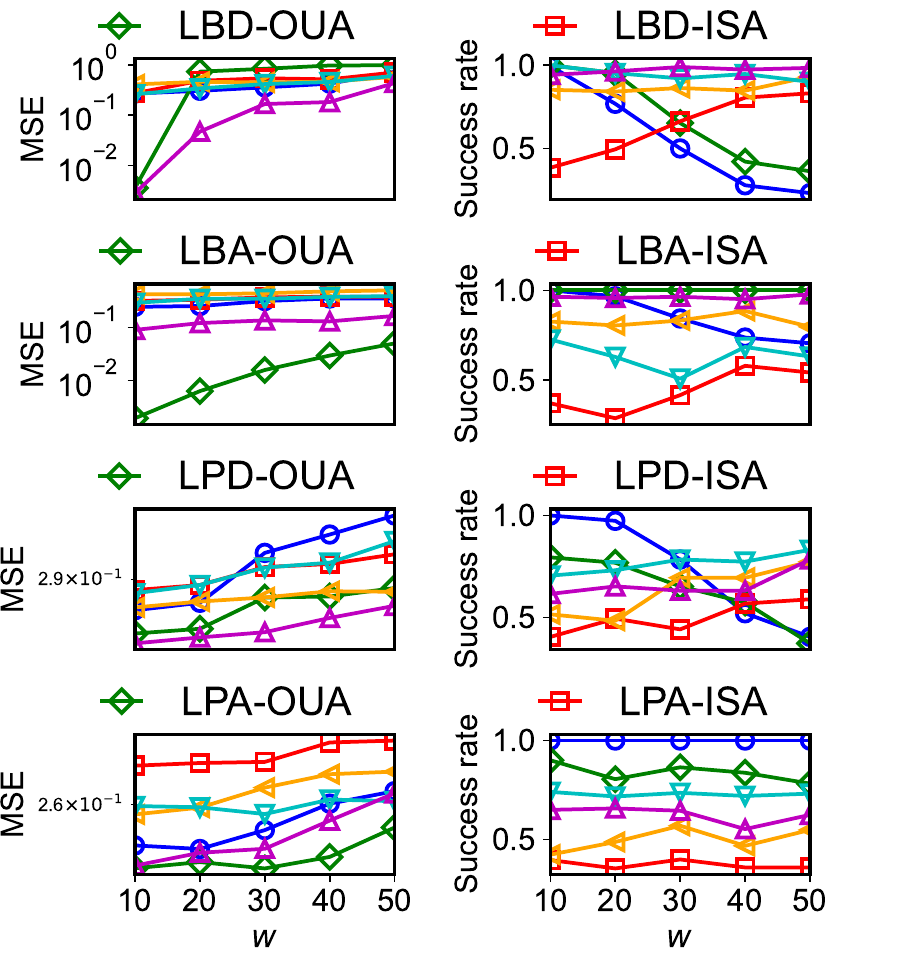}}\vspace{-0.13cm}\hspace{-4mm}
 	\subfigure[\textsf{Pulse} dataset, Gaussian $\mathbf{\tilde{f}}$]{
 		\includegraphics[width=0.24\textwidth]{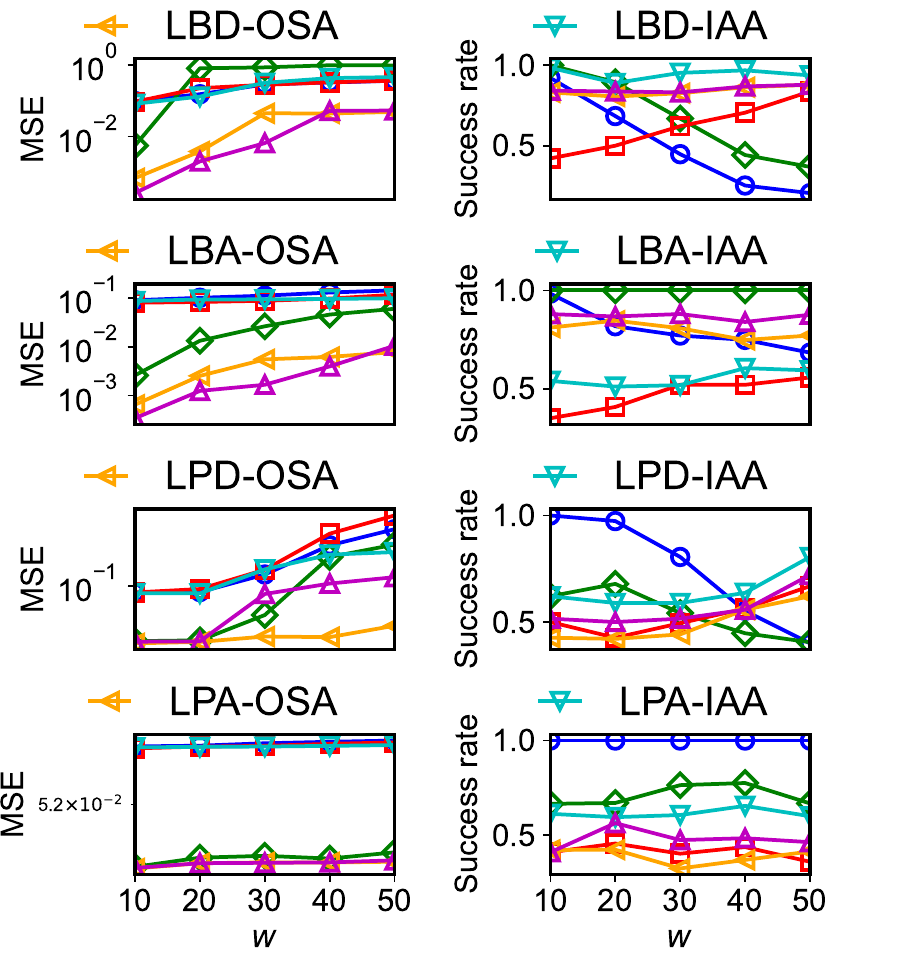}}\vspace{-0.13cm}\hspace{-4mm}
 	\subfigure[\textsf{Pulse} dataset, Sigmoid $\mathbf{\tilde{f}}$]{
 		\includegraphics[width=0.24\textwidth]{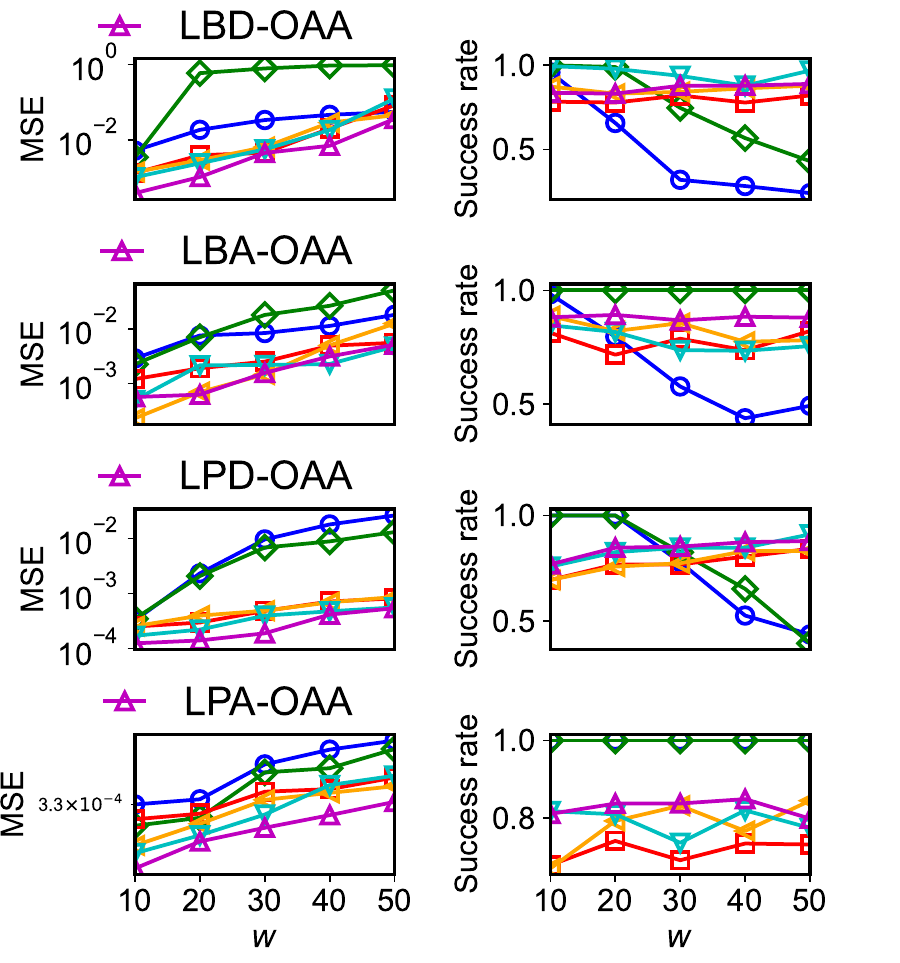}}
 	\caption{\small Attack effectiveness for Synthetic datasets, varying $w$.
  }\centering
 	\label{fig:w synthetic data} 
 	\vspace{-0.5cm}
 \end{figure*}

\begin{figure*}[htbp]
 	\centering	
         \subfigure[\textsf{LNS} dataset, Uniform $\mathbf{\tilde{f}}$]{
 		\includegraphics[width=0.24\textwidth]{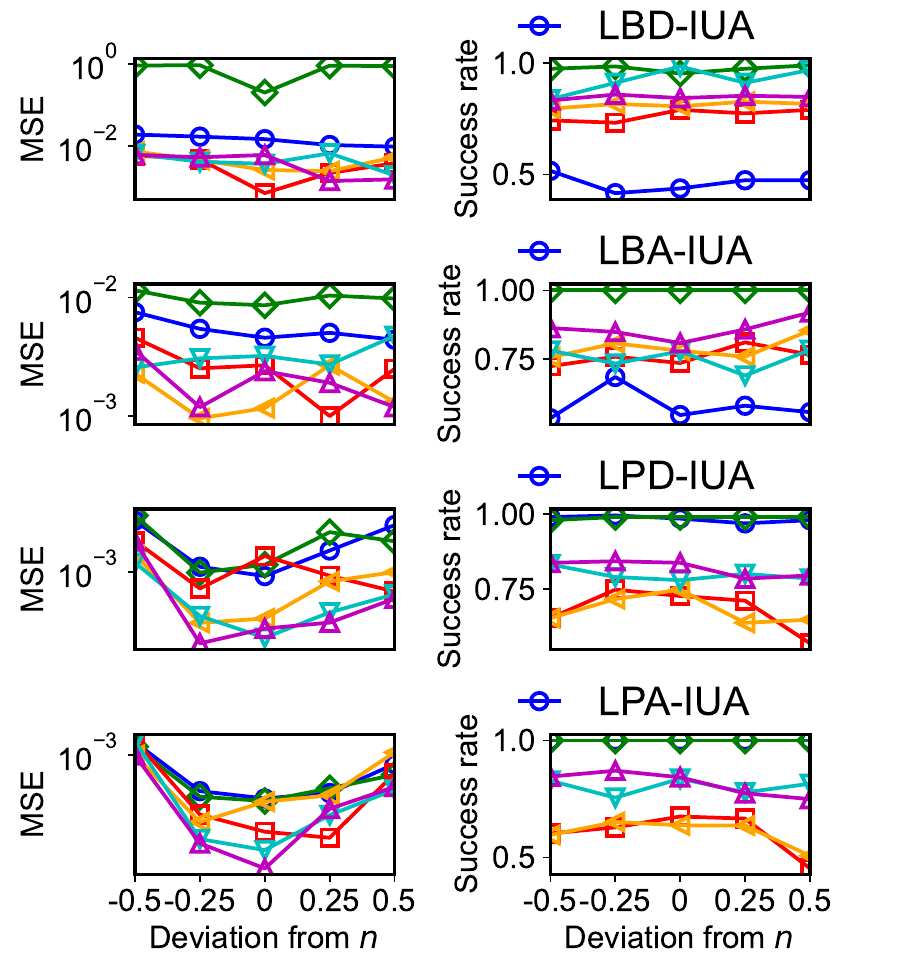}}\vspace{-0.06cm}\hspace{-4mm}
 	\subfigure[\textsf{LNS} dataset, Pulse $\mathbf{\tilde{f}}$]{
 		\includegraphics[width=0.24\textwidth]{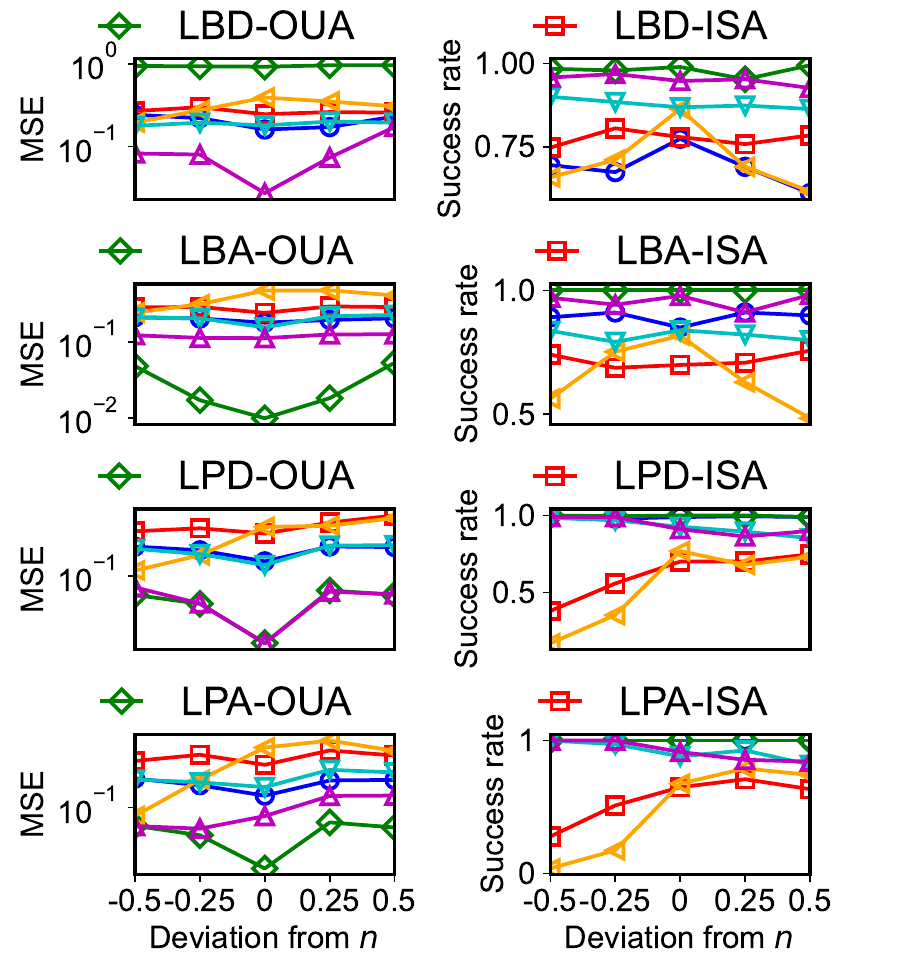}}\vspace{-0.06cm}\hspace{-4mm}
 	\subfigure[\textsf{LNS} dataset, Gaussian $\mathbf{\tilde{f}}$]{
 		\includegraphics[width=0.24\textwidth]{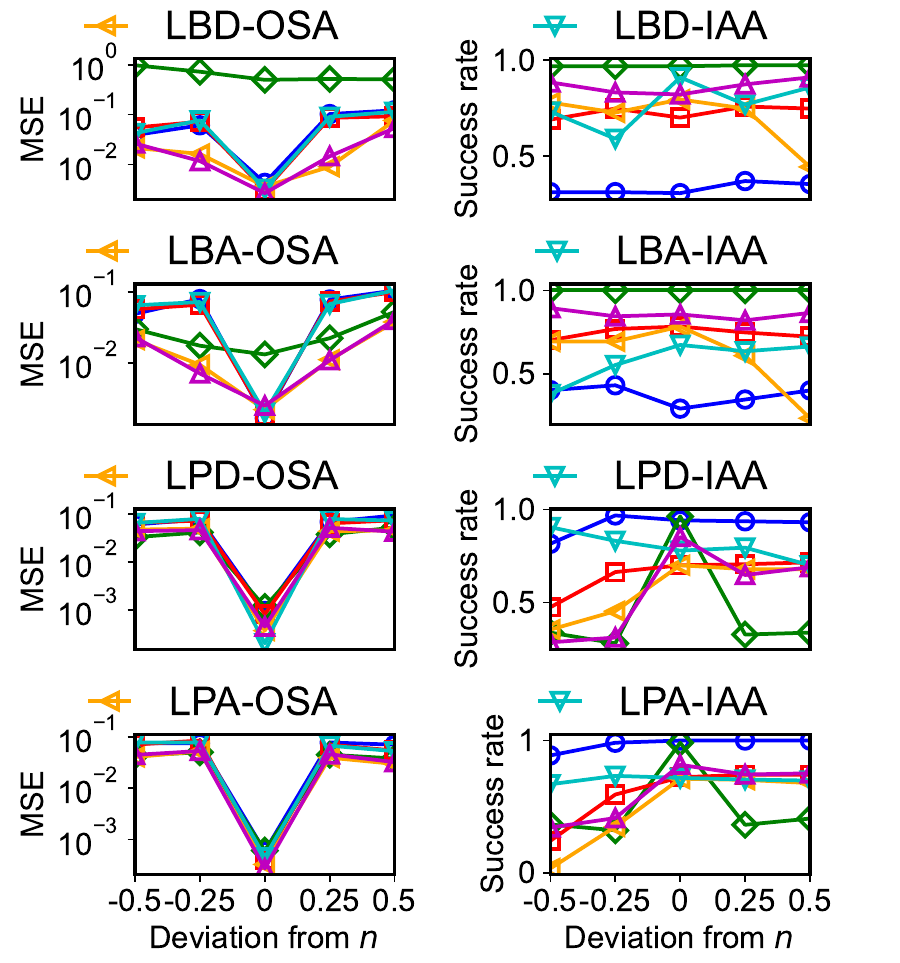}}\vspace{-0.06cm}\hspace{-4mm}
 	\subfigure[\textsf{LNS} dataset, Sigmoid $\mathbf{\tilde{f}}$]{
 		\includegraphics[width=0.24\textwidth]{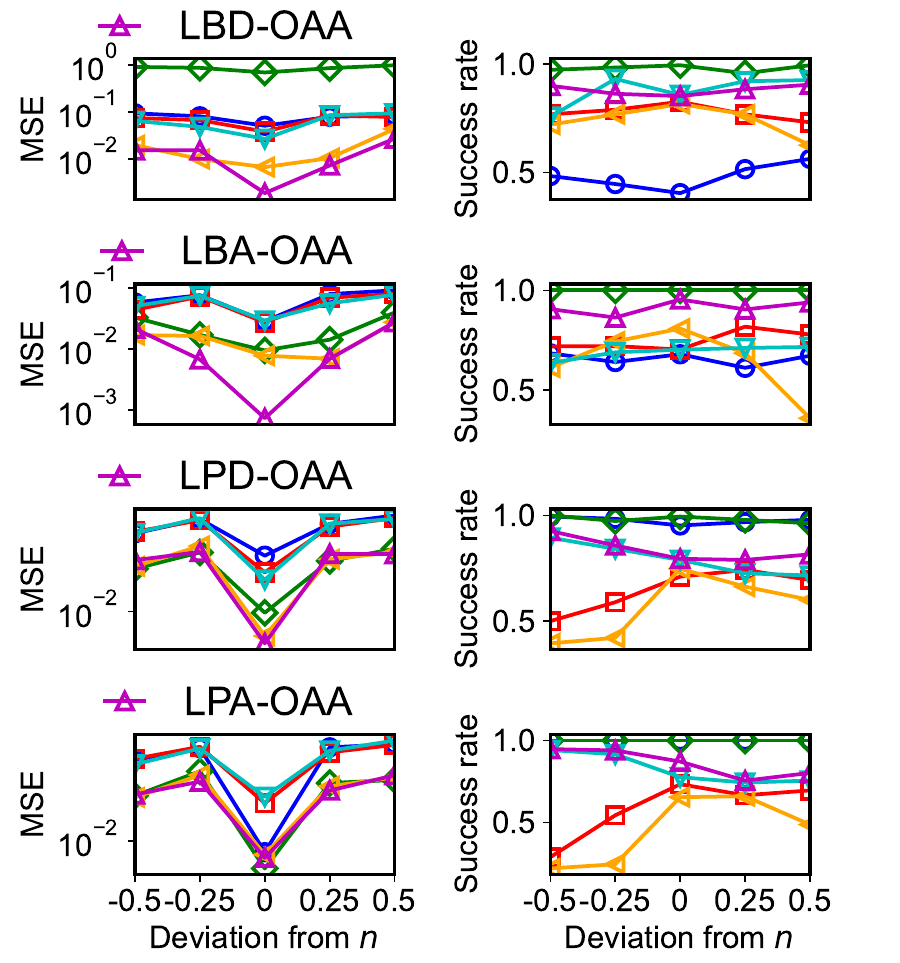}}\vspace{-0.06cm}\hspace{-4mm}
 	\\
 
         \subfigure[\textsf{Sin} dataset, Uniform $\mathbf{\tilde{f}}$]{
 		\includegraphics[width=0.24\textwidth]{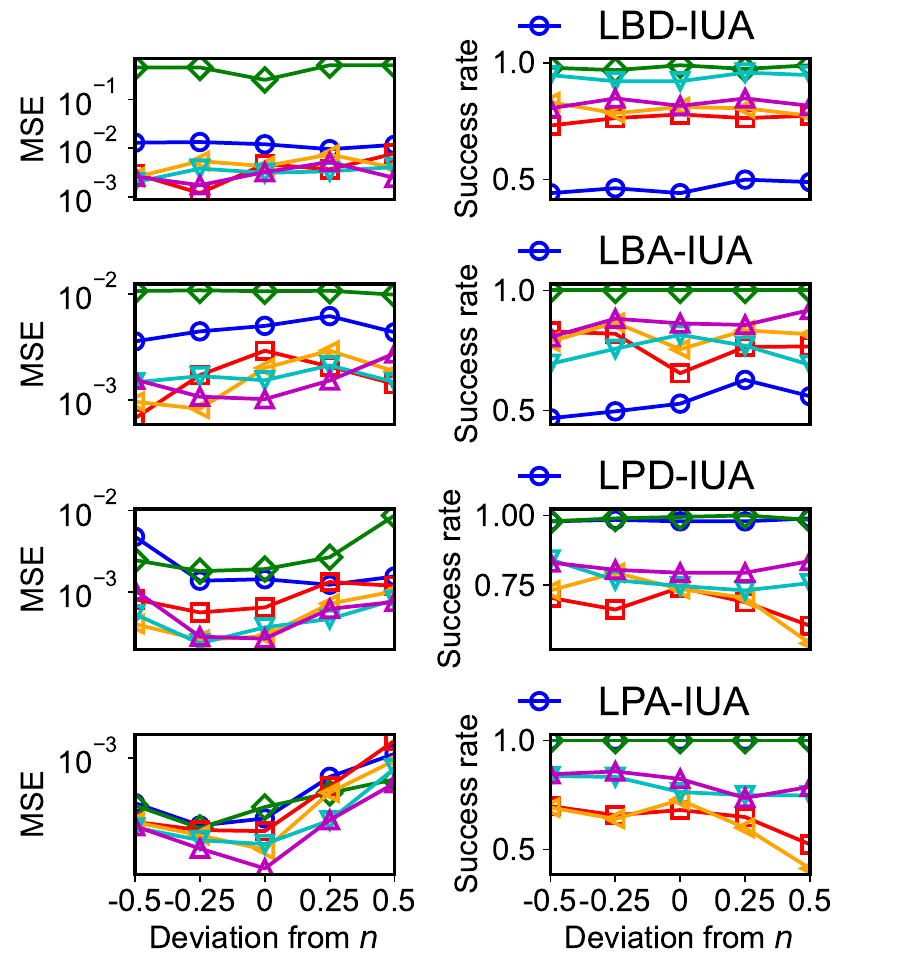}}\vspace{-0.06cm}\hspace{-4mm}
 	\subfigure[\textsf{Sin} dataset, Pulse $\mathbf{\tilde{f}}$]{
 		\includegraphics[width=0.24\textwidth]{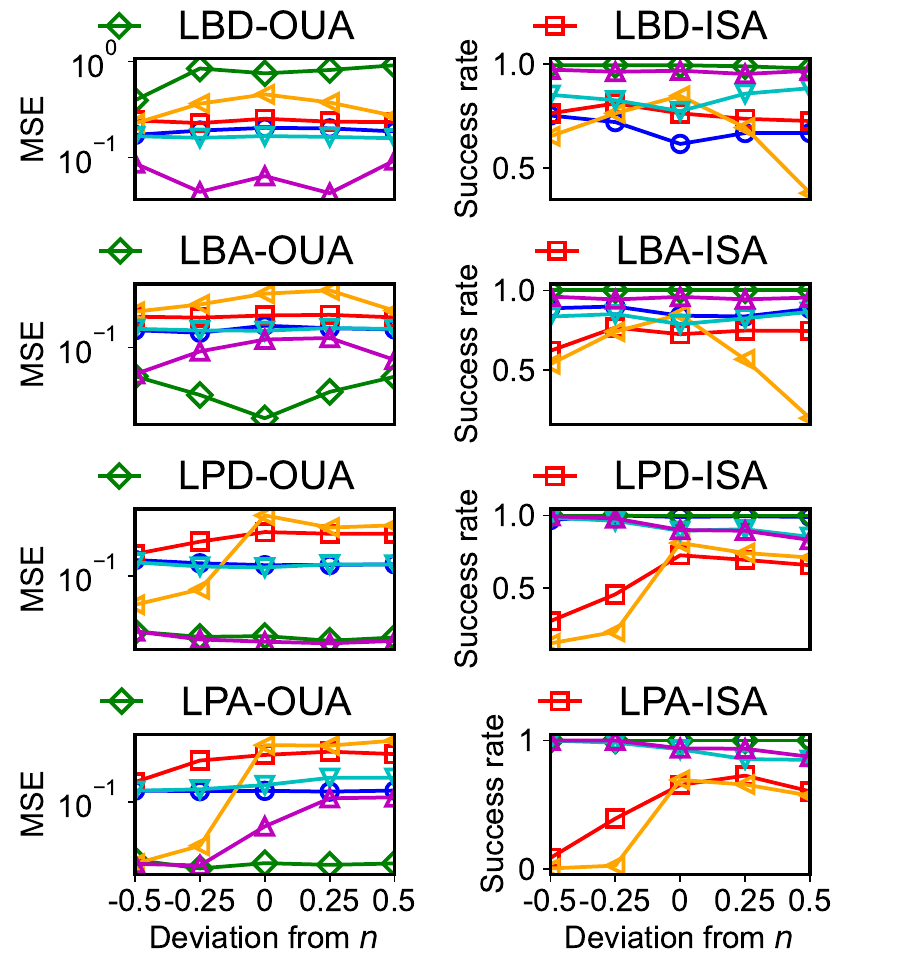}}\vspace{-0.06cm}\hspace{-4mm}
 	\subfigure[\textsf{Sin} dataset, Gaussian $\mathbf{\tilde{f}}$]{
 		\includegraphics[width=0.24\textwidth]{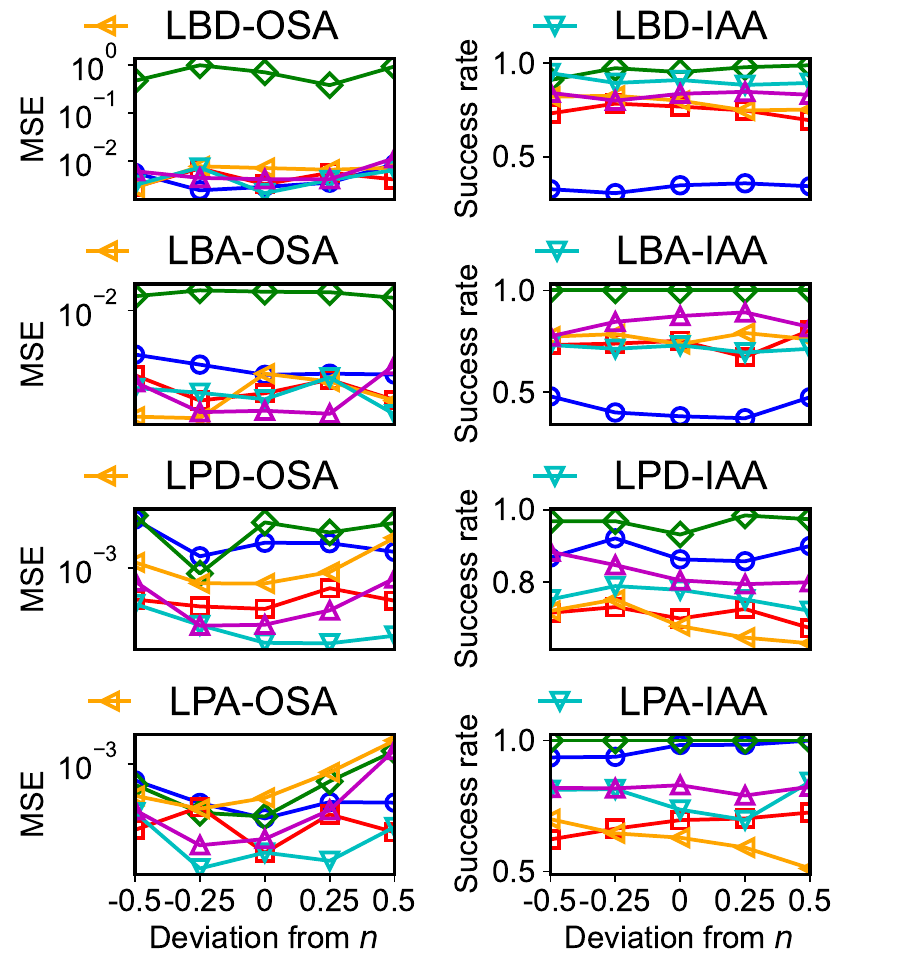}}\vspace{-0.06cm}\hspace{-4mm}
 	\subfigure[\textsf{Sin} dataset, Sigmoid $\mathbf{\tilde{f}}$]{
 		\includegraphics[width=0.24\textwidth]{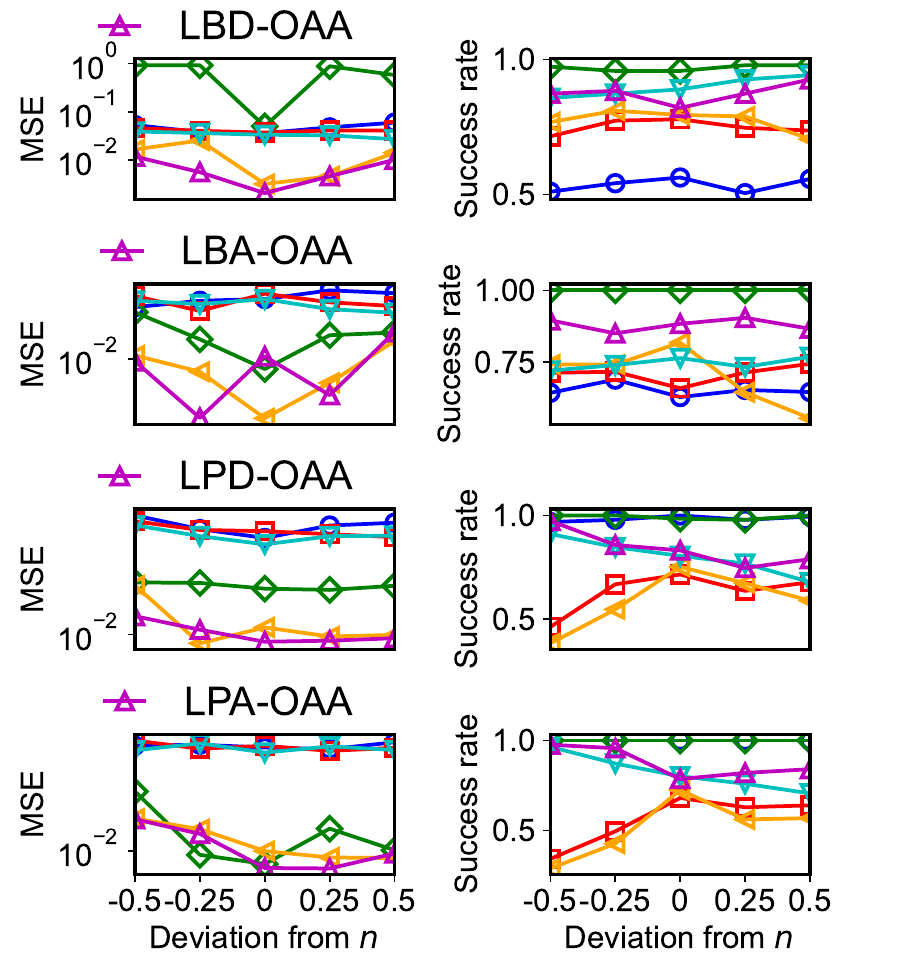}}\vspace{-0.06cm}\hspace{-4mm}
         \\
        
         \subfigure[\textsf{Log} dataset, Uniform $\mathbf{\tilde{f}}$]{
 		\includegraphics[width=0.24\textwidth]{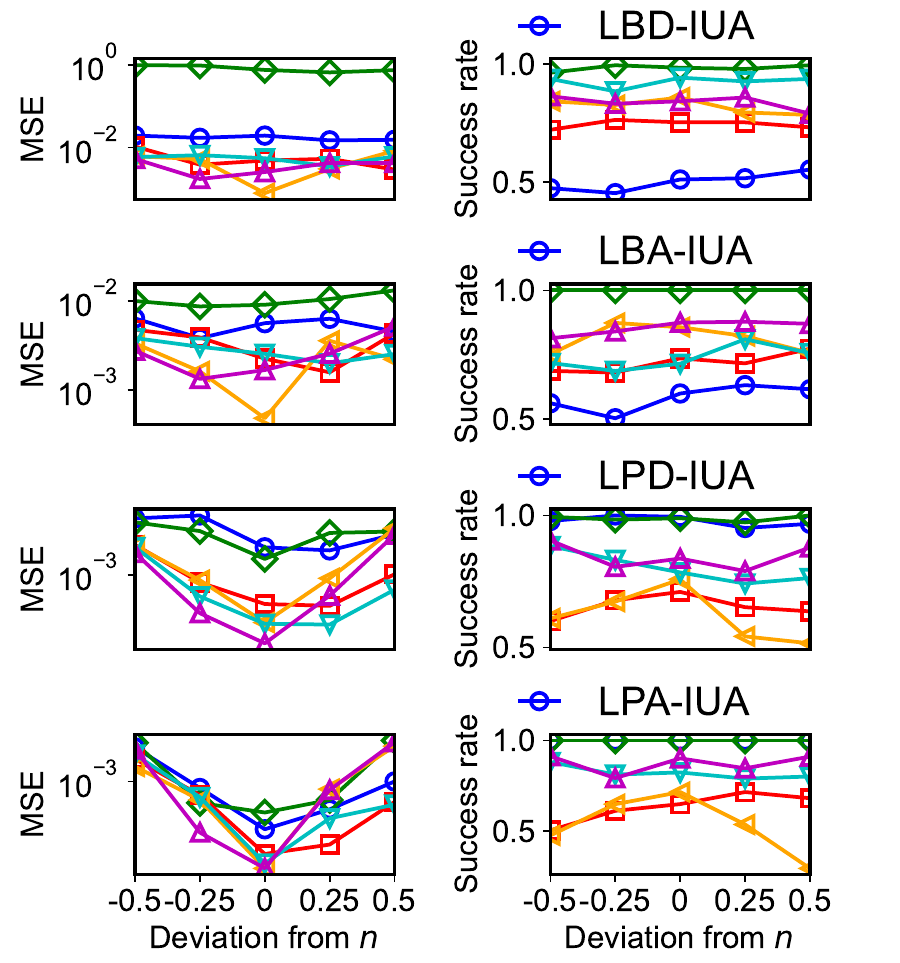}}\vspace{-0.05cm}\hspace{-4mm}
 	\subfigure[\textsf{Log} dataset, Pulse $\mathbf{\tilde{f}}$]{
 		\includegraphics[width=0.24\textwidth]{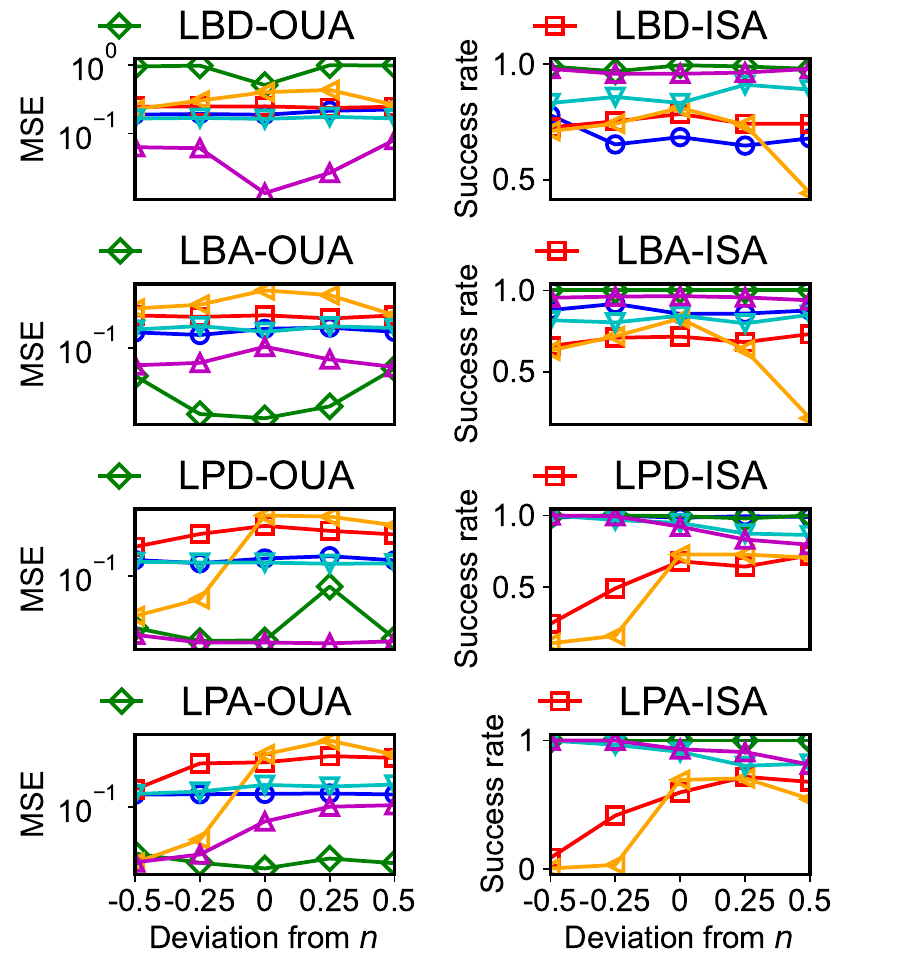}}\vspace{-0.06cm}\hspace{-4mm}
 	\subfigure[\textsf{Log} dataset, Gaussian $\mathbf{\tilde{f}}$]{
 		\includegraphics[width=0.24\textwidth]{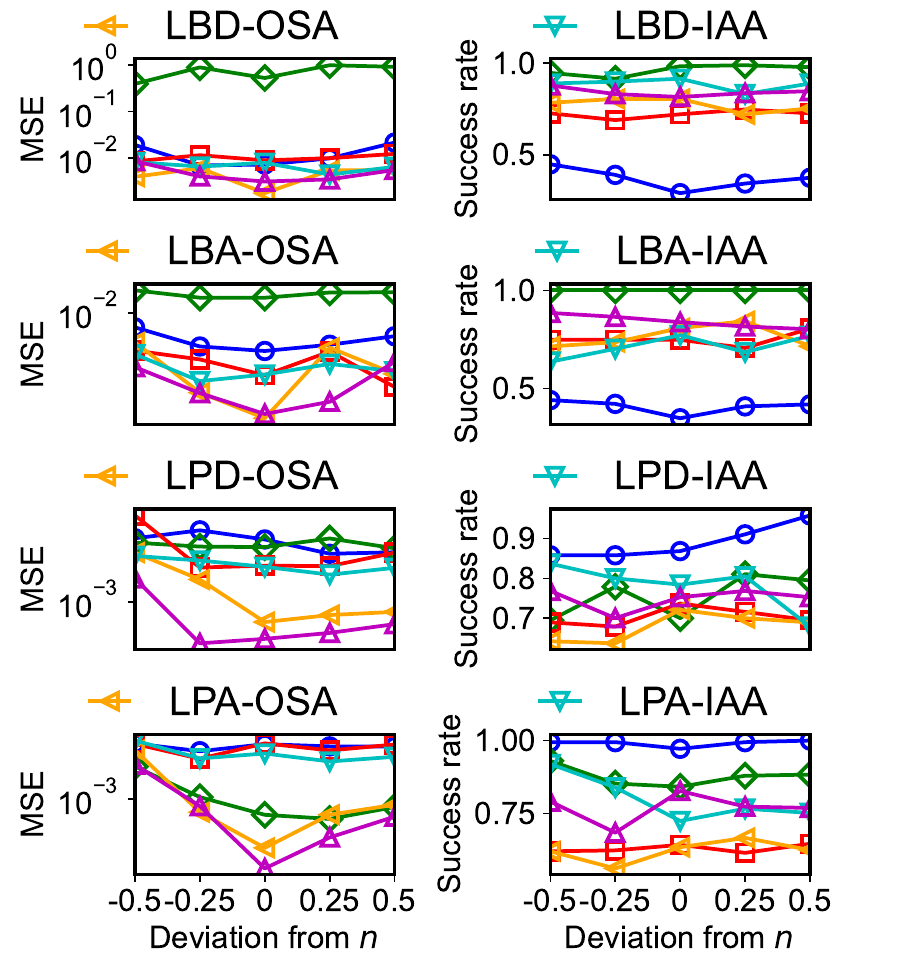}}\vspace{-0.06cm}\hspace{-4mm}
 	\subfigure[\textsf{Log} dataset, Sigmoid $\mathbf{\tilde{f}}$]{
 		\includegraphics[width=0.24\textwidth]{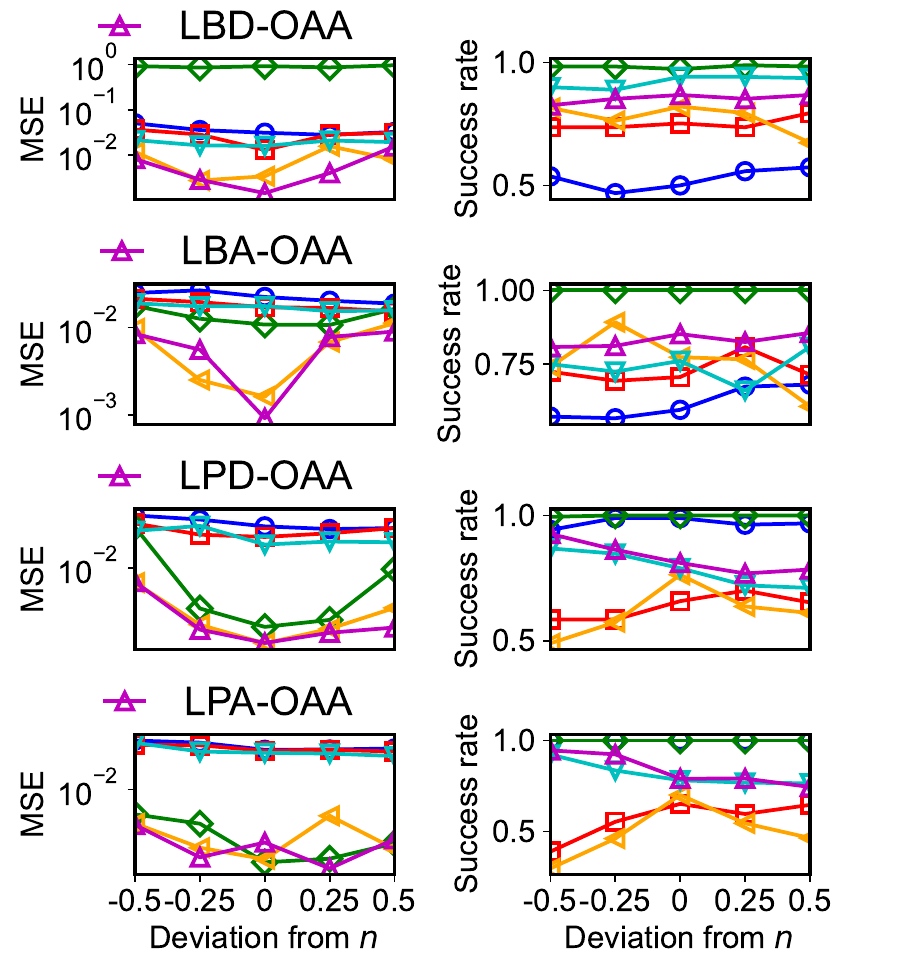}}\vspace{-0.06cm}\hspace{-4mm}
         \\
        
         \subfigure[\textsf{Pulse} dataset, Uniform $\mathbf{\tilde{f}}$]{
 		\includegraphics[width=0.24\textwidth]{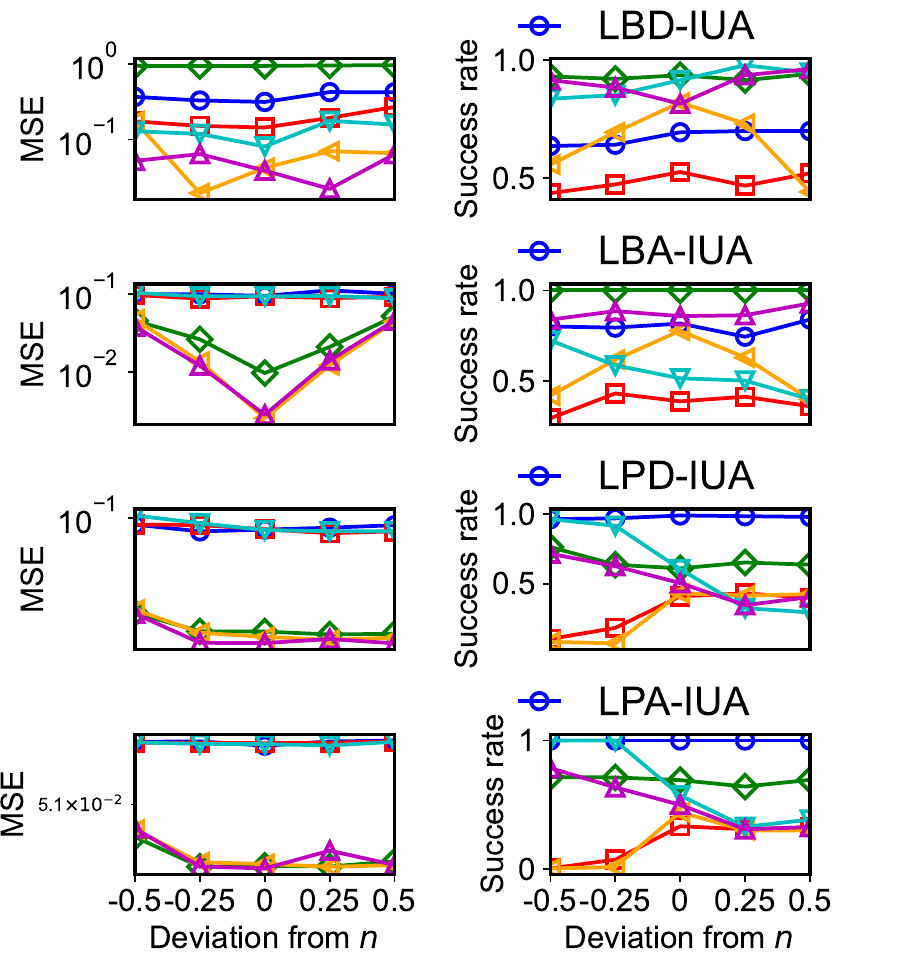}}\vspace{-0.13cm}\hspace{-4mm}
 	\subfigure[\textsf{Pulse} dataset, Pulse $\mathbf{\tilde{f}}$]{
 		\includegraphics[width=0.24\textwidth]{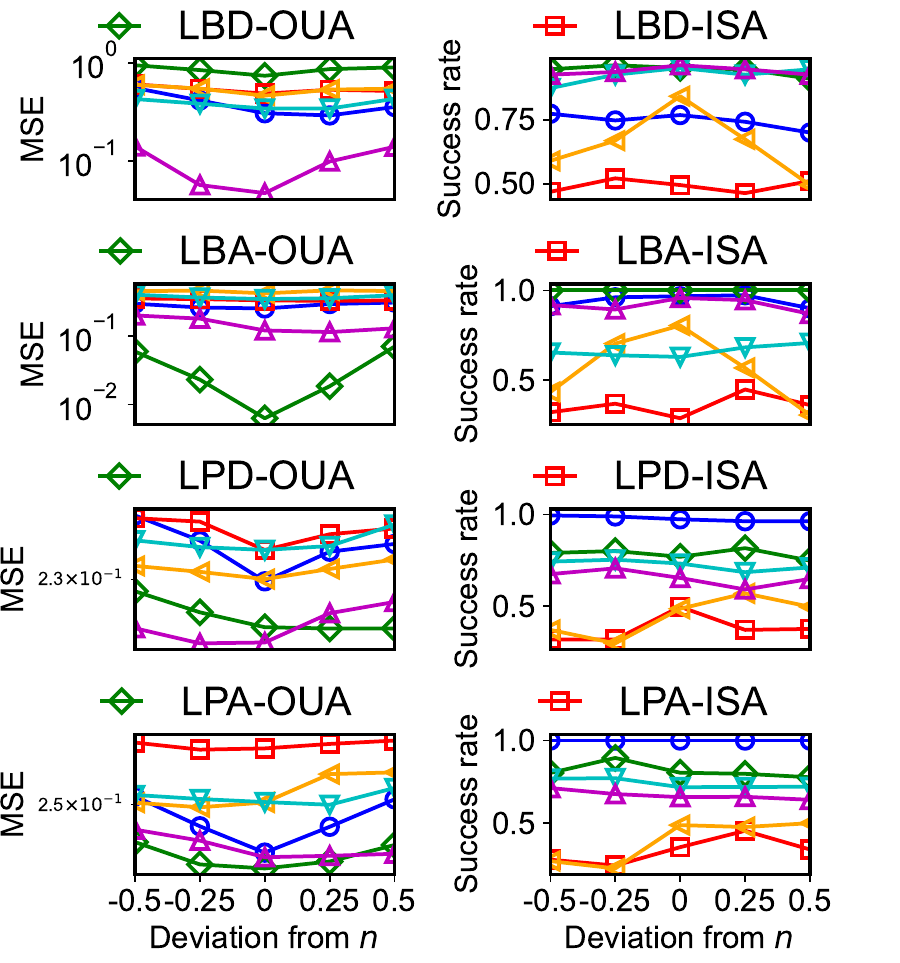}}\vspace{-0.13cm}\hspace{-4mm}
 	\subfigure[\textsf{Pulse} dataset, Gaussian $\mathbf{\tilde{f}}$]{
 		\includegraphics[width=0.24\textwidth]{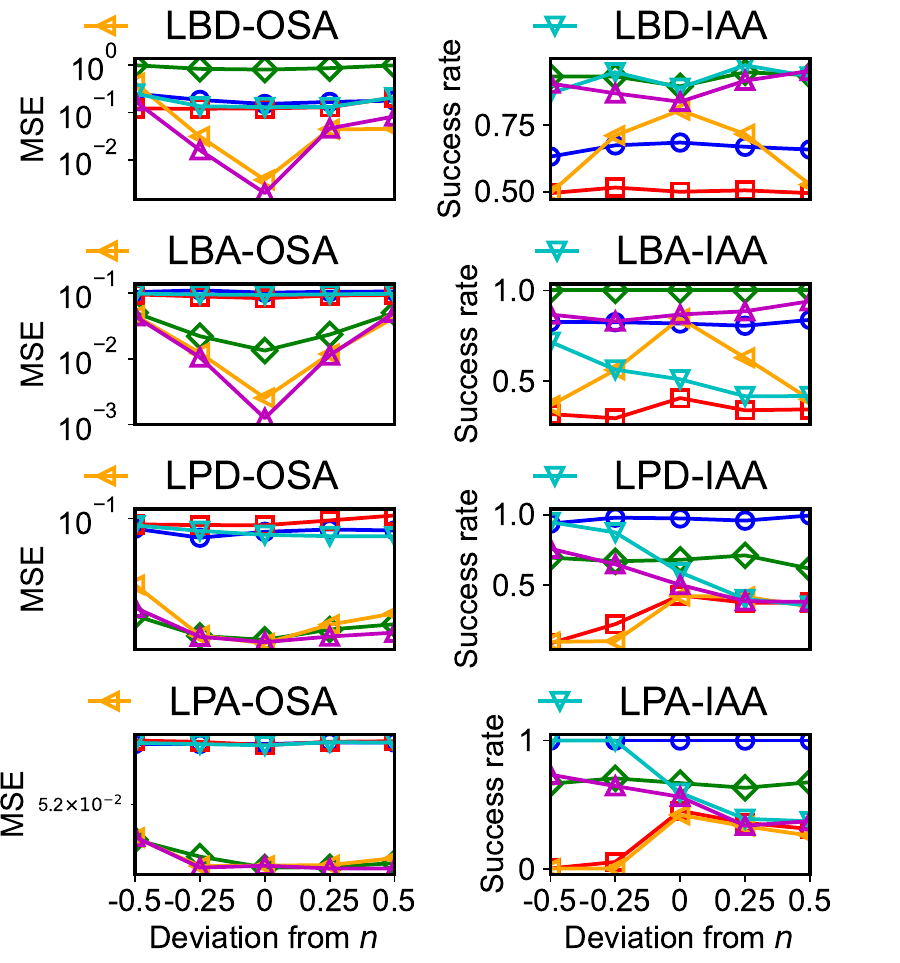}}\vspace{-0.13cm}\hspace{-4mm}
 	\subfigure[\textsf{Pulse} dataset, Sigmoid $\mathbf{\tilde{f}}$]{
 		\includegraphics[width=0.24\textwidth]{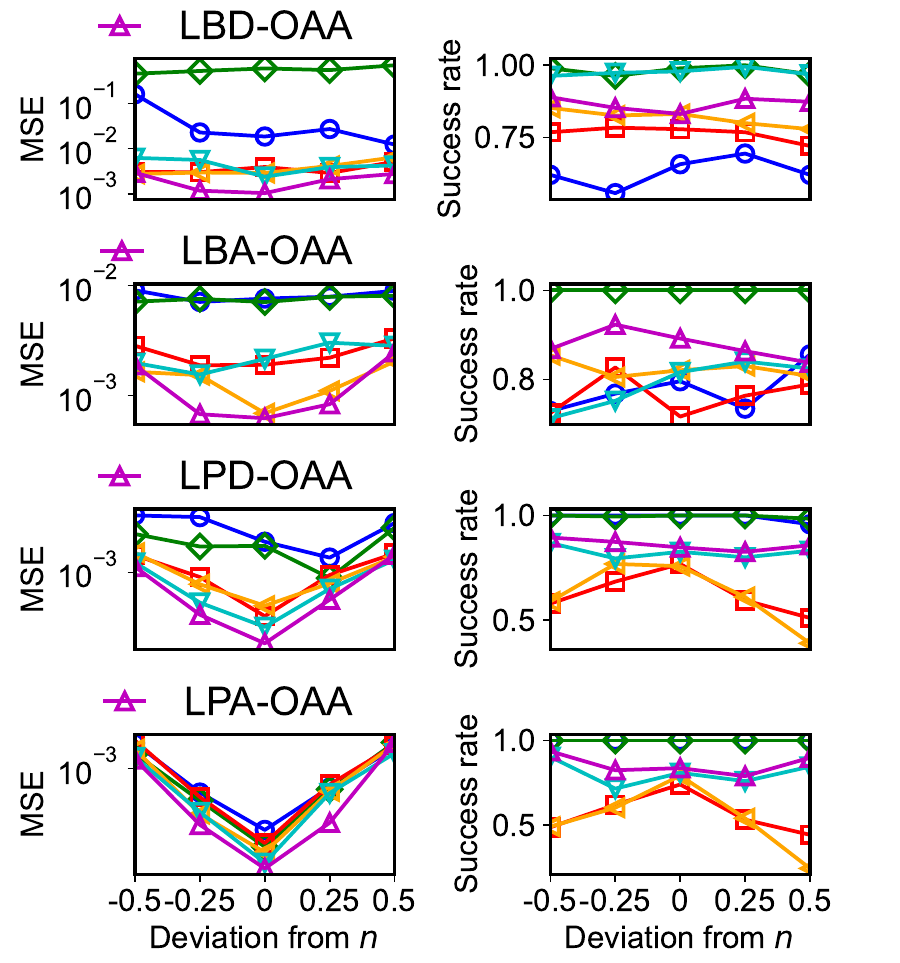}}
 	\caption{\small Attack effectiveness for Synthetic datasets, varying $n^e$. 
  }\centering
 	\label{fig:n^e synthetic data} 
 	\vspace{-0.5cm}
 \end{figure*}

\begin{figure*}[htbp]
 	\centering	
         \subfigure[\textsf{LNS} dataset, Uniform $\mathbf{\tilde{f}}$]{
 		\includegraphics[width=0.24\textwidth]{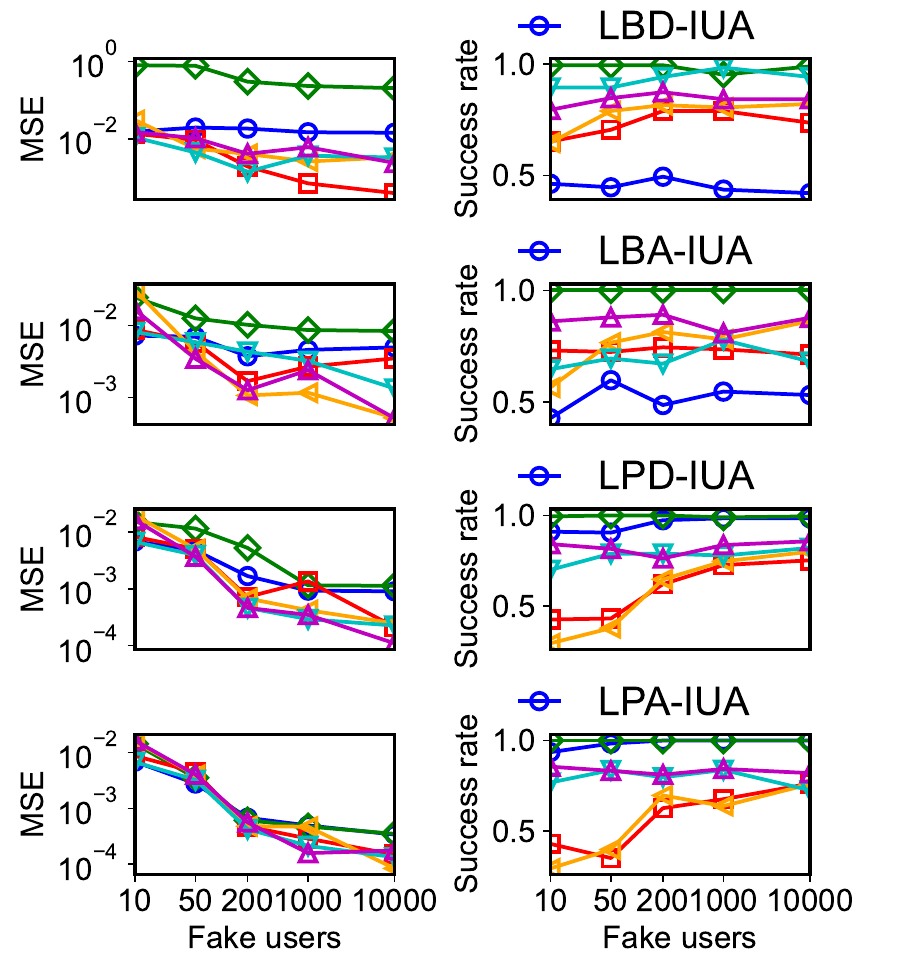}}\vspace{-0.06cm}\hspace{-4mm}
 	\subfigure[\textsf{LNS} dataset, Pulse $\mathbf{\tilde{f}}$]{
 		\includegraphics[width=0.24\textwidth]{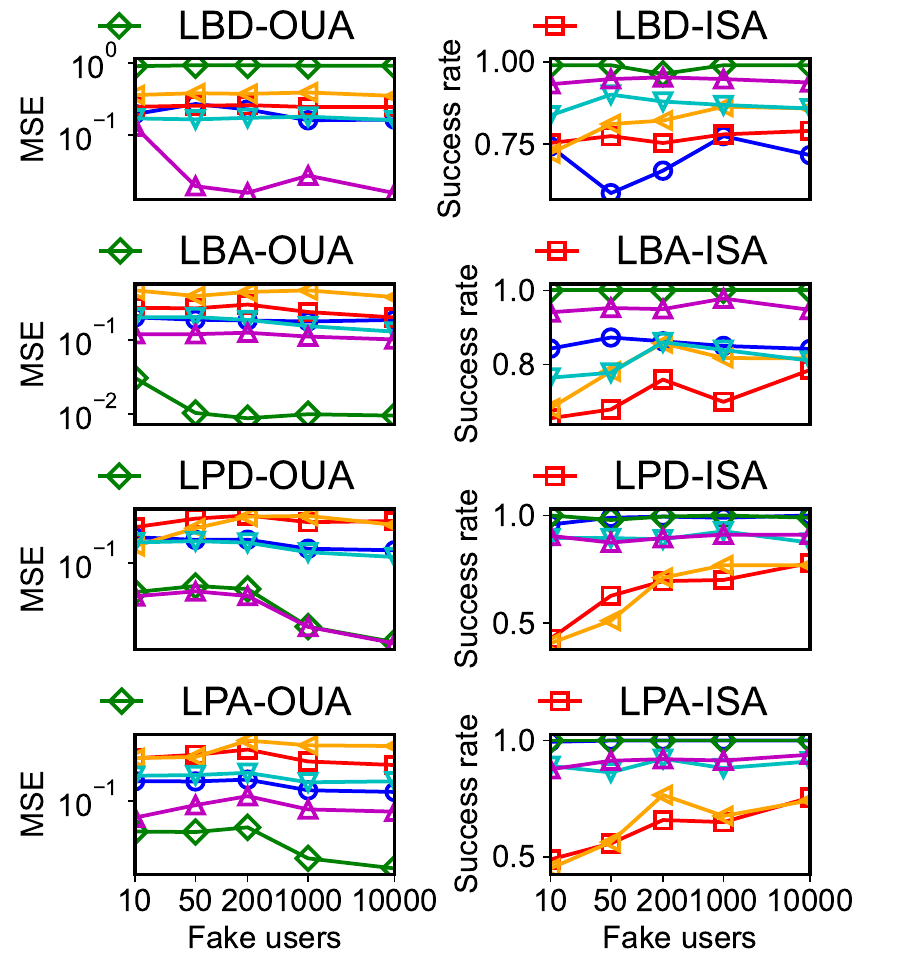}}\vspace{-0.06cm}\hspace{-4mm}
 	\subfigure[\textsf{LNS} dataset, Gaussian $\mathbf{\tilde{f}}$]{
 		\includegraphics[width=0.24\textwidth]{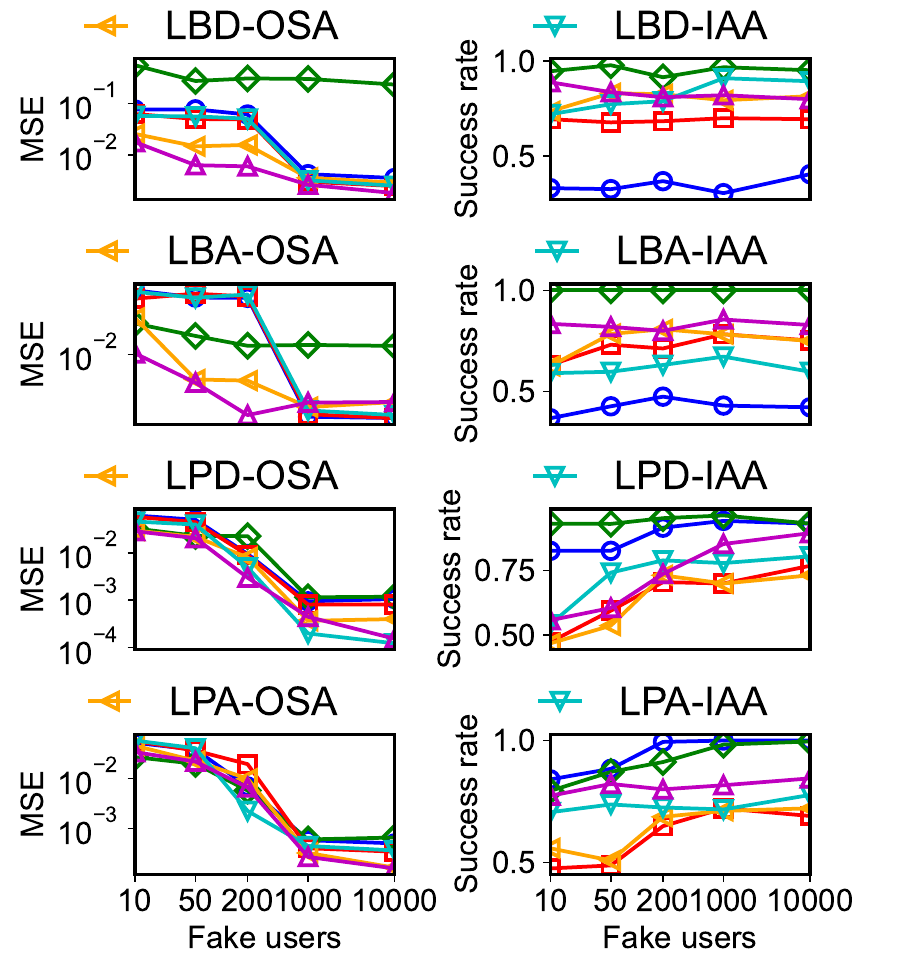}}\vspace{-0.06cm}\hspace{-4mm}
 	\subfigure[\textsf{LNS} dataset, Sigmoid $\mathbf{\tilde{f}}$]{
 		\includegraphics[width=0.24\textwidth]{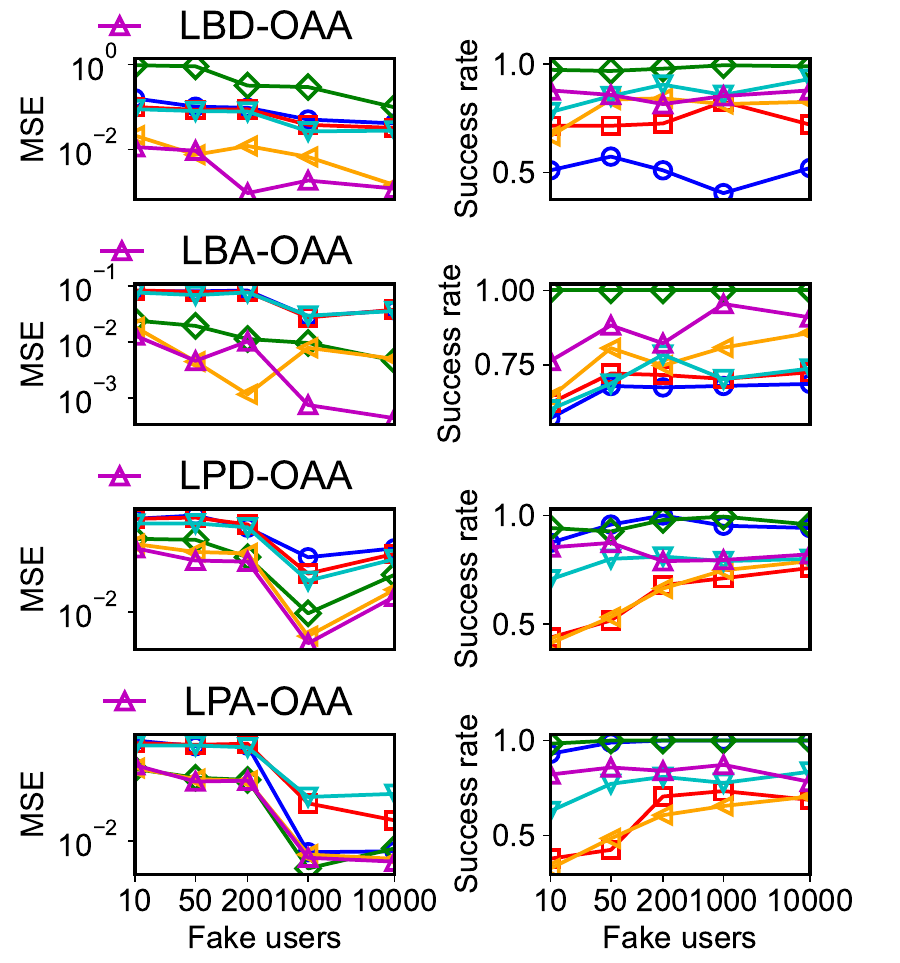}}\vspace{-0.06cm}\hspace{-4mm}
 	\\
 
         \subfigure[\textsf{Sin} dataset, Uniform $\mathbf{\tilde{f}}$]{
 		\includegraphics[width=0.24\textwidth]{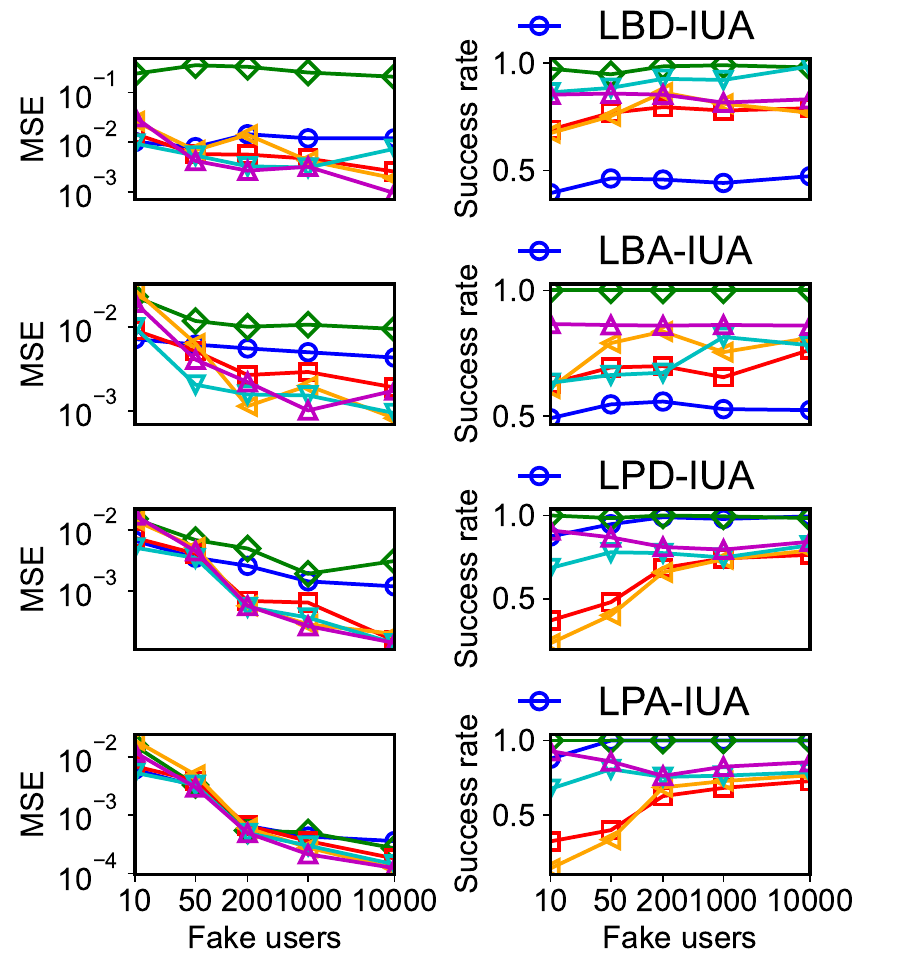}}\vspace{-0.06cm}\hspace{-4mm}
 	\subfigure[\textsf{Sin} dataset, Pulse $\mathbf{\tilde{f}}$]{
 		\includegraphics[width=0.24\textwidth]{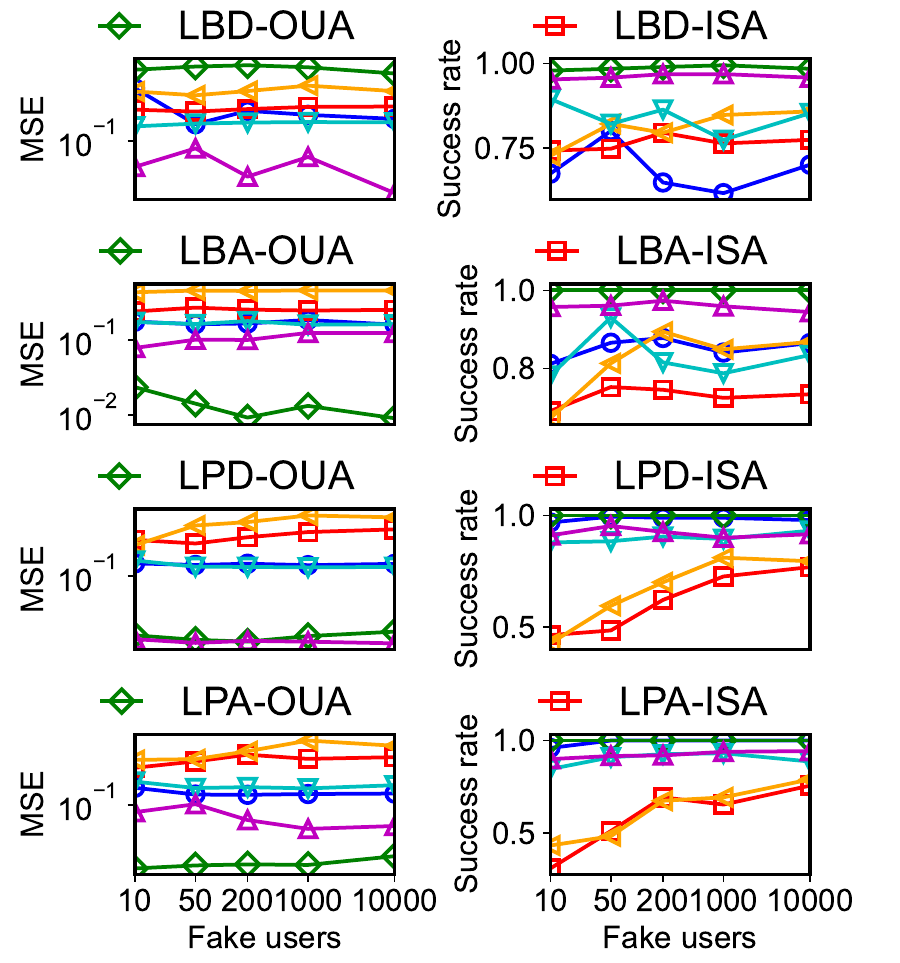}}\vspace{-0.06cm}\hspace{-4mm}
 	\subfigure[\textsf{Sin} dataset, Gaussian $\mathbf{\tilde{f}}$]{
 		\includegraphics[width=0.24\textwidth]{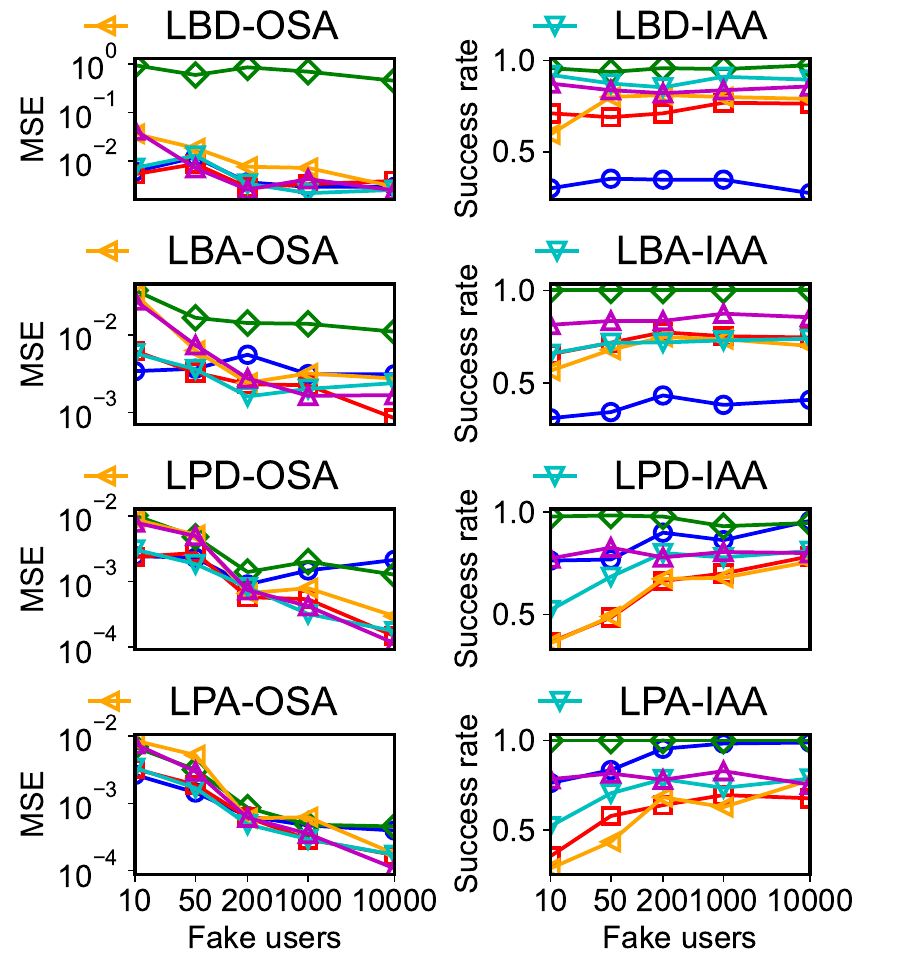}}\vspace{-0.06cm}\hspace{-4mm}
 	\subfigure[\textsf{Sin} dataset, Sigmoid $\mathbf{\tilde{f}}$]{
 		\includegraphics[width=0.24\textwidth]{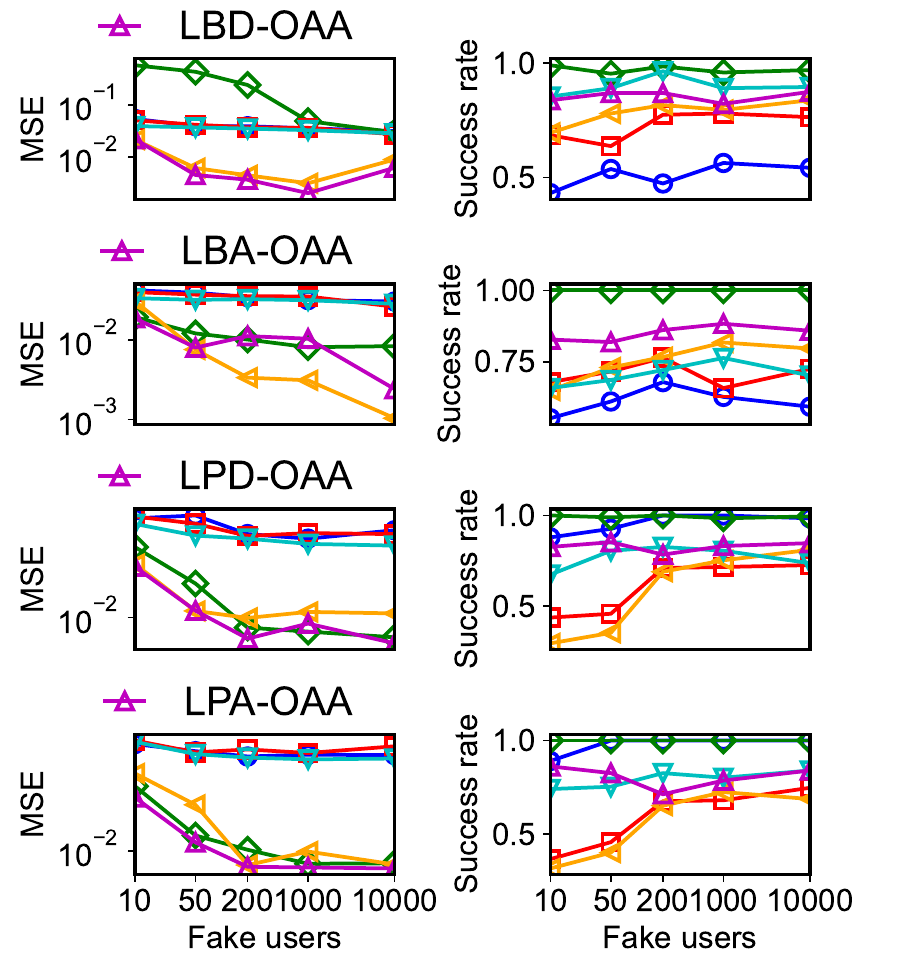}}\vspace{-0.06cm}\hspace{-4mm}
         \\
        
         \subfigure[\textsf{Log} dataset, Uniform $\mathbf{\tilde{f}}$]{
 		\includegraphics[width=0.24\textwidth]{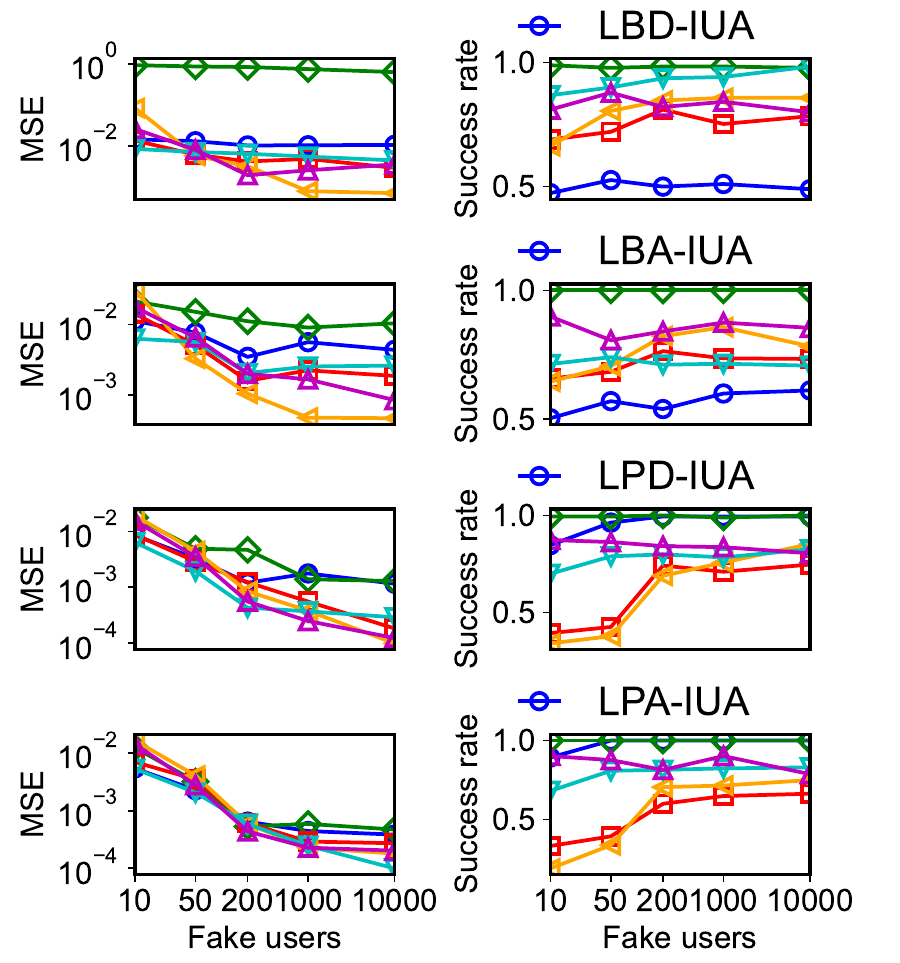}}\vspace{-0.05cm}\hspace{-4mm}
 	\subfigure[\textsf{Log} dataset, Pulse $\mathbf{\tilde{f}}$]{
 		\includegraphics[width=0.24\textwidth]{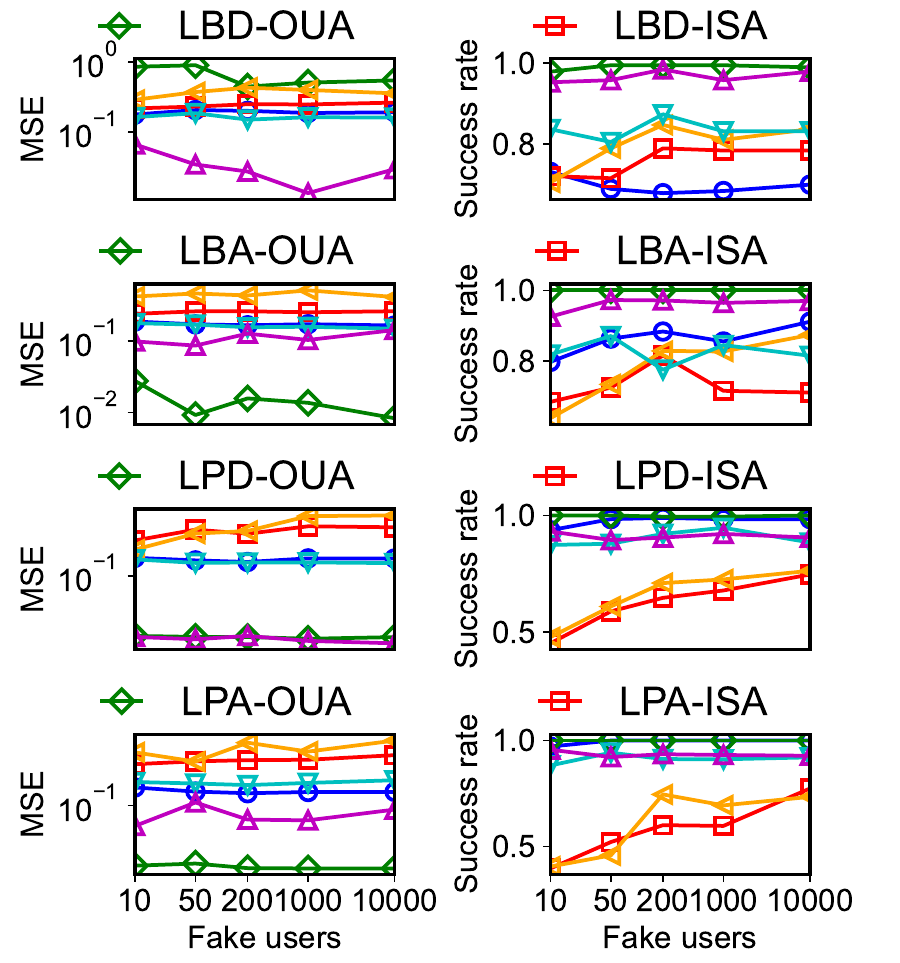}}\vspace{-0.06cm}\hspace{-4mm}
 	\subfigure[\textsf{Log} dataset, Gaussian $\mathbf{\tilde{f}}$]{
 		\includegraphics[width=0.24\textwidth]{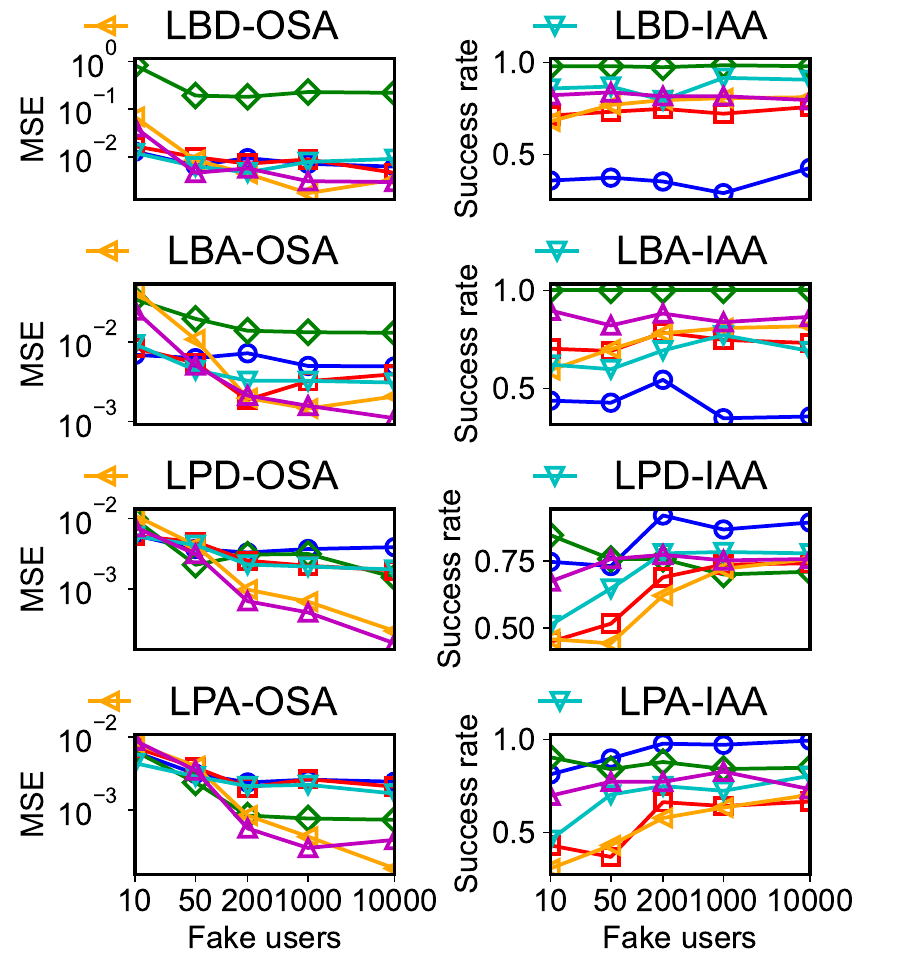}}\vspace{-0.06cm}\hspace{-4mm}
 	\subfigure[\textsf{Log} dataset, Sigmoid $\mathbf{\tilde{f}}$]{
 		\includegraphics[width=0.24\textwidth]{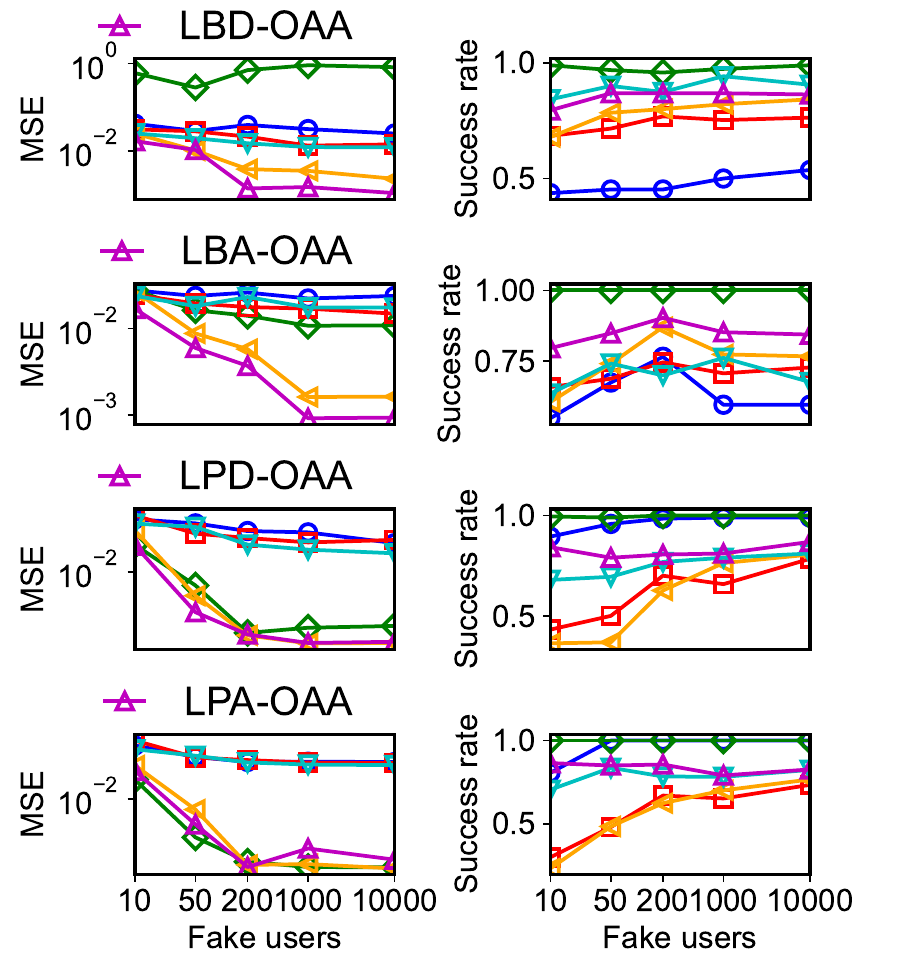}}\vspace{-0.06cm}\hspace{-4mm}
         \\
        
         \subfigure[\textsf{Pulse} dataset, Uniform $\mathbf{\tilde{f}}$]{
 		\includegraphics[width=0.24\textwidth]{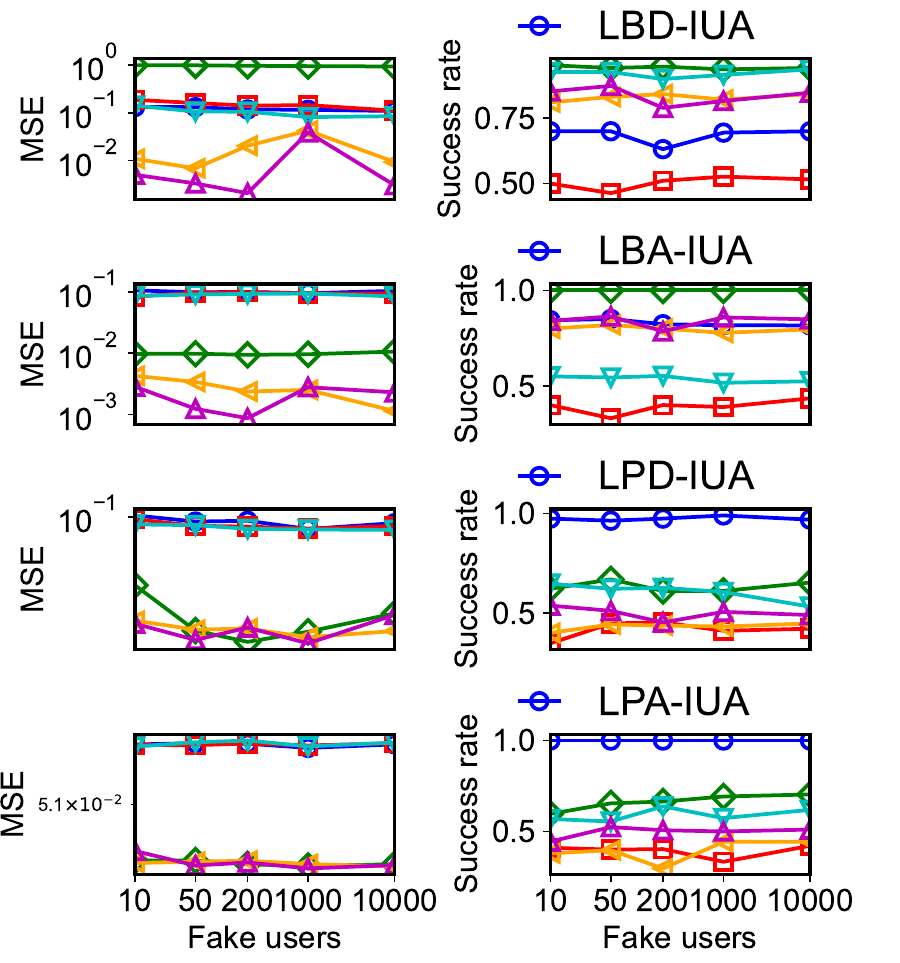}}\vspace{-0.13cm}\hspace{-4mm}
 	\subfigure[\textsf{Pulse} dataset, Pulse $\mathbf{\tilde{f}}$]{
 		\includegraphics[width=0.24\textwidth]{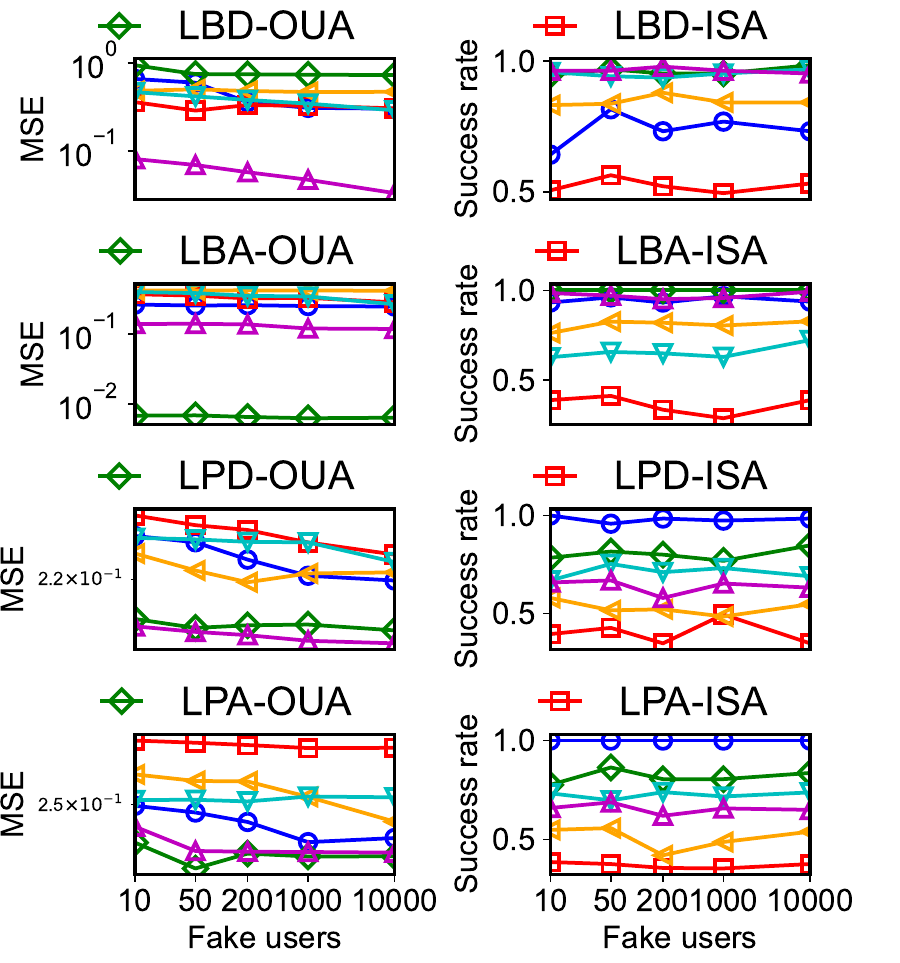}}\vspace{-0.13cm}\hspace{-4mm}
 	\subfigure[\textsf{Pulse} dataset, Gaussian $\mathbf{\tilde{f}}$]{
 		\includegraphics[width=0.24\textwidth]{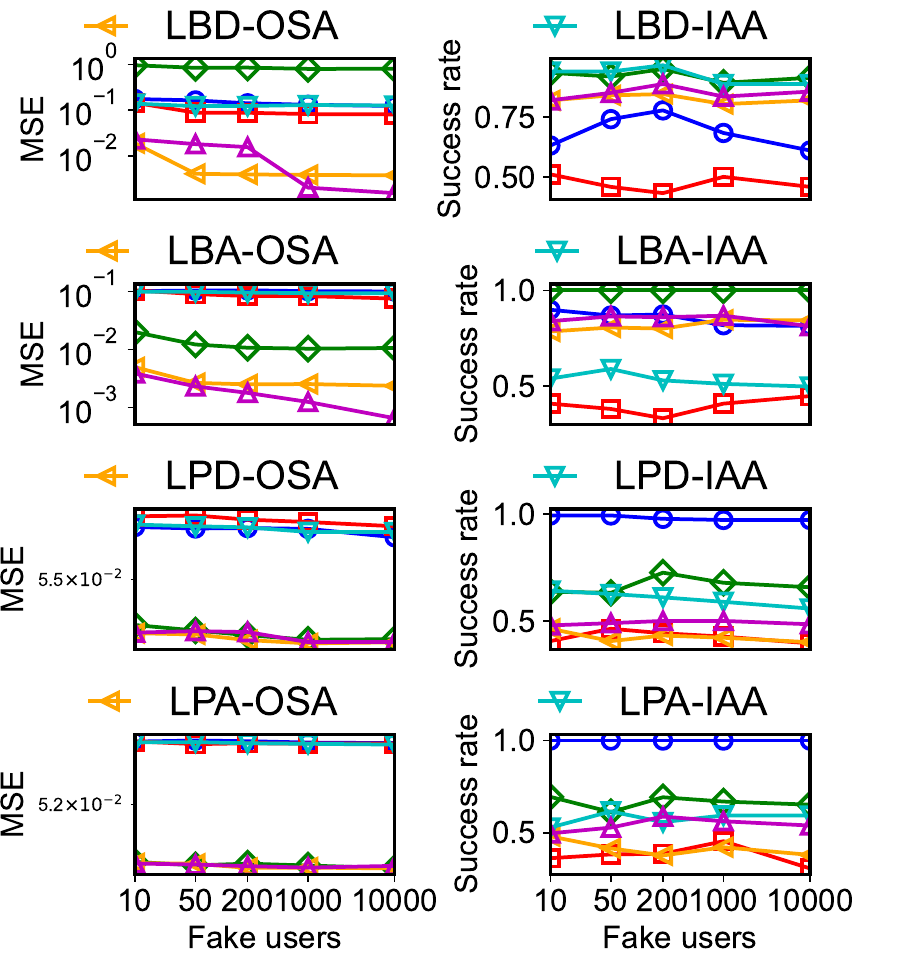}}\vspace{-0.13cm}\hspace{-4mm}
 	\subfigure[\textsf{Pulse} dataset, Sigmoid $\mathbf{\tilde{f}}$]{
 		\includegraphics[width=0.24\textwidth]{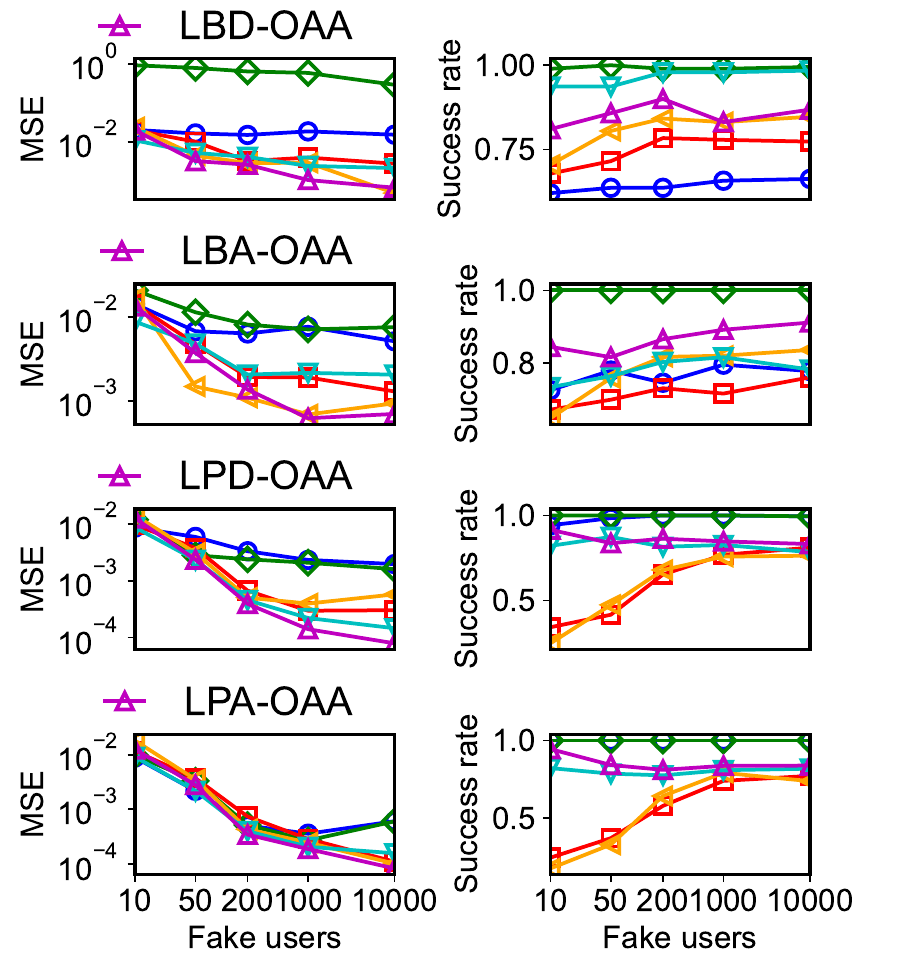}}
 	\caption{\small Attack effectiveness for Synthetic datasets, varying the number of fake users for $\mathbf{f}^e$ calculation. 
  }\centering
 	\label{fig:f_e synthetic data} 
 	\vspace{-0.5cm}
 \end{figure*}

Our experimental results on these datasets are depicted in Figs.~\ref{fig:beta synthetic data}, \ref{fig:epsilon synthetic data}, \ref{fig:w synthetic data}, \ref{fig:n^e synthetic data}, and \ref{fig:f_e synthetic data}, illustrating attack impacts by varying $\beta$, $\epsilon$, $w$, $n^e$, and $\mathbf{f}^e$.

\section{Attacking Mean Estimation over Numerical Domain}\label{sec: Attack Mean Estimation over Numerical Domain}



\textbf{Apply HM to LDP-IDS \cite{LDP_IDS}.} The Hybrid Mechanism \cite{Wang2019CollectingAA} is an LDP mean estimation technique merging Stochastic Rounding (SR) \cite{Duchi2016MinimaxOP} and Piecewise Mechanism (PM) \cite{Wang2019CollectingAA} for minimal error. For $\epsilon>0.61$, it employs PM with probability $1-e^{-\epsilon/2}$ and SR with $e^{-\epsilon/2}$. Below $\epsilon\le 0.61$, it solely uses SR. Considering that in practice users' private values $v\in\left[-1,1\right]$ are close to 0, the worst case variance of the perturbed value $y$ in HM is written as
\begin{align}
\vspace{-1mm}
\nonumber
    \text{Var}^\ast\left[y\right]=
    \left\{
    \begin{aligned}
      &\left(\frac{e^\epsilon+1}{e^\epsilon-1}\right)^2,\text{if} \ \epsilon\le 0.61\\
      &\frac{1}{e^{\epsilon/2}}\left[\left(\frac{e^\epsilon+1}{e^\epsilon-1}\right)^2+\frac{e^{\epsilon/2}+3}{3\left(e^{\epsilon/2}-1\right)^2}\right],\text{if} \ \epsilon>0.61
    \end{aligned}\right.
    \vspace{-1mm}
\end{align}
The variance of the mean estimate $f=\frac{1}{n}\sum_{j=1}^{n}y_j$ (denote the perturbed value of $j$-th user as $y_j$) can be calculated as $\text{Var}\left[f\right]=\frac{1}{n^2}\sum_{j=1}^{n}\text{Var}\left[y_j\right]$. Simplifying, we use $\text{Var}\left[f\right]=\frac{1}{n}\text{Var}^\ast\left[y\right]$, denoted as $\text{Var}(n,\epsilon)$. The HM mechanism can then replace FOs in the LDP-IDS framework for mean estimation.

\textbf{Attack HM-based LDP-IDS.}  \cite{Li2023Finegrained} introduced a fine-grained poisoning attack, Output Poisoning Attack (OPA), for SR and PM to manipulate estimated means and variances. For SR, the mean estimation manipulation gap of OPA is given by $\frac{2n-2(p-q)^2S^{(2)}}{(m+n)^2(p-q)^2}+\frac{S^{(2)}}{(m+n)^2}+(\frac{n^e-n}{m+n}\mu_t+\frac{S^{(1)}-S_e^{(1)}}{m+n})^2$ where $p=\frac{e^\epsilon}{1+e^\epsilon}$, $q=1-p$, and $S^{(1)}$, $S^{(2)}$ are sums of genuine user inputs and their squares, respectively, and $S_e^{(1)}$ is the attacker-estimated $S^{(1)}$. For PM, it's $(\frac{n^e-n}{m+n}\mu_t+\frac{S^{(1)}-S_e^{(1)}}{m+n})^2+\frac{2n(e^{\epsilon/2}+3)}{3(m+n)^2(e^{\epsilon/2}-1)^2}+\frac{(1+e^{\epsilon/2})S^{(2)}}{(m+n)^2(e^{\epsilon/2}-1)}$. OPA maintains security-privacy consistency~\cite{Li2023Finegrained} and improves as $m$ and $n$ scale equally, making it suitable for replacing PMAs in Adaptive Attack frameworks. For DMAs, when attackers want to minimize dissimilarity, they just need to set the target value as the last estimated mean and use OPA to attack. When attackers want to maximize dissimilarity, they can set the target as an extreme value (e.g., $-1$ or $1$ when data is normalized to $\left[-1,1\right]$) and use OPA to attack. Manipulation gaps are calculated using the formulas above, noting that only half the users are used for mean estimation in \cite{Li2023Finegrained}, requiring adjustment by doubling $m$ and $n$ for exclusive mean estimation.


\section{Attacking Other Streaming LDP Algorithms}\label{sec: Existing LDP Finite Data Stream Algorithms}

LDP protocols for data streams, which often employ methods like budget and population division for better utility, can be exploited by attackers who mimic these methodologies to optimize attack strategies for improved results, as discussed in Sec.~\ref{subsec: Arttack Strategy Determination}. This approach is effective because both data estimation and our attacks benefit from larger privacy budgets or populations, enhancing estimation and attack outcomes.

Besides LDP-IDS, streaming algorithms like event-level private PeGaSus \cite{Chen2017PeGaSusDD}, user-level private FAST \cite{Fan2014AnAA}, and $w$-event private $\text{DSAT}_w$ \cite{10.1145/2806416.2806441} and RescueDP \cite{Wang2016RescueDPRS} can be adapted to LDP settings. Our attack methods are also applicable to these protocols. We also consider two LDP streaming alogorithms over numerical domains, CGM~\cite{baocgm} and ToPL~\cite{wang2020continuous}. 

\textbf{PeGaSus.} \cite{Chen2017PeGaSusDD} describes an event-level CDP method with three components: perturber, grouper, and smoother. The perturber adds noise independently at each timestamp, while the grouper divides the stream into partitions, and the smoother post-processes these partitions. In event-level-LDP PeGaSus, the perturber allocates $\epsilon$ LDP budget to all users with FOs at each timestamp, generating an estimate ${\bar{\mathbf{f}}}_t$. The grouper then partitions the stream based on these estimates without additional budget, defining deviation as the squared distance within a partition. The smoother operates as in the CDP algorithm. Steps (2) and (3) are aggregator-handled, aiming for improved data stream estimation. Attackers focusing on Step (1) can manipulate the initial estimates by sending tailored data. Since PeGaSus, like LBU and LPU, performs data estimation at each timestamp, only Uniform Attacks can be adopted for attacks.

\textbf{FAST.} \cite{Fan2014AnAA} describes FAST, a user-level DP algorithm adaptable to user-event LDP settings, using PID controller-based adaptive sampling and Kalman-filter-based post-processing. In these settings, FAST aggregates data using LDP protocols from all users with a reduced privacy budget of $\epsilon/M$, generating an initial estimate ${\bar{\mathbf{f}}}_t$ with FOs, where $M$ is the number of sampling timestamps. A Kalman filter then refines this into a posterior estimate ${\hat{\mathbf{f}}}_t$, output at each sampling timestamp. FAST dynamically adjusts the sampling interval using the PID controller. Attacks on FAST can only target the initial estimate ${\bar{\mathbf{f}}}_t$ to influence the result. Like attacking LSP, attackers can only manipulate FAST at sampling timestamps. Thus, only Sampling Attacks are adopted.

\textbf{$\text{DSAT}_w$} is a $w$-event private adaptation of DSAT \cite{10.1145/2806416.2806441} for $w$-event LDP. It allocates a portion of the budget or population for distance computation (dissimilarity budget $\epsilon_1$ or population $n_1$) and the remainder (publication budget $\epsilon_2$ or population $n_2$) for data publication. Budget or population usage is tracked at each timestamp to avoid depletion. If depleted, $\text{DSAT}_w$ repeats the last release $\hat{\mathbf{f}}_t=\hat{\mathbf{f}}_{t-1}$. If not, it uses $\epsilon_1/C$ for all users or $\epsilon$ for $n_1/C$ sampled users for LDP reporting, estimating dissimilarity $\overline{dis}$. This $\overline{dis}$ is compared against an adaptively controlled threshold to decide on publication or approximation. Publication decisions trigger a new reporting round for an updated estimate. Attacks on $\text{DSAT}_w$ mirror those on LBD/LBA/LPD/LPA, exploiting the dual data collection per timestamp. Attackers target the statistics during LDP aggregation phases. For effective manipulation, both DMAs and PMAs should be designed to influence dissimilarity and control estimates, with mechanisms to decide on publishing or approximating to aid attacks.

\textbf{RescueDP.} \cite{Wang2016RescueDPRS} is a real-time spatio-temporal data release mechanism that extends FAST \cite{Fan2014AnAA} with $w$-event privacy for multi-dimensional data, adaptable to $w$-event LDP protocols when using FOs. Each timestamp begins with a KF-Prediction to get a prior estimate ${\check{\mathbf{f}}}_t$. If it’s not a sampling timestamp, this estimate is immediately released as ${\hat{\mathbf{f}}}_t$. At sampling timestamps, the aggregator calculates the budget or population to be used, then aggregates reports via LDP to derive ${\bar{\mathbf{f}}}_t$, and combines these with the prior ${\check{\mathbf{f}}}_t$ to produce and release a posterior estimate ${\hat{\mathbf{f}}}_t$ using a Kalman filter. Like FAST, RescueDP dynamically adjusts sampling intervals. Attacks on RescueDP, similar to those on FAST, can only target ${\bar{\mathbf{f}}}_t$ to influence outputs. Thus, only Sampling Attack are adopted.

\textbf{CGM.} ~\cite{baocgm} proposes a novel Correlated Gaussian Mechanism (CGM) for enforcing ($\epsilon$, $\delta$)-LDP on streaming data collection over numerical domains. At each timestamp, CGM injects temporally corelated gaussian noise, computed through an optimization program that takes into account the given autocorrelation pattern, data value range, and utility metric, to the original data for publication. It is noted that CGM is used for processing only one user’s data stream. It needs to be run $n$ times to process $n$ users’ data streams. We consider the situation that aggregator utilizes CGM for mean value estimation, i.e., the aggregator collects every user’s CGM-processed data stream and releases the mean value over all users at each timestamp. For attacks, attackers also hope to manipulate the released mean value to be close to the target mean value $\mathbf{\tilde{f}}_t$ at every timestamp. CGM publishes at every timestamp, causing only Uniform Attacks to be adopted. Assume that attackers estimate the number of users and the real mean value as $n^e$ and $\mathbf{f}^e_t$ and fake users report $z_j$, $j \in [1,...,m]$ to the aggregator at each timestamp, bypassing the noise injection of CGM. Considering the expectation of gaussian noise injected to genuine users is 0, attackers only need to solve $z_j$, which satisfies $\frac{n^e \cdot \mathbf{f}^e_t + \sum_{j=1}^{m} z_j}{n^e+m}=\mathbf{\tilde{f}}_t $ for attack.

\textbf{ToPL.} ~\cite{wang2020continuous} describes ToPL method for outputting streaming data in event-level LDP setting over numerical domains. ToPL consists of two parts, Threshold Optimizer and Perturber. ToPL first utilizes Threshold Optimizer to cash a period of reports to privately find the optimal threshold $\theta$ by minimizing overall estimation errors. After obtaining $\theta$, the sever sends it to all users. When a user reports a value, it will first be truncated to be no lager than $\theta$. The user than adopts the Hybrid Mechanism (HM)~\cite{Wang2019CollectingAA}, which combines SR~\cite{Duchi2016MinimaxOP} and PM~\cite{Wang2019CollectingAA}, as the Perturber to report the truncated value and release the estimated mean values. For attacks, considering that Threshold Optimizer adopts Square Wave (SW)~\cite{li2020estimating} which utilizes Expectation Maximization algorithm to estimate distribution iteratively, we only heuristically  adopt  IPMA for attack. For Perturber, OPA~\cite{Li2023Finegrained}, which is specifically designed for attacking mean value estimation, is used to perform each timestamp attacks on HM. ToPL publishes at every timestamp like LBU/LPU. Thus, only Uniform Attacks are adopted.


\end{document}